\documentclass[aps,prd,showpacs,superscriptaddress,showpacs,amsmath,amssymb,preprintnumbers]{revtex4-1}
\usepackage[dvipdfmx]{graphicx}
\usepackage[colorlinks, linkcolor=blue, citecolor=blue, dvipdfmx]{hyperref}
\usepackage{aas_macros,multirow}

\newcommand{\calH}{\mathcal{H}}
\newcommand{\calO}{\mathcal{O}}
\newcommand{\calR}{\mathcal{R}}

\newcommand{\bfq}{\mbox{\boldmath$q$}}
\newcommand{\bfx}{\mbox{\boldmath$x$}}
\newcommand{\bfk}{\mbox{\boldmath$k$}}

\newcommand{\be}{\begin{equation}}
\newcommand{\ee}{\end{equation}}
\newcommand{\ba}{\begin{eqnarray}}
\newcommand{\ea}{\end{eqnarray}}

\newcommand{\ltsim}{\protect\raisebox{-0.5ex}
  {$\:\stackrel{\textstyle <}{\sim}\:$}}
\newcommand{\gtsim}{\protect\raisebox{-0.5ex}
  {$\:\stackrel{\textstyle >}{\sim}\:$}}

\begin{document}

\preprint{IPMU14-0105}

\title{%
Understanding higher-order nonlocal halo bias at large scales\\
by combining the power spectrum with the bispectrum 
}%
\author{%
Shun Saito
}%
\email[E-mail me at: ]{shun.saito@ipmu.jp}
\affiliation{%
Kavli Institute for the Physics and Mathematics of the Universe (WPI), 
Todai Institutes for Advanced Study, 
The University of Tokyo, Chiba 277-8582, Japan
}%
\author{%
Tobias Baldauf
}%
\affiliation{%
School of Natural Sciences, Institute for Advanced Study,
1 Einstein Drive, Princeton, NJ 08540, USA
}%
\author{%
Zvonimir Vlah
}%
\affiliation{%
Physik Institut, University of Z\"{u}rich, Winterthurerstrasse 190, 
CH-8057 Z\"{u}rich, Switzerland
}%
\author{%
Uro\v{s} Seljak
}%
\affiliation{%
Department of Physics, University of California Berkeley, CA 94720, USA
}%
\affiliation{%
Lawrence Berkeley National Laboratory, Physics Department, Berkeley, CA 94720, USA
}%
\author{%
Teppei Okumura
}
\affiliation{%
Kavli Institute for the Physics and Mathematics of the Universe (WPI), 
Todai Institutes for Advanced Study, 
The University of Tokyo, Chiba 277-8582, Japan
}%
\author{%
Patrick McDonald
}%
\affiliation{%
Lawrence Berkeley National Laboratory, Physics Department, Berkeley, CA 94720, USA
}%

\date{\today}

\begin{abstract}
Understanding the relation between underlying matter distribution and biased tracers such as galaxies or 
dark matter halos is essential to extract cosmological information from ongoing or future galaxy redshift surveys. 
At sufficiently large scales such as the Baryon Acoustic Oscillation (BAO) scale,  
a standard approach for the bias problem on the basis of the perturbation theory (PT)
is to assume the `local bias' model in which the density field of biased tracers is deterministically expanded 
in terms of matter density field at the same position. 
The higher-order bias parameters are then determined by combining the power spectrum with higher-order 
statistics such as the bispectrum.\par 
As is pointed out by recent studies, however, nonlinear gravitational evolution naturally induces nonlocal bias 
terms even if initially starting only with purely local bias. As a matter of fact, previous works showed that 
the second-order nonlocal bias term, which corresponds to the gravitational tidal field, is important to explain 
the characteristic scale-dependence of the bispectrum. 
In this paper we extend the nonlocal bias term up to third order, and investigate whether the PT-based model 
including nonlocal bias terms can simultaneously explain the power spectrum and the bispectrum of 
simulated halos in $N$-body simulations. 
The bias renormalization procedure ensures that only one additional term is necessary to be introduced to 
the power spectrum as a next-to-leading order correction, even if third-order nonlocal bias terms are taken 
into account. We show that the power spectrum, including density and momentum, and the bispectrum 
between halo and matter in $N$-body simulations can be simultaneously well 
explained by the model including up to third-order nonlocal bias terms at $k\ltsim 0.1h$/Mpc. 
Also, the results are in a good agreement with theoretical predictions of a simple coevolution picture, 
although the agreement is not perfect. These trend can be found for a wide range of halo mass, 
$0.7\ltsim M_{\rm halo}\,[10^{13}M_{\odot}/h]\ltsim 20$ at various redshifts, $0 \le z\le 1$.
These demonstrations clearly show a failure of the local bias model even at such large scales, and 
we conclude that nonlocal bias terms should be consistently included in order to accurately model 
statistics of halos. 
\end{abstract}


\maketitle

\section{Introduction}
\label{sec: Introduction}
Precise observation of the early universe has been well established by measurements of 
temperature and polarization anisotropy of the Cosmic Microwave Background (CMB) 
such as Wilkinson Microwave Anisotropy Probe (WMAP) 
\cite{Komatsu:2009qy,Komatsu:2011rs,Hinshaw:2013rt} or Planck \cite{Planck-Collaboration:2013fr}. 
Now we enter a new era of precision cosmology by getting in hand various kinds of large-scale 
structure measurements in late-time universe, mainly aiming at unveiling dark universe 
(see \cite{Weinberg:2012uq} for a recent review). 
In particular, clustering of galaxies in a three dimensional map of the universe 
offers us a lot of fruitful cosmological information via the Baryon Acoustic Oscillations (BAOs), 
Redshift-Space Distortion (RSD), or the shape of galaxy clustering statistics such as 
the power spectrum and the bispectrum (for an encompassing review, see \cite{Bernardeau:2002lr}). 
As a matter of fact, recent works by Baryon Oscillation Spectroscopic Survey (BOSS) \cite{Schlegel:2009uq}
in Sloan Digital Sky Survey III (SDSS-III) \cite{Eisenstein:2011ve} or WiggleZ survey \cite{Drinkwater:2010rz} 
have already accomplished very accurate measurements of such signals 
\cite{Anderson:2013qy,Reid:2012ly,Anderson:2013qq,Samushia:2013lr,Beutler:2013kx,Beutler:2014ty,Zhao:2013fk,
Reid:2014qy, Blake:2011fj,Blake:2011jf,Blake:2011kc,Blake:2011zv,Blake:2012nx,Contreras:2013ys,Scrimgeour:2012pd}. 
Planned or near-future galaxy redshift surveys, which include 
Subaru Prime Focus Spectrograph (PFS) Survey \cite{Takada:2012ww}, 
Hobby-Eberly Telescope Dark Energy Experiment (HETDEX) \cite{Adams:2011ak}, 
Dark Energy Spectroscopic Instrument (DESI) \cite{Levi:2013ly} and 
Euclid \cite{Laureijs:2011jw}, will continue to improve measurement accuracy at various redshift and scales.\par 

In order to unlock full potential of cosmological information in the galaxy clustering, it is essential to understand 
the relation between underlying matter distribution and galaxies, known as the so-called galaxy bias problem. 
It is often assumed that galaxy distribution well traces underlying matter distribution which can be directly probed 
by cosmological $N$-body simulations. 
Given the fact that we do not have complete knowledge of galaxy formation scenario in nonlinear structure formation, 
it is a common practice to connect observed galaxy 
distribution to simulated dark matter halos. This approach is based on the halo model 
\cite{Cooray:2002lr,Seljak:2000mz}, and its associated techniques such as Halo Occupation Distribution (HOD) 
and Subhalo Abundance Matching (SHAM) (e.g., \cite{Behroozi:2010lr,Reddick:2012qy,Hearin:2013vn,Zentner:2013lq}) 
are applied to somewhat small-scale galaxy clustering (typically $\sim\mathcal{O}(\text{0.1-10})\,$Mpc)  
(see e.g., \cite{Leauthaud:2011fk,Tinker:2013fk,Guo:2014fj} and references therein for recent studies).\par 

Even though dark matter halos can be easily constructed in $N$-body simulations, it is important to theoretically 
understand clustering of the halos, or halo bias, especially at large scales around BAOs ($\sim 150\,$Mpc), 
because the halo clustering is sensitive to underlying cosmology at the regimes (where, in other words, 
the two-halo term is dominant in the halo-model context). Some authors tried to formulate the halo or galaxy 
bias in parametric ways (see e.g., \cite{Cole:2005fp,Tegmark:2006pl,Yamamoto:2010kj,Oka:2014aa}) and showed 
a successful performance depending on their specific purpose, although it might be hard to be justified 
in more general situations. It is therefore desirable to develop an analytic formulation to describe the halo 
clustering in a physically-well motivated way. Perturbation theory (PT) is a natural approach along this direction, 
and, in the PT approach, the so-called `local bias' model \cite{Kaiser:1984bh,Fry:1993lq} in which the density field 
of halos is deterministically Tailor-expanded in terms of matter density field at the same position as
\be
 \delta_{\rm h}(\bfx) = \sum_{n}\frac{b_{n}}{n!}\delta_{\rm m}(\bfx)^{n}, 
\ee
where $b_{n}$ is the bias coefficient at $n$-th order, and $\delta_{\rm h}$ and $\delta_{\rm m}$ describes 
density fields of halos and matter, respectively. It is well known that the local bias model works 
well at linear order to some extent \cite{Kaiser:1984bh}, 
and the fitting formula for the halo mass function is calibrated so that it 
also consistently reproduces the linear bias value in simulations \cite{Sheth:1999dq,Tinker:2008fk}.  
However, a couple of issues in the model have been recently pointed out. 
First of all, the model prefers different values of nonlinear bias parameter like $b_{2}$ for the halo 
power spectrum and the bispectrum \cite{Pollack:2012kh,Pollack:2013fk}, although the model looks well 
fitted to the spectra by properly choosing nonlinear bias parameters (see e.g. 
\cite{Jeong:2009xe,Saito:2011yq,Nishizawa:2012lr}). 
In addition, the authors \cite{McDonald:2009lr,Matsubara:2011qy,Baldauf:2012lr,Chuen-Chan:2012fk} 
show that nonlinear gravitational evolution naturally induces nonlocal terms, and there are clear evidences  
of such a term at least at second-order perturbation observed in the bispectrum in simulations 
\cite{Baldauf:2012lr,Chuen-Chan:2012fk}. These caveats clearly warn adopting the local bias model 
from a physical point of view.\par 

In this paper we continue to study how well the bias model including nonlocal terms performs against 
the halo statistics in $N$-body simulations. In particular, we focus on how well such a model can 
simultaneously explain the power spectrum as well as the bispectrum which again cannot be realized 
in the simple local bias model. 
While the leading-order (i.e., tree-level) bispectrum requires only up to second-order perturbation, 
it is necessary to consider up to third order as a next-to-leading order correction in the power spectrum. 
The author \cite{McDonald:2006lr} showed that the bias renormalization procedure allows us to write down 
a {\it physical} expression for the halo statistics and the third-order local bias term is absorbed into 
the linear bias. As we will revisit later, Ref.~\cite{McDonald:2009lr} shows that all the correction terms 
associated with the third-order nonlocal bias can be summarized into only one term. 
This bias renormalization approach has been recently readdressed in terms of the effective 
field theory by \cite{Assassi:2014lr}, and they also reached the same conclusion 
(see also \cite{Kehagias:2013gf}). 
Thus we have in hand a very simple bias model on the basis of PT even if considering 
all the local and nonlocal terms up to third order. Then the natural question that arises is whether 
the simple bias model can well explain the simulated halo power spectrum as well, and also 
whether the fitted value of the bias parameter is consistent with what is physically expected. 
In order to answer these questions, we study the halo-matter statistics in a standard 
$\Lambda$CDM universe at a various halo-mass range and redshift. 
We jointly fit the PT model to the power spectrum together with the bispectrum. 
An advantage of focusing on the halo-matter statistics is that it is free from the stochastic bias 
\cite{Seljak:2009nl,Kitaura:2014wo} and the velocity bias 
\cite{Percival:2008fk,Desjacques:2010qy,Desjacques:2010fk,Baldauf:2014xy}. 
We also investigate the cross spectrum between halo density and matter momentum 
which was recently studied in modeling the RSDs in the Distribution Function approach 
(see \cite{Seljak:2011dw,Okumura:2012gf,Okumura:2012rq,Vlah:2012fr,Vlah:2013qy,Blazek:2013ys} 
for a series of papers) and should be simultaneously explained by the same bias values if the model works. 
\par 

The outline of this paper is as follows. In \S.~\ref{sec: Formulation}, we first revisit the argument in \cite{McDonald:2009lr} 
and summarize a model to describe the halo-matter power spectrum and the bispectrum including nonlocal terms up to 
third order. In particular, we extend the model to the cross power spectrum between halo density and matter momentum 
which can be easily measured from the simulations and can be used to study the bias model as well. 
In addition, we study a simple coevolution picture of dark matter and halo fluids and derive a third-order solution.  
In \S.~\ref{sec: simulation} we describe our simulation details and fitting procedure. We then show our results 
in \S.~\ref{sec: results} in which the bias model is compared with halo-matter power spectra in detail. 
Finally we make a summary and conclusion in in \S.~\ref{sec: summary}. 

\section{The halo-matter cross statistics in the presence of nonlocal bias terms}
\label{sec: Formulation}

In this section we explicitly write down expressions for the halo-matter power spectrum 
on the basis of the perturbation theory (PT), including nonlocal bias terms. 
For this purpose we revisit an procedure proposed by \cite{McDonald:2009lr} in which 
all the possible bias terms are introduced by symmetry arguments and can be properly 
renormalized. After we review exactly the same procedure in \cite{McDonald:2009lr} 
for the matter-density and halo-density power spectrum, we will extend it 
to the matter-momentum and halo-density cross spectrum in a similar manner.
We also discuss the bispectrum and bias renormalization \cite{Baldauf:2012lr}. 
For readers unfamiliar with PT, we refer to Appendix.~\ref{sec: PT basics} where  
basic equations in the PT formalism and our notations are summarized. 
While we here focus on the cross power spectrum, we present expressions for 
the auto correlators in Appendix.~\ref{sec: PT basics} as well.\par 

\subsection{The halo-matter density power spectrum}
\label{subsec: P^hm_00}

Starting from Eq.~(\ref{eq: MR ansatz}) which includes all possible perturbations up to third order 
for the next-to-leading order calculation of the power spectrum (e.g., see \cite{Bernardeau:2002lr}), 
the matter-halo density power spectrum is written as 
\ba
 P^{\rm hm}_{\,00}(k) & = & \;\;
 c_{\delta}P(k) +  c_{\delta}P^{(13)}_{\delta\delta}(k) + c_{\delta}P^{(22)}_{\delta\delta}(k)\nonumber\\
 && + \frac{34}{21}c_{\delta^{2}}\sigma^{2}P(k) + \frac{1}{2}c_{\delta^{3}}\sigma^{2}P(k) + \frac{1}{3}c_{\delta s^{2}}\sigma^{2}P(k)
 +\frac{1}{2}c_{\delta\epsilon^{2}}\sigma^{2}_{\epsilon}P(k)\nonumber\\
 && + c_{\delta^{2}} \int\frac{d^{3}q}{(2\pi)^{3}}\,P(q)P(|\bfk-\bfq|)F^{(2)}_{\rm S}(\bfq,\bfk-\bfq)\nonumber\\
 && + c_{s^{2}} \int\frac{d^{3}q}{(2\pi)^{3}}\,P(q)P(|\bfk-\bfq|)F^{(2)}_{\rm S}(\bfq,\bfk-\bfq)S^{(2)}(\bfq,\bfk-\bfq)\nonumber\\
 && + 2c_{s^{2}}P(k)\int\frac{d^{3}q}{(2\pi)^{3}}\,P(q)F^{(2)}_{\rm S}(-\bfq,\bfk)S^{(2)}(\bfq,\bfk-\bfq)\nonumber\\
 && + 2c_{st}P(k)\int\frac{d^{3}q}{(2\pi)^{3}}\,P(q)D^{(2)}_{\rm S}(-\bfq,\bfk)S^{(2)}(\bfq,\bfk-\bfq)\nonumber\\
 && + 2c_{\psi}P(k)\int\frac{d^{3}q}{(2\pi)^{3}}\,P(q)\left[\frac{3}{2}D^{(3)}_{\rm S}(\bfq,-\bfq,-\bfk)
 -2F^{(2)}_{\rm S}(-\bfq,\bfk)D^{(2)}_{\rm S}(\bfq,\bfk-\bfq)\right], 
 \label{eq: dd_org}
\ea
where the superscript `h' stands for a quantity for halos and the subscript `0' stands for 
the zeroth moment of mass-wighted velocity. 
All the bias coefficients, $c_{n}$, are bare bias parameters, and do not necessarily have 
clear physical meaning as explained later.   
$P(k)$ denotes the linear matter power spectrum, and the r.m.s of the fluctuated matter field, 
$\sigma^{2}$, is defined by $\sigma^{2}\equiv \int q^{2}dq\,P(q)/(2\pi^{2})$. 
Note that the term involving the third-order tidal term, $s^{3}$, vanishes in this case. 
Ref.~\cite{McDonald:2006lr} argued that the first and second lines in Eq.~(\ref{eq: dd_org}) can be renormalized 
to a {\it physical} linear bias as follows: in the limit of $k\to 0$, one finds 
\ba
 &&c_{\delta}P(k) +  c_{\delta}P^{(13)}_{\delta\delta}(k) + c_{\delta}P^{(22)}_{\delta\delta}(k)
 +\frac{34}{21}c_{\delta^{2}}\sigma^{2}P(k) + \frac{1}{2}c_{\delta^{3}}\sigma^{2}P(k) + \frac{1}{3}c_{\delta s^{2}}\sigma^{2}P(k)
 +\frac{1}{2}c_{\delta\epsilon^{2}}\sigma^{2}_{\epsilon}P(k)\nonumber\\
 & \xrightarrow[k\to 0]{} & 
 \left(c_{\delta}+\frac{34}{21}c_{\delta^{2}}\sigma^{2}+\frac{1}{2}c_{\delta^{3}}\sigma^{2}+ \frac{1}{3}c_{\delta s^{2}}\sigma^{2}
 +\frac{1}{2}c_{\delta\epsilon^{2}}\sigma^{2}_{\epsilon}\right)P(k).  
 \label{eq: linear bias renormalization 1}
\ea
In the limit of $k\to 0$, all the terms proportional to $P(k)$ should 
behave as the linear bias parameter times the linear power spectrum $P(k)$, 
which means that all the terms in the bracket can be interpreted as a renormalized linear bias. 
The third-order local bias term, $c_{\delta^{3}}$, is thus renormalized into the linear bias and 
not necessary to be considered. 
Ref.~\cite{McDonald:2009lr} further found that the fifth, sixth and seventh lines in Eq.~(\ref{eq: dd_org}), 
whose origins are the third-order nonlocal terms, can be renormalized in a similar manner into 
a linear bias and {\it just one} additional bias parameter. In order to see this, let us first separate 
out $k\to 0$ limit of these terms, 
\ba
 && \int\frac{d^{3}q}{(2\pi)^{3}}\,P(q)F^{(2)}_{\rm S}(-\bfq,\bfk)S^{(2)}(\bfq,\bfk-\bfq) \to \frac{34}{63}\sigma^{2},\\
 && \int\frac{d^{3}q}{(2\pi)^{3}}\,P(q)D^{(2)}_{\rm S}(-\bfq,\bfk)S^{(2)}(\bfq,\bfk-\bfq) \to  -\frac{8}{63}\sigma^{2},\\
 && \int\frac{d^{3}q}{(2\pi)^{3}}\,P(q)\left[\frac{3}{2}D^{(3)}_{\rm S}(\bfq,-\bfq,-\bfk)
 -2F^{(2)}_{\rm S}(-\bfq,\bfk)D^{(2)}_{\rm S}(\bfq,\bfk-\bfq)\right] \to 0. 
\ea
These terms thus behaves as constants at $k \to 0$ and hence can be renormalized to linear bias 
parameters just as Eq.~(\ref{eq: linear bias renormalization 1}). 
In addition, Ref.~\cite{McDonald:2009lr} found that these integrals exactly match each other 
and behaves as a filter function, once constants in $k\to 0$ limit are separated out 
and normalization factors are properly chosen, 
\ba
 && \int\frac{d^{3}q}{(2\pi)^{3}}\,P(q)F^{(2)}_{\rm S}(-\bfq,\bfk)S^{(2)}(\bfq,\bfk-\bfq) 
 = -\frac{8}{21}\sigma^{2}_{3}(k)+\frac{34}{63}\sigma^{2},
 \label{eq: PFS}\\
 && \int\frac{d^{3}q}{(2\pi)^{3}}\,P(q)D^{(2)}_{\rm S}(-\bfq,\bfk)S^{(2)}(\bfq,\bfk-\bfq) 
 = \frac{16}{105}\sigma^{2}_{3}(k)-\frac{8}{63}\sigma^{2},\\
 && \int\frac{d^{3}q}{(2\pi)^{3}}\,P(q)\left[\frac{3}{2}D^{(3)}_{\rm S}(\bfq,-\bfq,-\bfk)
 -2F^{(2)}_{\rm S}(-\bfq,\bfk)D^{(2)}_{\rm S}(\bfq,\bfk-\bfq)\right] 
 = \frac{256}{2205}\sigma^{2}_{3}(k), 
\ea
where we define $\sigma^{2}_{3}(k)$ as
\be
 \sigma^{2}_{3}(k) \equiv k^{3}\int \frac{r^{2}dr}{2\pi^{2}}P(kr)I_{\rm R}(r). 
\ee
For instance in the case of Eq.~(\ref{eq: PFS}), $I_{\rm R}(r)$ is described as, 
\be
 I_{R}(r)=\frac{5}{128r^4}(1+r^2)(-3+14r^2-3r^4)+\frac{3(r^2-1)^4}{256r^5}\ln\left| \frac{1+r}{1-r} \right|. 
\ee
$I_{\rm R}(r)$ is the filtering function satisfying $I_{\rm R}(r) \to 1$ at $r\to 0$ and $I_{\rm R}(r) \to 0$ 
at $r\to \infty$ (see Fig.2 in \cite{McDonald:2009lr}). 
Again, these three terms end up with a constant plus the $\sigma^{2}_{3}(k)$ term even though 
the functional forms of this filtering function for each term are all different. 
Based upon the considerations above all, one finds an expression for the halo-matter density power 
spectrum
\ba
 P^{\rm hm}_{\,00}(k) & = & \;\;
 \left(c_{\delta}+\frac{34}{21}c_{\delta^{2}}\sigma^{2}
 +\frac{1}{2}c_{\delta^{3}}\sigma^{2}+ \frac{1}{3}c_{\delta s^{2}}\sigma^{2}
 +\frac{1}{2}c_{\delta\epsilon^{2}}\sigma^{2}_{\epsilon}
 +\frac{68}{63}c_{s^{2}}\sigma^{2}-\frac{16}{63}c_{st}\sigma^{2}\right)P^{\rm NL}_{\delta\delta}(k)\nonumber\\
 && + c_{\delta^{2}} \int\frac{d^{3}q}{(2\pi)^{3}}\,P(q)P(|\bfk-\bfq|)F^{(2)}_{\rm S}(\bfq,\bfk-\bfq)\nonumber\\
 && + c_{s^{2}} \int\frac{d^{3}q}{(2\pi)^{3}}\,P(q)P(|\bfk-\bfq|)F^{(2)}_{\rm S}(\bfq,\bfk-\bfq)S^{(2)}(\bfq,\bfk-\bfq)\nonumber\\
 && + \left(-\frac{16}{21}c_{s^{2}}+\frac{32}{105}c_{st}+ \frac{512}{2205}c_{\psi}\right)\sigma^{2}_{3}(k)P(k)\nonumber\\
 & = & b_{1}P^{\rm NL}_{\delta\delta}(k) + b_{2}P_{b2,\delta}(k) 
 + b_{s^{2}}P_{bs2,\delta}(k) + b_{3{\rm nl}}\,\sigma^{2}_{3}(k)P(k), 
 \label{eq: dd_fin}
\ea
where we redefine the bias parameters as 
\ba
 b_{1}  & = &  c_{\delta}+\frac{34}{21}c_{\delta^{2}}\sigma^{2}+\frac{1}{2}c_{\delta^{3}}\sigma^{2}+ \frac{1}{3}c_{\delta s^{2}}\sigma^{2}
 +\frac{1}{2}c_{\delta\epsilon^{2}}\sigma^{2}_{\epsilon}
 +\frac{68}{63}c_{s^{2}}\sigma^{2}-\frac{16}{63}c_{st}\sigma^{2},\label{eq: c1 to b1}\\
 b_{2} & = &  c_{\delta^{2}},\\
 b_{s^{2}} & = & c_{s^{2}},\\ 
 b_{\rm 3nl} & = &  -\frac{16}{21}c_{s^{2}}+\frac{32}{105}c_{st}+ \frac{512}{2205}c_{\psi},
\ea
and terms associated with these bias parameters are defined as
\ba
 P_{b2,\delta}(k)  & \equiv & \int\frac{d^{3}q}{(2\pi)^{3}}\,P(q)P(|\bfk-\bfq|)F^{(2)}_{\rm S}(\bfq,\bfk-\bfq),\\
 P_{bs2,\delta}(k) & \equiv & \int\frac{d^{3}q}{(2\pi)^{3}}\,P(q)P(|\bfk-\bfq|)F^{(2)}_{\rm S}(\bfq,\bfk-\bfq)
 S^{(2)}(\bfq,\bfk-\bfq).
\ea 
Thus all the third-order nonlocal bias terms can be grouped into only one bias parameter, $b_{\rm 3nl}$. 
The main purpose of this paper is to investigate whether the $b_{\rm 3nl}$ term is important to explain the 
halo-matter power spectrum in $N$-body simulations. 

\subsection{The cross power spectrum between halo density and matter momentum}
\label{subsec: P^hm_01}

Let us next extend to the case of the cross spectrum between halo density and matter momentum. 
Higher-order nonlocal bias could also affect the cross spectrum between halo density and matter momentum. 
An advantage of the momentum spectrum is that it can be easily measured from $N$-body simulations 
without any ambiguity in interpolating the velocity divergence field \cite{Okumura:2012rq}.  
Also, since the momentum spectrum is an essential ingredient in predicting the nonlinear 
RSDs (see \cite{Seljak:2011dw,Okumura:2012gf,Okumura:2012rq,Vlah:2012fr,Vlah:2013qy,Blazek:2013ys}), 
it would be important to see an impact of the nonlocal bias terms on the momentum spectrum. 
Here we derive an explicit formula including the nonlocal bias terms up to third order and 
show that it can be renormalized in a similar manner to the case of halo and matter density 
correlation. 

The cross spectrum between halo density and matter momentum, $P^{\rm hm}_{\,01}(k)$ is given by 
\ba
 P^{\rm hm}_{\,01}(\bfk)(2\pi)^{3}\delta_{D}(\bfk+\bfk')
 & = & \langle T_{\parallel}^{\rm h,0}(\bfk)T_{\parallel}^{\rm m,1}(\bfk') \rangle \nonumber\\
 & = & if\frac{\mu}{k} \langle \delta_{\rm h}(\bfk)\theta(\bfk') \rangle
  +if \int\,\frac{d^{3}q}{(2\pi)^{3}}\,\frac{q_{\parallel}}{q^{2}} \langle \delta_{\rm h}(\bfk)\theta(-\bfq)\delta(\bfk'+\bfq) \rangle, 
  \label{eq: start of P^hm_01}
\ea
where $f$ is the growth parameter defined by $f\equiv d\ln D/d\ln a$ with $D$ and $a$ being 
the linear growth rate and scale factor of the universe, respectively, and 
$\mu$ is cosine of the angle between wavevector and line of sight. We define an isotropic part, 
$P^{\rm hm}_{\,01}(k)$, as
$P^{\rm hm}_{\,01}(\bfk)=i\mu\,P^{\rm hm}_{\,01}(k)/k$, motivated by the fact that it reduces to 
\be
 P^{\rm hm}_{\,01}(\bfk) = i\frac{\mu}{k}c_{\delta}fP(k), 
\ee
in linear regime. The bispectrum term in Eq.~(\ref{eq: start of P^hm_01}) is not affected by the third-order perturbations, 
while the first term Eq.~(\ref{eq: start of P^hm_01}) is. We then redo the similar renormalization procedure in the first 
term, i.e., the cross spectrum between halo density and matter velocity fields which becomes 
\ba
 P_{\rm \delta_{\rm h}\theta}(k) & = & \;\;
 c_{\delta}P(k) +  c_{\delta}P^{(13)}_{\delta\theta}(k) + c_{\delta}P^{(22)}_{\delta\theta}(k)\nonumber\\
 && + \frac{34}{21}c_{\delta^{2}}\sigma^{2}P(k) + \frac{1}{2}c_{\delta^{3}}\sigma^{2}P(k) + \frac{1}{3}c_{\delta s^{2}}\sigma^{2}P(k)
 +\frac{1}{2}c_{\delta\epsilon^{2}}\sigma^{2}_{\epsilon}P(k)\nonumber\\
 && + c_{\delta^{2}} \int\frac{d^{3}q}{(2\pi)^{3}}\,P(q)P(|\bfk-\bfq|)G^{(2)}_{\rm S}(\bfq,\bfk-\bfq)\nonumber\\
 && + c_{s^{2}} \int\frac{d^{3}q}{(2\pi)^{3}}\,P(q)P(|\bfk-\bfq|)G^{(2)}_{\rm S}(\bfq,\bfk-\bfq)S^{(2)}(\bfq,\bfk-\bfq)\nonumber\\
 && + 2c_{s^{2}}P(k)\int\frac{d^{3}q}{(2\pi)^{3}}\,P(q)F^{(2)}_{\rm S}(-\bfq,\bfk)S^{(2)}(\bfq,\bfk-\bfq)\nonumber\\
 && + 2c_{st}P(k)\int\frac{d^{3}q}{(2\pi)^{3}}\,P(q)D^{(2)}_{\rm S}(-\bfq,\bfk)S^{(2)}(\bfq,\bfk-\bfq)\nonumber\\
 && + 2c_{\psi}P(k)\int\frac{d^{3}q}{(2\pi)^{3}}\,P(q)\left[\frac{3}{2}D^{(3)}_{\rm S}(\bfq,-\bfq,-\bfk)
 -2F^{(2)}_{\rm S}(-\bfq,\bfk)D^{(2)}_{\rm S}(\bfq,\bfk-\bfq)\right]. 
 \label{eq: dv_org}
\ea
Since the last three lines are exactly same with the terms in the halo-density and matter-density spectrum, we confirm 
that $P_{\delta_{\rm h}\theta}(k)$ can be similarly renormalized as 
\ba
 P_{\delta_{\rm h}\theta}(k) = 
 b_{1}P^{\rm NL}_{\delta\theta}(k) 
 + b_{2}P_{b2,\theta}(k) 
 + b_{s^{2}}P_{bs2,\theta}(k) 
 + b_{3{\rm nl}}\,\sigma^{2}_{3}(k)P(k), 
 \label{eq: dv_fin}
\ea
where we define the terms associated with the second-order bias as 
\ba
 P_{b2,\theta}(k)  & \equiv & \int\frac{d^{3}q}{(2\pi)^{3}}\,P(q)P(|\bfk-\bfq|)G^{(2)}_{\rm S}(\bfq,\bfk-\bfq),\\
 P_{bs2,\theta}(k) & \equiv & \int\frac{d^{3}q}{(2\pi)^{3}}\,P(q)P(|\bfk-\bfq|)G^{(2)}_{\rm S}(\bfq,\bfk-\bfq)
 S^{(2)}(\bfq,\bfk-\bfq).
\ea 
A symmetric structure in integrations of the bispectrum allows us to write down 
the second term in Eq.~(\ref{eq: start of P^hm_01}) as \cite{Matsubara:2008vp,Taruya:2010lr}:
\ba
  \int\,\frac{d^{3}q}{(2\pi)^{3}}\,\frac{q_{\parallel}}{q^{2}} \langle \delta_{\rm h}(\bfk)\theta(-\bfq)\delta(\bfk'+\bfq) \rangle
  & = & \frac{\mu}{k}\left\{c_{\delta}B_{b1}(k) + c_{\delta^{2}}B_{b2}(k) + c_{s^{2}}B_{bs2}(k)\right\}
  (2\pi)^{3}\delta_{D}(\bfk+\bfk') \label{eq: dvd c1}\\
  & \simeq & \frac{\mu}{k}\left\{ b_{1}B_{b1}(k) + b_{2} B_{b2}(k) + b_{s^{2}} B_{bs2}(k)\right\}
  (2\pi)^{3}\delta_{D}(\bfk+\bfk'), 
  \label{eq: dvd to Bk}
\ea
where $B_{b1}(k)$, $B_{b2}(k)$ and $B_{bs2}(k)$ are expressed as follows:
\ba
 \frac{\mu}{k}B_{b1}(k) & \equiv & \int\,\frac{d^{3}q}{(2\pi)^{3}}\,\frac{q_{\parallel}}{q^{2}}\,
 2\left\{
 P(q)P(|\bfk-\bfq|)F^{(2)}_{\rm S}(\bfq,\bfk-\bfq)\right.\nonumber\\
 &&
 \left.\ \ \ \ \ \ \ \ \ \ \ \ \ \ \ \ \ \ \ +P(q)P(k)F^{(2)}_{\rm S}(\bfq,-\bfk)
 +P(|\bfk-\bfq|)P(k)G^{(2)}_{\rm S}(\bfk-\bfq,-\bfk)
 \right\},\\
 \frac{\mu}{k}B_{b2}(k) & \equiv & \int\,\frac{d^{3}q}{(2\pi)^{3}}\,\frac{q_{\parallel}}{q^{2}}\,P(q)P(|\bfk-\bfq|),\\
 \frac{\mu}{k}B_{bs2}(k) & \equiv & \int\,\frac{d^{3}q}{(2\pi)^{3}}\,\frac{q_{\parallel}}{q^{2}}\,P(q)P(|\bfk-\bfq|)
 S^{(2)}(\bfq,\bfk-\bfq). 
\ea
Collecting all the terms in Eqs.~(\ref{eq: dv_fin}) and (\ref{eq: dvd to Bk}), we finally obtain 
\ba
 P^{\rm hm}_{\,01}(k) & = &  b_{1}
 \left\{P^{\rm NL}_{\delta\theta}(k) + B_{b1}(k)\right\} 
 + b_{2}\left\{P_{b2,\theta}(k) + B_{b2}(k)\right\}\nonumber\\ 
 &&+ b_{s^{2}}\left\{P_{bs2,\theta}(k) + B_{bs2}(k)\right\} 
 + b_{3{\rm nl}}\,\sigma^{2}_{3}(k)P(k).
 \label{eq: P^hm_01 fin}
\ea
Thus the cross spectrum between halo density and matter momentum also includes only the $b_{\rm 3nl}$ term 
as a third-order nonlocal bias. 
Note that the first bracket, $ \left\{P^{\rm NL}_{\delta\theta} + B_{b1}(k)\right\} $, is nothing but the cross spectrum 
between matter density and momentum, $P^{\rm mm}_{\,01}(k)$, which is easily measured from simulations.\par 

\begin{figure}[t]
\begin{center}
\includegraphics[width=0.48\textwidth]{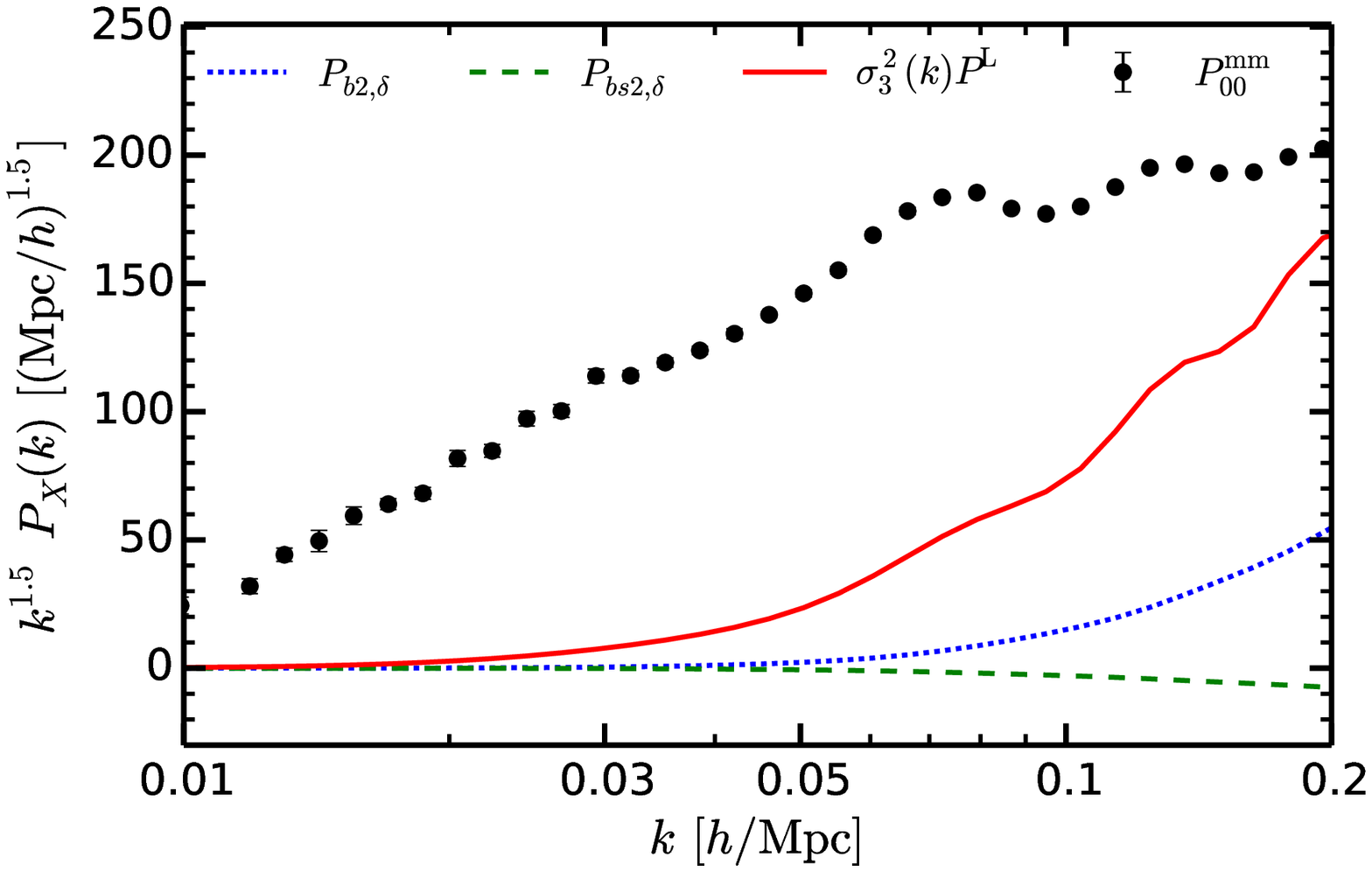}
\includegraphics[width=0.48\textwidth]{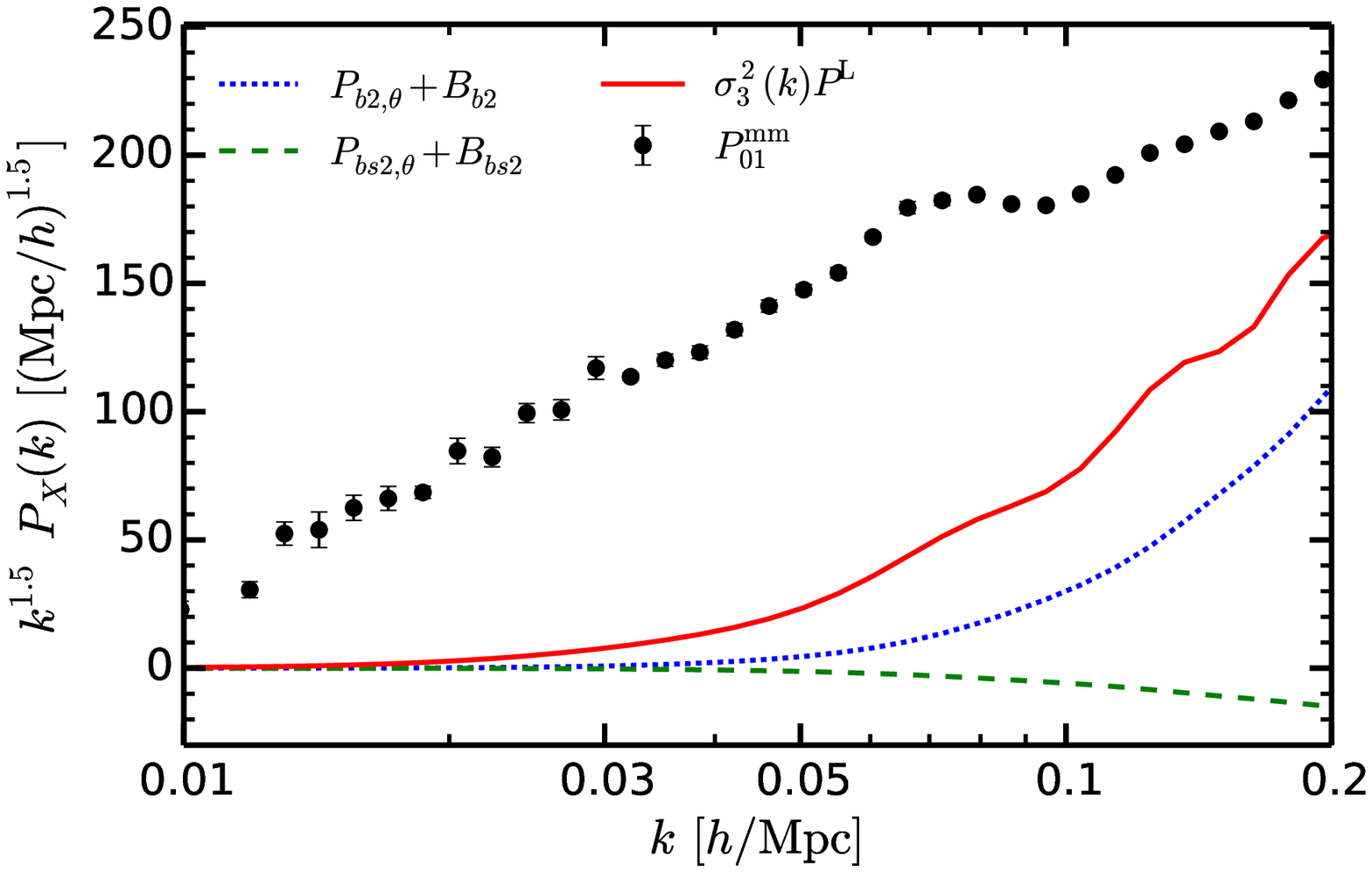}
\end{center}
\vspace*{-2em}
\caption{ 
A comparison of the PT correction terms at $z=0$ for $P^{\rm hm}_{\,00}$ ({\it left}) 
and $P^{\rm hm}_{\,01}$ ({\it right}). 
The data points are the nonlinear matter power spectrum directly measured from 
our simulations described in Sec.~\ref{sec: simulation}.
}  
\label{fig: PT comparison}
\end{figure}

In summary, we show that we only need four physical and renormalized bias parameters to describe 
the halo-matter spectra; 
the renormalized linear bias parameter, $b_{1}$, the second-order local bias parameter, $b_{2}$,
the second-order nonlocal bias parameter, $b_{s^{2}}$, and the third-order nonlocal bias parameter, $b_{3\rm nl}$.
We show the shape of each terms discussed so far in Fig.~\ref{fig: PT comparison}, together with the nonlinear 
matter power spectra in our simulations. Each line corresponds to the case in which the bias parameter is equal 
to be unity. As shown in the figures, the third-order nonlocal bias terms can dominate over the second-order local 
and nonlocal terms. As we will confirm later, the third-order nonlocal bias term becomes more significant 
than the second-order terms especially as long as the $b_{2}$ term is sufficiently small. 
This is not the case at massive halos with $M_{\rm halo}\gtsim 5\times 10^{13}\,[M_{\odot}/h]$ 
where $b_{2}$ becomes large enough to dominate over the $b_{3\rm nl}$ term. 

\subsection{The bispectrum and the bias renormalization}
\label{subsec: B^hmm_000}
So far we have observed that the four bias parameters, 
i.e., $(b_{1},b_{2},b_{s^{2}},b_{3\rm nl})$ are introduced to describe 
the cross power spectrum between halo density and matter density, or the one  
between halo density and matter momentum at the next-to-leading order 
when the nonlocal bias terms are considered. 
As is discussed in Ref.~\cite{Baldauf:2012lr}, the bispectrum at the lowest order 
(i.e., at tree level) demands perturbations only up to the second order, described as 
\be
 B^{\rm hmm}_{\,000}(\bfk_{1},\bfk_{2},\bfk_{3}) =  b_{1}B^{\rm mmm}_{\,000}(\bfk_{1},\bfk_{2},\bfk_{3})
 + P(k_{1})P(k_{2})\left[b_{2}+b_{s^{2}}\left(\mu_{k_{1},k_{2}}^{2}-\frac{1}{3}\right)\right], 
 \label{eq: B^hmm_000}
\ee
where $\mu_{k_{1},k_{2}}$ is cosine of the angle between $\bfk_{1}$ and $\bfk_{2}$, 
and the three arguments satisfy $\bfk_{1}+\bfk_{2}+\bfk_{3}=0$. 
In order to derive this as well as Eq.~(\ref{eq: dvd to Bk}) starting from Eq.~(\ref{eq: MR ansatz}), 
one may notice that a nontrivial approximation has been introduced, namely, $b_{1}\simeq c_{1}$. 
However, Ref.~\cite{McDonald:2006lr} argued that this is not the case. As we have seen in the 
renormalization procedure, all the renormalized terms originate from those in the limit of $k\to 0$. 
This fact means that a {\it physical} biased field should be defined so that homogeneous mean density 
is recovered at $k\to 0$. In other words, we should start with 
\ba
 \delta_{\rm h}(\bfx) & = & b_{1}\delta_{\rm m}(\bfx) 
 + \frac{1}{2}b_{2}\left[ \delta_{\rm m}(\bfx)^{2} - \langle \delta_{\rm m}(\bfx)^{2} \rangle \right] 
 + \frac{1}{2}b_{s^{2}}\left[s(\bfx)^{2} - \langle s(\bfx)^{2} \rangle \right]
 + \dots,
 \label{eq: physical bias}
\ea
rather than Eq.~(\ref{eq: MR ansatz}), and hence Eqs.~(\ref{eq: dvd to Bk}) and (\ref{eq: B^hmm_000}) 
are naturally derived. The same argument can be found in \cite{Assassi:2014lr} as well. 

Ref.~\cite{Baldauf:2012lr} shows that the specific $\mu_{k_{1},k_{2}}$ dependence 
in Eq.~(\ref{eq: B^hmm_000}) enables us to reliably determine both of the second-order bias parameters, 
$b_{2}$ and $b_{s^{2}}$ at the same time from the large-scale bispectrum. 
In later section we are going to simultaneously fit the power spectrum as well as the bispectrum, while 
Ref.~\cite{Baldauf:2012lr} fit the bispectrum with a prior on the linear bias $b_{1}$ determined from 
the halo-matter power spectrum only at $z=0$. In Appendix~\ref{sec: bispectrum only}, we present 
the results when we fit solely to the bispectrum with $b_{1}$ treated as free. 
In short, the differences in two approaches are generally small especially 
for $b_{s^{2}}$, indicating that $b_{s^{2}}$ is essentially determined by the characteristic $\mu_{k_{1},k_{2}}$
dependence.


\subsection{Coevolution of halos and dark matter up to third order}
\label{subsec: co-evolution}
So far we have discussed what kind of nonlocal bias terms are allowed in terms of symmetry in the fields 
set by gravity. Another way of studying the nonlocal bias terms induced by nonlinear gravitational evolution 
is to perturbatively solve the coupled equations between halos and dark matter under fluid approximation. 
This coevolution picture was first introduced by \cite{Fry:1996lr}, followed by e.g., 
\cite{Taruya:2000uq,Hui:2008qy,Chuen-Chan:2012fk,Baldauf:2012lr}. 
Here we simply assume the initial condition is purely local in the Lagrangian space, and thus 
this simple coevolution approach corresponds to the local Lagrangian evolution model. 
Assuming no velocity bias and 
a conservation of halo number, the continuity and the Euler equations combined with 
the Poisson equation for a matter-halo system are given by
\ba
 \delta_{\rm h}(\bfk,y)'-\theta(\bfk,y) 
 & = & \int \frac{d^{3}q}{(2\pi)^{3}}\alpha(\bfq,\bfk-\bfq)\theta(\bfq,y)\delta_{\rm h}(\bfk-\bfq,y),\\
 \delta_{\rm m}(\bfk,y)'-\theta(\bfk,y) 
 & = & \int \frac{d^{3}q}{(2\pi)^{3}}\alpha(\bfq,\bfk-\bfq)\theta(\bfq,y)\delta_{\rm m}(\bfk-\bfq,y),\\
 \{f\theta(\bfk,\eta)\}'
 +\left(1+\frac{\calH'}{\calH^{2}}\right)\theta(\bfk,y)
 -\frac{3}{2 f}\Omega_{\rm m}(y)\delta_{\rm m}(\bfk,y)
 & = &  f \int \frac{d^{3}q}{(2\pi)^{3}}\beta(\bfq,\bfk-\bfq)\theta(\bfq,\eta)\theta(\bfk-\bfq,\eta),
\ea
where we introduce $y\equiv \ln D(\eta)$ as a time variable rather than the conformal time $\eta$, 
and the prime denotes derivative w.r.t $y$. The Hubble parameter $\calH$ is defined by $\calH=da/(ad\eta)$.  
The linear-order solutions for this system are give by 
$\delta^{(1)}_{\rm m}(\bfk,y)  = e^{y}\delta_{0}(\bfk,y_{i})$, $\theta^{(1)}(\bfk,y)=\delta^{(1)}(\bfk,y)$,  
and $\delta^{(1)}_{\rm h}(\bfk,y)=b^{\rm E}_{1}(y)e^{y}\delta_{0}(\bfk,y_{i})$ where 
\ba
 \frac{b^{\rm L}_{1}(y)}{b^{\rm L}_{1}(y_{i})}=\frac{b^{\rm E}_{1}(y)-1}{b^{\rm E}_{1}(y_{i})-1}
 =\frac{e^{y_{i}}}{e^{y}}.
 \label{eq: b1L-b1E}
\ea
As is shown in Ref.~\cite{Catelan:1998qy, Baldauf:2012lr, Chuen-Chan:2012fk}, the second-order solution 
for halos is written by 
\ba
 \delta^{(2)}_{\rm h}(\bfk,y)  & = & 
 b^{\rm E}_{1}(y)\int \frac{d^{3}q}{(2\pi)^{3}}F^{(2)}_{\rm S}(\bfq,\bfk-\bfq)
 \delta^{(1)}_{\rm m}(\bfq,y)\delta^{(1)}_{\rm m}(\bfk-\bfq,y)\nonumber\\
 && \ +\left\{\frac{1}{2}b^{\rm L}_{2}(y)+\frac{4}{21}b^{\rm L}_{1}(y)\right\}
 \int \frac{d^{3}q}{(2\pi)^{3}}\delta^{(1)}_{\rm m}(\bfq,\eta)
 \delta^{(1)}_{\rm m}(\bfk-\bfq,y)\nonumber\\
 && \ -\frac{2}{7}b^{\rm L}_{1}(y)\int \frac{d^{3}q}{(2\pi)^{3}}S^{(2)}(\bfq,\bfk-\bfq)\delta^{(1)}_{\rm m}(\bfq,y)
 \delta^{(1)}_{\rm m}(\bfk-\bfq,y), 
 \label{eq: co-evolution second solution}
\ea
where we used the fact that $b_{n}^{\rm L}(y)=b_{n}^{\rm L}(y_{i})e^{n(y-y_{i})}$. 
Hence a correspondence of the local and nonlocal bias terms at second order to 
Eq.~(\ref{eq: full expression for third order}) is clearly found, and it shows that 
the tidal field is allowed to be a source of the nonlocal bias at second order: 
\ba
 b^{\rm coev}_{2} & = & b^{\rm L}_{2}(y)+\frac{8}{21}b^{\rm L}_{1}(y), \label{eq: b2L-b2E}\\
 b^{\rm coev}_{s^{2}} & = & -\frac{4}{7}b^{\rm L}_{1}(y) = -\frac{4}{7}(b^{\rm E}_{1}(y)-1). 
\ea
Continuing to this exercise to third order, we find the solution as 
\ba
 \delta^{(3)}_{\rm h}(\bfk,y) & = & \delta^{(3)}_{\rm h}(\bfk,y_{i})
 +\frac{1}{3}\int \frac{d^{3}q_1}{(2\pi)^{3}}\frac{d^{3}q_2}{(2\pi)^{3}}
 G^{(3)}_{\rm S}(\bfq_{1},\bfq_{2},\bfk-\bfq_{1}-\bfq_{2})\,
 \delta^{(1)}_{\rm m}(\bfq_{1},y)\delta^{(1)}_{\rm m}(\bfq_{2},y)\delta^{(1)}_{\rm m}(\bfk-\bfq_{1}-\bfq_{2},y)\nonumber\\
 && \ +\left(\frac{1}{2}b_{1}^{\rm L}(y)+\frac{1}{3} \right)\int \frac{d^{3}q_1}{(2\pi)^{3}}\frac{d^{3}q_2}{(2\pi)^{3}}
 \left[ \alpha(\bfq_{1},\bfq_{2}+\bfq_{3})F^{(2)}_{\rm S}(\bfq_{2},\bfq_{3}) \right]_{\rm sym}\,\delta^{3}\nonumber\\
 &&\ + \left[\frac{1}{2}b^{\rm L}_{2}(y)+\frac{2}{21}b^{\rm L}_{1}(y)\right]\int \frac{d^{3}q_1}{(2\pi)^{3}}\frac{d^{3}q_2}{(2\pi)^{3}}
 \left[ \alpha(\bfq_{1},\bfq_{2}+\bfq_{3})\right]_{\rm sym}\,\delta^{3}\nonumber\\
 &&\ - \frac{1}{4}b_{s^{2}}\int \frac{d^{3}q_1}{(2\pi)^{3}}\frac{d^{3}q_2}{(2\pi)^{3}}
 \left[ \alpha(\bfq_{1},\bfq_{2}+\bfq_{3})S^{(2)}(\bfq_{2},\bfq_{3}) \right]_{\rm sym}\,\delta^{3}\nonumber\\
 &&\ +\left(\frac{1}{2}b_{1}^{\rm L}(y)+\frac{1}{3} \right)\int \frac{d^{3}q_1}{(2\pi)^{3}}\frac{d^{3}q_2}{(2\pi)^{3}}
 \left[ \alpha(\bfq_{2}+\bfq_{3},\bfq_{1})G^{(2)}_{\rm S}(\bfq_{2},\bfq_{3}) \right]_{\rm sym}\,\delta^{3}. 
 \label{eq: co-evolution third solution}
\ea
Although the third-order solution looks somewhat complicated, it is useful 
to isolate its contribution to the matter-halo power spectrum, i.e., 
$\langle \delta^{(3)}_{\rm h}(\bfk,y) \delta^{(1)}_{\rm m}(\bfk,y)\rangle$. 
Subtracting out the terms proportional to the linear bias, we find 
\ba
 P^{\rm hm,(31)}_{\rm coev}(k)-b^{\rm E}_{1}P^{(31)}(k) = 
 \frac{32}{315}b^{\rm L}_{1}\sigma^{2}_{3}P(k) 
 +\left(
 b^{\rm L}_{2}+\frac{1}{2}b^{\rm L}_{3}
 \right)\sigma^{2}P(k).
\ea 
Now it is straightforward to correspond this formula to Eq.~(\ref{eq: dd_fin}):
\ba
 b^{\rm coev}_{3{\rm nl}} & = & \frac{32}{315}b^{\rm L}_{1}(y) = \frac{32}{315}(b^{\rm E}_{1}(y)-1),\label{eq: LLB b3nl prediction}\\
 b_{1} & = & b_{1}^{\rm E}+\left(b^{\rm L}_{2}+\frac{1}{2}b^{\rm L}_{3}\right)\sigma^{2}. \label{eq: LLB b1 prediction}
\ea
Thus the nonlocal bias term at third order which we discussed in the previous section 
can be related to the linear Lagrangian bias in this specific way. 
We will compare this prediction with our $b_{\rm 3nl}$ measurement from simulations 
in the following sections. 


\section{$N$-body simulations and the fitting methodology}
\label{sec: simulation}

\subsection{$N$-body simulation detail}
\label{subsec: matter-halo}
We performed a suite of $N$-body simulations using the publicly available {\tt Gadget2} code 
\cite{Springel:2005uq} to make 14 realizations at $z=0$, $0.5$ and $1$ 
with cosmological parameters in a flat $\Lambda$CDM model preferred by the WMAP results 
\cite{Komatsu:2009qy}, i.e., a mass density parameter $\Omega_{\rm m} = 0.272$, 
a baryon density parameter $\Omega_{\rm b} = 0.0455$, a Hubble constant $h = 0.704$, 
a spectral index $n_{\rm s} = 0.967$, and a normalization of the curvature perturbations 
of $\Delta^{2}_{\calR}= 2.42\times 10^{-9}$ at the pivot scale of $k = 0.002\,{\rm Mpc}^{-1}$, 
giving $\sigma_{8}=0.81$. 
The total simulation volume is $47.25\,[({\rm Gpc}/h)^{3}]$ which is 
larger roughly by a factor of ten than the current galaxy survey like BOSS. 
We generated initial conditions at $z=99$ using the second-order Lagrangian perturbation theory 
to initialize the second order growth correctly and allow for a reliable bispectrum extraction at low redshift.
The box size and number of particles are $L=1500\,{\rm Mpc}/h$ and $N_{\rm particle}=1024^{3}$, 
respectively, yielding a particle mass resolution of $2.37\times 10^{11}\,M_{\odot}/h$.\par 
We identify halos using the Friends-of-Friends finder with a linking length of 0.2 times 
the mean inter particle spacing. 
We only take halos which contain more than 20 particles, and hence our minimum halo 
mass is approximately $4.74\times 10^{12}\,M_{\odot}/h$. 
We divide the halo catalog into several mass bins at each redshift slice, whose detail is summarized 
in Table.~\ref{table: halo mass bin}. 
Note that $\sim 10^{13}M_{\odot}/h$ halo roughly corresponds to a typical host halo in which 
observed galaxies live. 
In order to estimate the power spectrum and the bispectrum, the particles are assigned on 
a $N_{\rm c}=1024$ grid with the Cloud-in-Cell algorithm, and the gridded density field is 
properly corrected by the window function. 
We also estimate the power spectrum of mass-weighted momentum of matter, following 
the method in \cite{Okumura:2012gf,Okumura:2012rq}. Note that our simulation is different 
from that used in \cite{Desjacques:2009qy,Baldauf:2012lr,Okumura:2012gf,Okumura:2012rq}. 
We mainly focus on combined measurement using the power spectrum and the bispectrum 
but will present results in the case of the bispectrum only in Appendix~\ref{sec: bispectrum only}. 
 
The errors of the power spectrum are estimated by the standard deviation among 14 realizations. 
Strictly speaking, it might be necessary to evaluate the covariance matrix to take account for 
the off-diagonal correlation among different modes. However, we neglect the correlation between 
different modes, since we focus on somewhat large scales, $k\ltsim 0.1\,h/{\rm Mpc}$. 
This part can be definitely improved by a proper treatment of the covariance matrix with 
larger number of realizations.  

\begin{table}[t]
{\tabcolsep = 3mm
 \renewcommand\arraystretch{1.3}
\begin{tabular}{ccccccc}
\hline
\hline
	redshift & mass bin & $\overline{M}_{\rm halo}\,[10^{13}\,M_\odot/h]$ & $b_1$ & $b_{2}$ & $b_{s^{2}}$ & $b_{3\rm nl}$\\
\hline \multirow{3}{*}{1} 
		&  I	&  0.763 & $2.0419\pm 0.0089$ & $-0.168\pm 0.027$ & $-1.099\pm 0.064$ & $0.211\pm 0.074$\\ 
		&  II	&  2.24   & $2.7957\pm 0.0114$ & $1.766\pm 0.039$ & $-1.409\pm 0.094$ &  $0.133\pm 0.100$\\ 
		&  III	&  6.50   & $4.0294\pm 0.0170$ & $8.0362\pm 0.062$ & $-1.708\pm 0.165$ &  $0.245\pm 0.150$\\ 
\hline \multirow{4}{*}{0.5}
		&  I	&  0.769 & $1.4426\pm 0.0057$ & $-0.792\pm 0.018$ & $-0.469\pm 0.038$ & $0.153\pm 0.030$\\ 
		&  II	&  2.29   & $1.9033\pm 0.0078$ & $-0.394\pm 0.024$ & $-0.785\pm 0.052$ & $0.170\pm 0.043$\\ 
		&  III	&  6.75   & $2.7005\pm 0.0115$ & $1.586\pm 0.035$ & $-1.286\pm 0.080$ &  $0.268\pm 0.061$\\ 
		&  IV	&  19.3   & $4.1349\pm 0.0204$ & $8.650\pm 0.066$ & $-1.837\pm 0.155$ &  $-0.294\pm 0.112$\\
\hline \multirow{4}{*}{0}
		&  I	&  0.773 & $1.0488\pm 0.0048$ & $-0.777\pm 0.013$ & $-0.099\pm 0.026$ & $0.092\pm 0.019$\\ 
		&  II	&  2.33   & $1.3094\pm 0.0062$ & $-0.873\pm 0.018$ & $-0.267\pm 0.035$ & $0.132\pm 0.023$\\ 
		&  III	&  6.92   & $1.7977\pm 0.0087$ & $-0.462\pm 0.025$ & $-0.514\pm 0.051$ & $0.193\pm 0.035$\\ 
		&  IV	&  20.1   & $2.6741\pm 0.0136$ & $1.500 \pm 0.040$ & $-1.028\pm 0.086$ & $0.105\pm 0.053$\\
\hline
\hline 
\end{tabular}
}
\caption{Summary of halo catalogs used in this paper. 
We also show the best-fitting values of four bias parameters determined by our fitting from the power spectrum 
and the bispectrum. The fitting range, $k_{\rm max}$, depends on redshift 
(see text on how to choose $k_{\rm max}$ in detail): 
$(k_{{\rm max},P(k)},k_{{\rm max},B(k)})=(0.08, 0.065)$, $(0.10, 0.075)$, and $(0.125, 0.075)$  at 
$z=0$, $0.5$ and $1$, respectively. 
Note that the definition of second-order bias parameters in \cite{Baldauf:2012lr} differs by a factor of two. 
}
\label{table: halo mass bin}
\end{table}

\subsection{Fitting procedure}
\label{subsec: fitting}
Let us briefly summarize how we determine the bias parameters from the simulated power spectra. 
As explained in the previous section, we have four bias parameters as free, i.e.,  two local bias 
parameters, $b_{1}$ and $b_{2}$, and second- and third-order nonlocal bias ones, $b_{s^{2}}$ and $b_{3\rm nl}$. 
When we fit the bias model to the halo-matter density power spectrum only, the fitted bias parameters are 
estimated so that they minimize 
\ba
 \chi^{2}_{P00} = \sum_{k_{i}\le k_{{\rm max},P(k)}}
 \frac{\left[P^{\rm hm}_{\,00}(k_{i})-\hat{P}^{\rm hm}_{\,00}(k_{i})\right]^{2}}{\Delta P^{\rm hm}_{\,00}(k_{i})^{2}}. 
 \label{eq: chi2_00}
\ea
Here theoretical template of $P^{\rm hm}_{\,00}$ at $k=k_{i}$ is given by Eq.~(\ref{eq: dd_fin}), 
$\hat{P}^{\rm hm}_{\,00}$ denotes the spectrum measured from the simulations, 
$\Delta P^{\rm hm}_{\,00}$ denotes the error of the spectrum amplitude, 
and $k_{{\rm max},P(k)}$ is the maximum wavenumber in the power spectrum analysis. 
Likewise we apply exactly the same procedure for the cross power spectrum between 
halo-density and matter-momentum by replacing `00' with `01' in the subscript in Eq.~(\ref{eq: chi2_00}). 
Note that we always insert the measured spectra from the simulation for nonlinear matter part, 
$P^{\rm NL}_{\delta\delta}$ for $P^{\rm mm}_{\,00}$ and $\{P^{\rm NL}_{\delta\theta}+B_{b1}\}$ for $P^{\rm mm}_{\,01}$. 
We also note that we use the power spectra averaged over 14 realizations rather than one in each realization. 
This is the reason why we will observe somewhat low values of reduced $\chi^{2}$, and hence this is not an overfitting issue. 
When we quote `00 only' (`01 only'), we simply use 
$\chi^{2}_{P(k)}= \chi^{2}_{P00}$ ($\chi^{2}_{P(k)}= \chi^{2}_{P01}$). 
When we include both $P^{\rm hm}_{\,00}$ and $P^{\rm hm}_{\,01}$, we assume they are independent 
and simply add two $\chi^{2}$ by neglecting the correlation between two, i.e., 
$\chi^{2}_{P(k)}= \chi^{2}_{P00}+ \chi^{2}_{P01}$. In principle, we could estimate 
the covariance matrix which includes correlation between both signals but 
the number of our realizations would not be sufficient to properly estimate it 
(see e.g. \cite{Percival:2013fj} for a recent study in such a direction).\par 

The similar procedure is adopted for the bispectrum as well. We search the best-fitting values of 
the bias parameters for the bispectrum so that they minimize
\ba
 \chi^{2}_{B(k)} = \sum_{k_{i,j}\le k_{{\rm max},B(k)}}
 \frac{\left[B^{\rm hmm}_{\,000}(k_{i},k_{j},\mu_{ij})
 -\hat{B}^{\rm hmm}_{\,000}(k_{i},k_{j},\mu_{ij})\right]^{2}}{\Delta B^{\rm hmm}_{\,000}(k_{i},k_{j},\mu_{ij})^{2}}, 
 \label{eq: chi2_Bk}
\ea
where $\mu_{i,j}$ is the cosine between $\bfk_{1}$ and $\bfk_{2}$, 
the theoretical template of $B^{\rm hmm}_{\,000}$ is given by Eq.~(\ref{eq: B^hmm_000}), 
$\Delta B^{\rm hm}_{\,00}$ denotes the error of the bispectrum amplitude, 
and $k_{{\rm max},B(k)}$ is the maximum wavenumber in the bispectrum analysis. 
Notice that the bispectrum depends only on three bias parameters, $b_{1}$, $b_{2}$ and $b_{s^2}$. 
We distinguish the maximum wavenumber range in the power spectrum case from
that in the bispectrum. It is not entirely clear if higher-order PT terms for different statistics 
become dominant at the same wavenumber. Our main purpose is to investigate how large 
the their-order contribution is, and hence we fix $k_{{\rm max},B(k)}$ to 0.065 (0.075)$h/$Mpc 
at $z=0$ ($z=0.5$ or $1$) in the following analysis. These choices are based on our fitting results 
to the bispectrum only, presented in Appendix~\ref{sec: bispectrum only}. 
Thus, we adopt $\chi^{2}=\chi^{2}_{P(k)}+\chi^{2}_{B(k)}$ when we jointly fit the PT model to
the power spectrum and the bispectrum.\par 

In order to fully investigate the probability distribution of preferred values of the bias parameters, 
we adopt the Markov chain Monte Carlo (MCMC) technique, assuming the Gaussian likelihood, i.e., 
$\mathcal{L} \propto \exp(-\chi^{2}/2)$. For this end, we modify the COSMOMC code  \cite{Lewis:2002lr}, 
considering future applications of the code to the actual galaxy sample. We ensure convergence of 
each chain, imposing $R<0.003$ where $R$ is the standard Gelman-Rubin criteria.\par 

\begin{figure}[t]
\begin{center}
\includegraphics[width=0.32\textwidth]{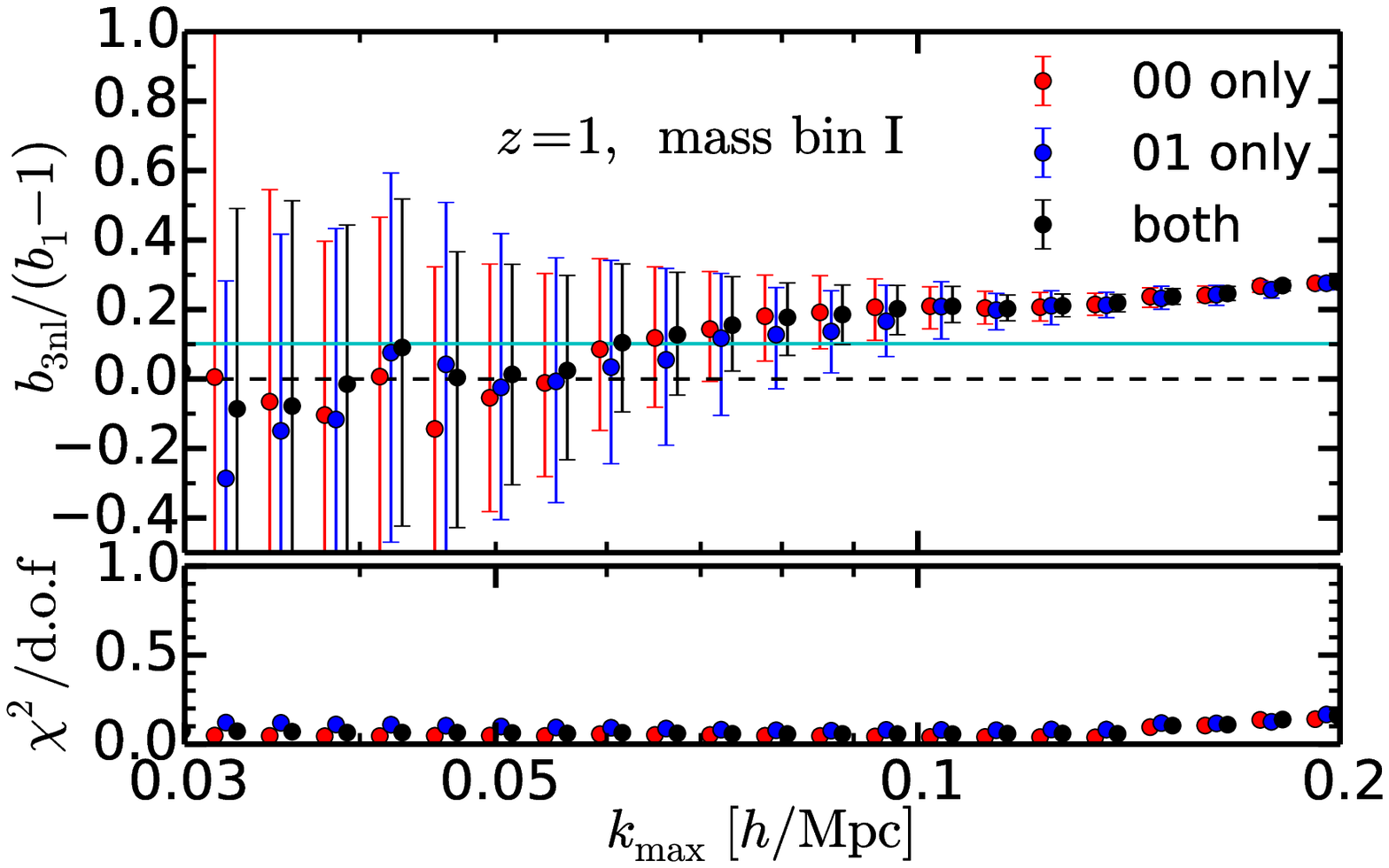}
\includegraphics[width=0.32\textwidth]{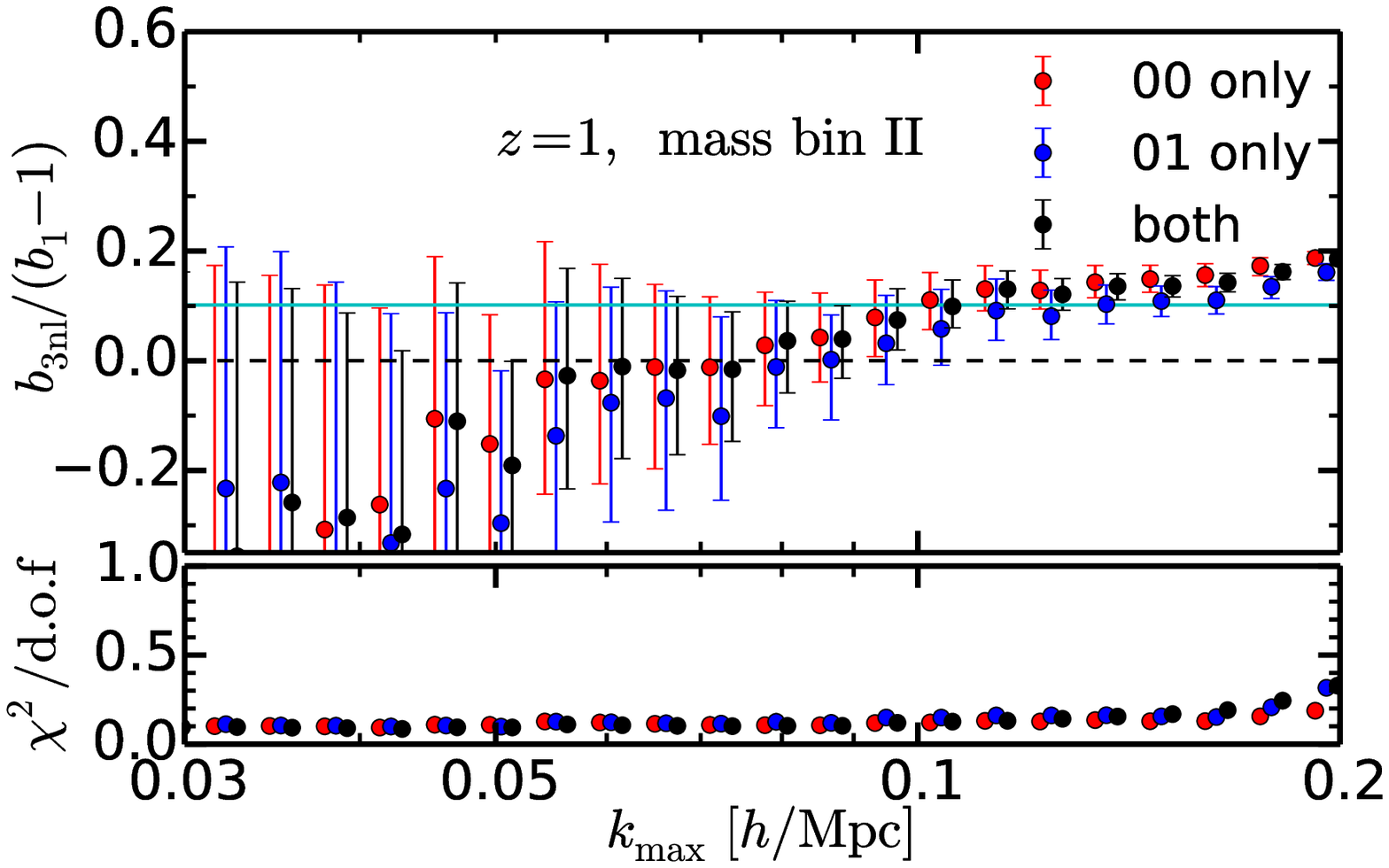}
\includegraphics[width=0.32\textwidth]{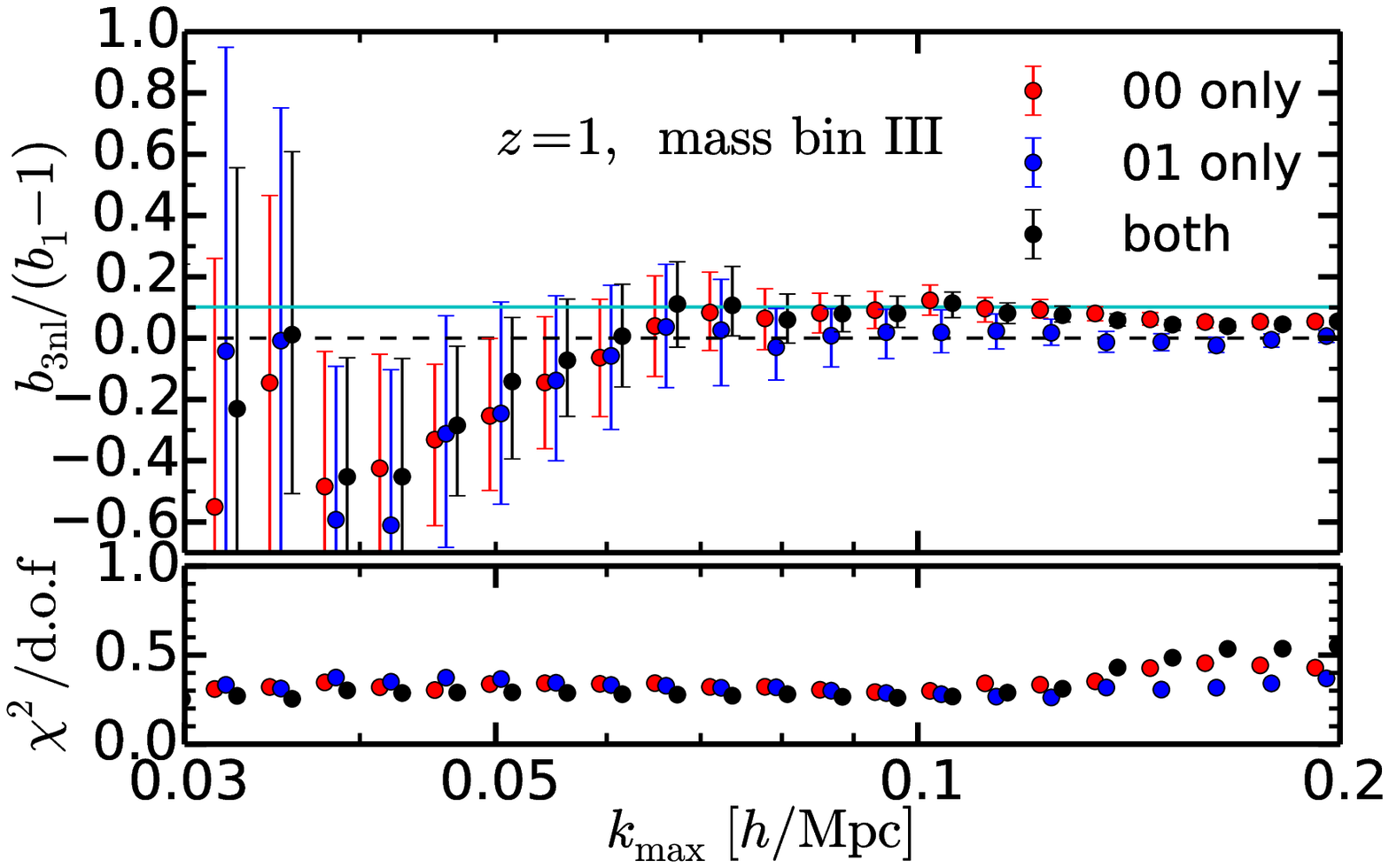}\\
\includegraphics[width=0.32\textwidth]{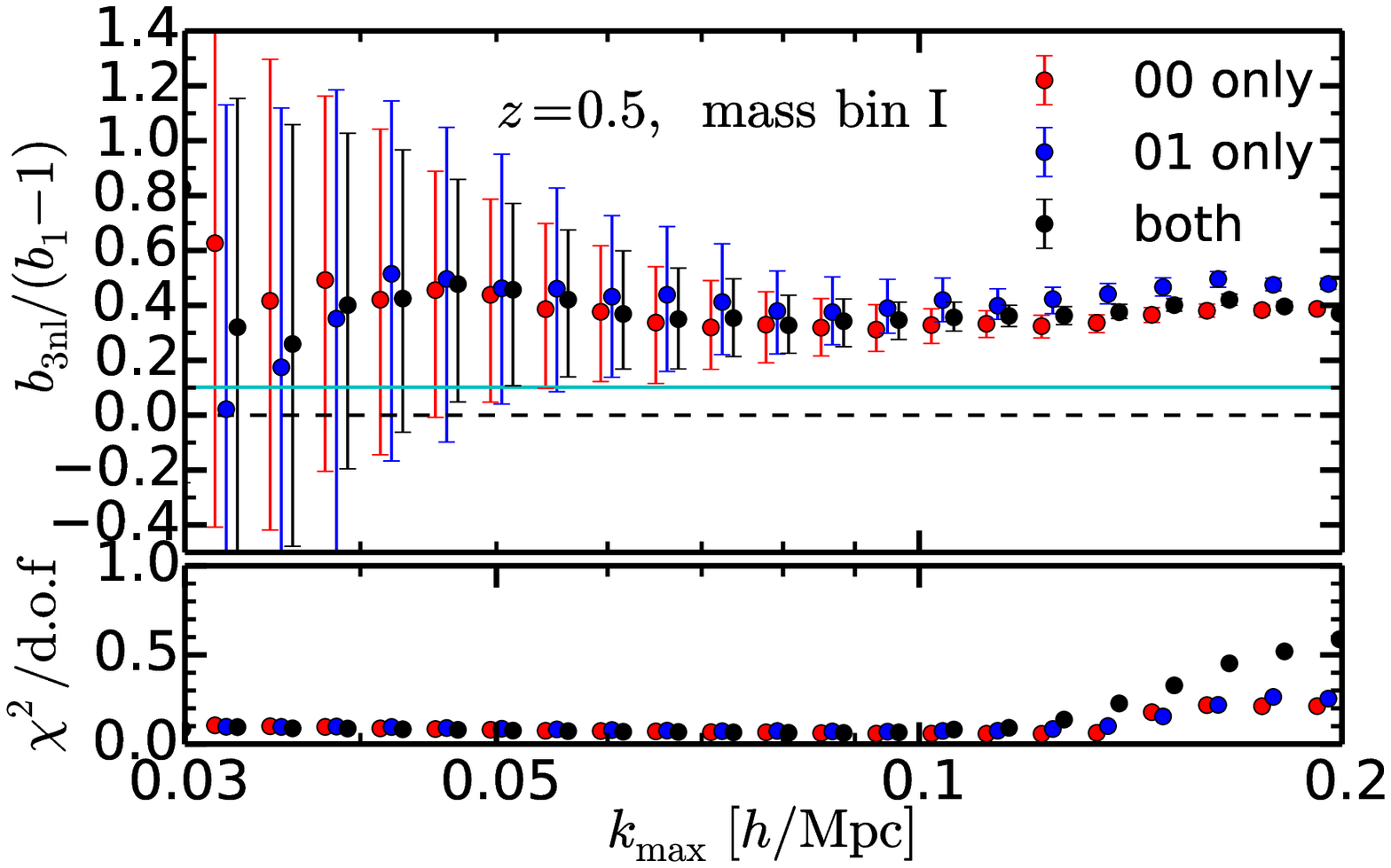}
\includegraphics[width=0.32\textwidth]{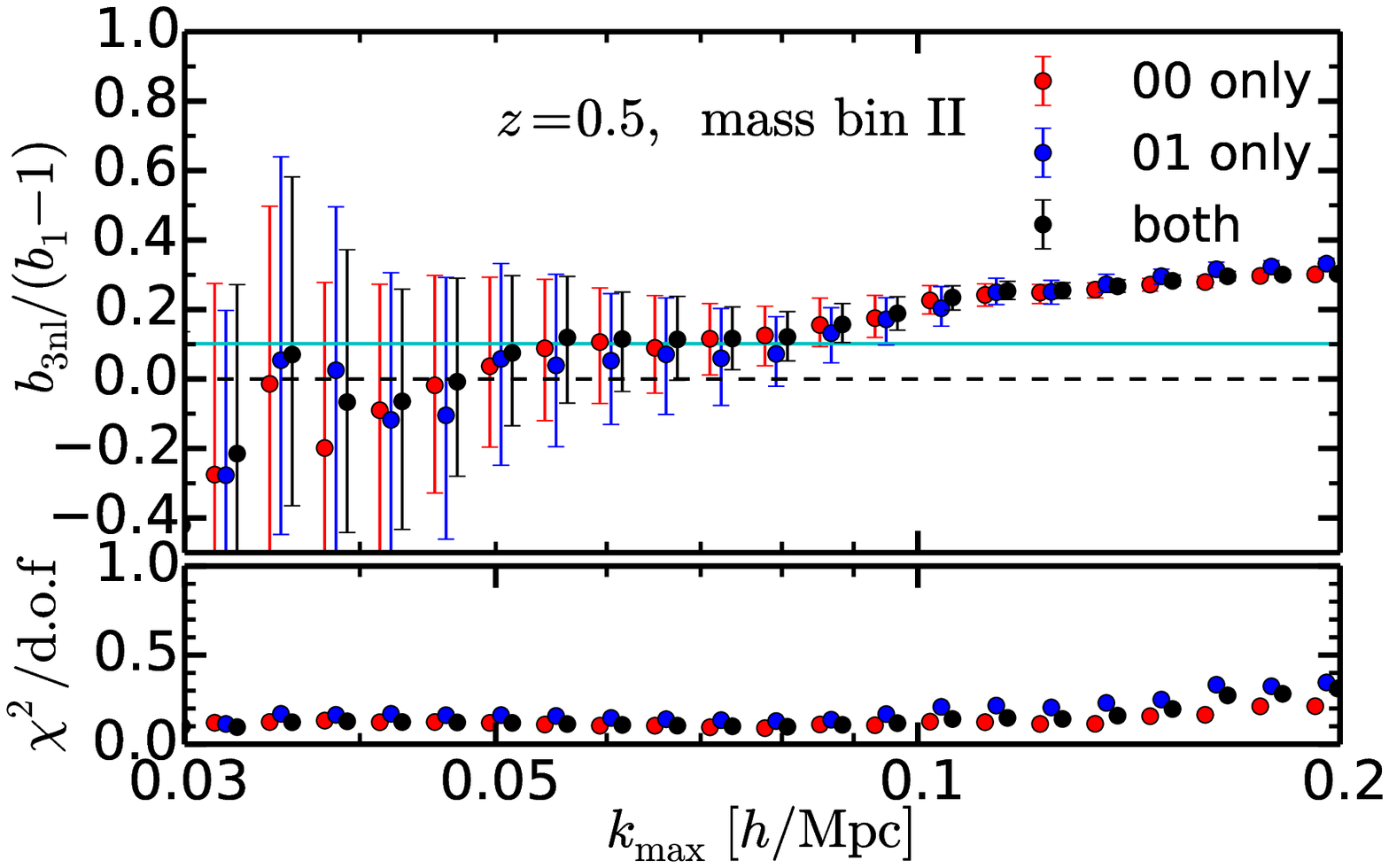}\\
\includegraphics[width=0.32\textwidth]{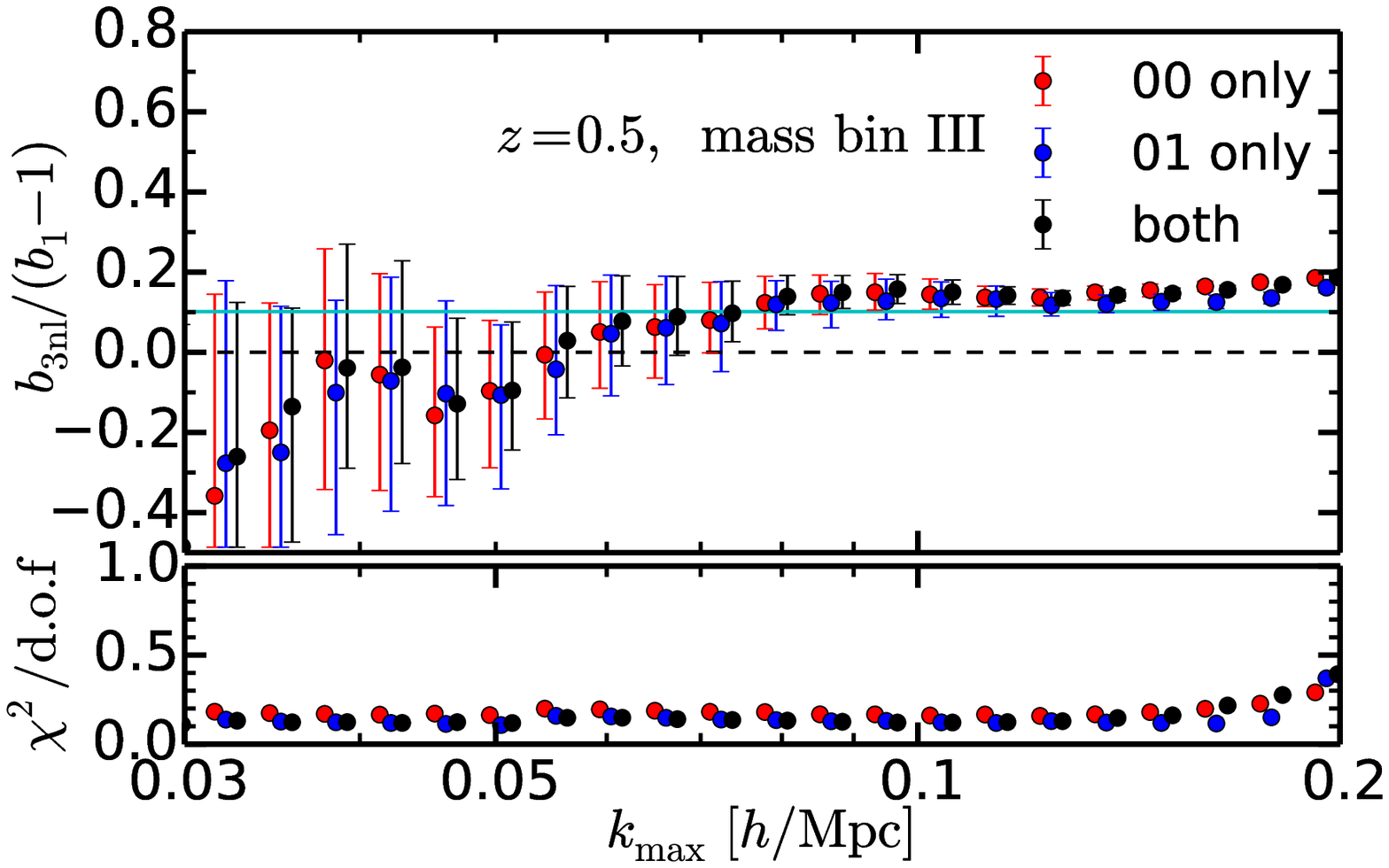}
\includegraphics[width=0.32\textwidth]{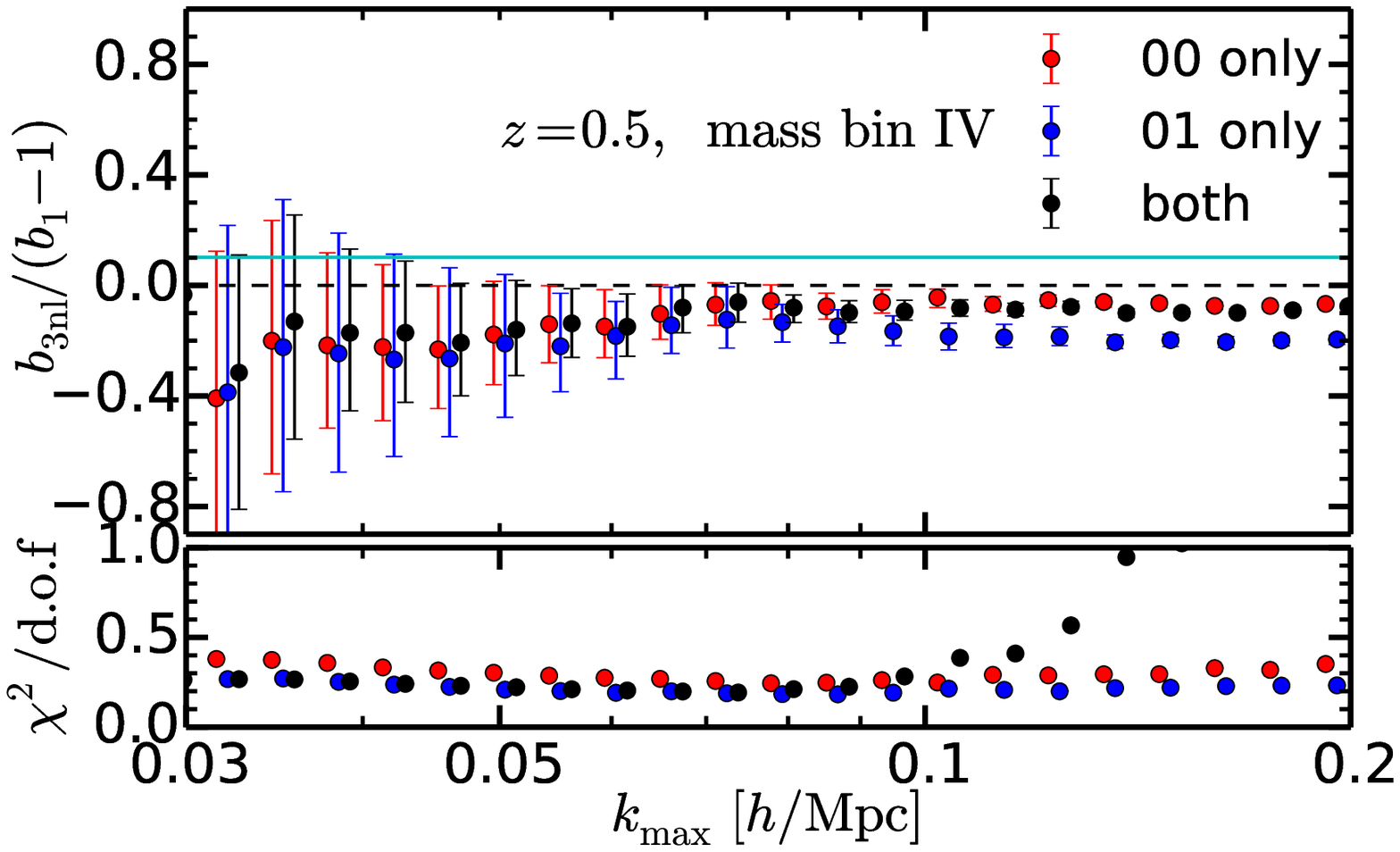}\\
\includegraphics[width=0.32\textwidth]{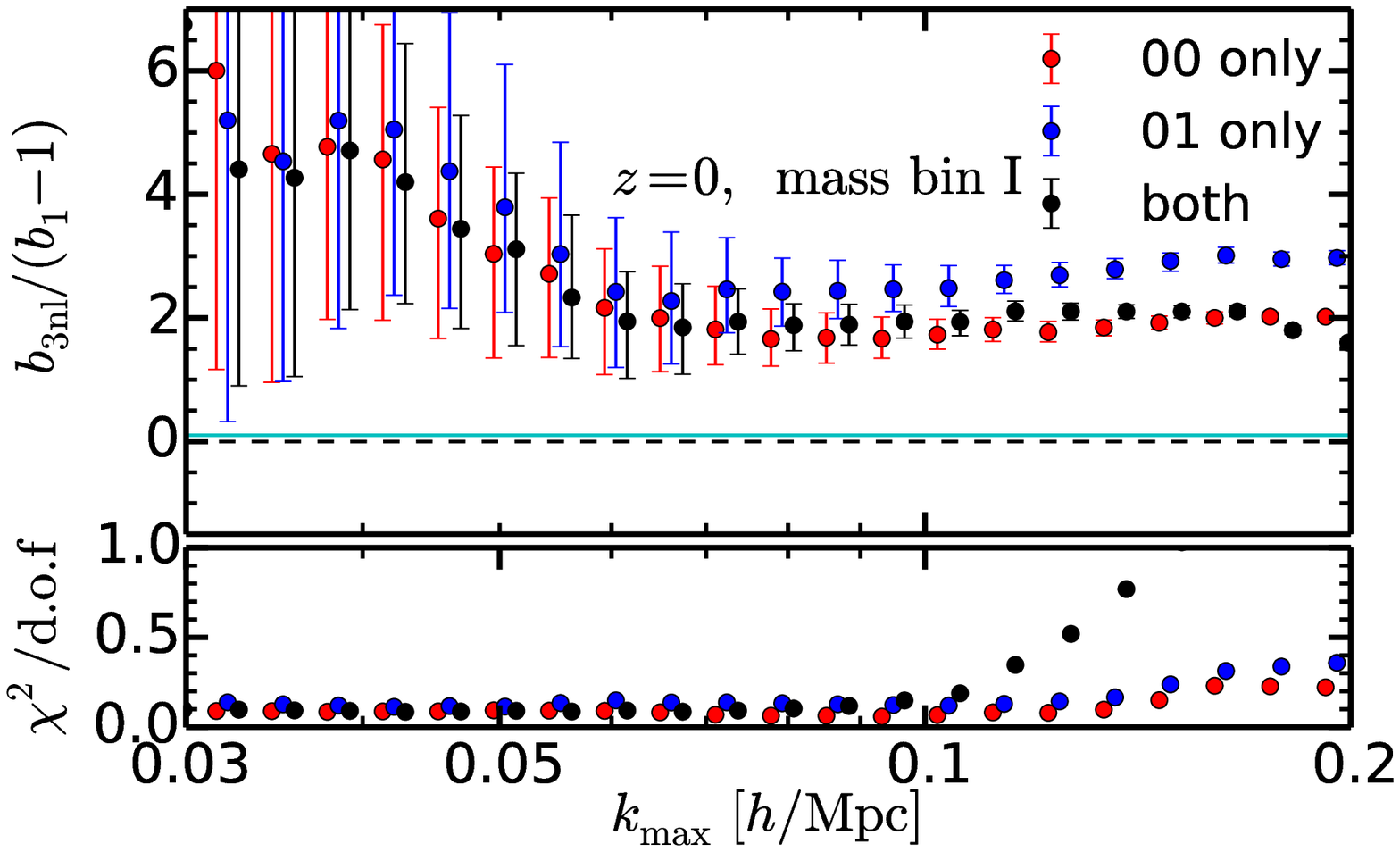}
\includegraphics[width=0.32\textwidth]{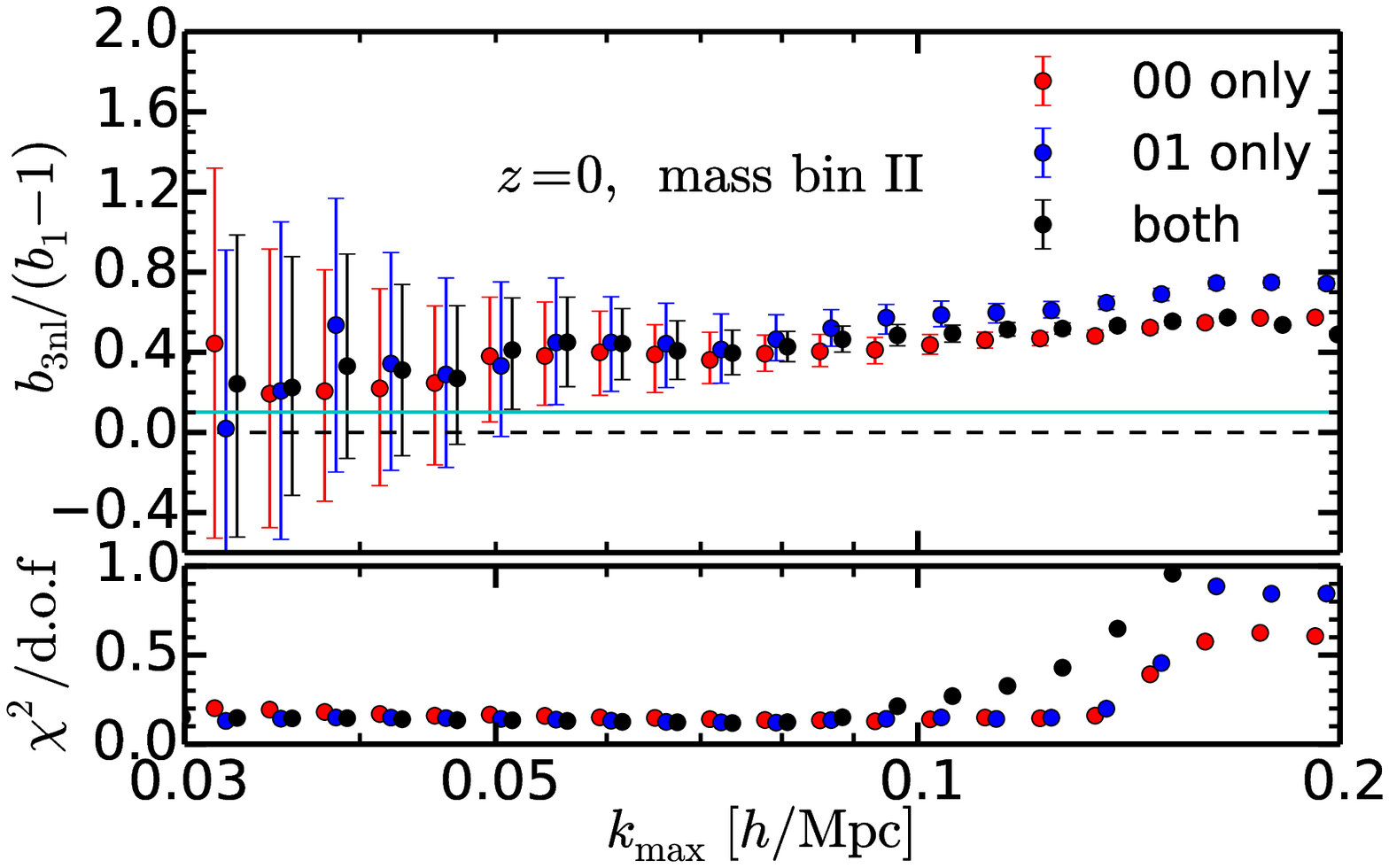}\\
\includegraphics[width=0.32\textwidth]{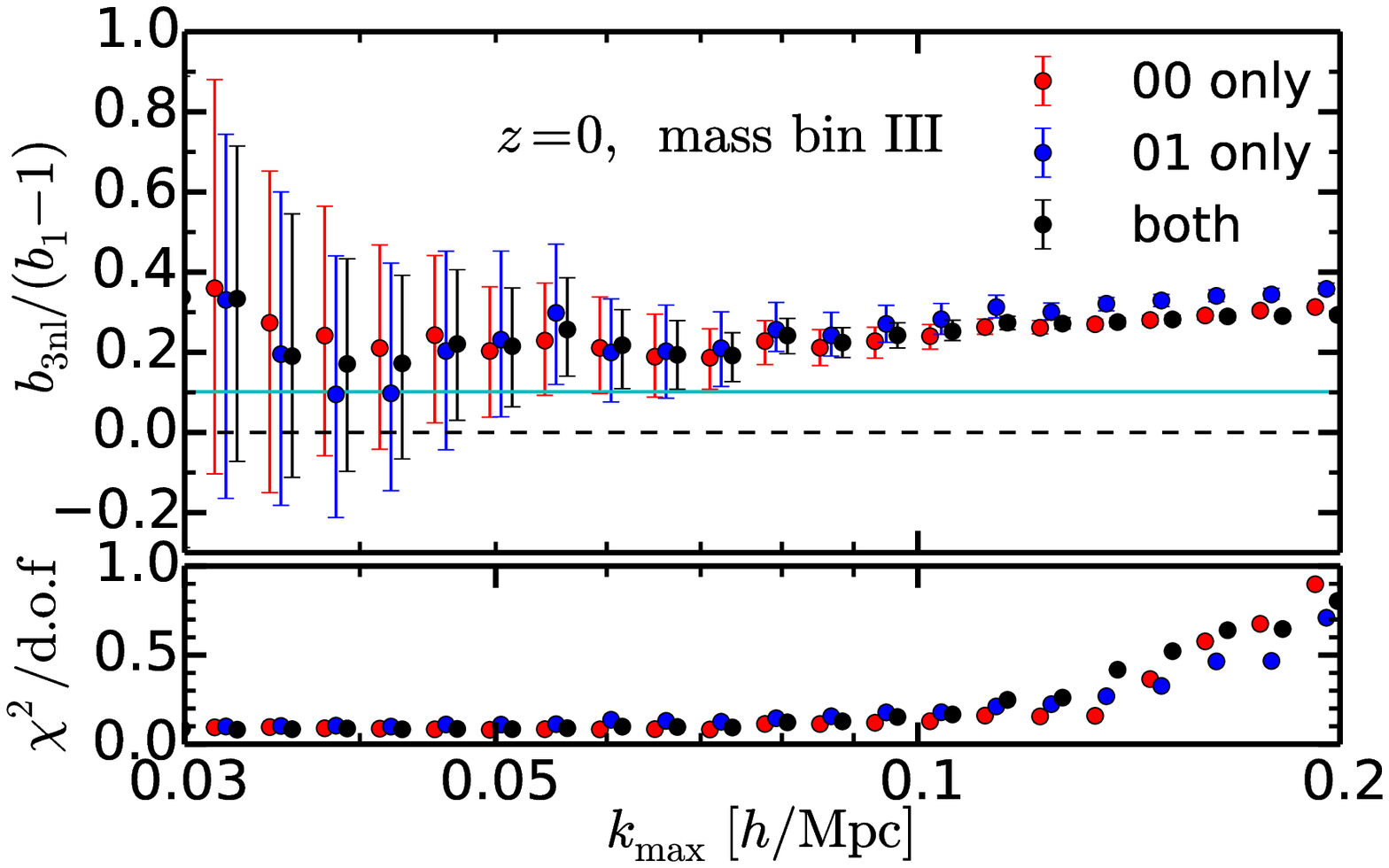}
\includegraphics[width=0.32\textwidth]{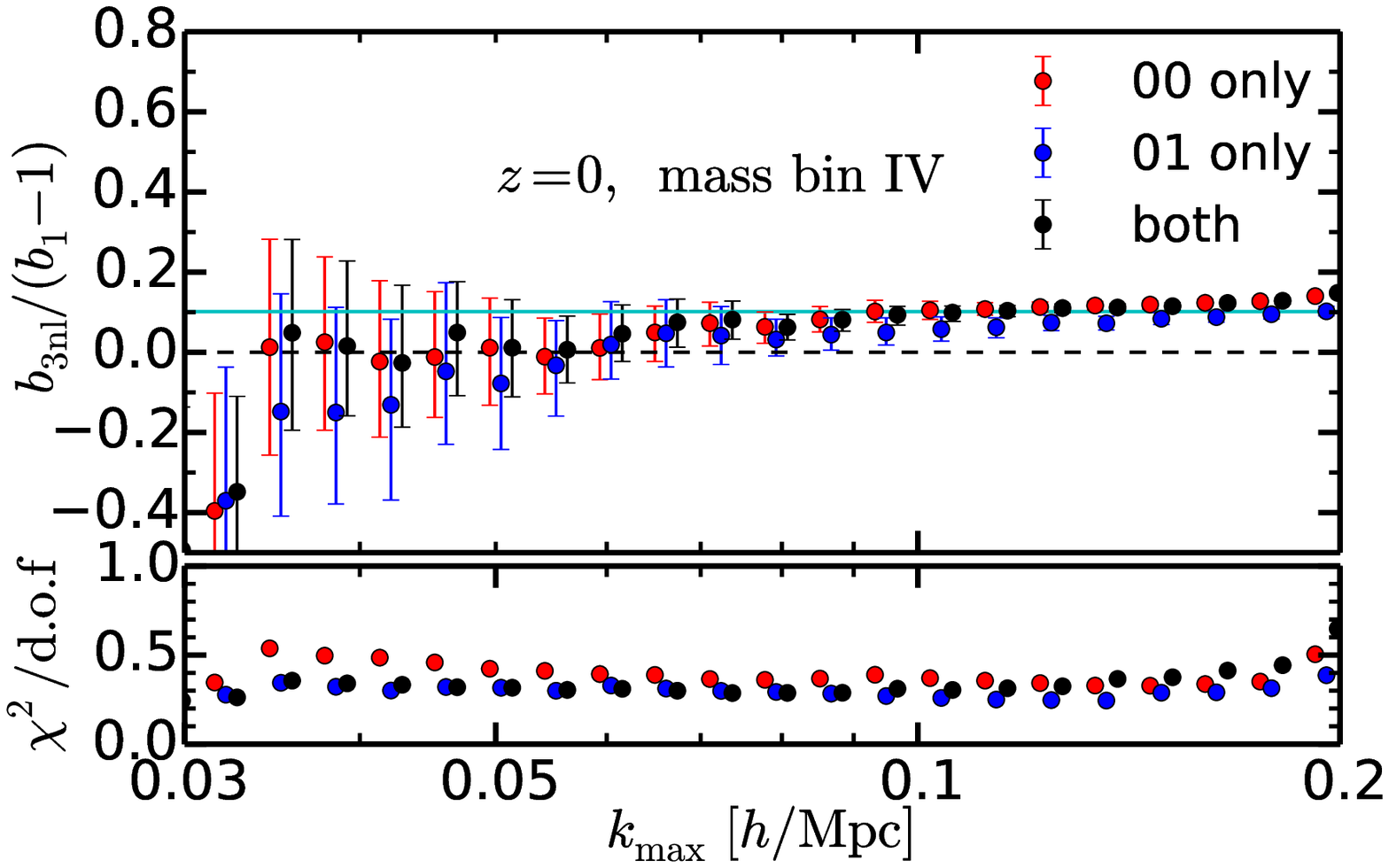}
\end{center}
\vspace*{-1em}
\caption{ 
The best-fitting values of $b_{3{\rm nl}}$ as a function of $k_{\rm max}$. 
We present results at $z=1$ ({\it top three}), at $z=0.5$ ({\it middle four}), and at 
$z=0$ ({\it bottom four}), for light to heavy (from I to IV) halo mass bins.  
In each panel, we show results in the case of $P^{\rm hm}_{\,00}$ only (red), 
$P^{\rm hm}_{\,01}$ only (blue), and both of two (black). 
The goodness of fit, $\chi^{2}_{P(k)}/{\rm d.o.f}$, is also plotted in the lower part 
of each panel. Note that we jointly fit the bispectrum together with the power 
spectrum. 
For comparison, the prediction from the coevolution picture (local Lagrangian bias model), 
$32/315$, is indicated by the horizontal line (cyan solid). 
}  
\label{fig: estimate b3nl}
\end{figure}

\begin{figure}[t]
\begin{center}
\includegraphics[width=0.4\textwidth]{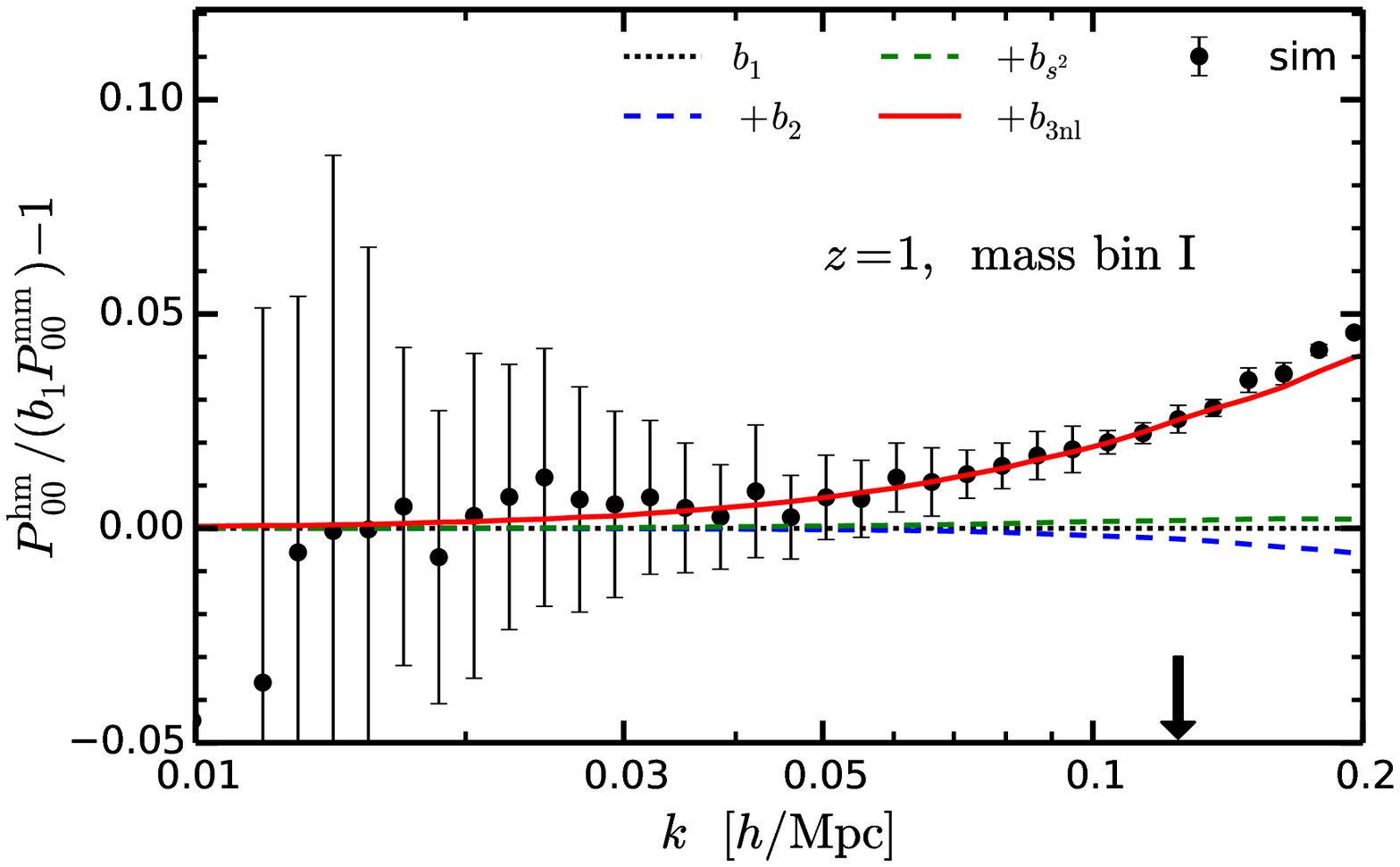}
\includegraphics[width=0.4\textwidth]{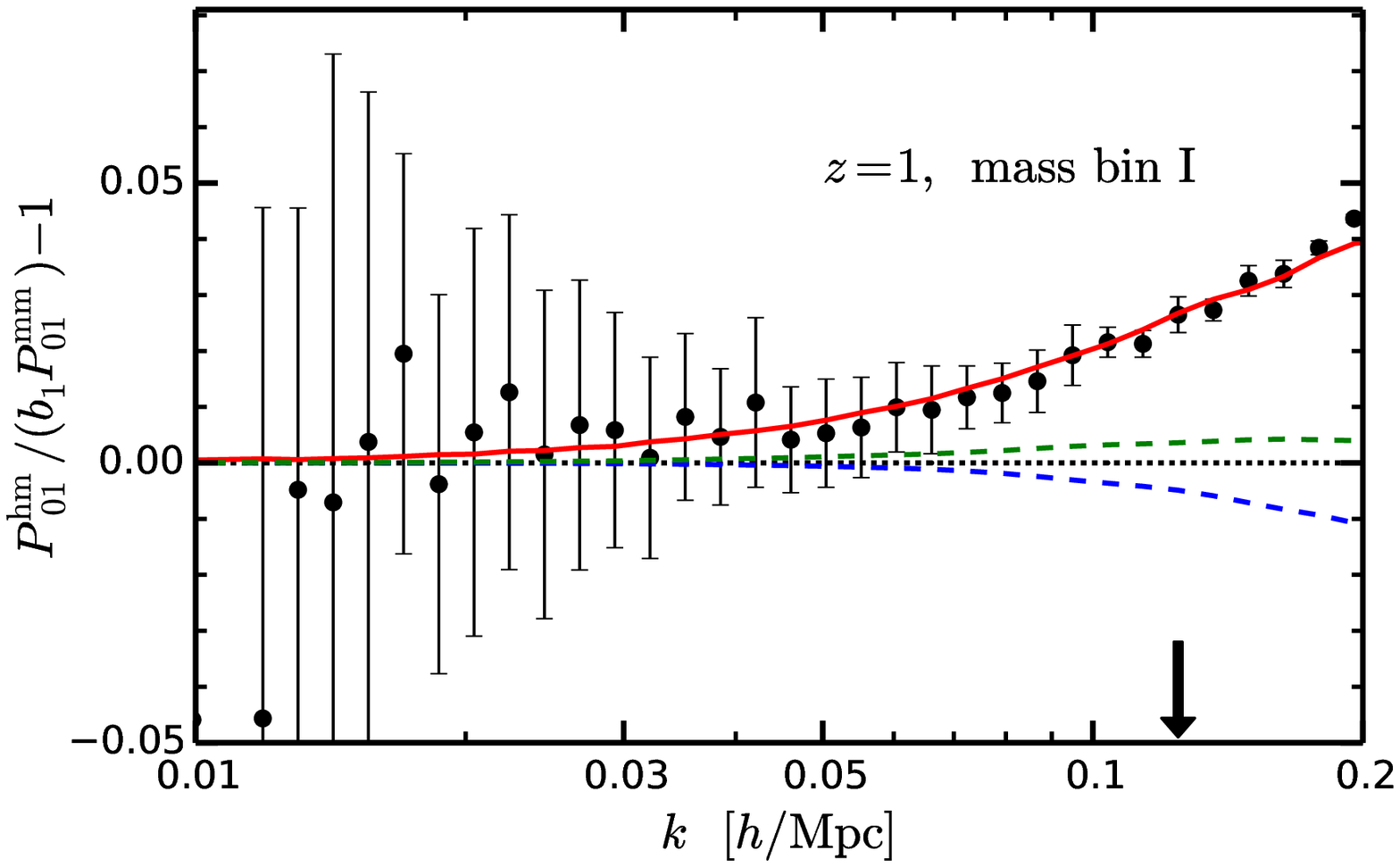}\\
\includegraphics[width=0.4\textwidth]{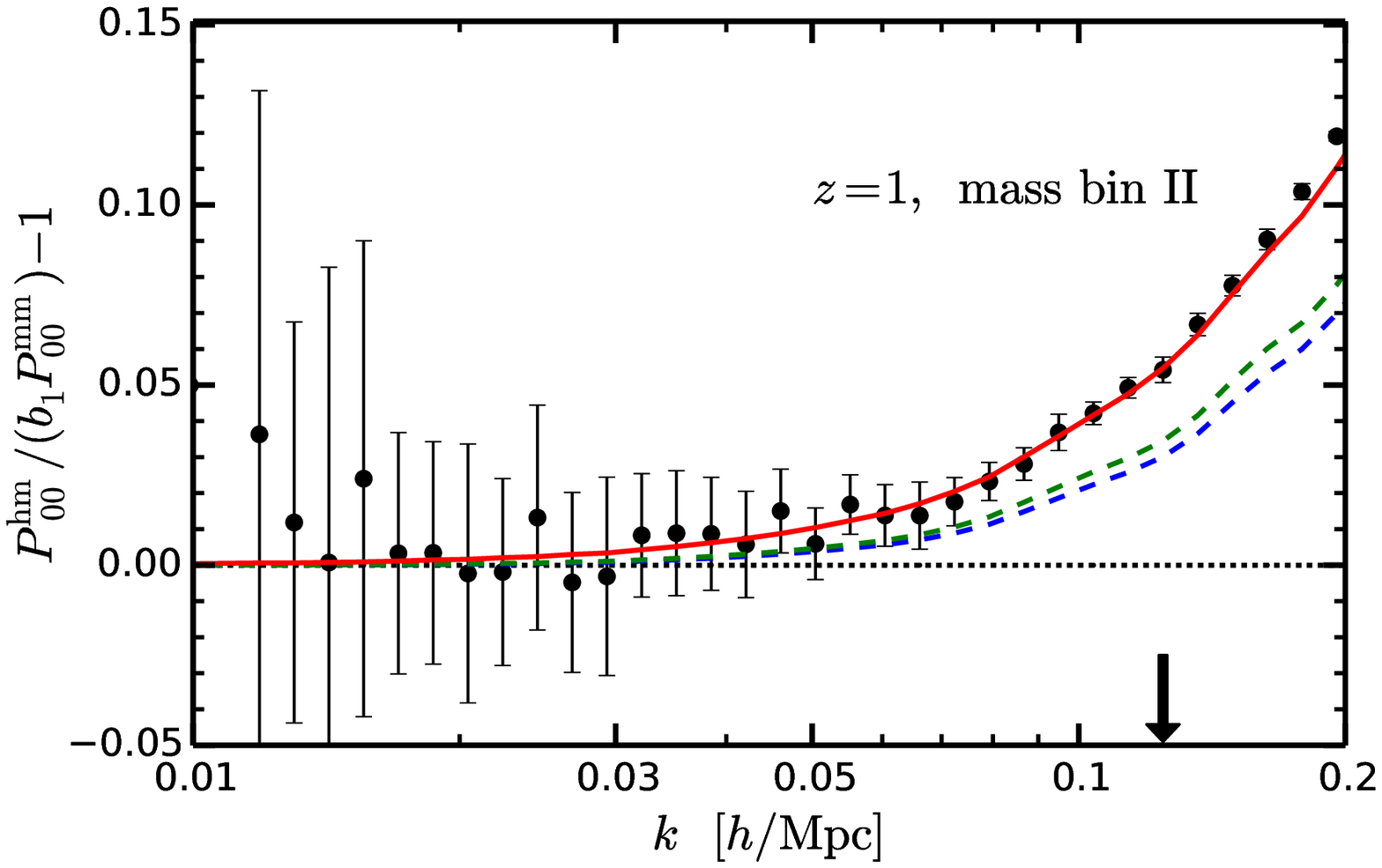}
\includegraphics[width=0.4\textwidth]{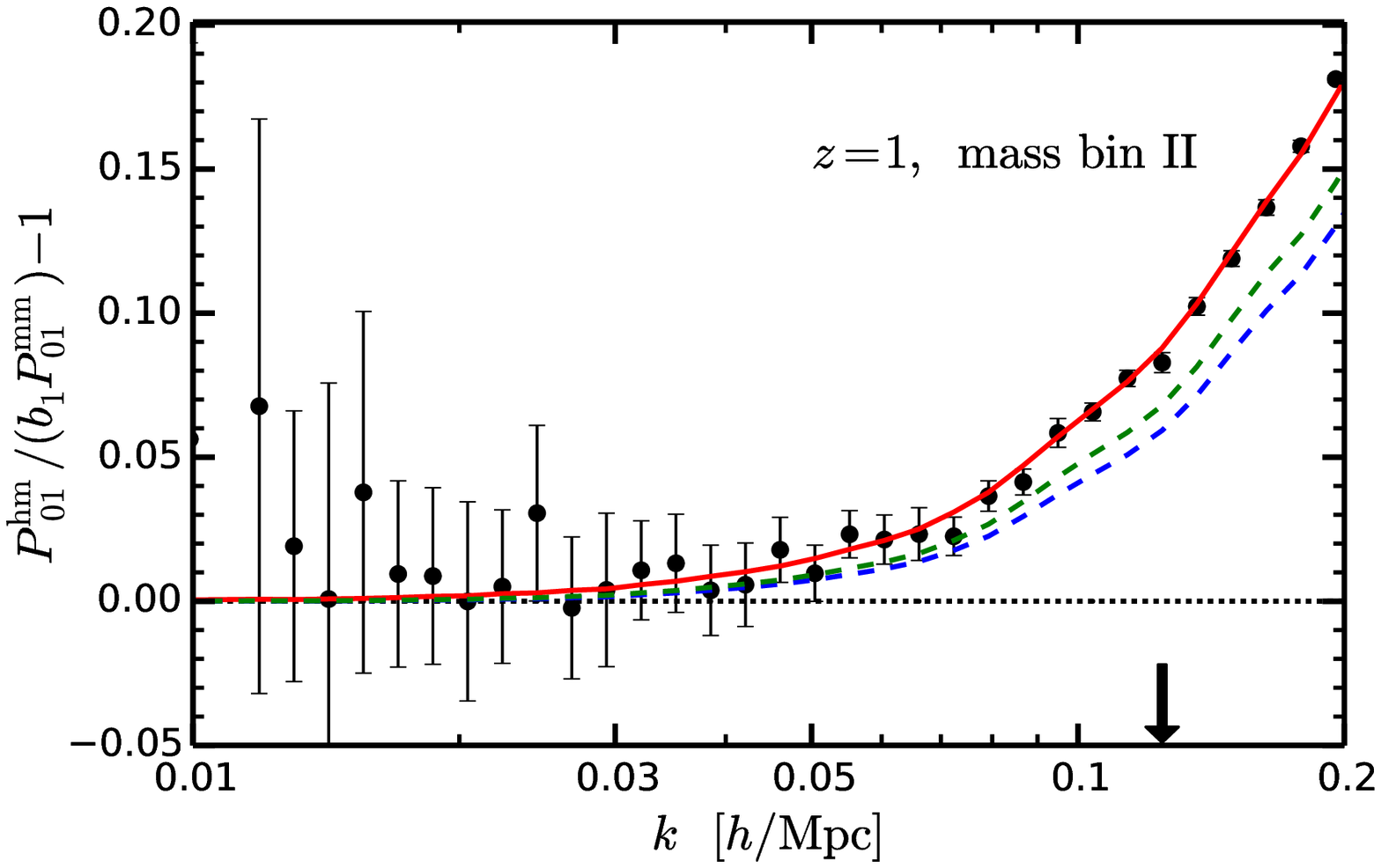}\\
\includegraphics[width=0.4\textwidth]{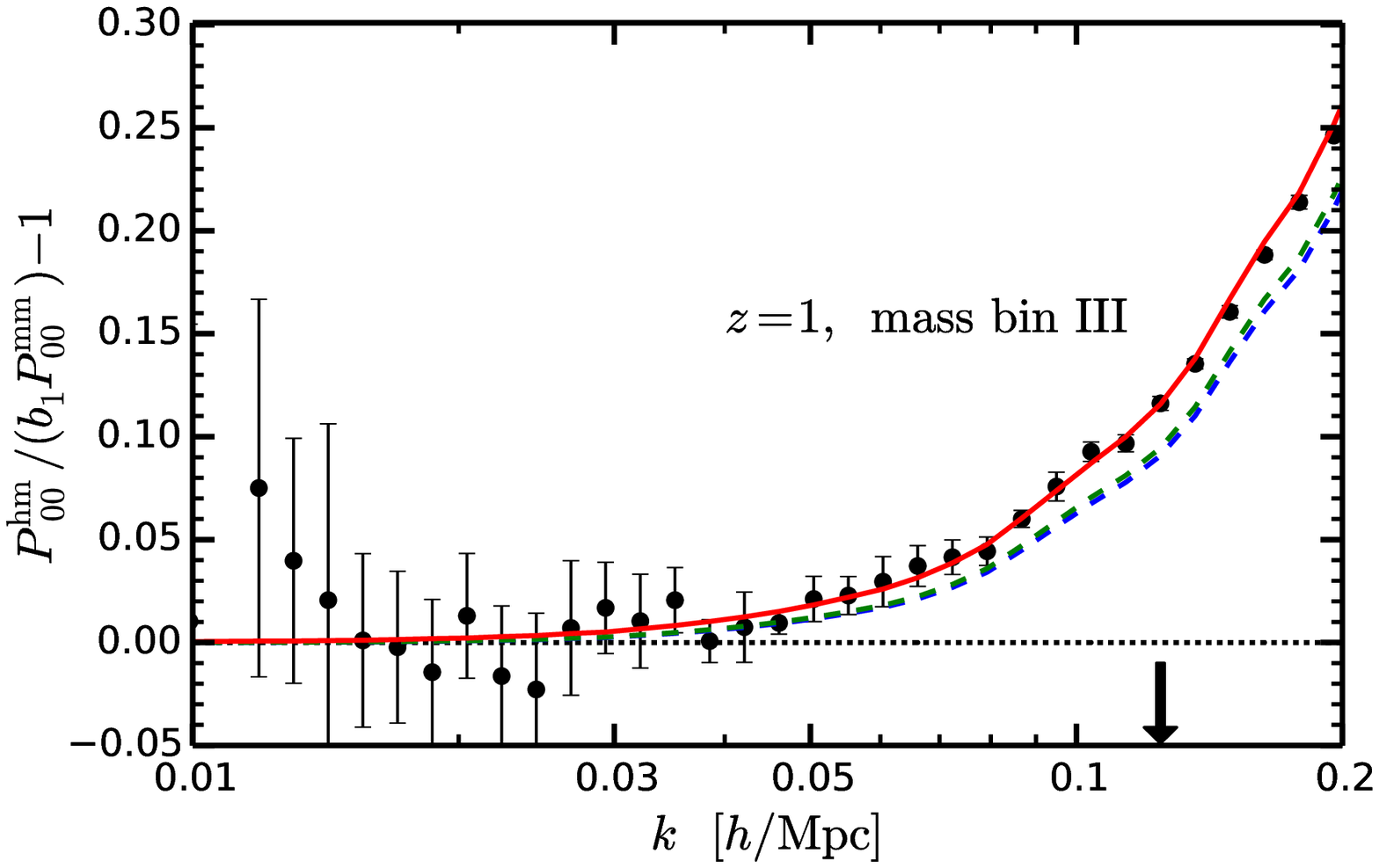}
\includegraphics[width=0.4\textwidth]{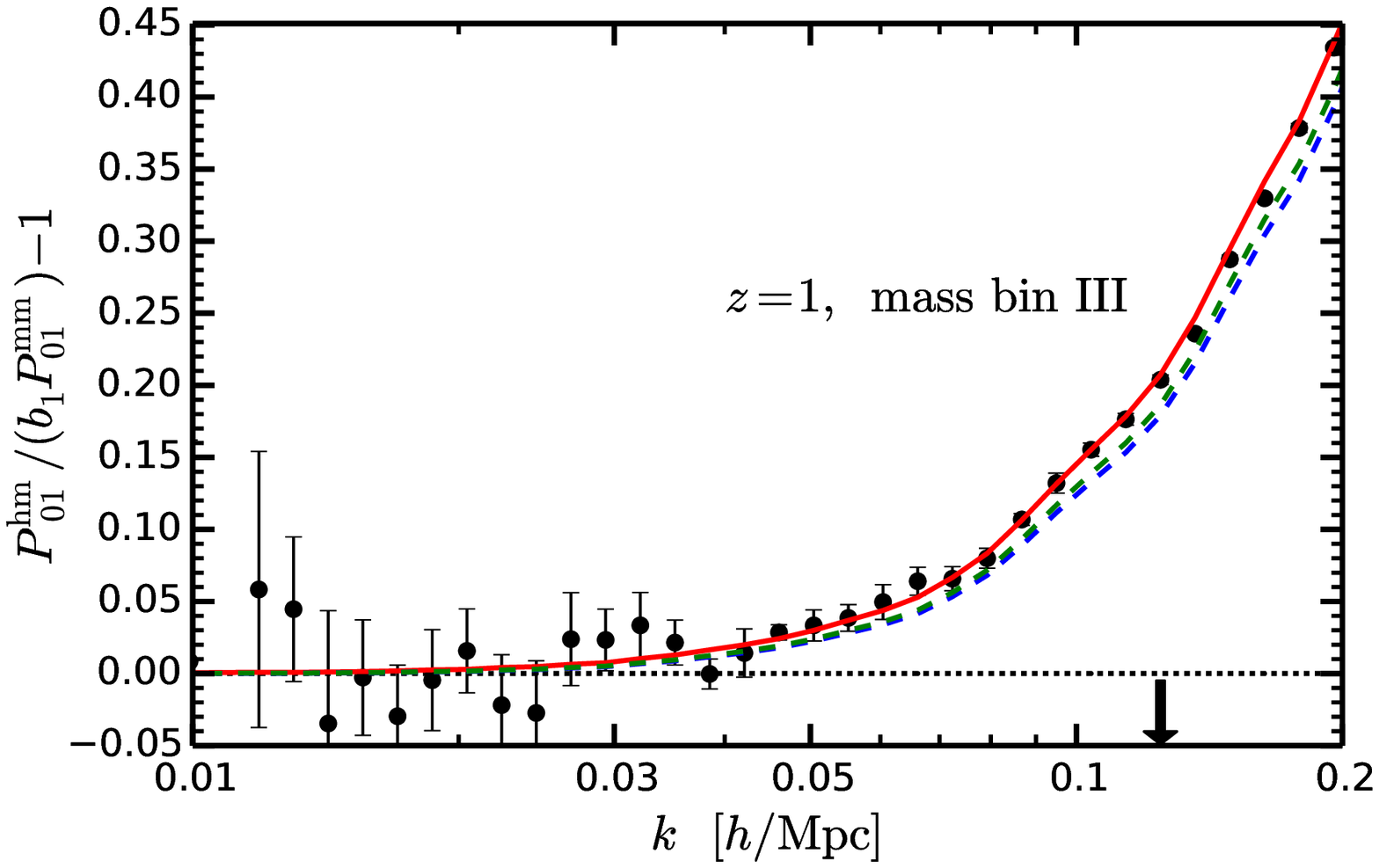}
\end{center}
\vspace*{-2em}
\caption{ 
The power spectra with best-fitting bias parameters at $z=1$. 
We here plot $P^{\rm hm}_{X}(k)/(b_{1}P^{\rm mm}_{X}(k))-1$ 
where $X$ is `00' ({\it left}) or `01' ({\it right}) with the best-fitting values of $b_{1}$ and $b_{\rm 3nl}$ 
at $k_{\rm max}=0.125\,h$/Mpc (specified as an arrow). 
Namely, zero values ({\it black dotted}) mean it matches to the linear bias term, any deviation from zero 
represents deviation from the linear bias model. 
The {\it red solid} line corresponds to the case including all contributions. The {\it blue dashed} line 
includes only local bias terms up to second order, while the {\it green dashed} line includes 
local and nonlocal bias terms up to second order. 
}  
\label{fig: estimate b3nl vs Pk z=1}
\end{figure}

\begin{figure}[t]
\begin{center}
\includegraphics[width=0.4\textwidth]{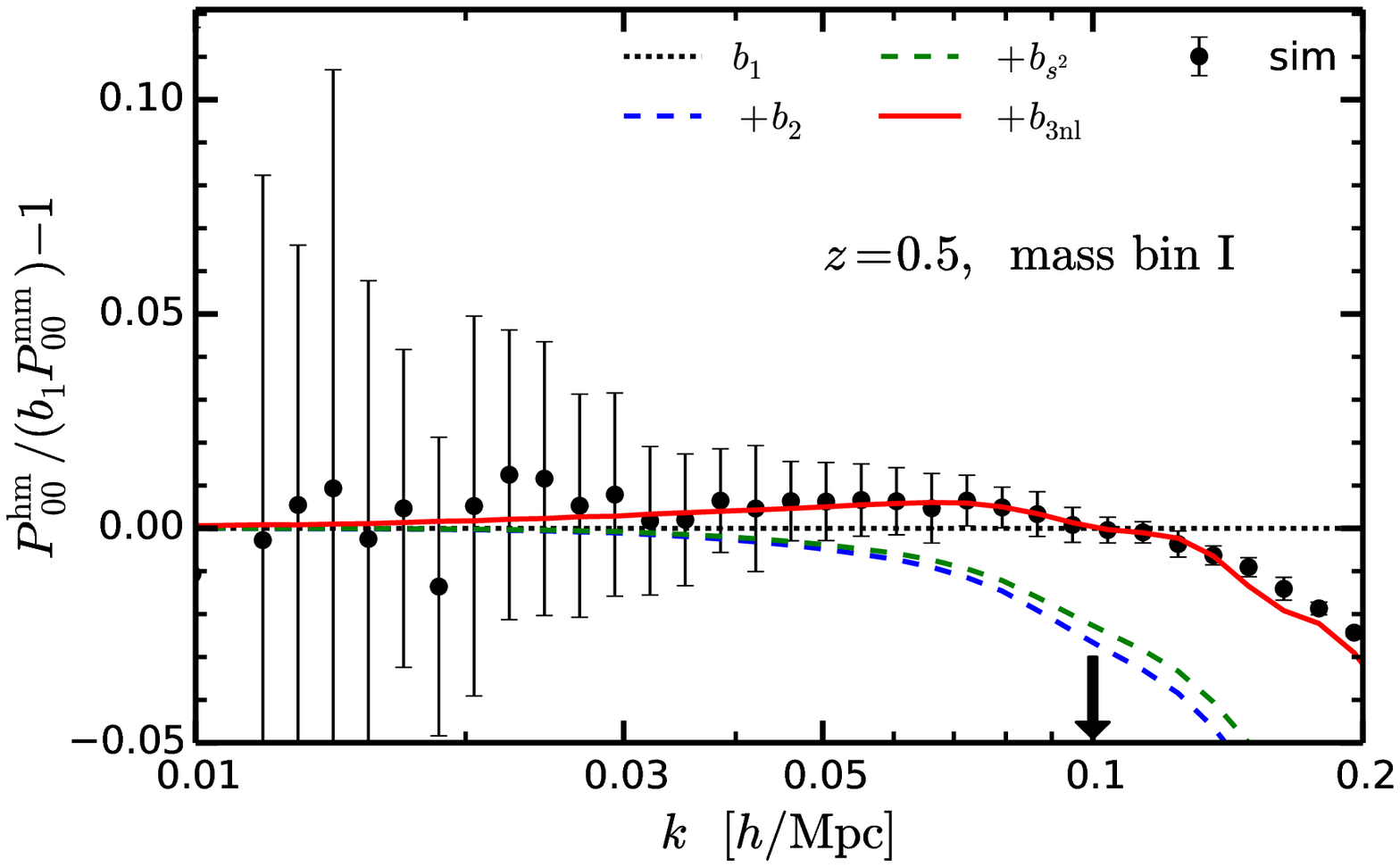}
\includegraphics[width=0.4\textwidth]{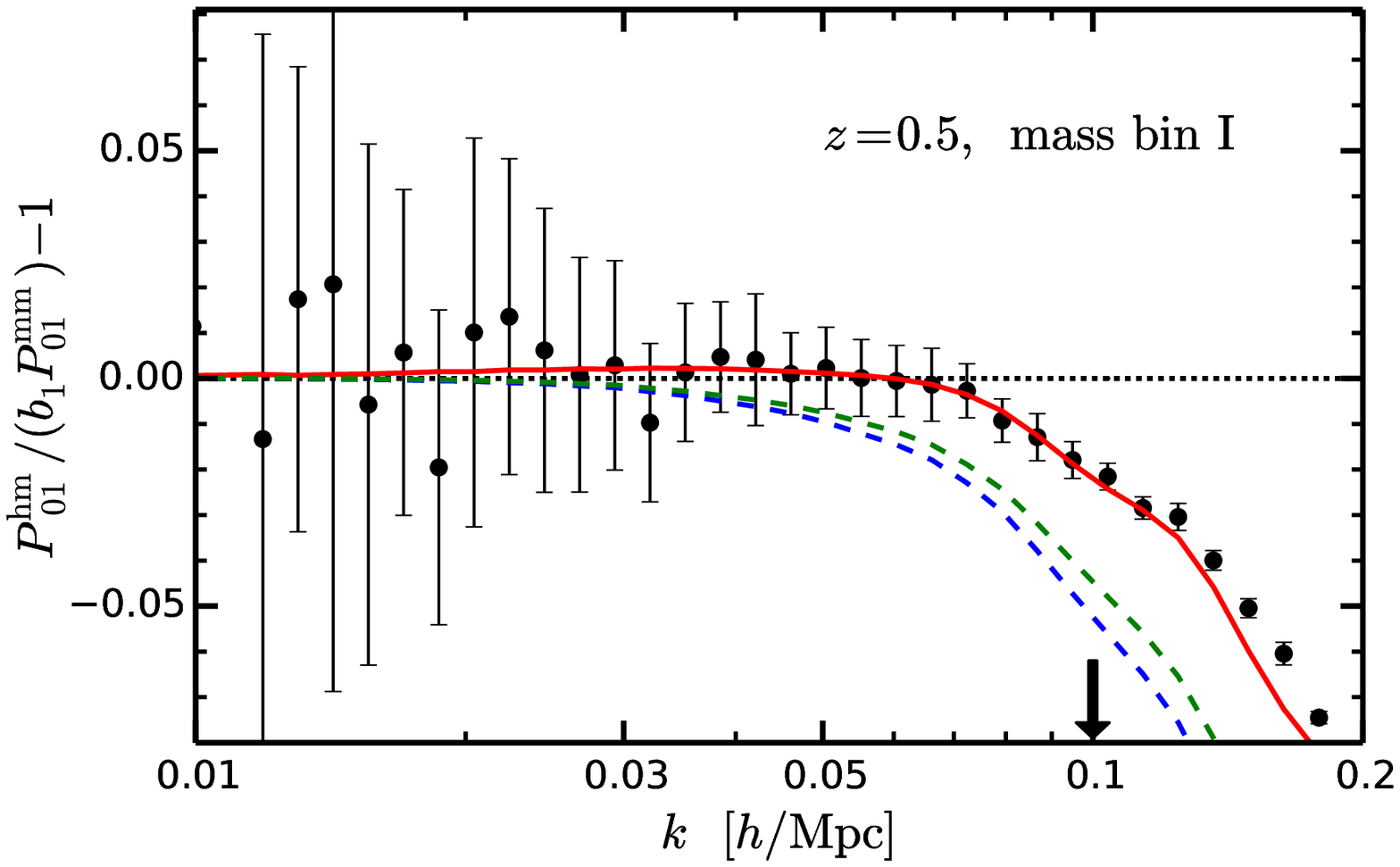}\\
\includegraphics[width=0.4\textwidth]{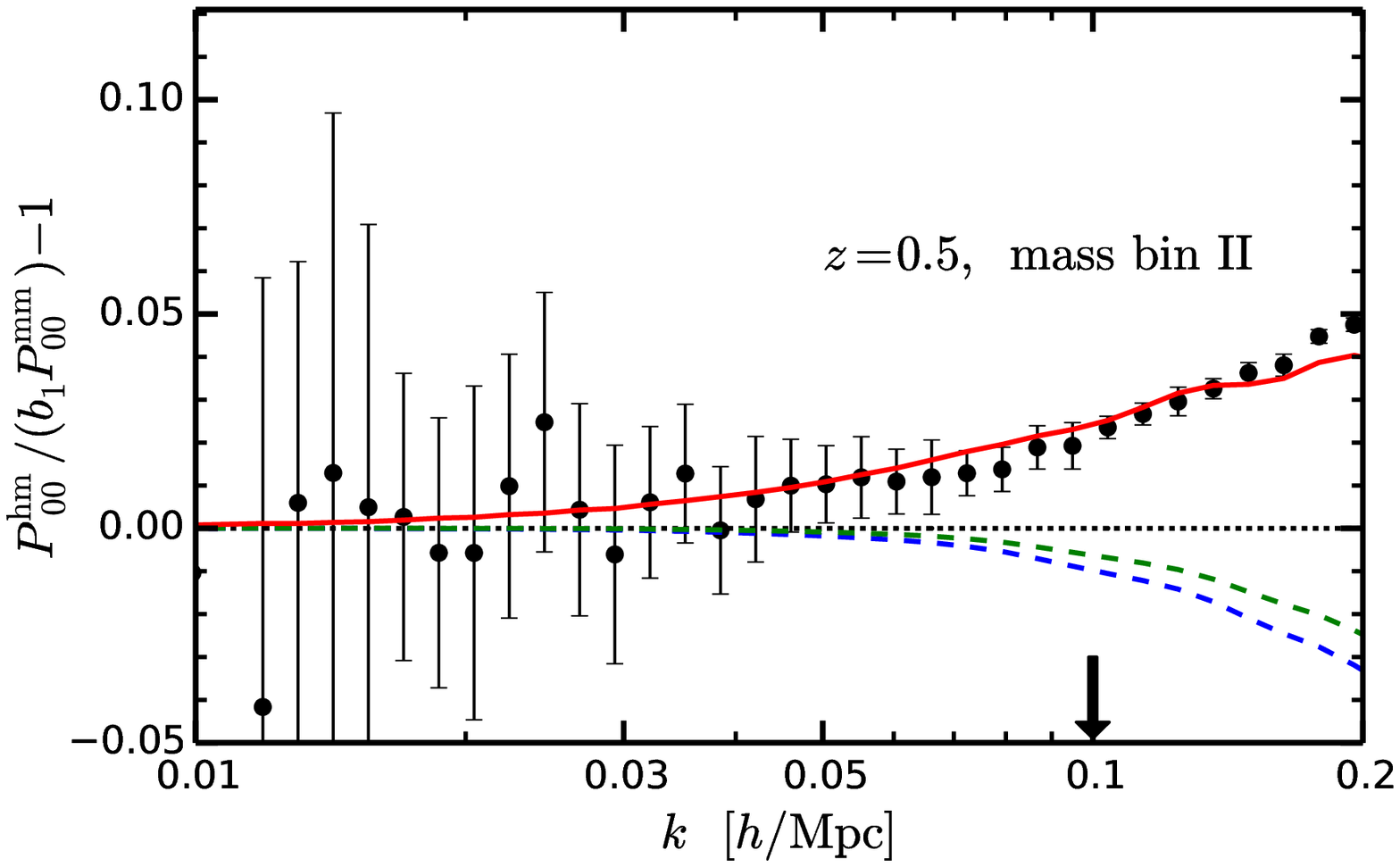}
\includegraphics[width=0.4\textwidth]{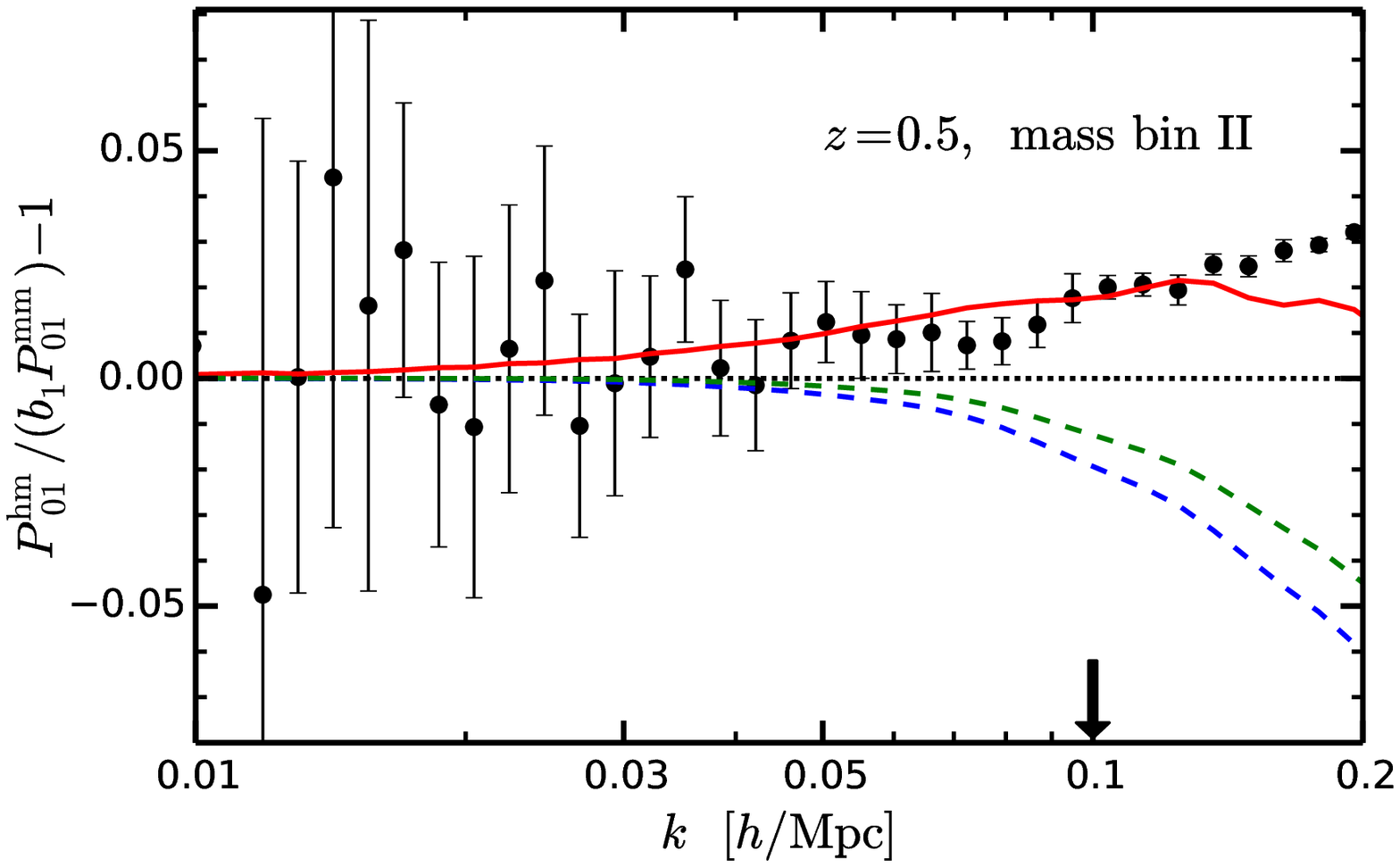}\\
\includegraphics[width=0.4\textwidth]{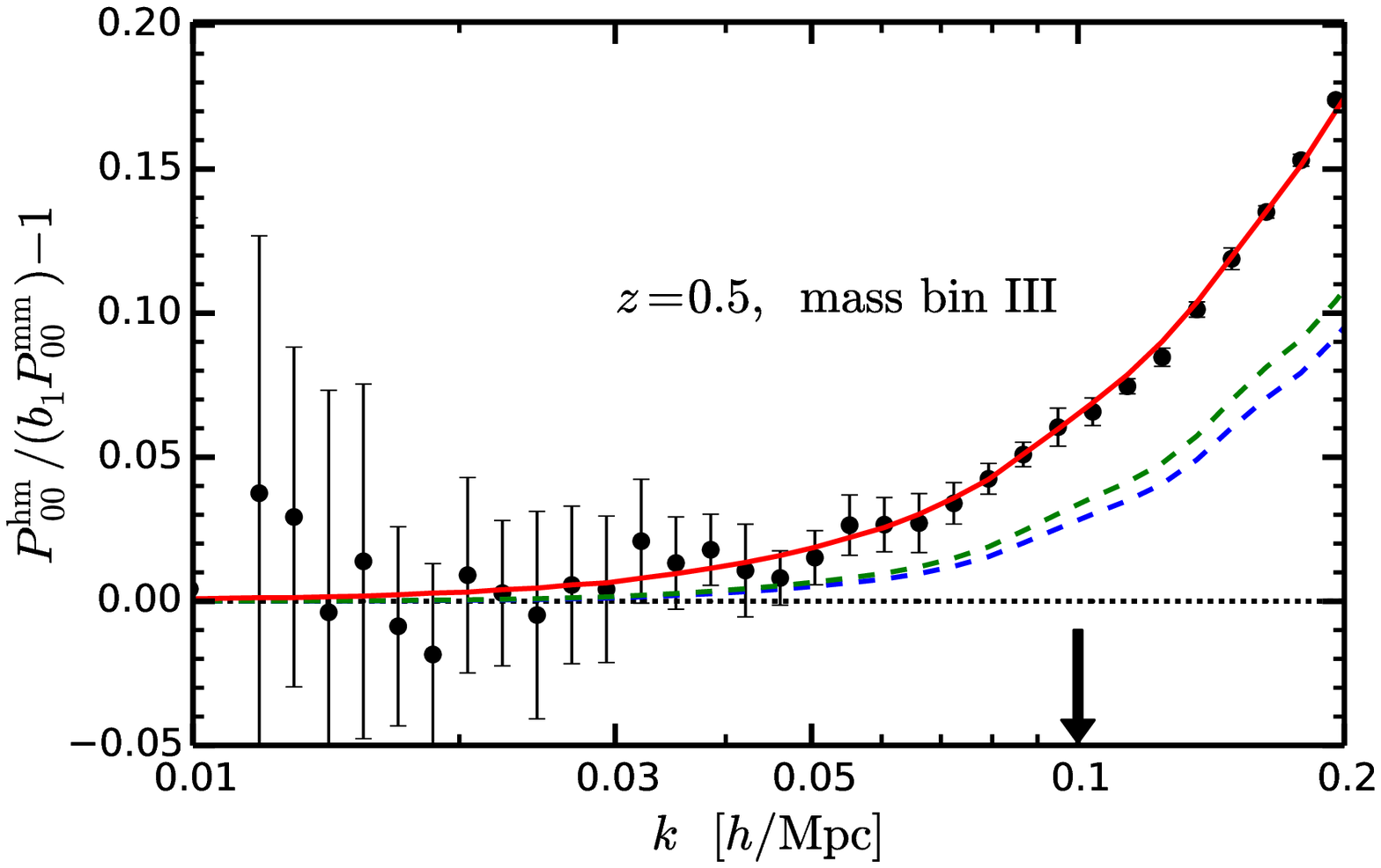}
\includegraphics[width=0.4\textwidth]{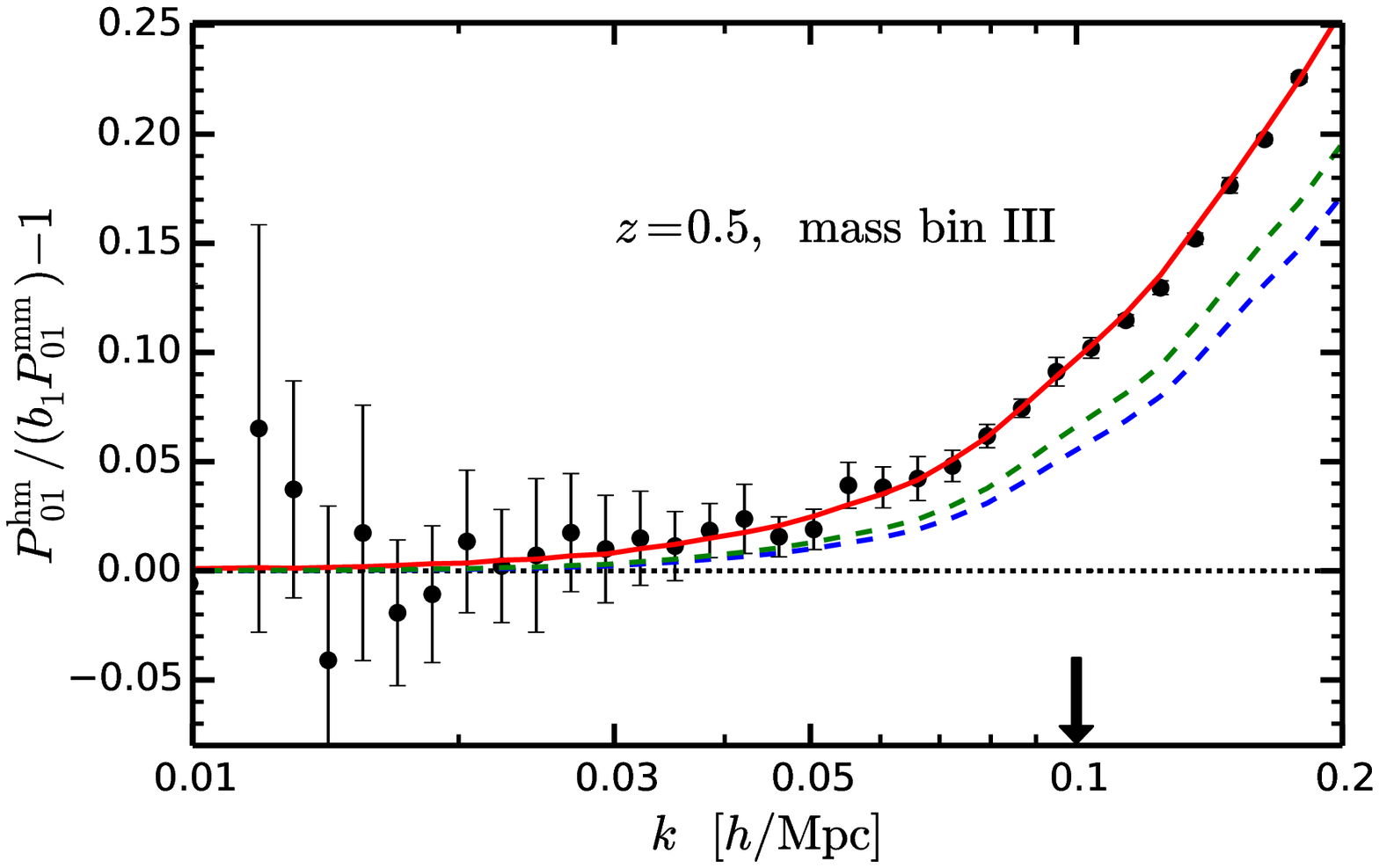}\\
\includegraphics[width=0.4\textwidth]{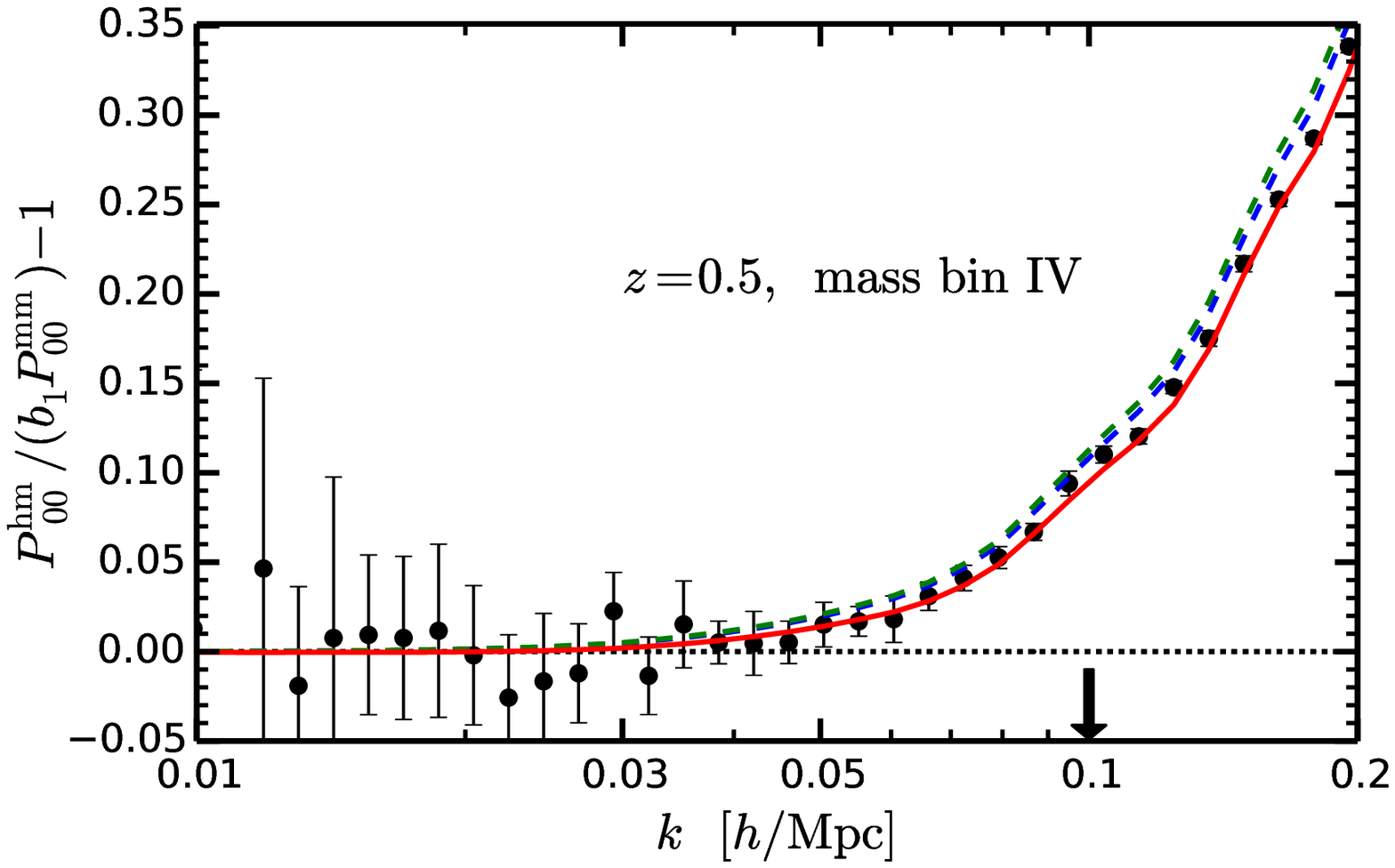}
\includegraphics[width=0.4\textwidth]{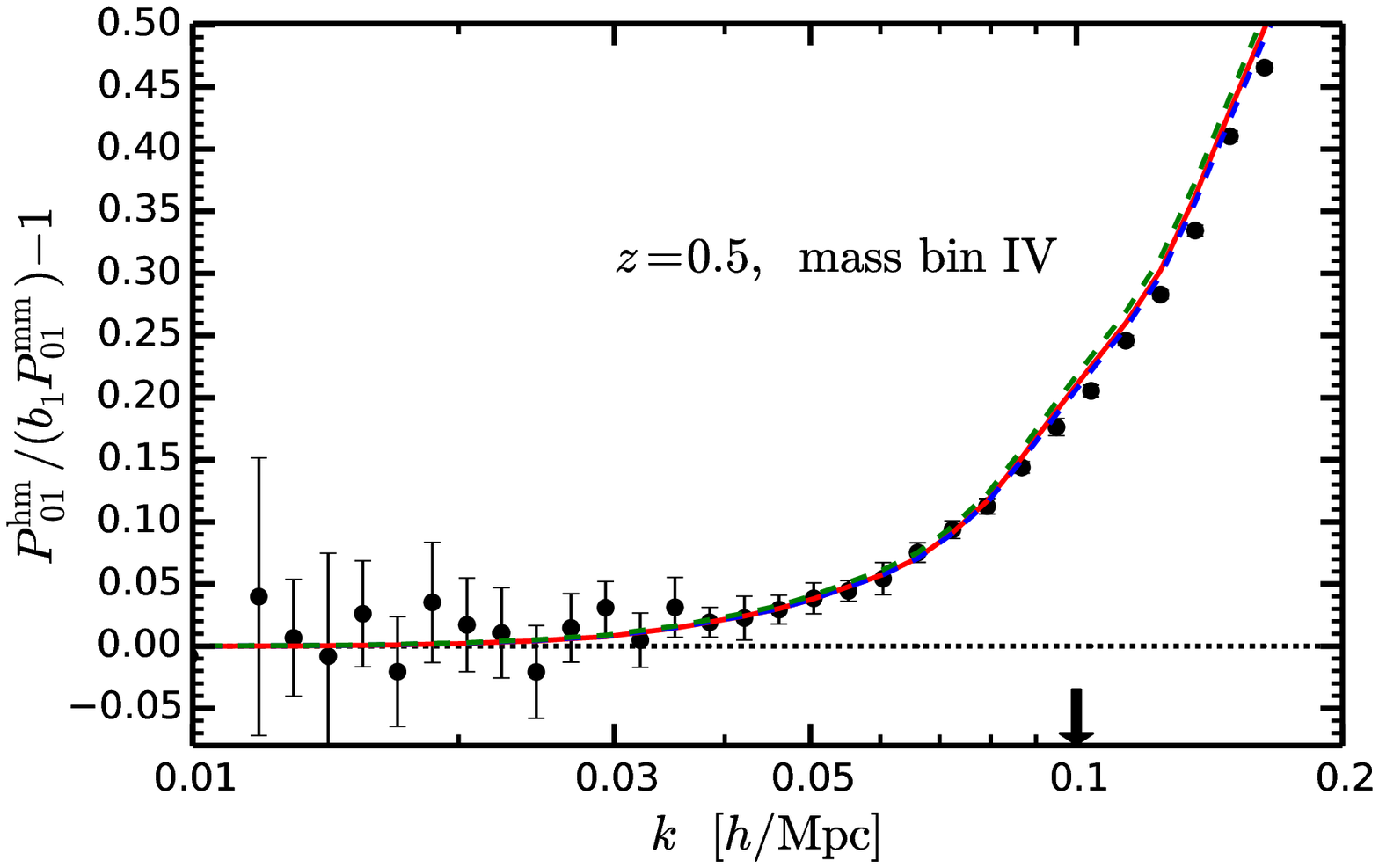}
\end{center}
\vspace*{-2em}
\caption{ 
Same as Fig.~\ref{fig: estimate b3nl vs Pk z=1}, but at $z=0.5$. 
The best-fitting values are derived at $k_{\rm max}=0.1\,h$/Mpc (specified as an arrow). 
}  
\label{fig: estimate b3nl vs Pk z=0.5}
\end{figure}

\begin{figure}[t]
\begin{center}
\includegraphics[width=0.4\textwidth]{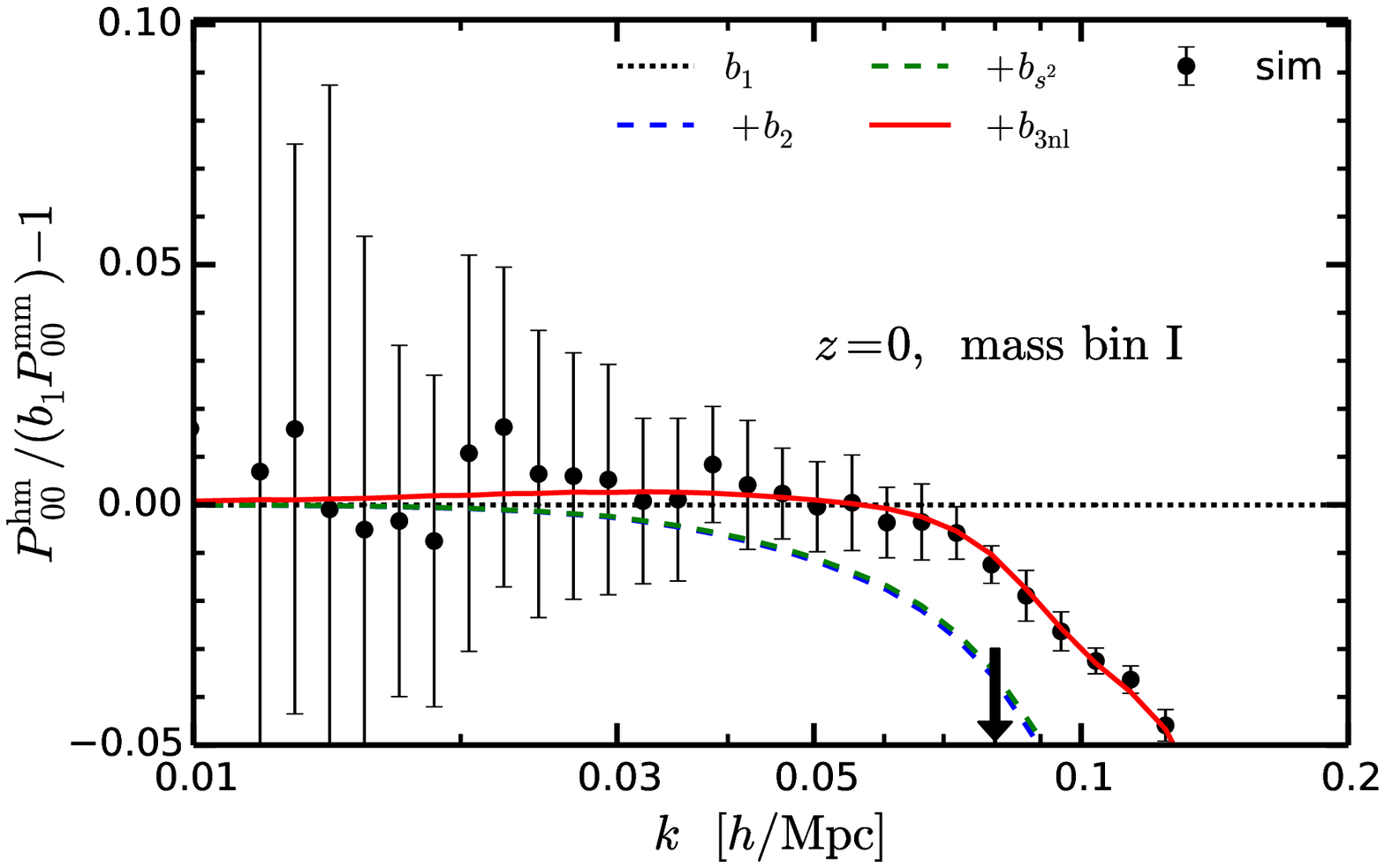}
\includegraphics[width=0.4\textwidth]{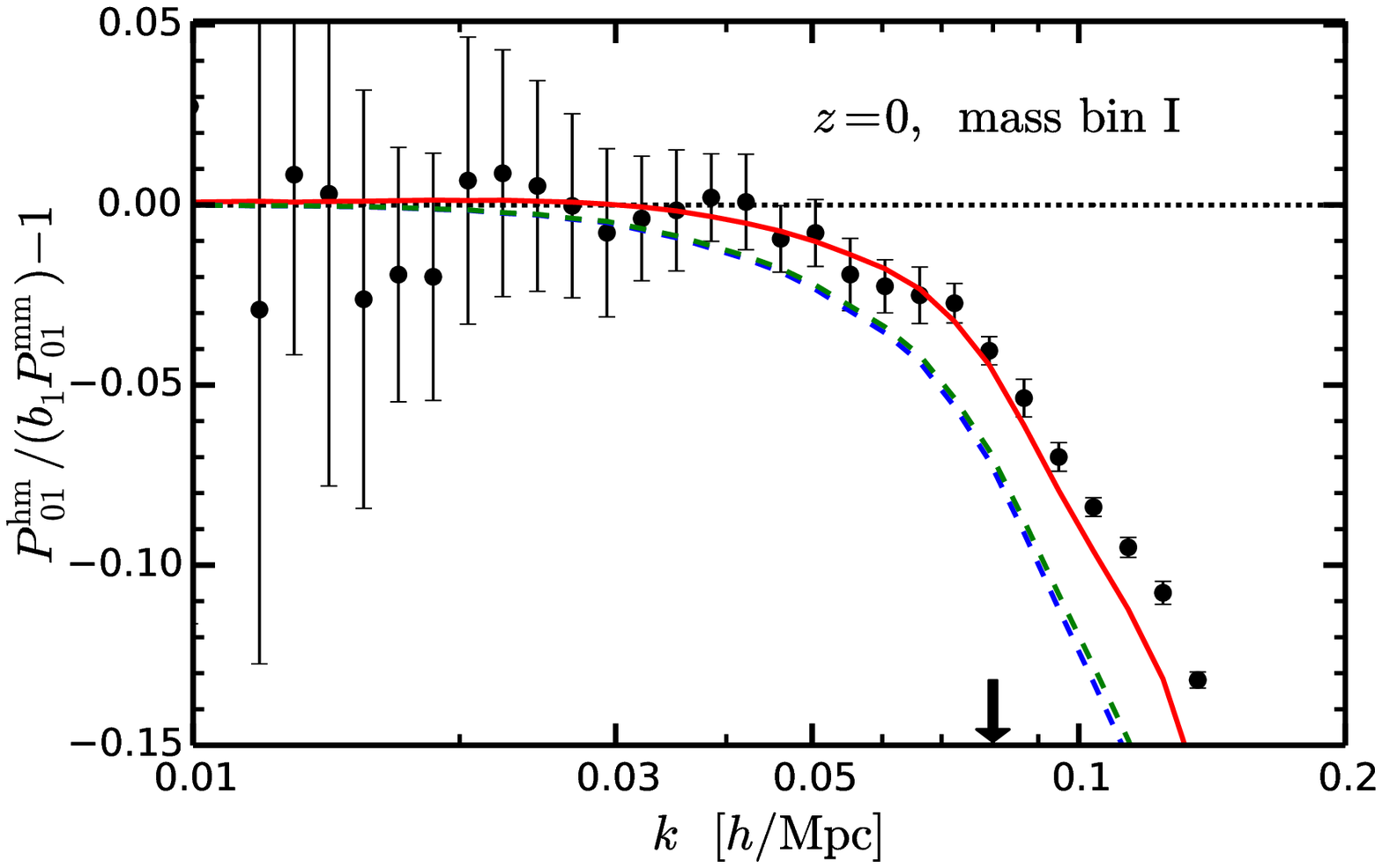}\\
\includegraphics[width=0.4\textwidth]{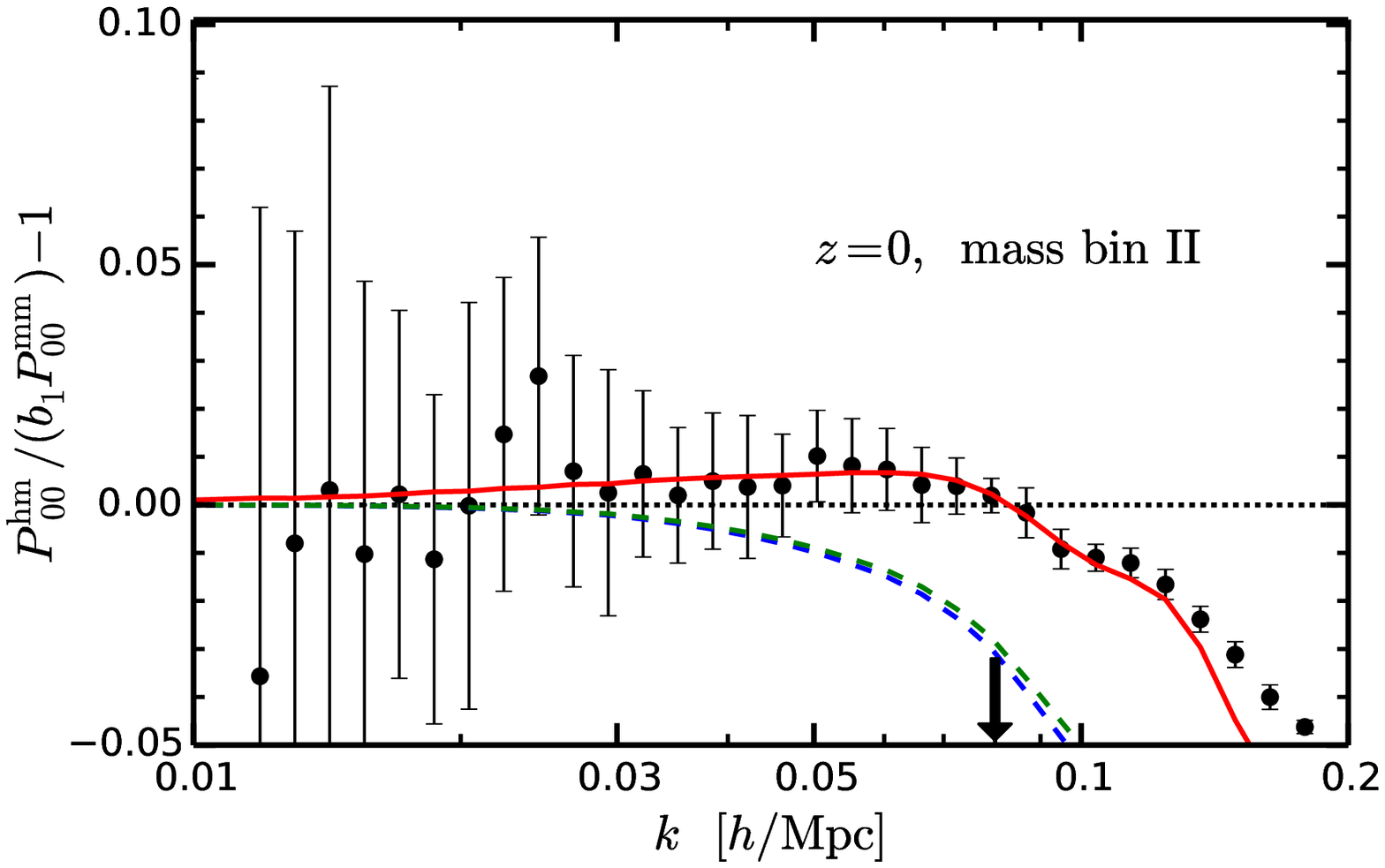}
\includegraphics[width=0.4\textwidth]{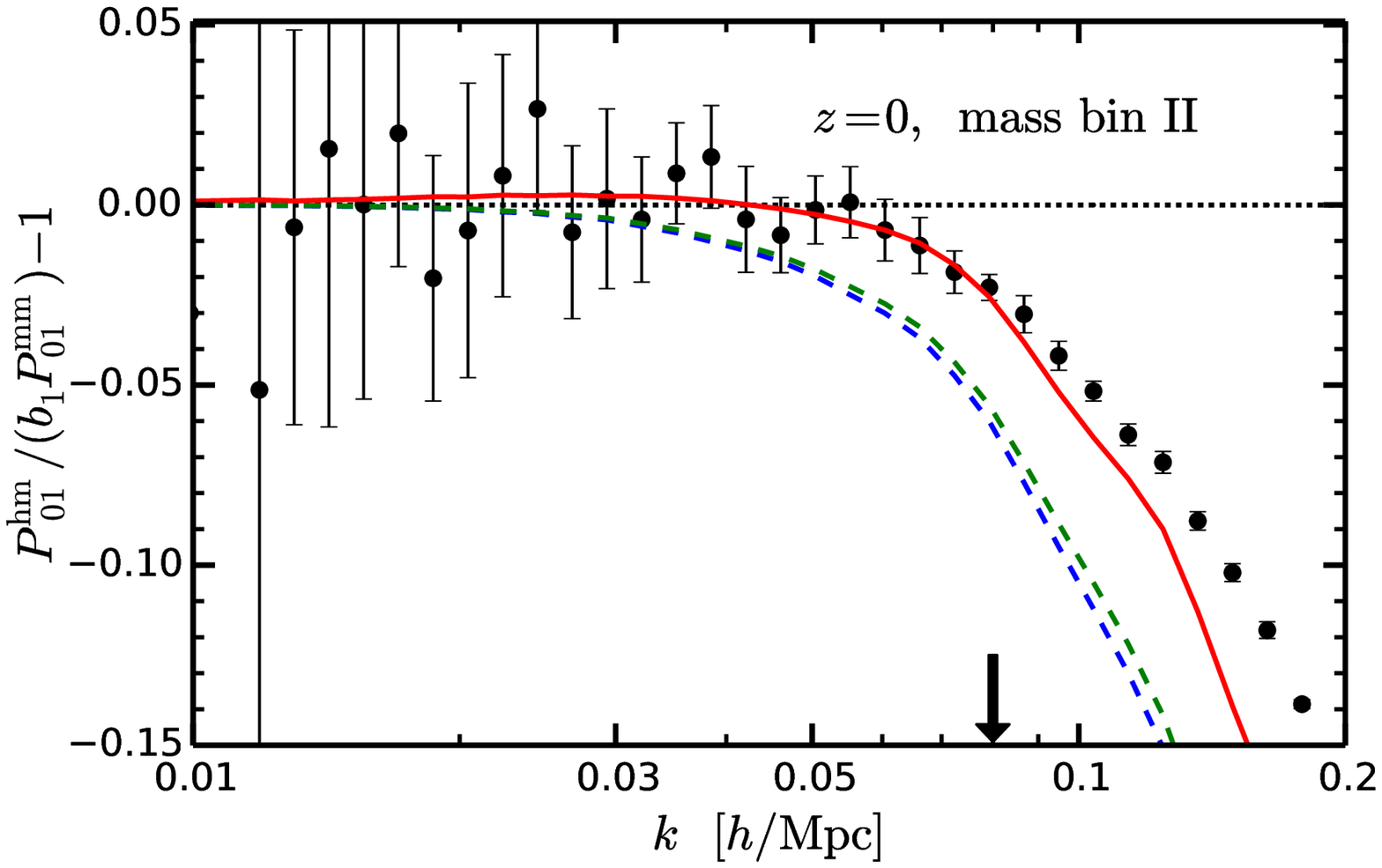}\\
\includegraphics[width=0.4\textwidth]{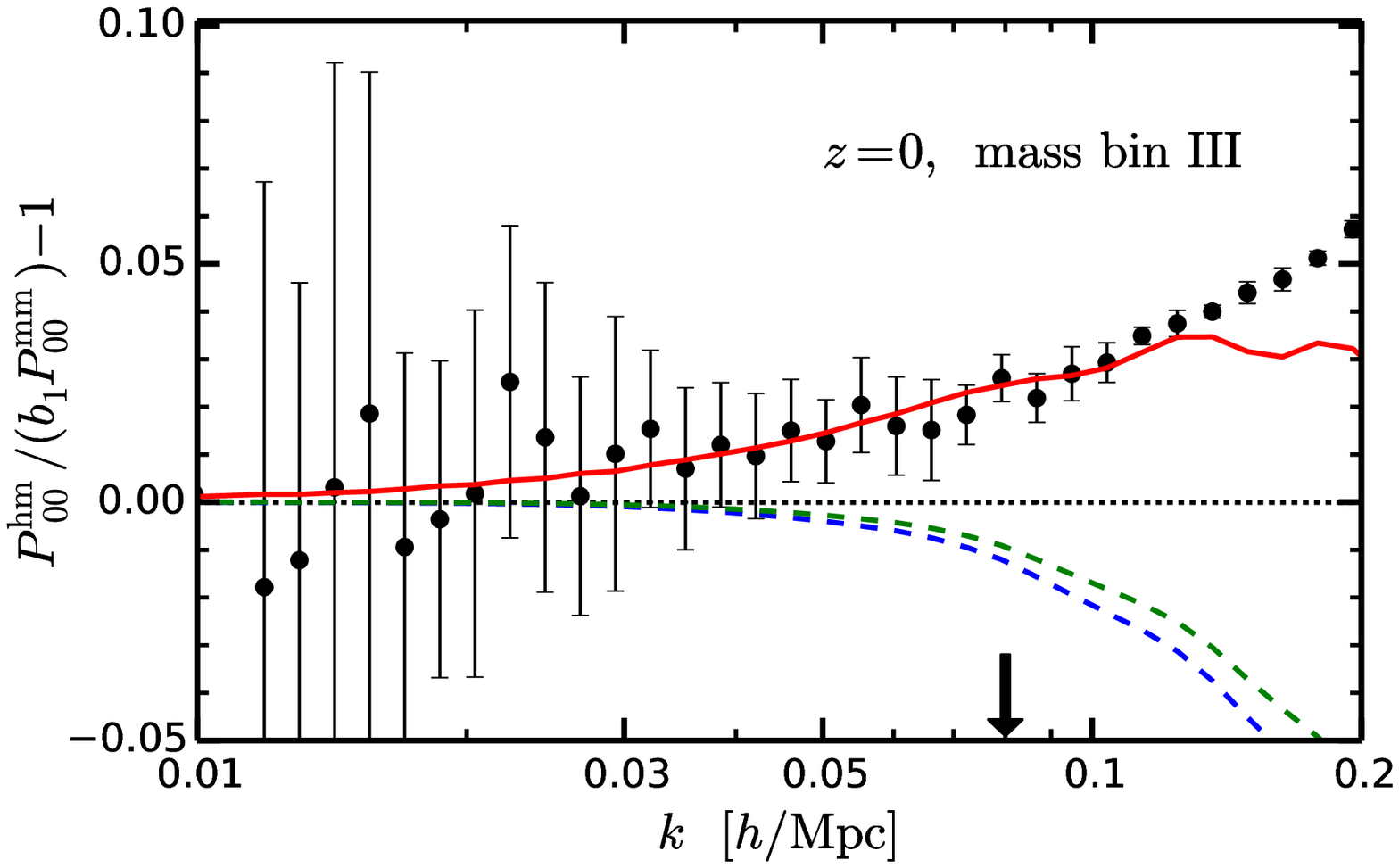}
\includegraphics[width=0.4\textwidth]{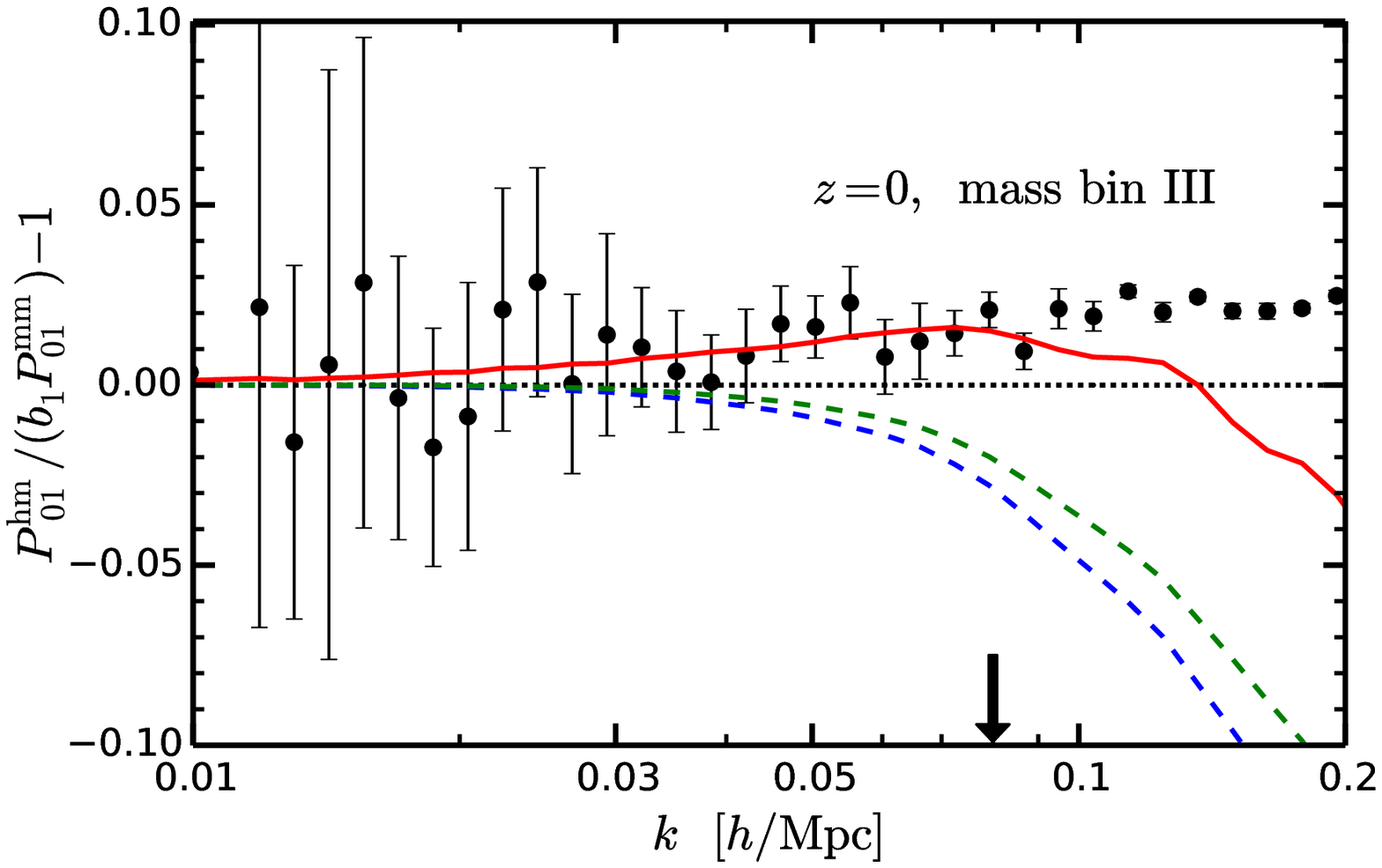}\\
\includegraphics[width=0.4\textwidth]{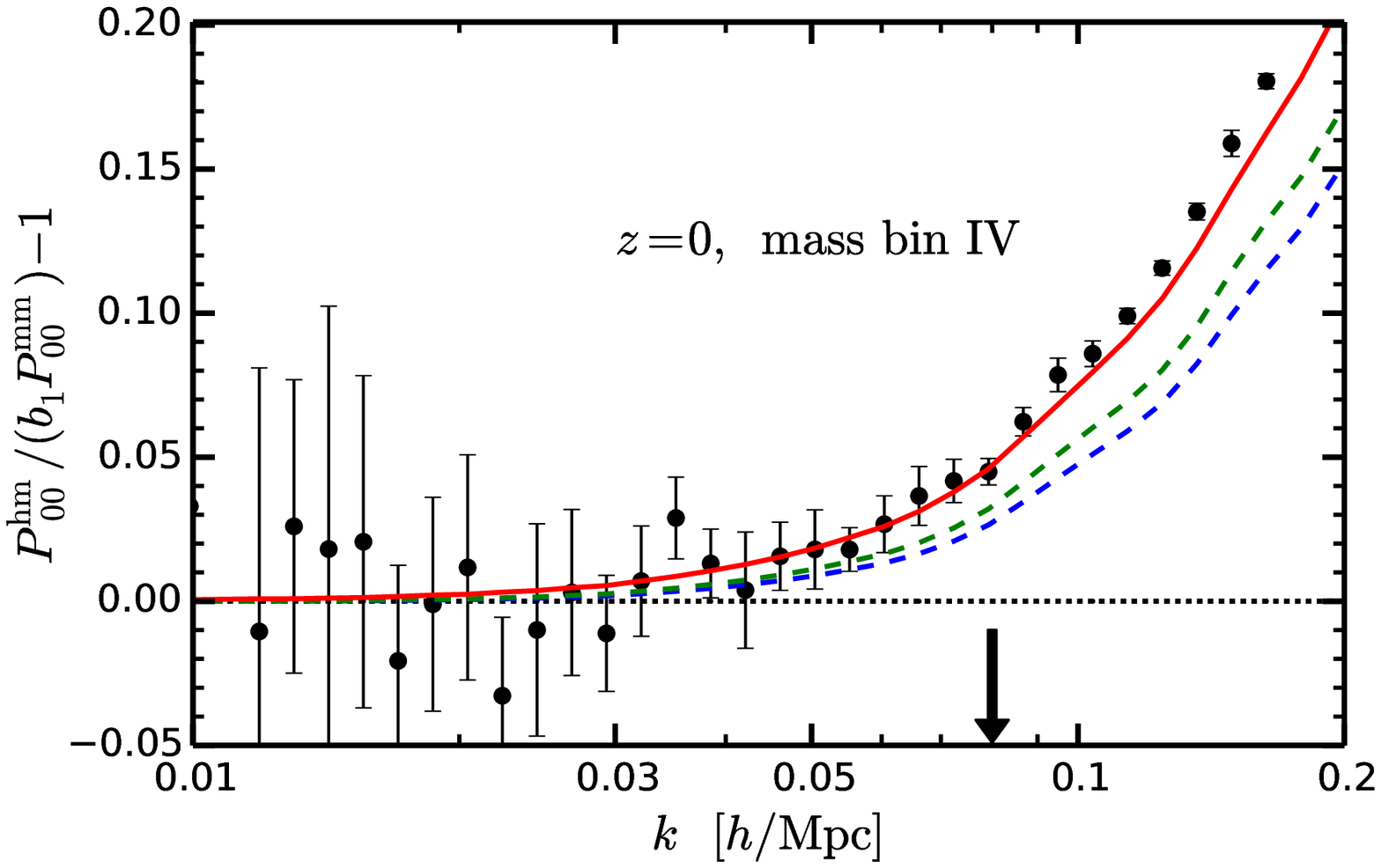}
\includegraphics[width=0.4\textwidth]{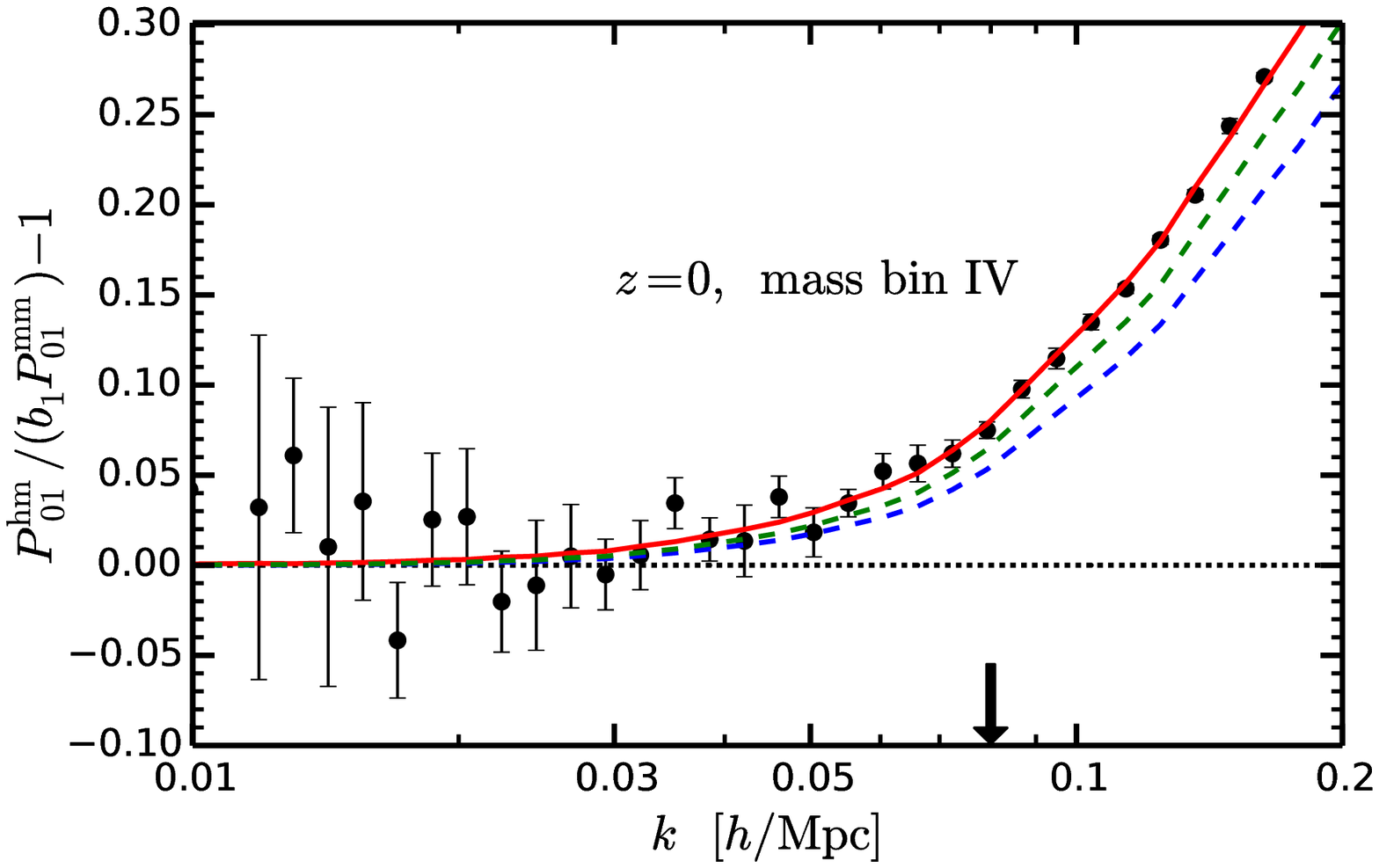}
\end{center}
\vspace*{-2em}
\caption{ 
Same as Fig.~\ref{fig: estimate b3nl vs Pk z=1}, but at $z=0$. 
The best-fitting values are derived at $k_{\rm max}=0.08\,h$/Mpc (specified as an arrow). 
}  
\label{fig: estimate b3nl vs Pk z=0}
\end{figure}
%

\begin{figure}[t]
\begin{center}
\includegraphics[width=0.35\textwidth]{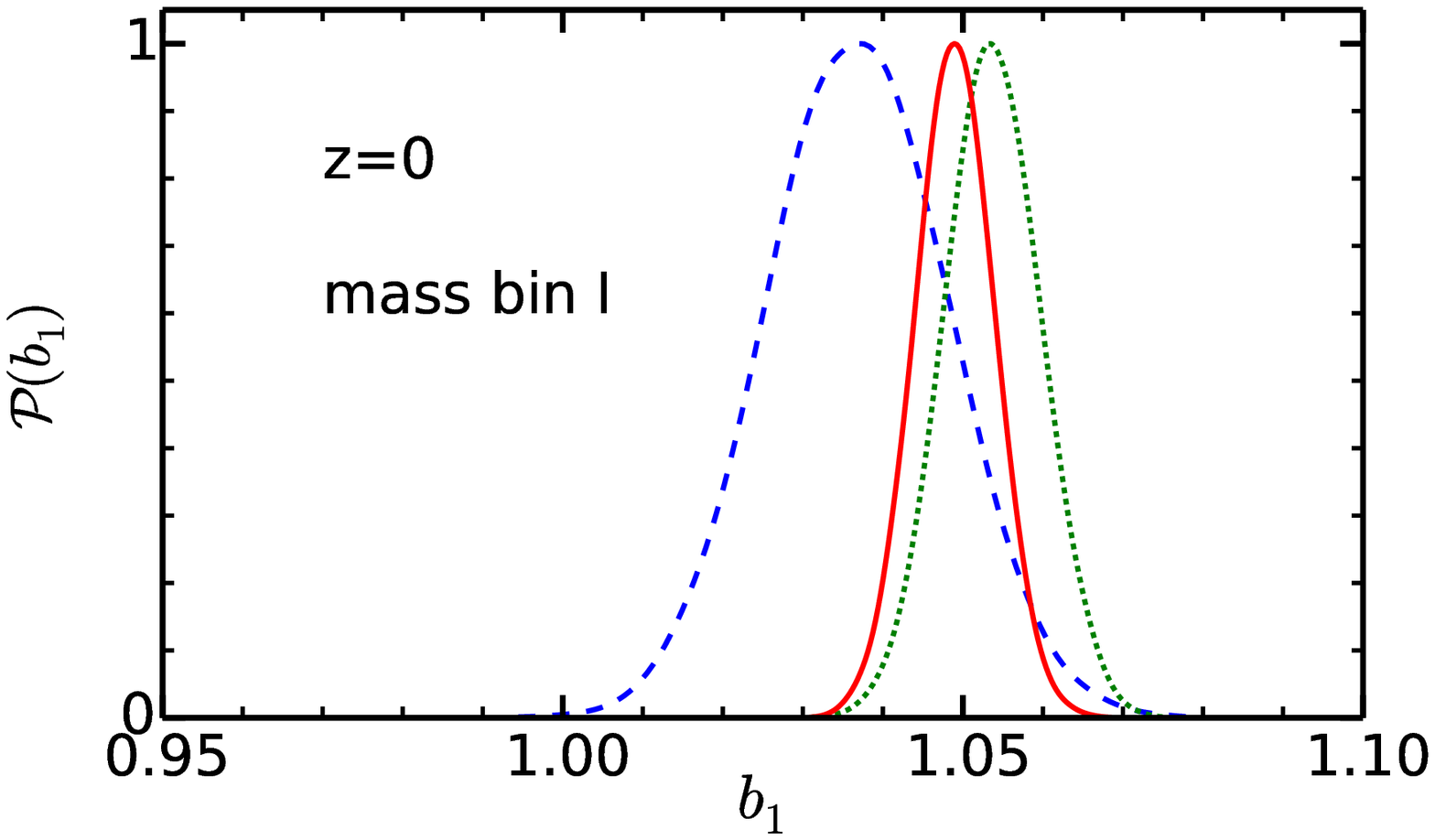}
\includegraphics[width=0.35\textwidth]{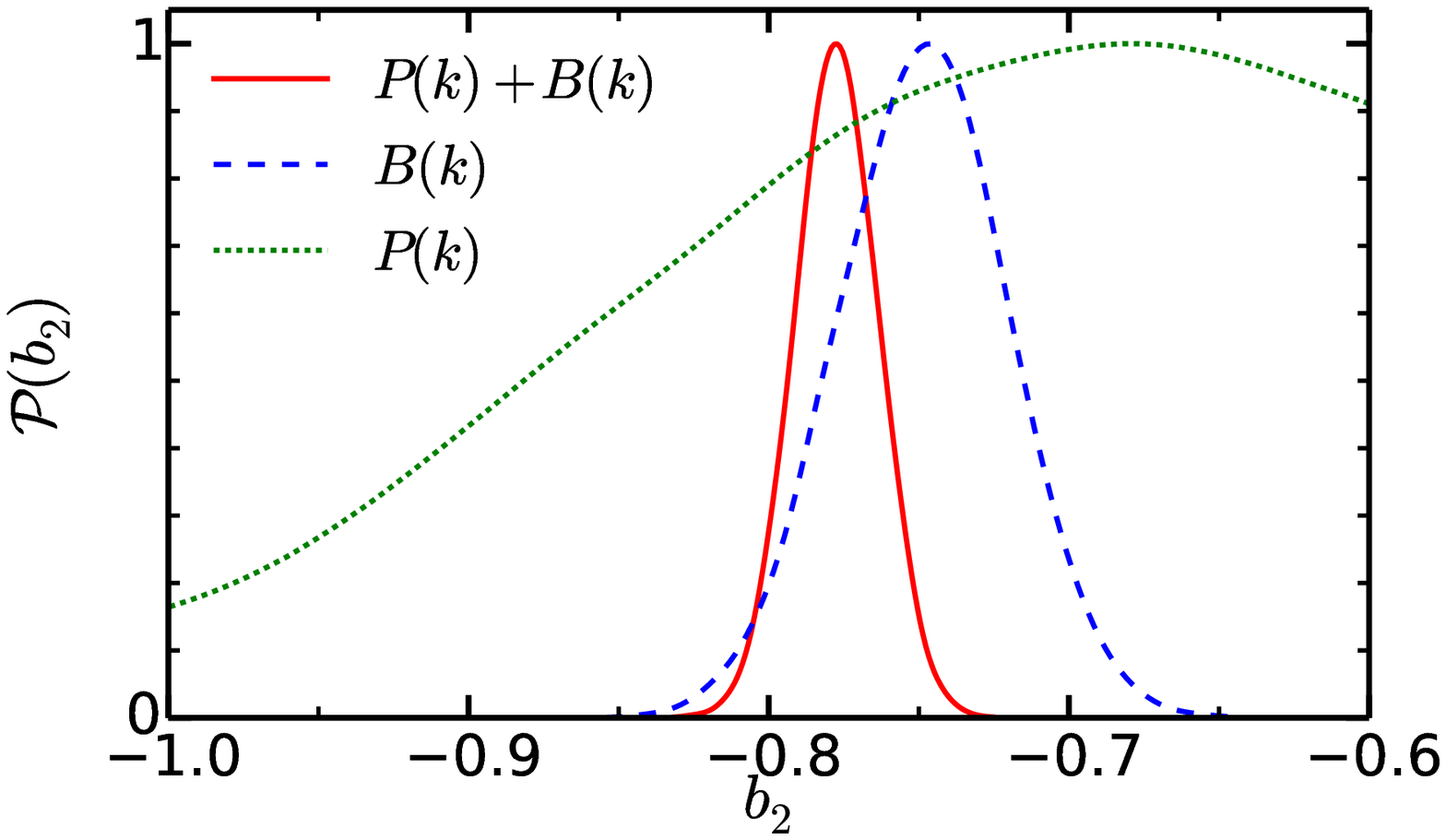}\\
\includegraphics[width=0.35\textwidth]{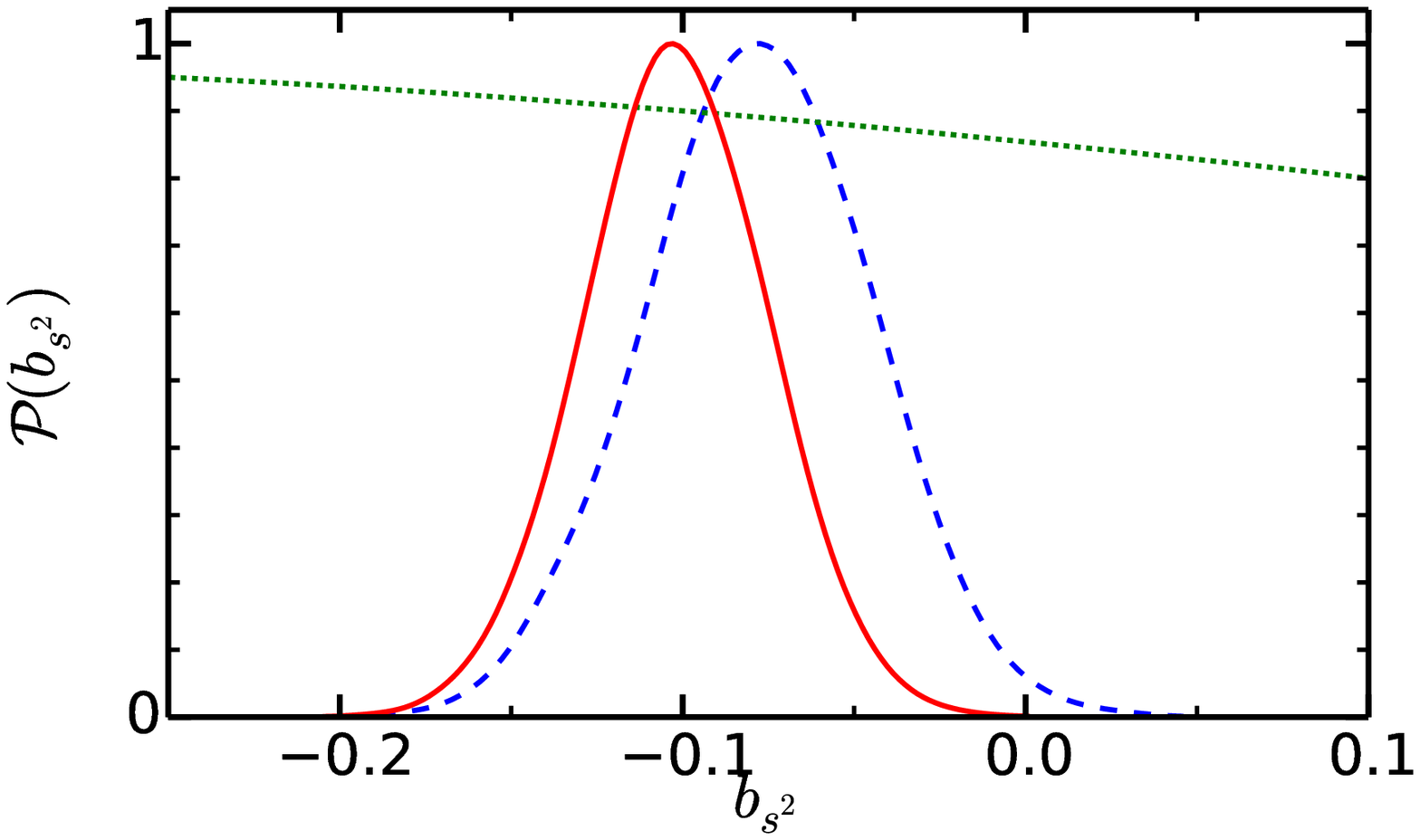}
\includegraphics[width=0.35\textwidth]{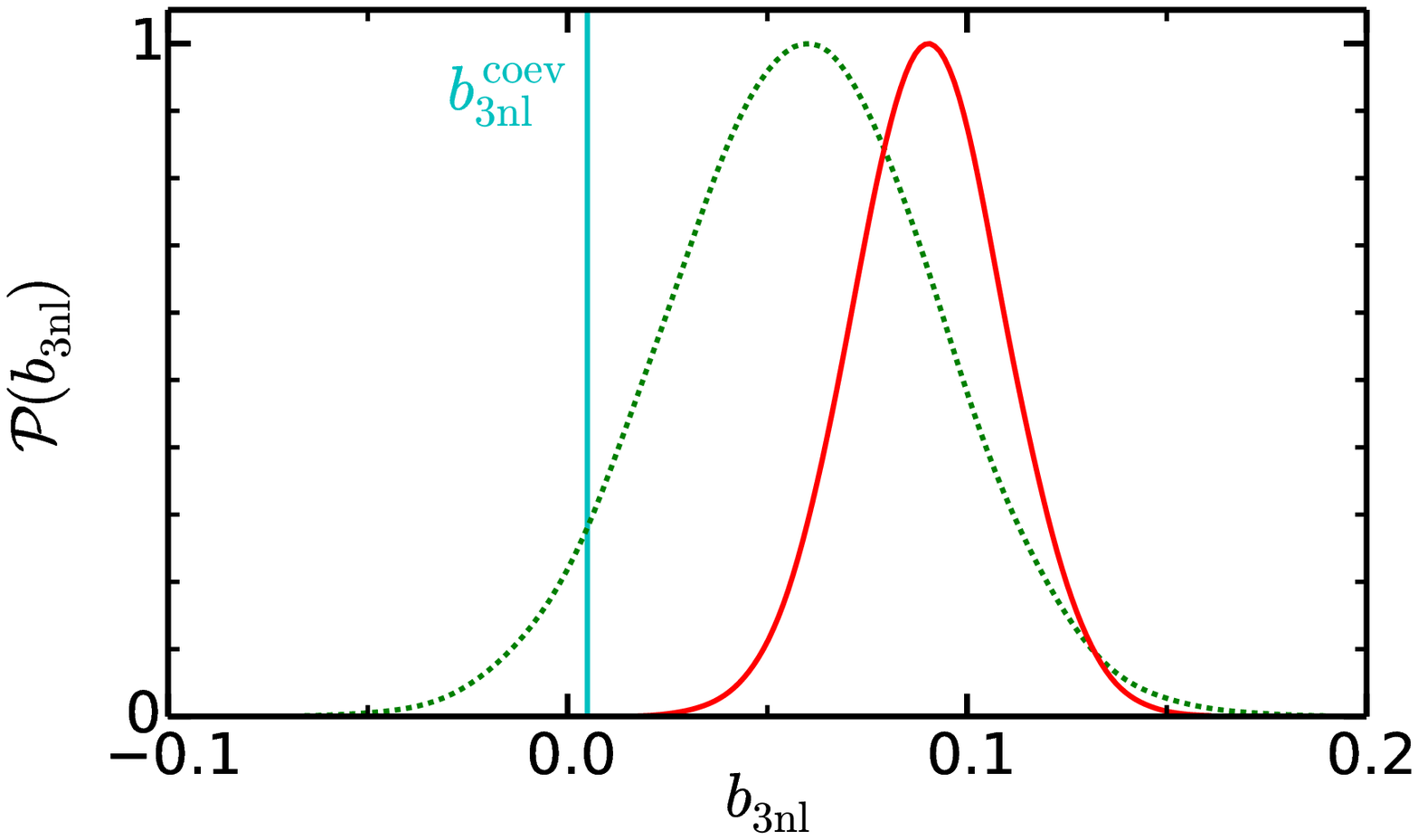}\\
\includegraphics[width=0.35\textwidth]{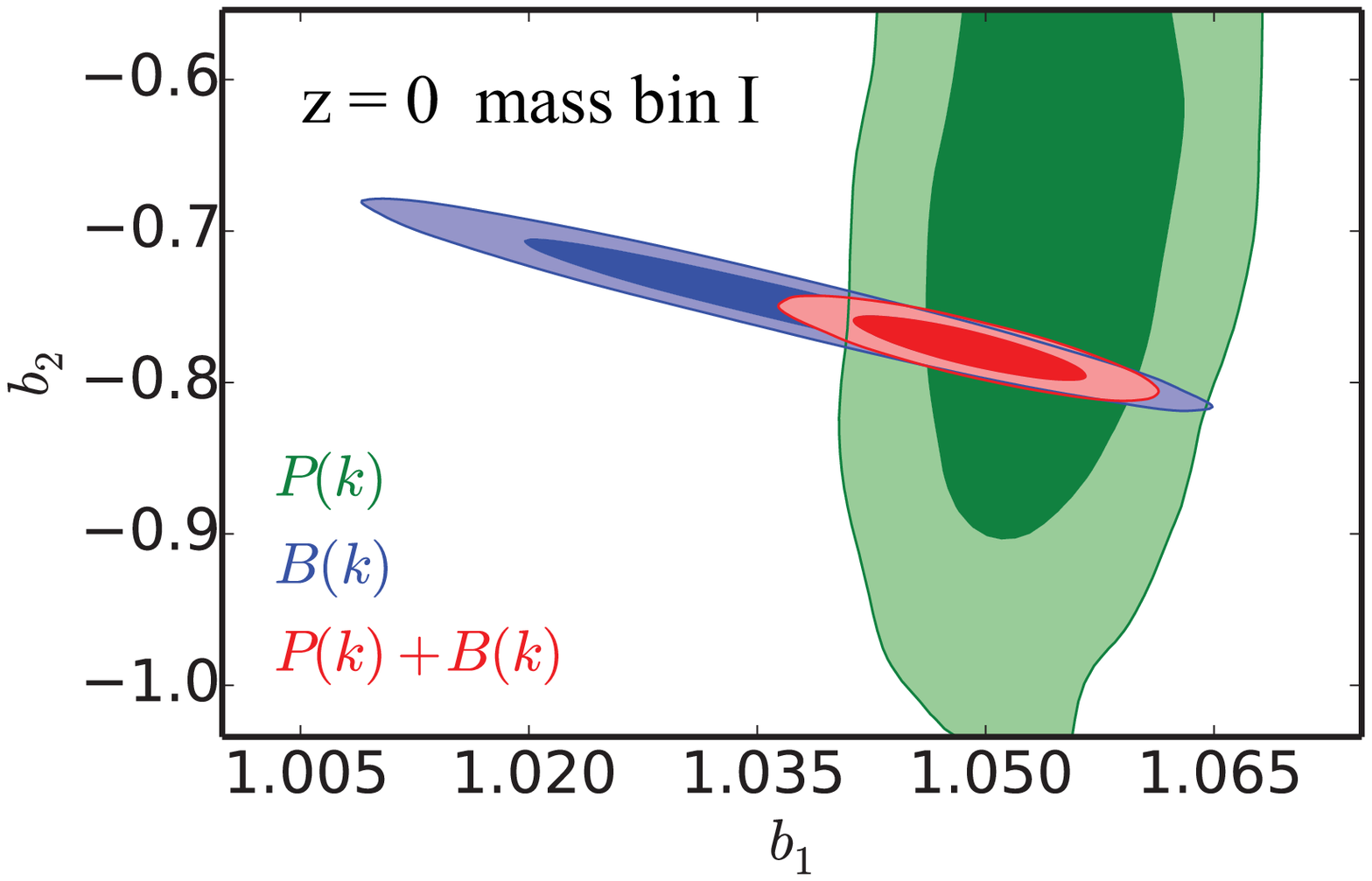}
\includegraphics[width=0.35\textwidth]{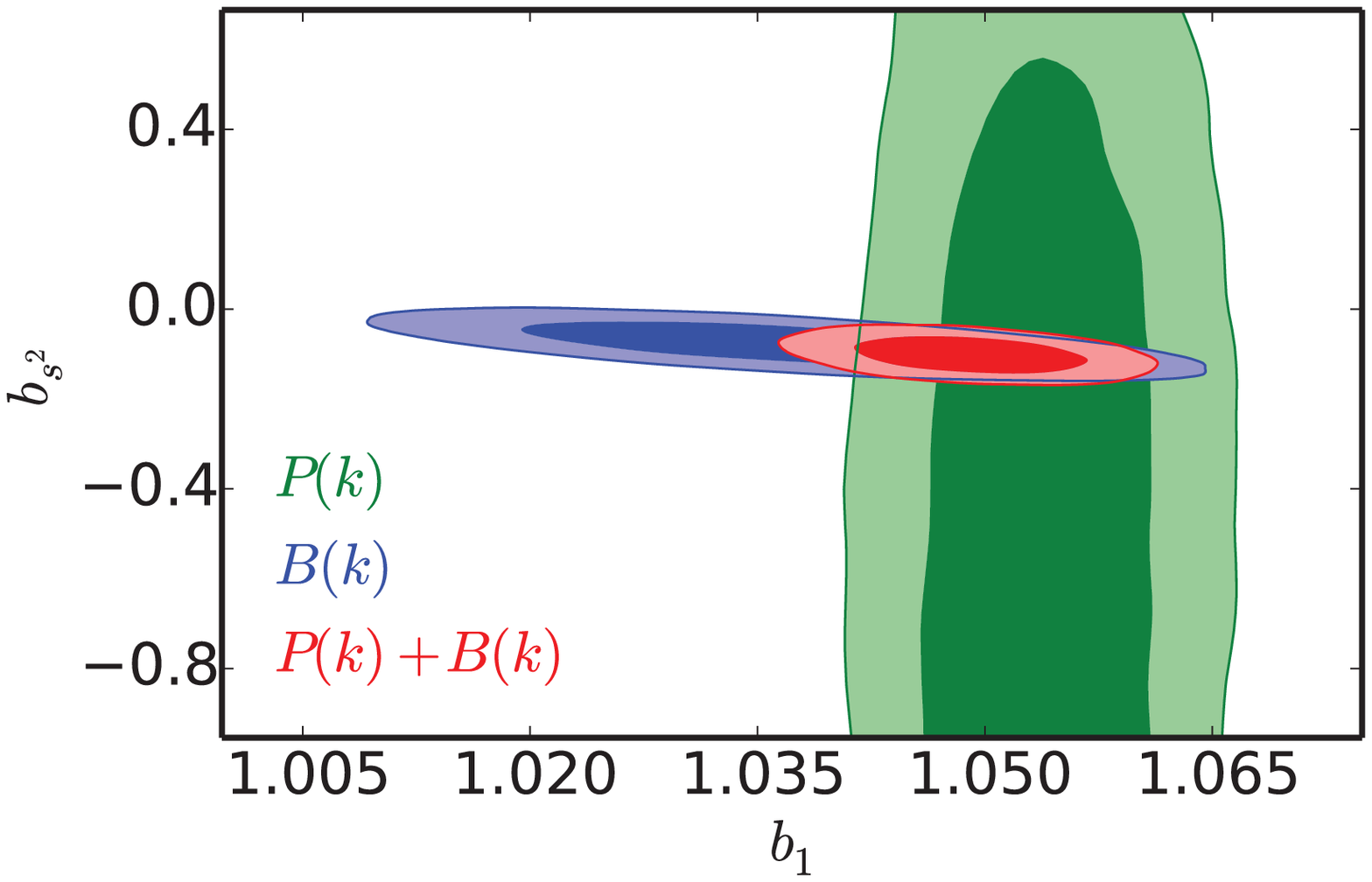}\\
\includegraphics[width=0.35\textwidth]{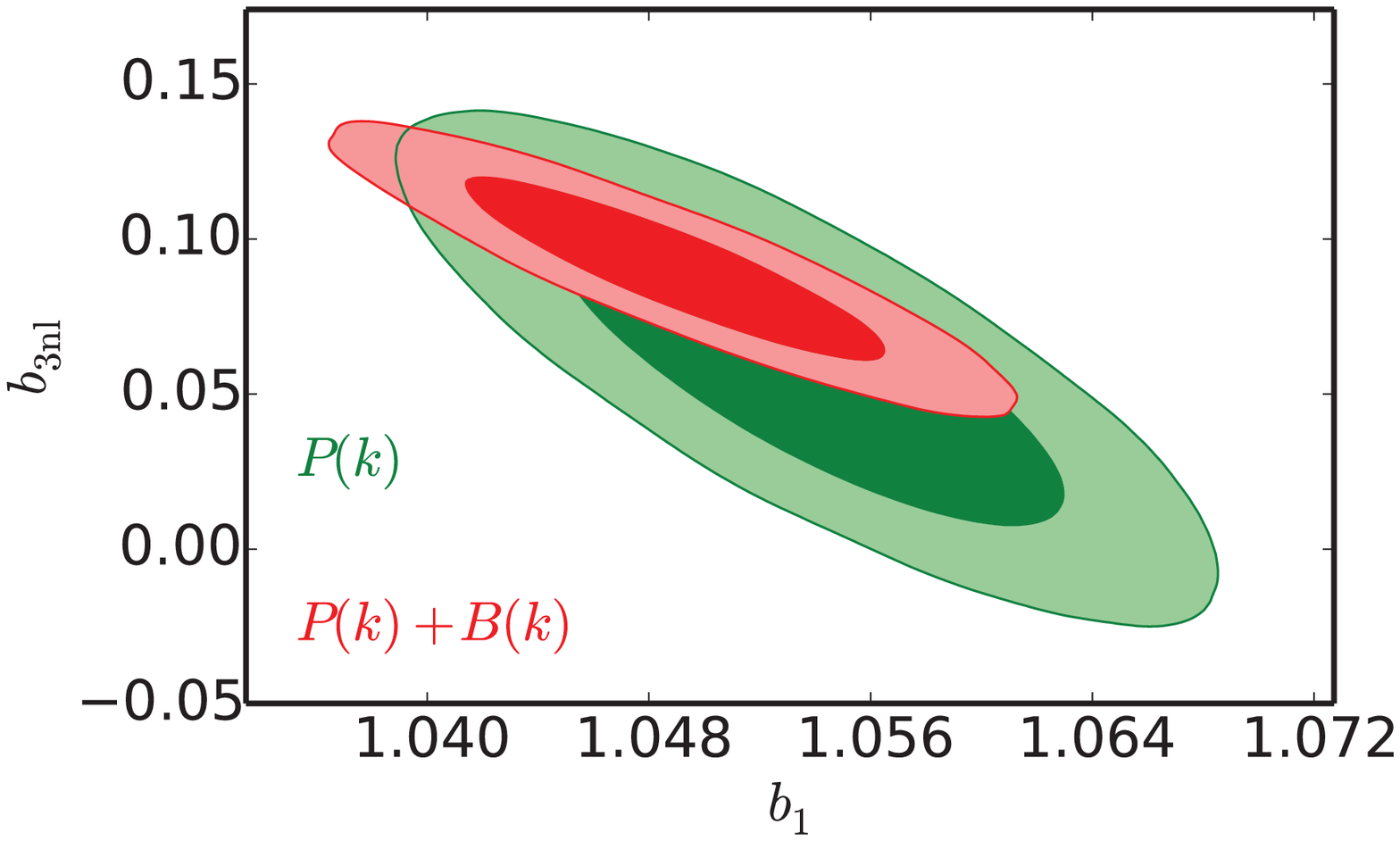}
\includegraphics[width=0.35\textwidth]{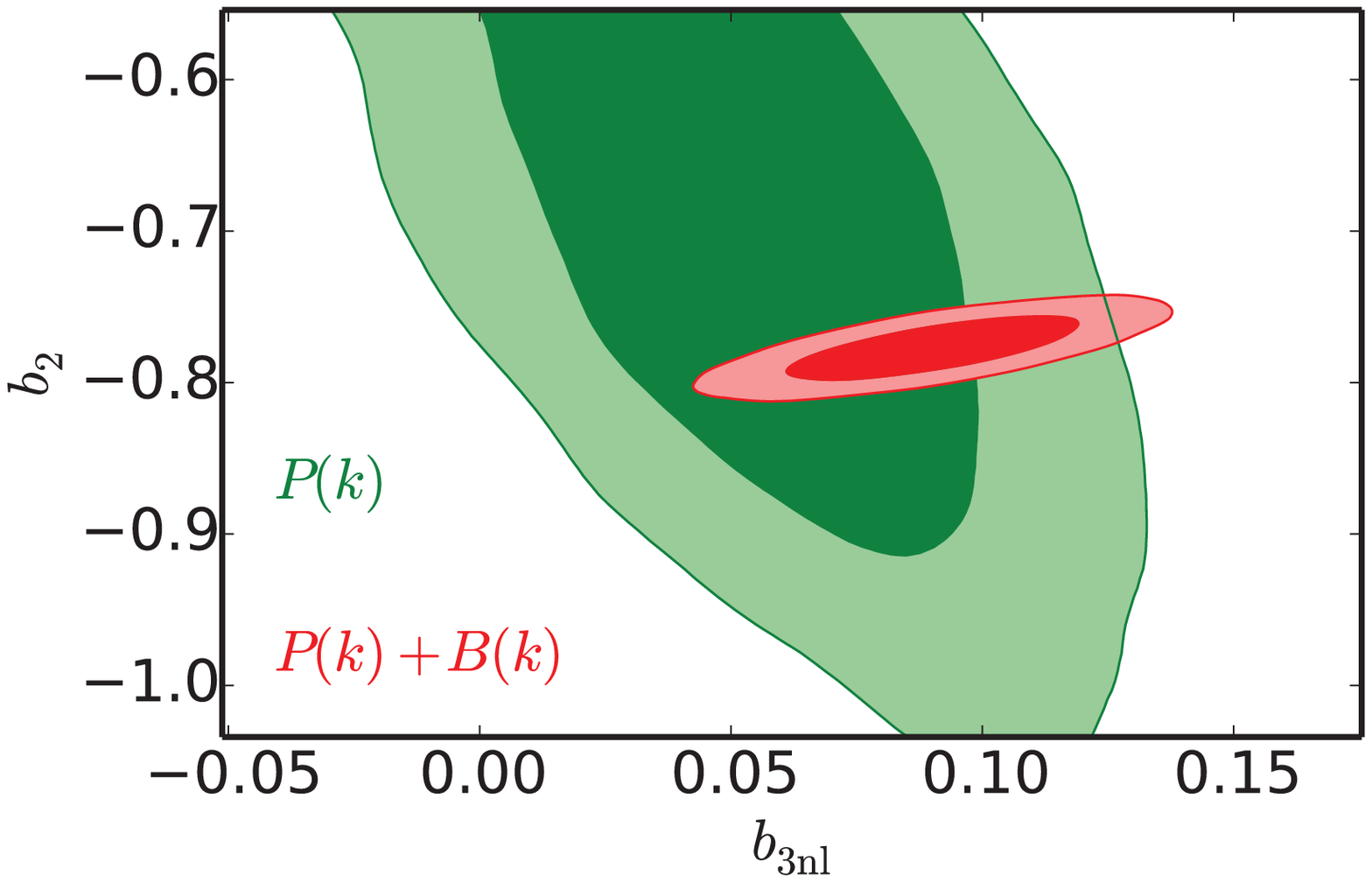}
\end{center}
\vspace*{-2em}
\caption{ 
({\it Upper four panels}) comparison between cases with and without combining 
the bispectrum information for mass bin I at $z=0$. 
We show the marginalized probability distribution for each bias parameter in the 
cases of the power spectrum combined with the bispectrum ({\it red solid}), 
the bispectrum only ({\it blue dashed}), and the power spectrum only ({\it green dotted}). 
As a reference, we show the prediction from the coevolution $b^{\rm coev}_{\rm 3nl}$ 
assuming $b_{1}\simeq b^{\rm E}_{1}$ is equal to the value obtained by joint fitting 
({\it cyan}). 
Note that, in the case of the power spectrum, we use both density-density and 
density-momentum power spectrum with $k_{{\rm max},P(k)}=0.08h$/Mpc. 
({\it Lower four panels}) the marginalized two-dimensional contours 
(68\% and 95\% C.L.) among the nonlinear bias parameters. 
}  
\label{fig: Pk only z0mbin1}
\end{figure}

\begin{figure}[t]
\begin{center}
\includegraphics[width=0.35\textwidth]{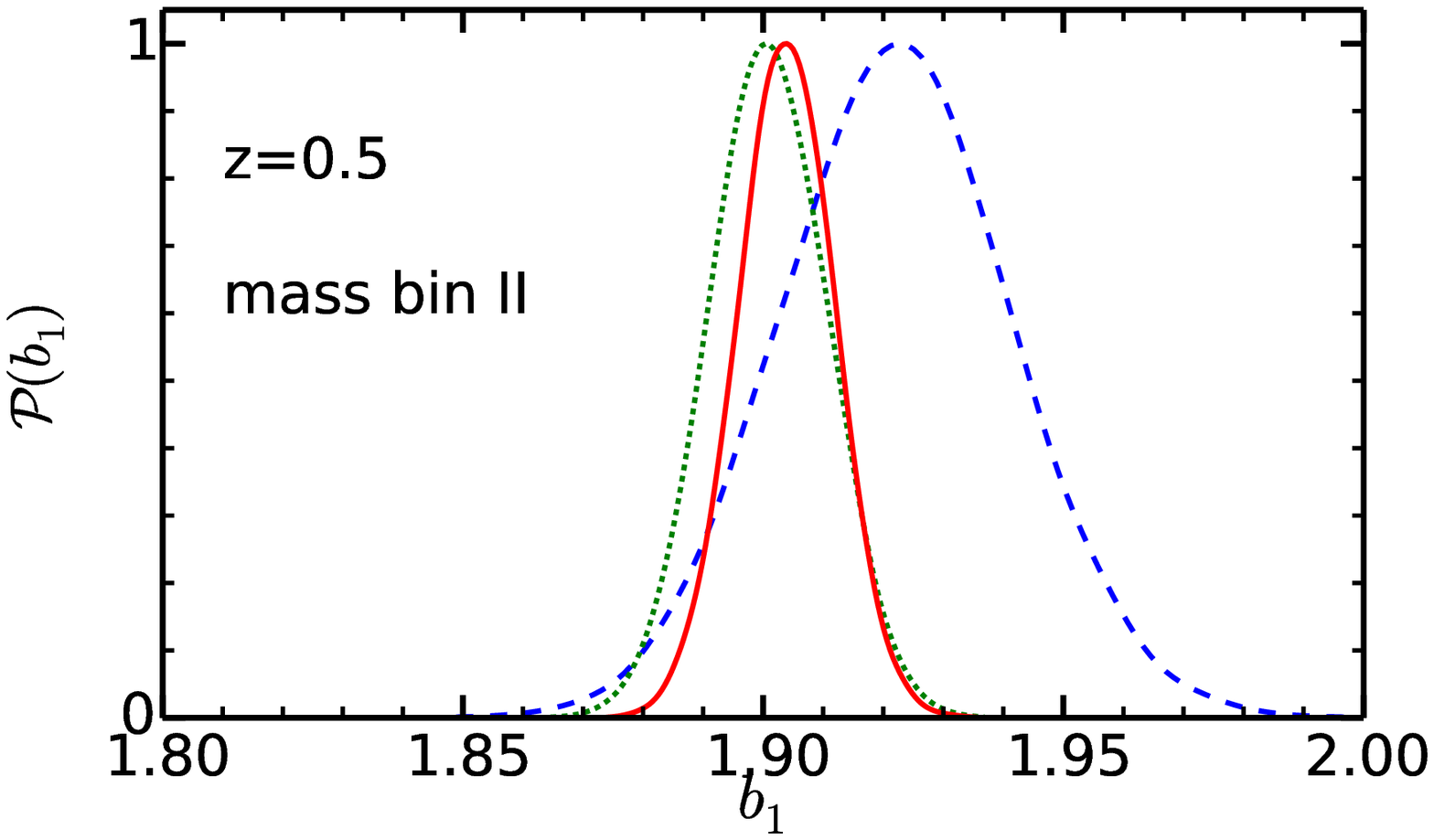}
\includegraphics[width=0.35\textwidth]{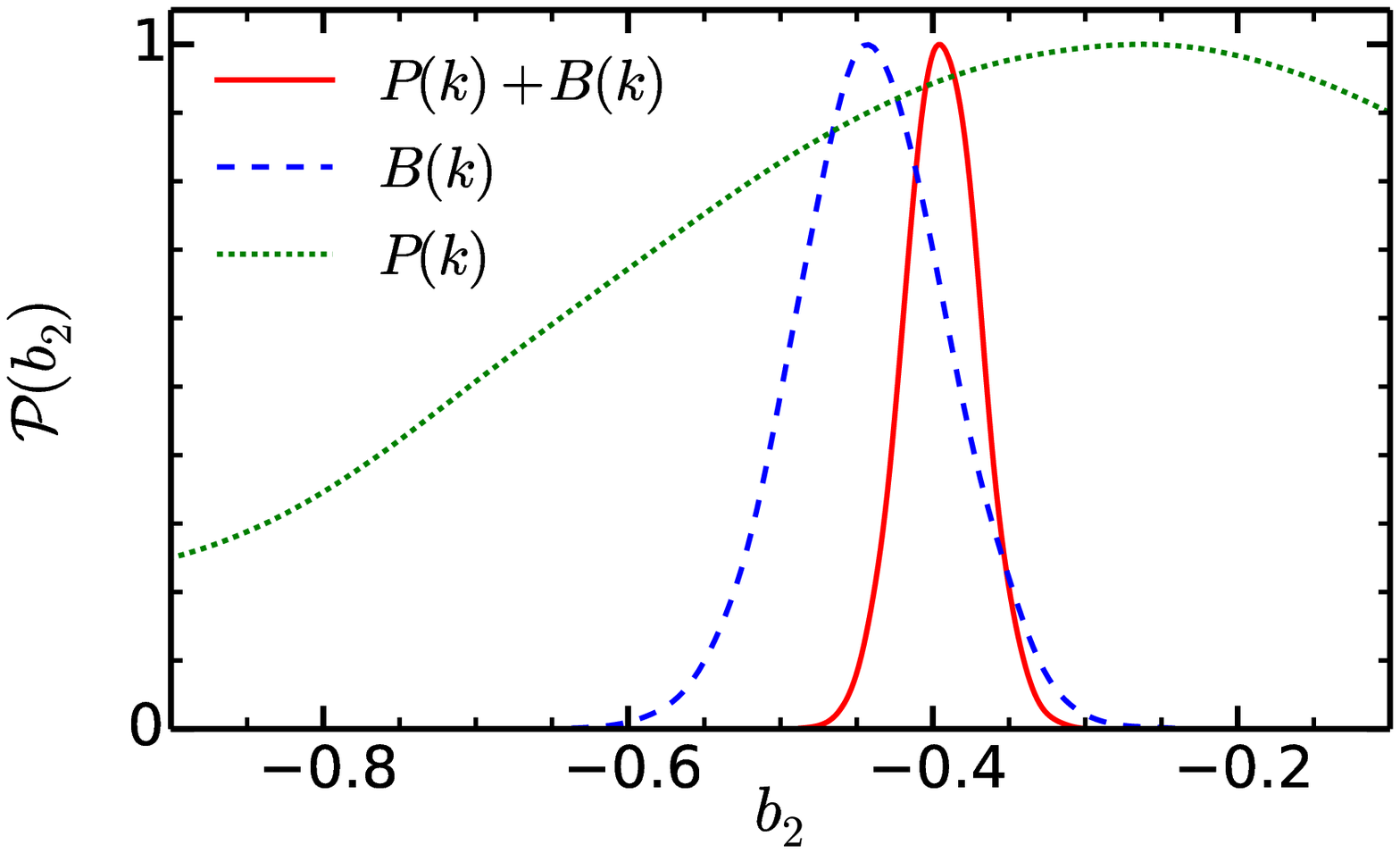}\\
\includegraphics[width=0.35\textwidth]{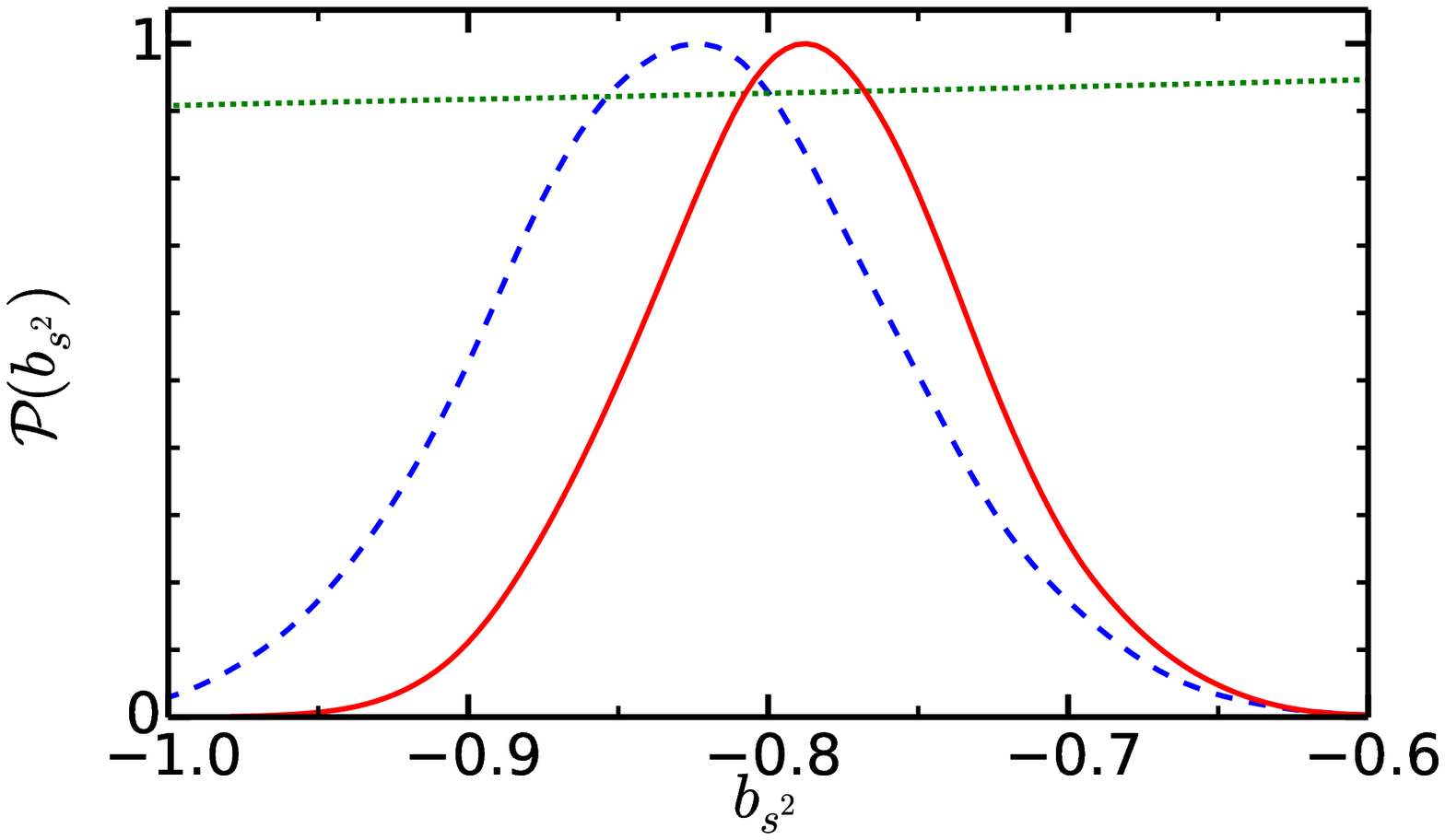}
\includegraphics[width=0.35\textwidth]{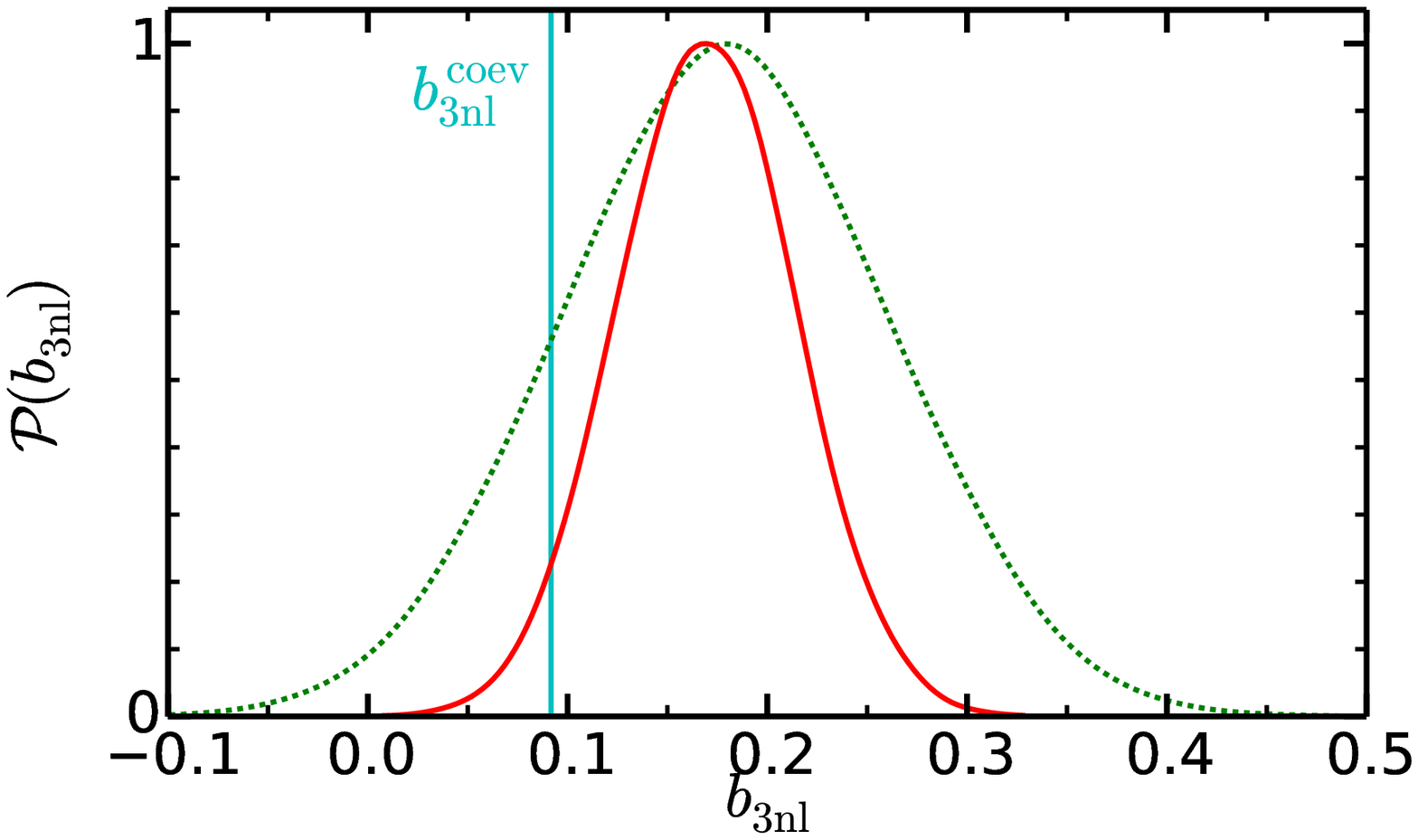}\\
\includegraphics[width=0.35\textwidth]{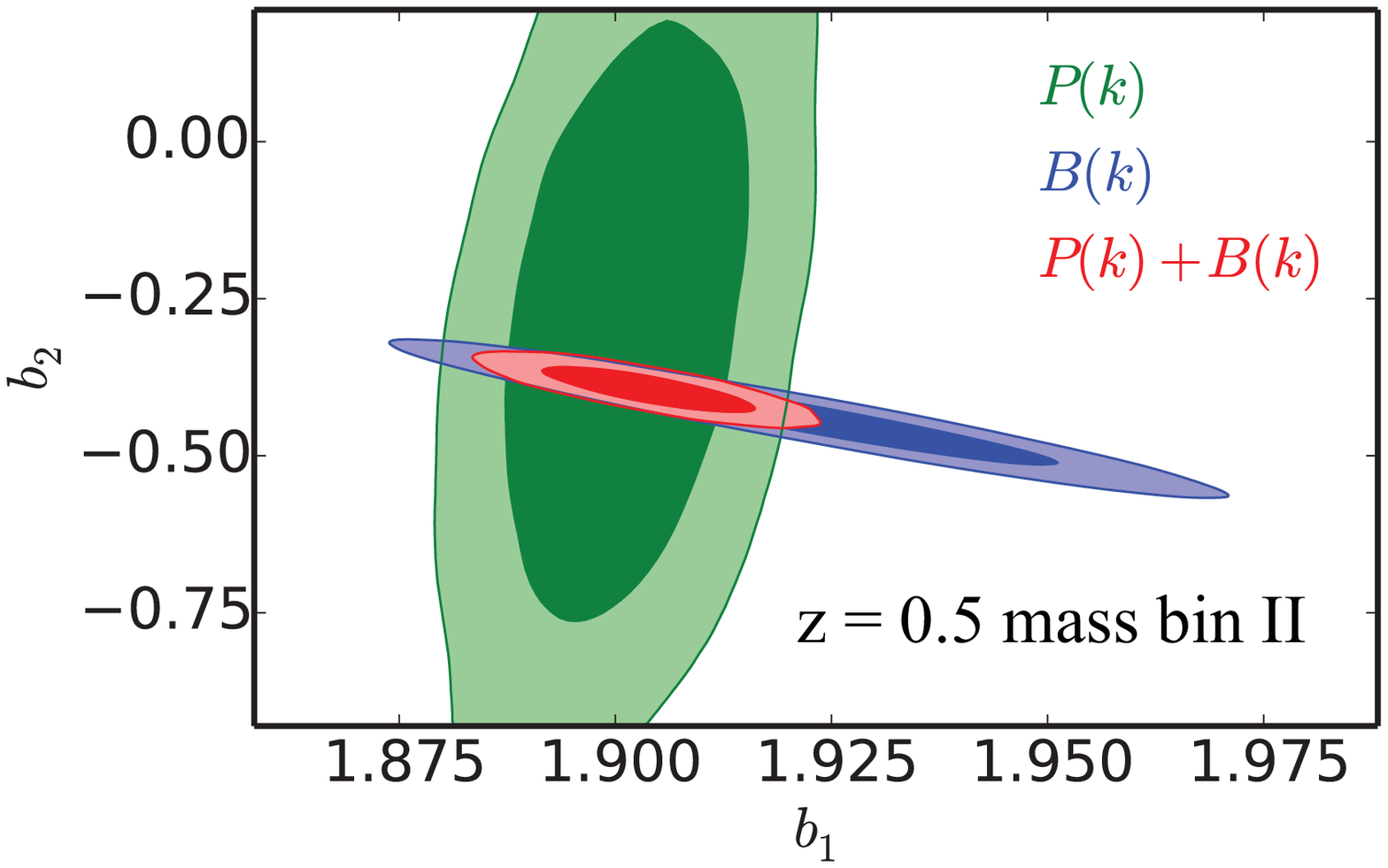}
\includegraphics[width=0.35\textwidth]{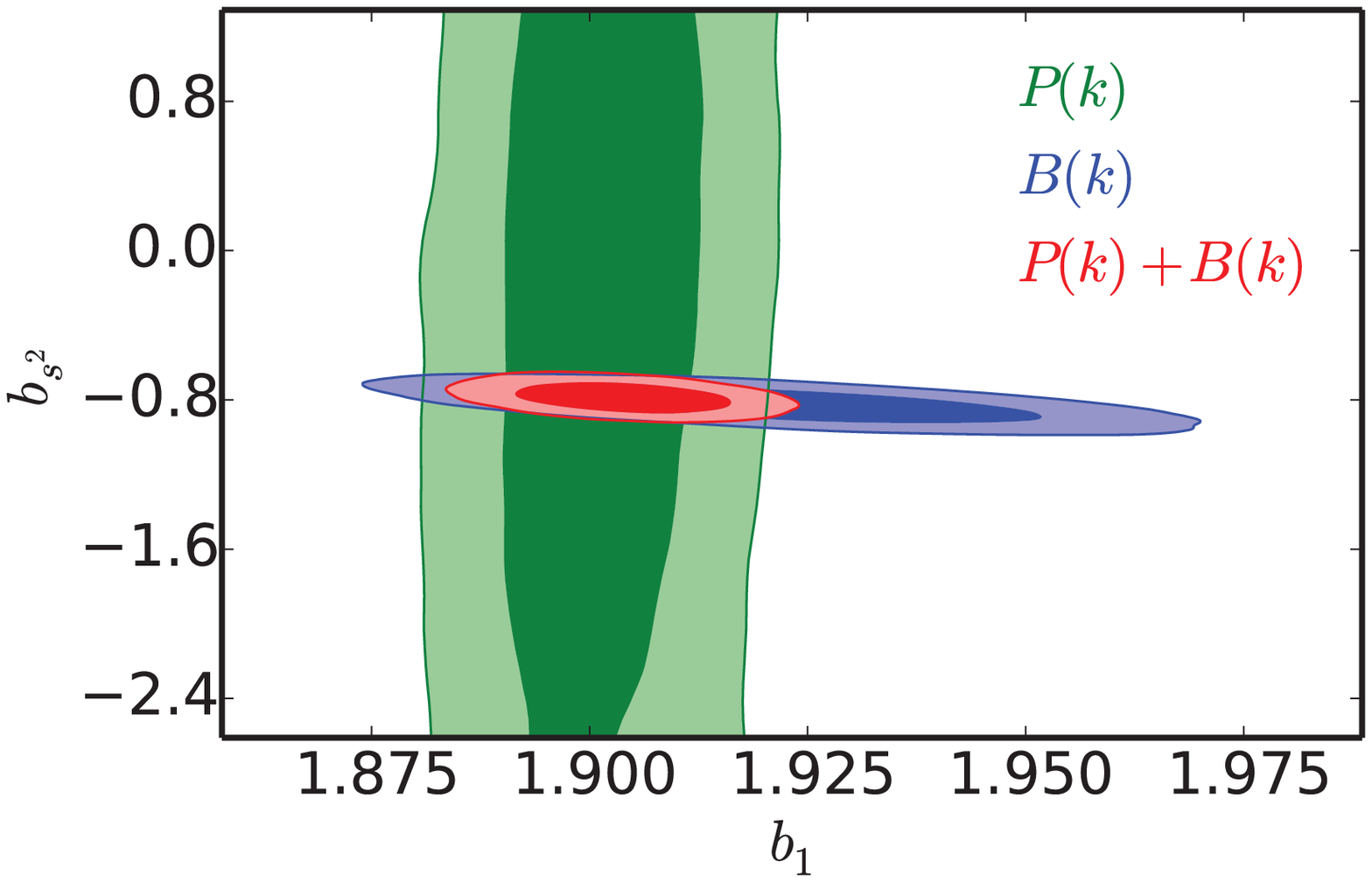}\\
\includegraphics[width=0.35\textwidth]{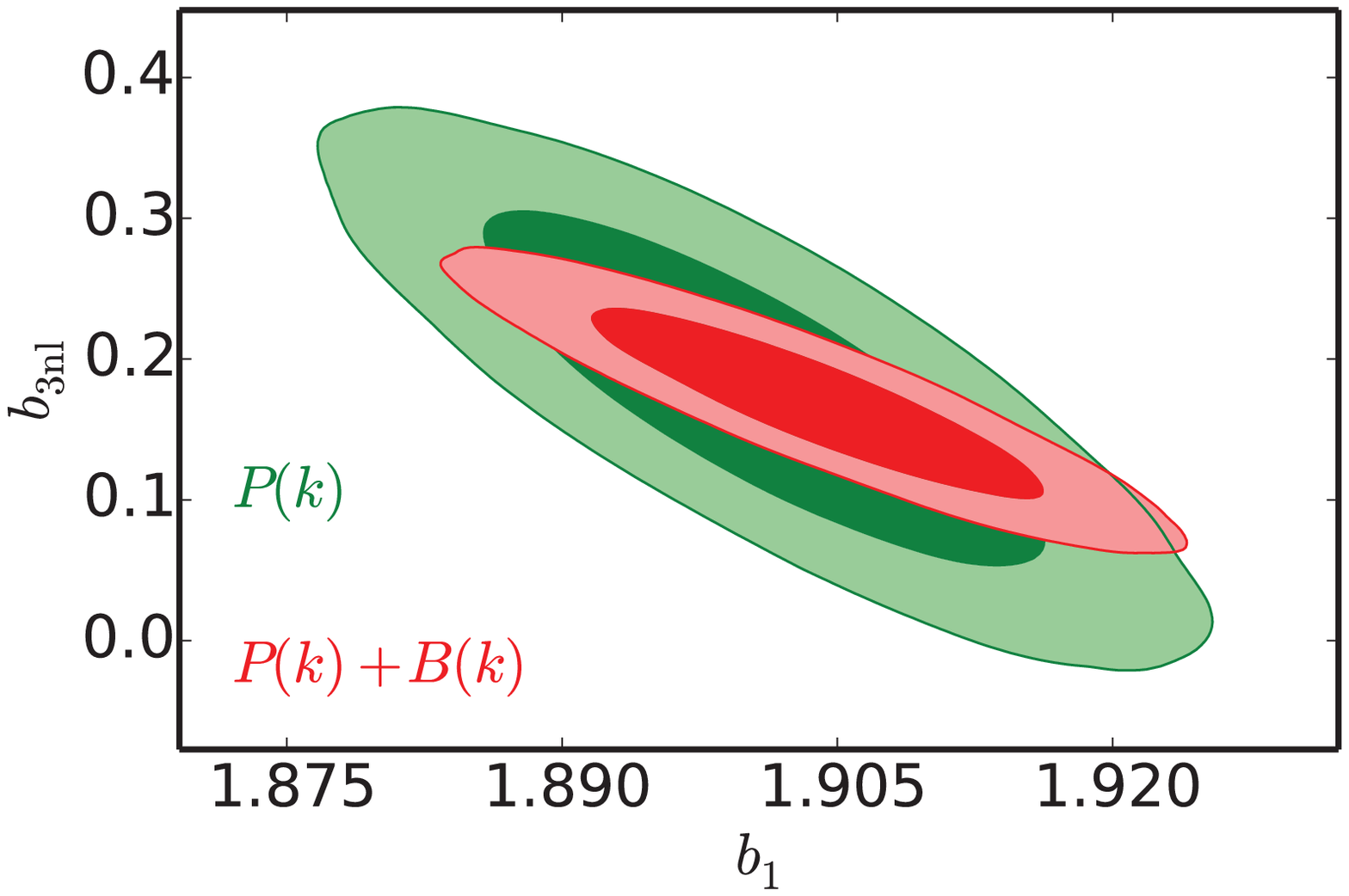}
\includegraphics[width=0.35\textwidth]{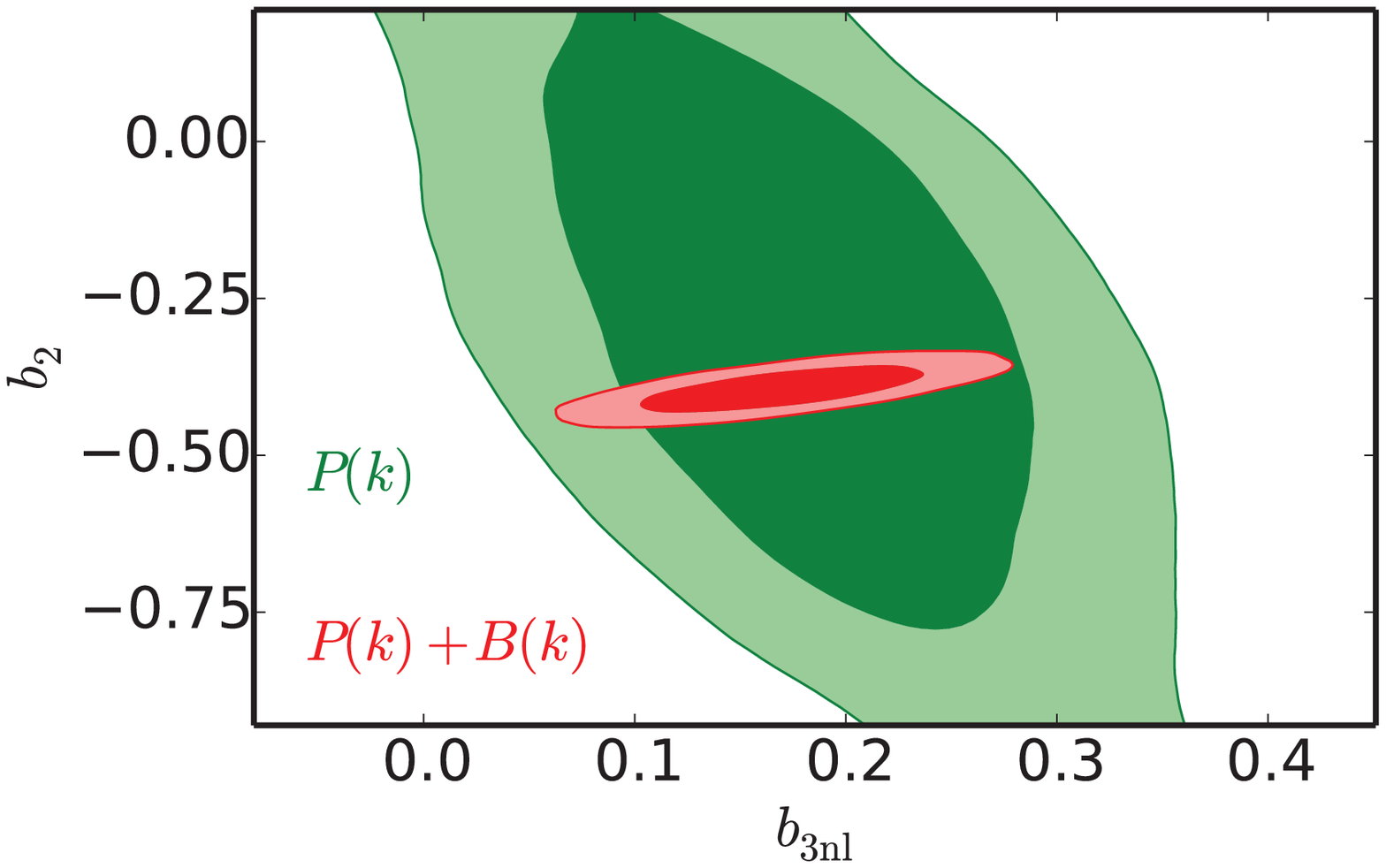}
\end{center}
\vspace*{-2em}
\caption{ 
Same as Fig.~\ref{fig: Pk only z0mbin1}, but for mass bin II at $z=0.5$. 
Note that, in the case of the power spectrum, we use both density-density and 
density-momentum power spectrum with $k_{{\rm max},P(k)}=0.1h$/Mpc.
}  
\label{fig: Pk only z05mbin2}
\end{figure}

\begin{figure}[t]
\begin{center}
\includegraphics[width=0.35\textwidth]{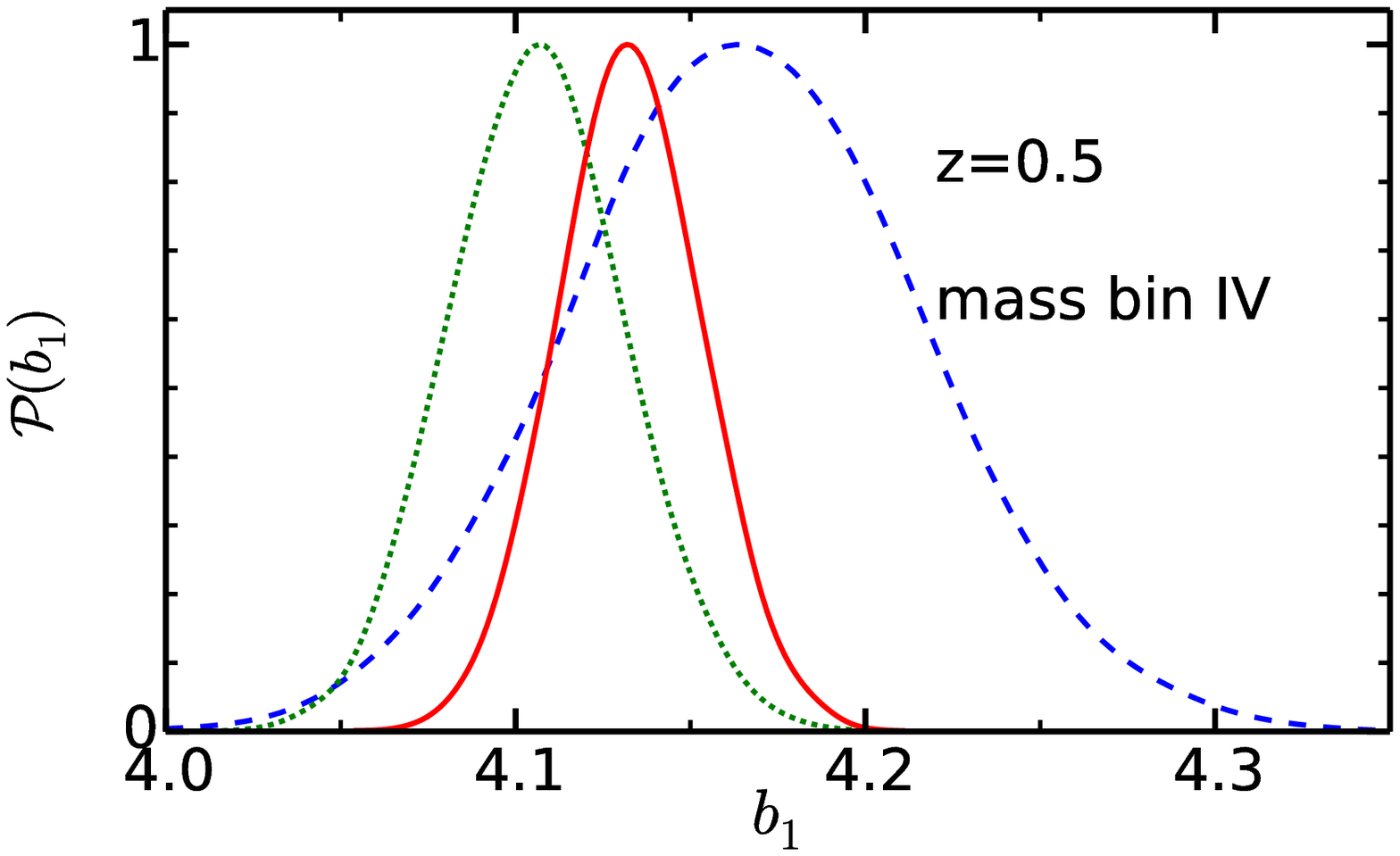}
\includegraphics[width=0.35\textwidth]{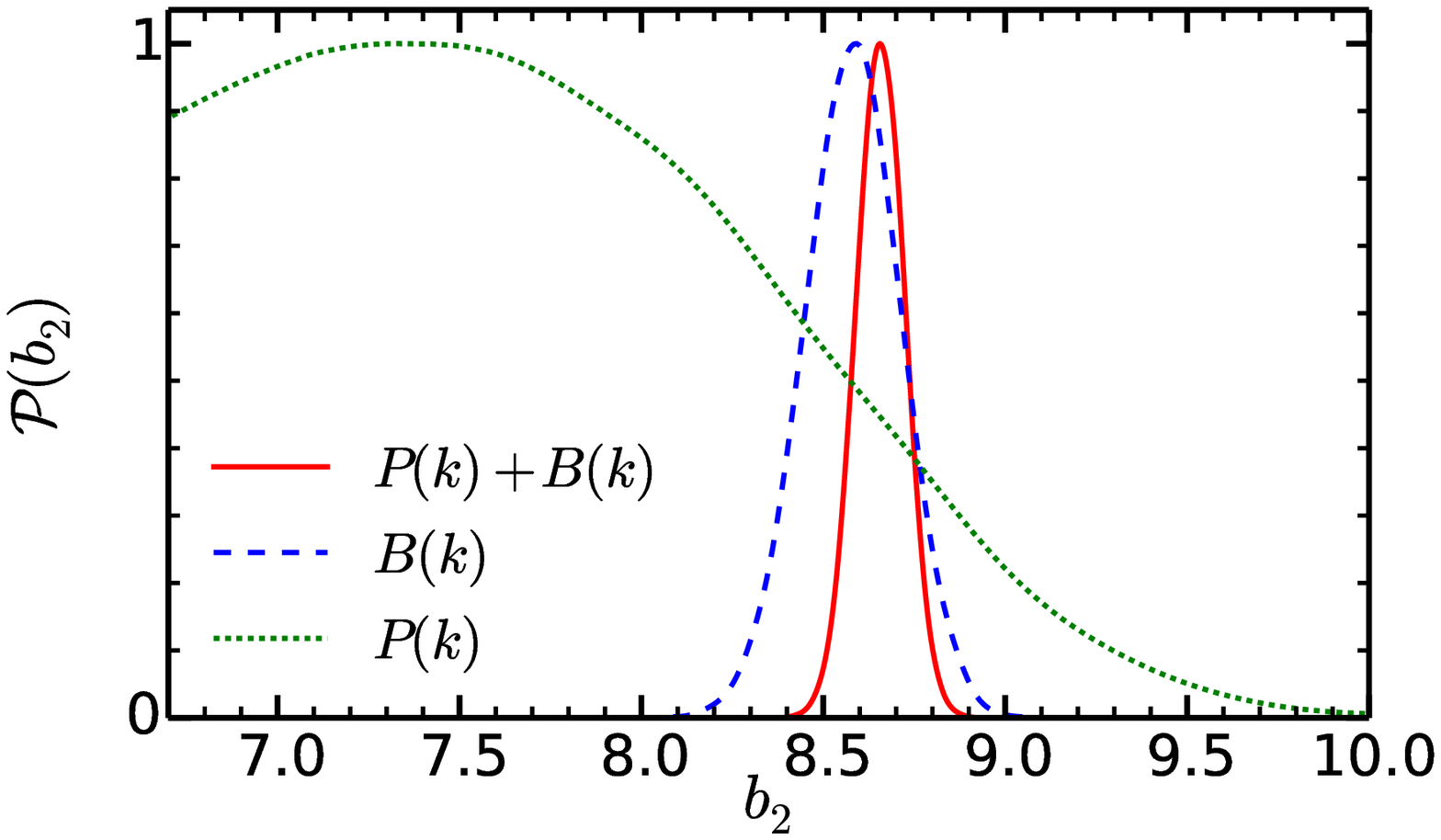}\\
\includegraphics[width=0.35\textwidth]{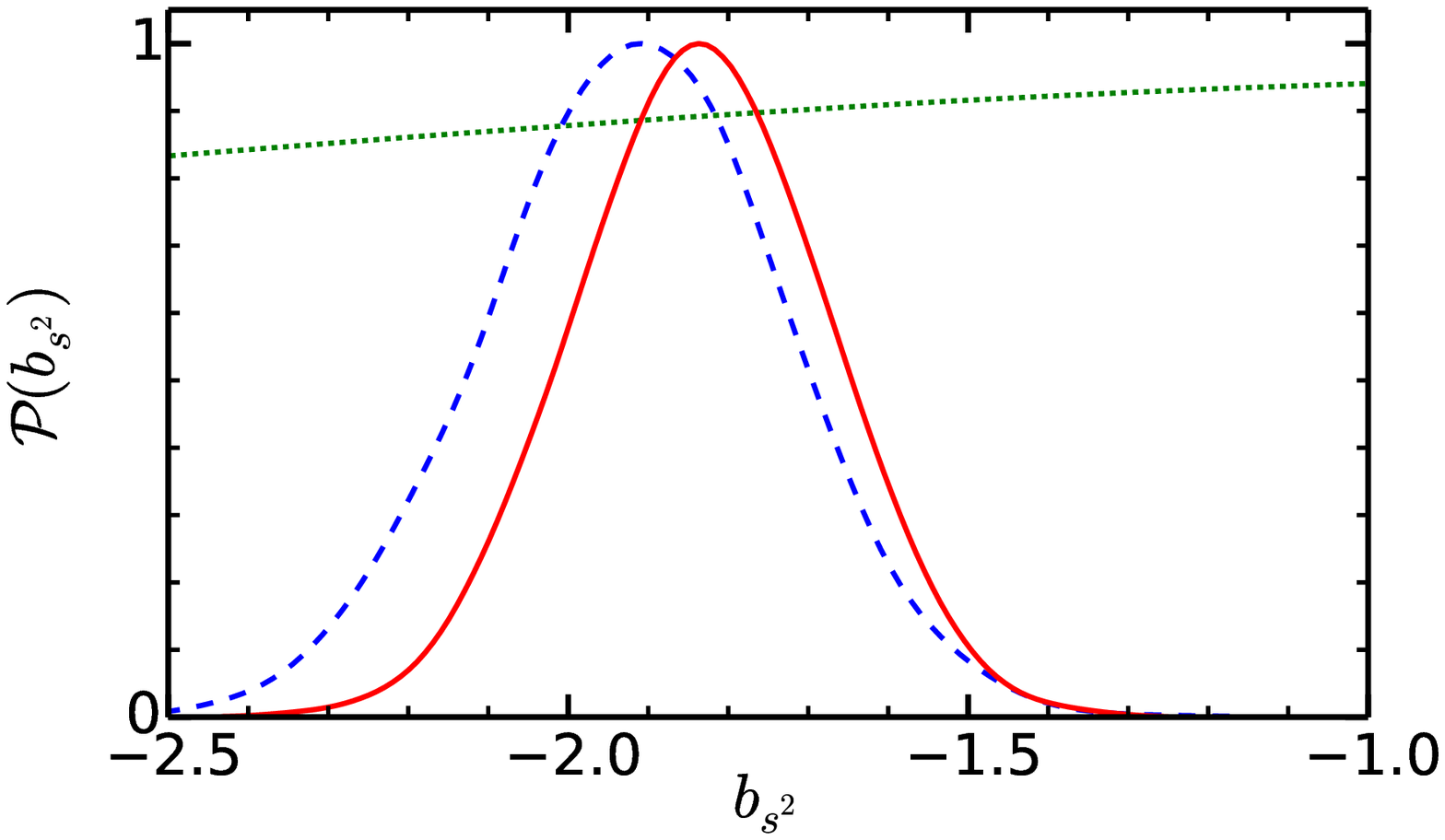}
\includegraphics[width=0.35\textwidth]{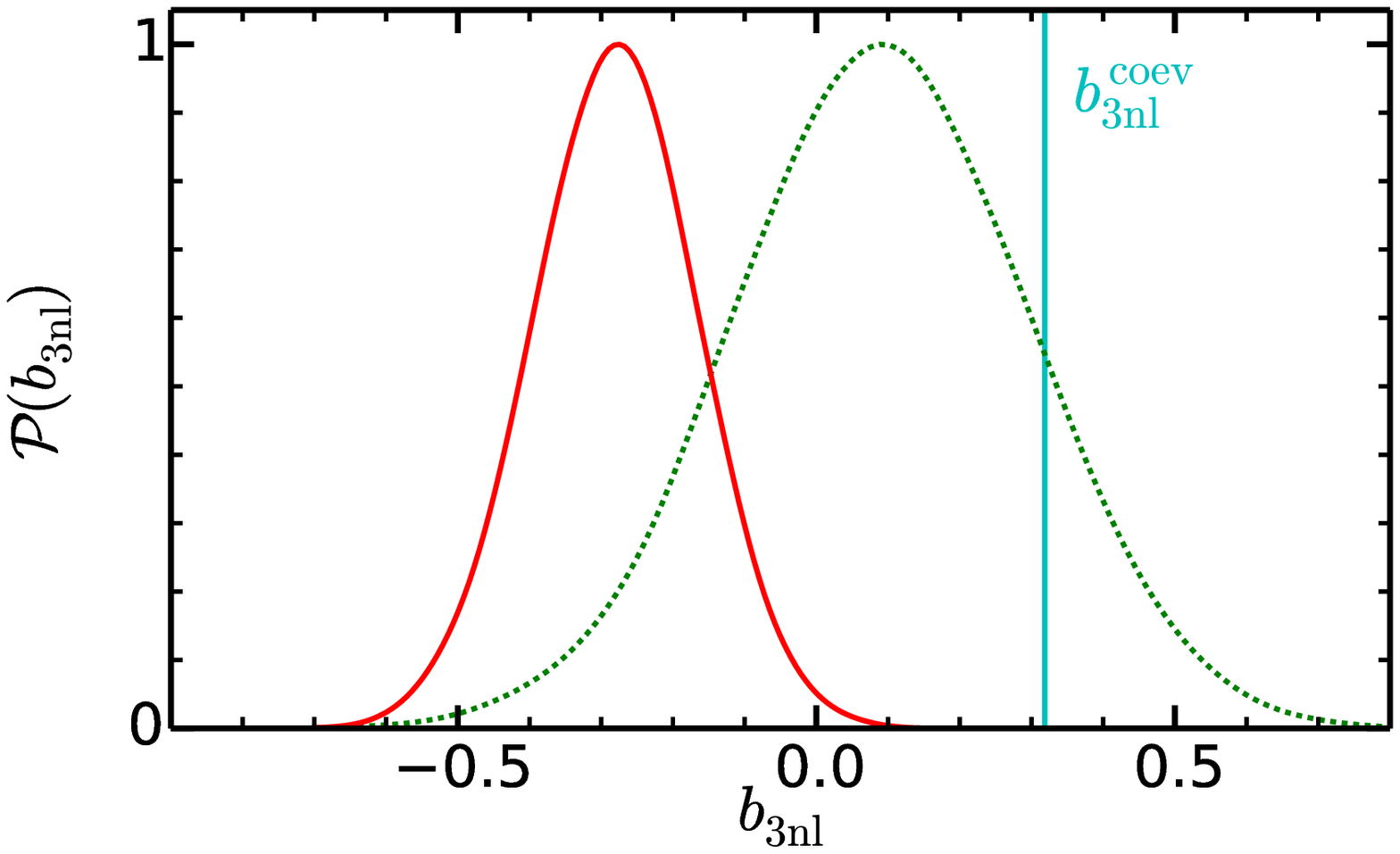}\\
\includegraphics[width=0.35\textwidth]{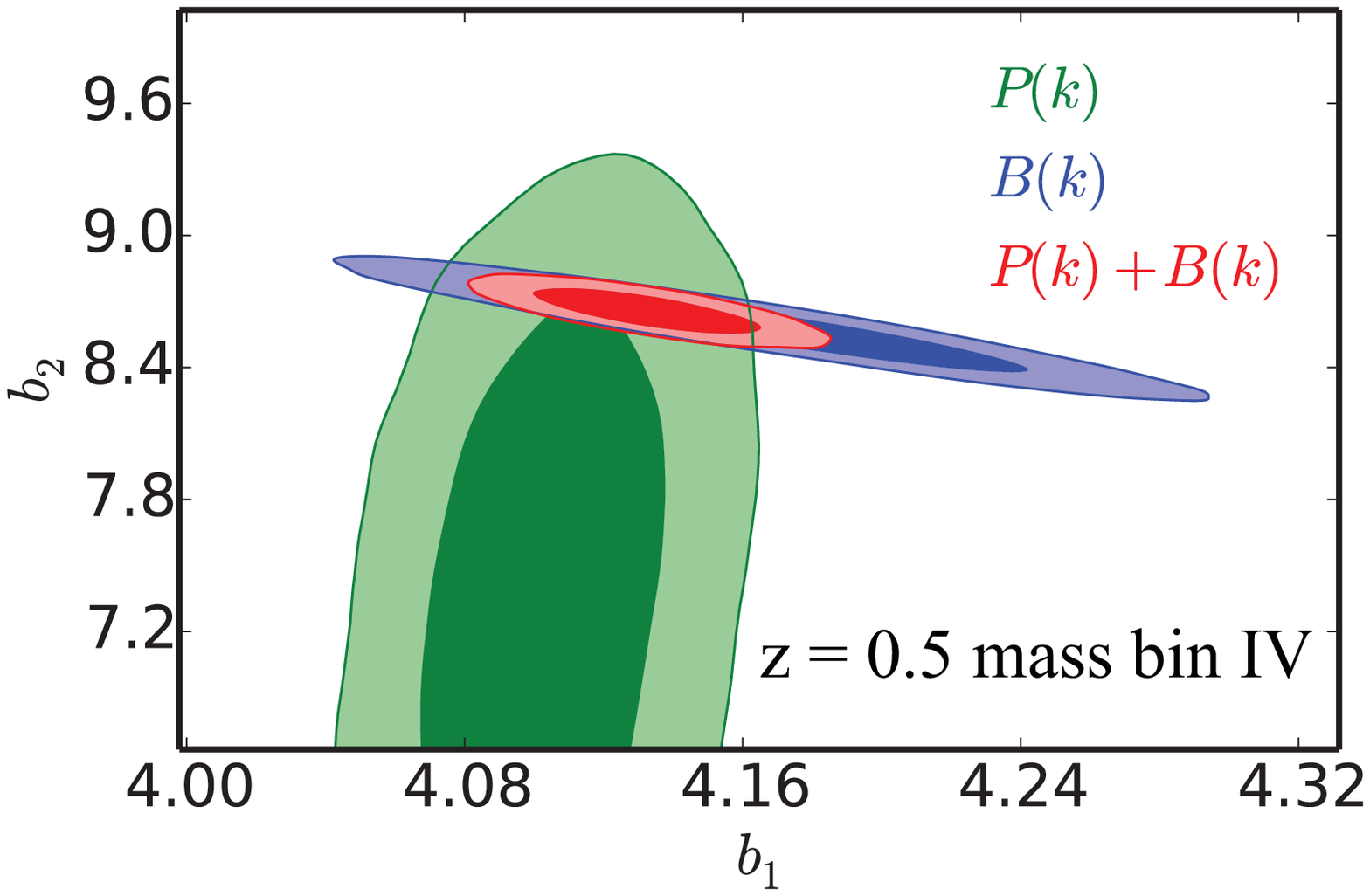}
\includegraphics[width=0.35\textwidth]{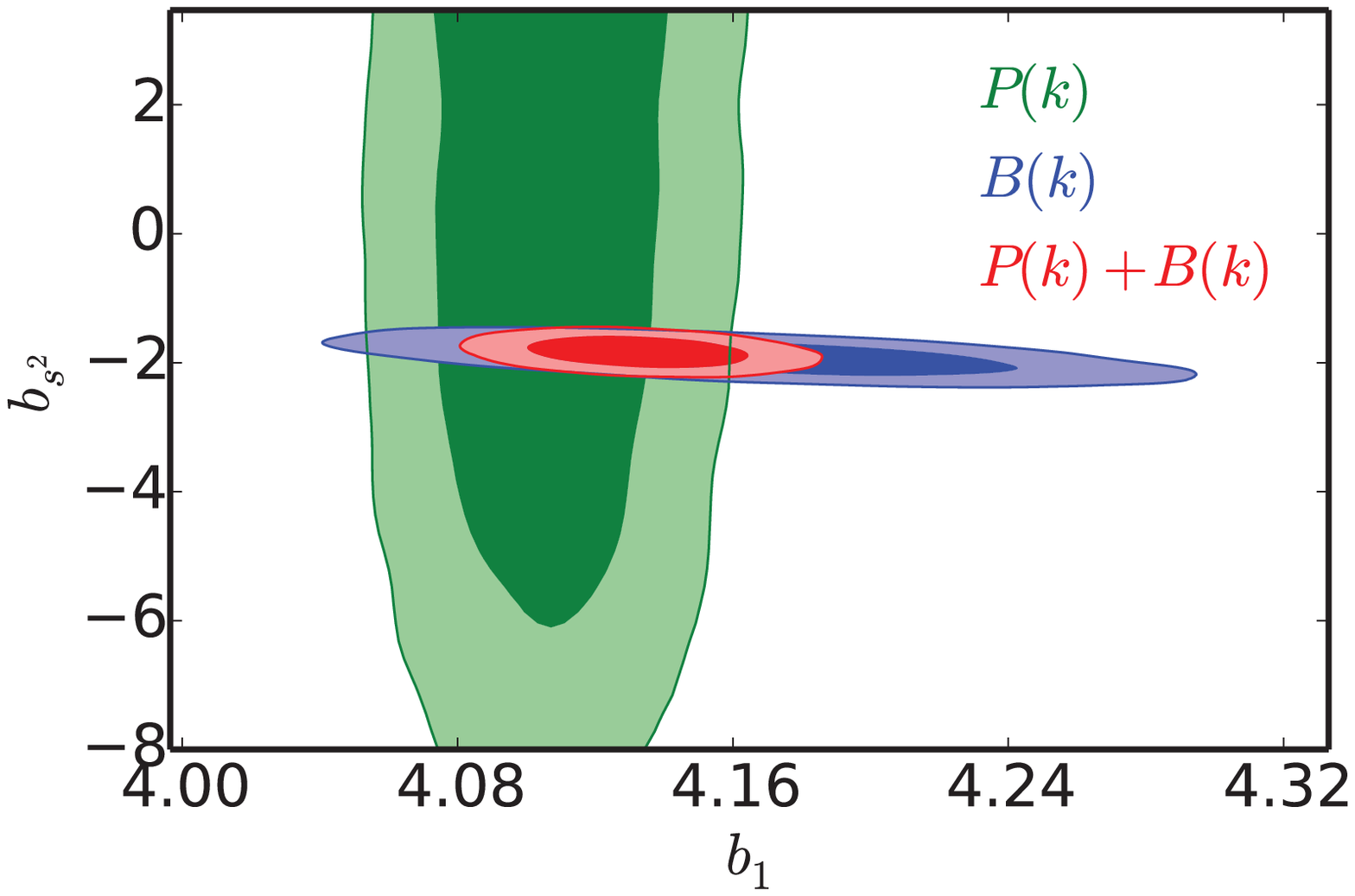}\\
\includegraphics[width=0.35\textwidth]{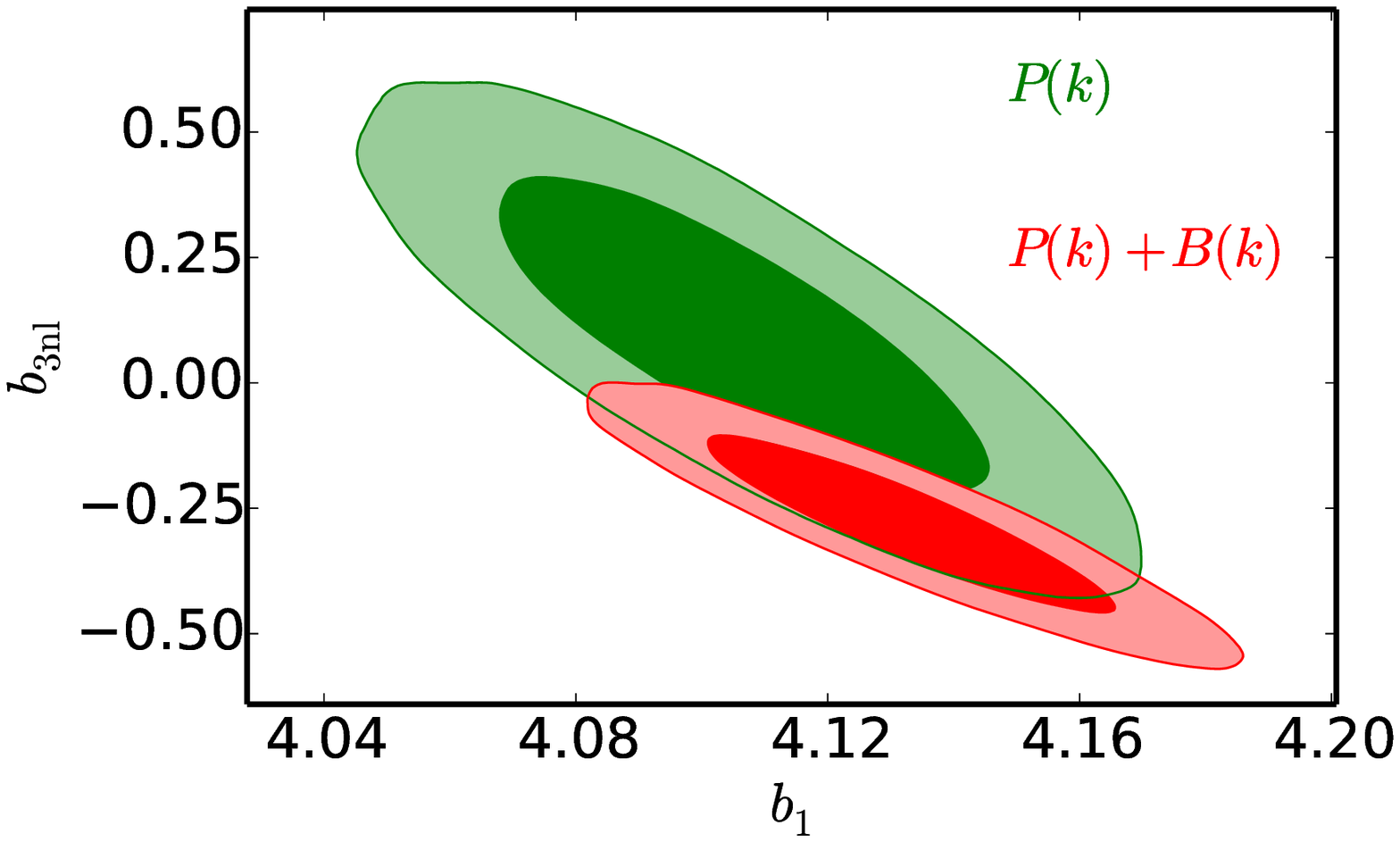}
\includegraphics[width=0.35\textwidth]{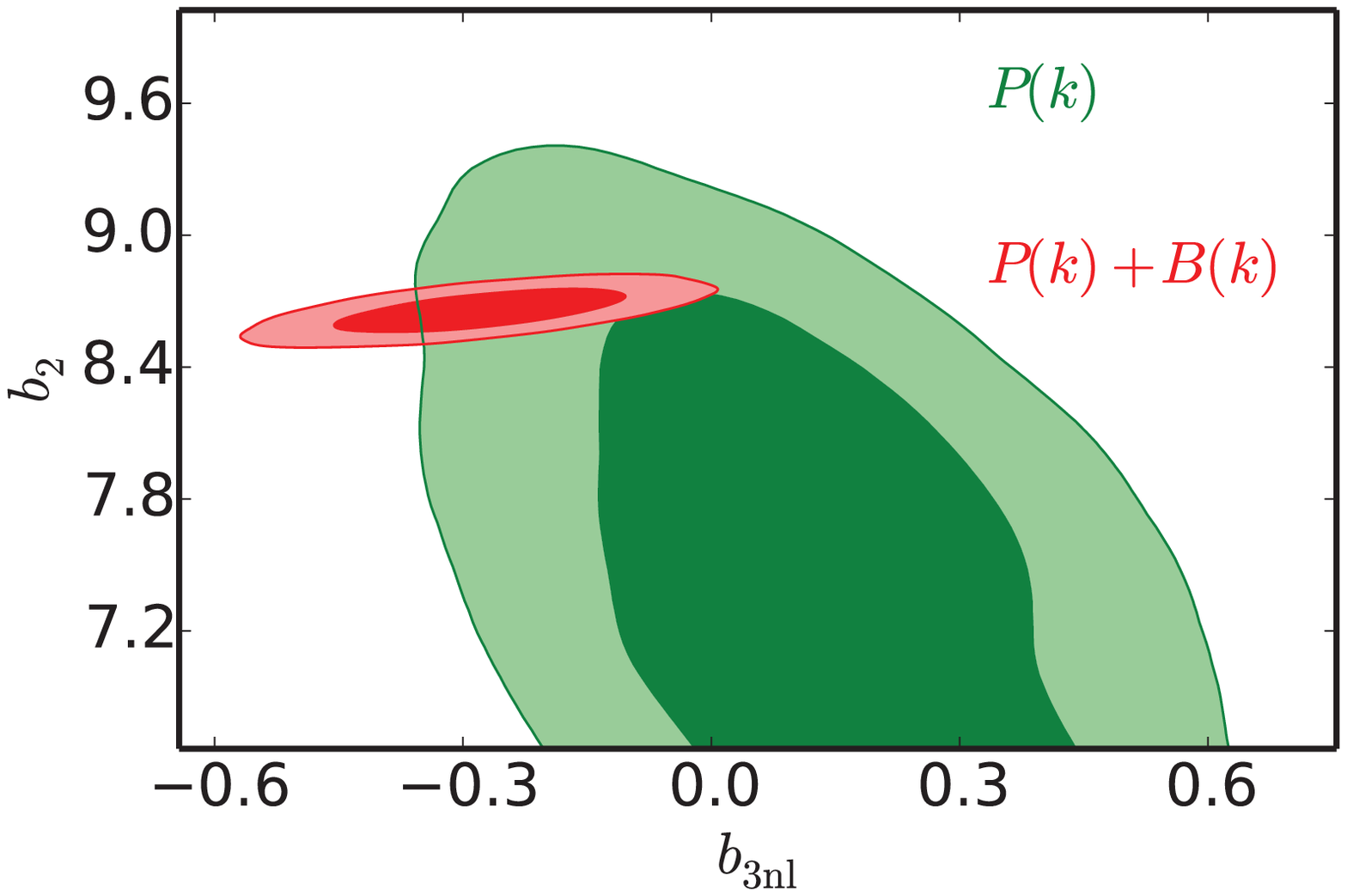}
\end{center}
\vspace*{-2em}
\caption{ 
Same as Fig.~\ref{fig: Pk only z0mbin1}, but for mass bin IV at $z=0.5$. 
Note that, in the case of the power spectrum, we use both density-density and 
density-momentum power spectrum with $k_{{\rm max},P(k)}=0.1h$/Mpc.
}  
\label{fig: Pk only z05mbin4}
\end{figure}

\begin{figure}[t]
\begin{center}
\includegraphics[width=0.48\textwidth]{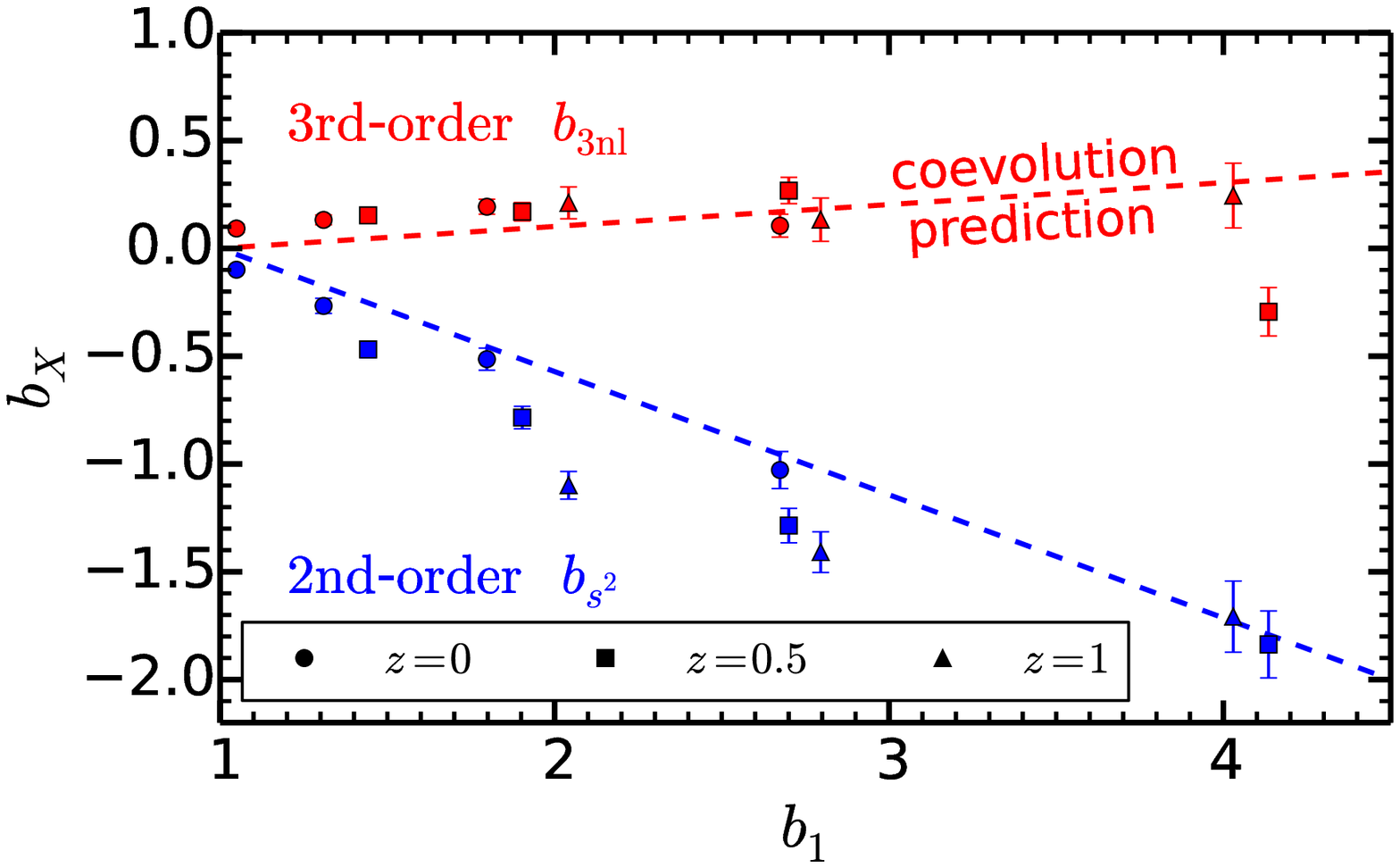}
\includegraphics[width=0.48\textwidth]{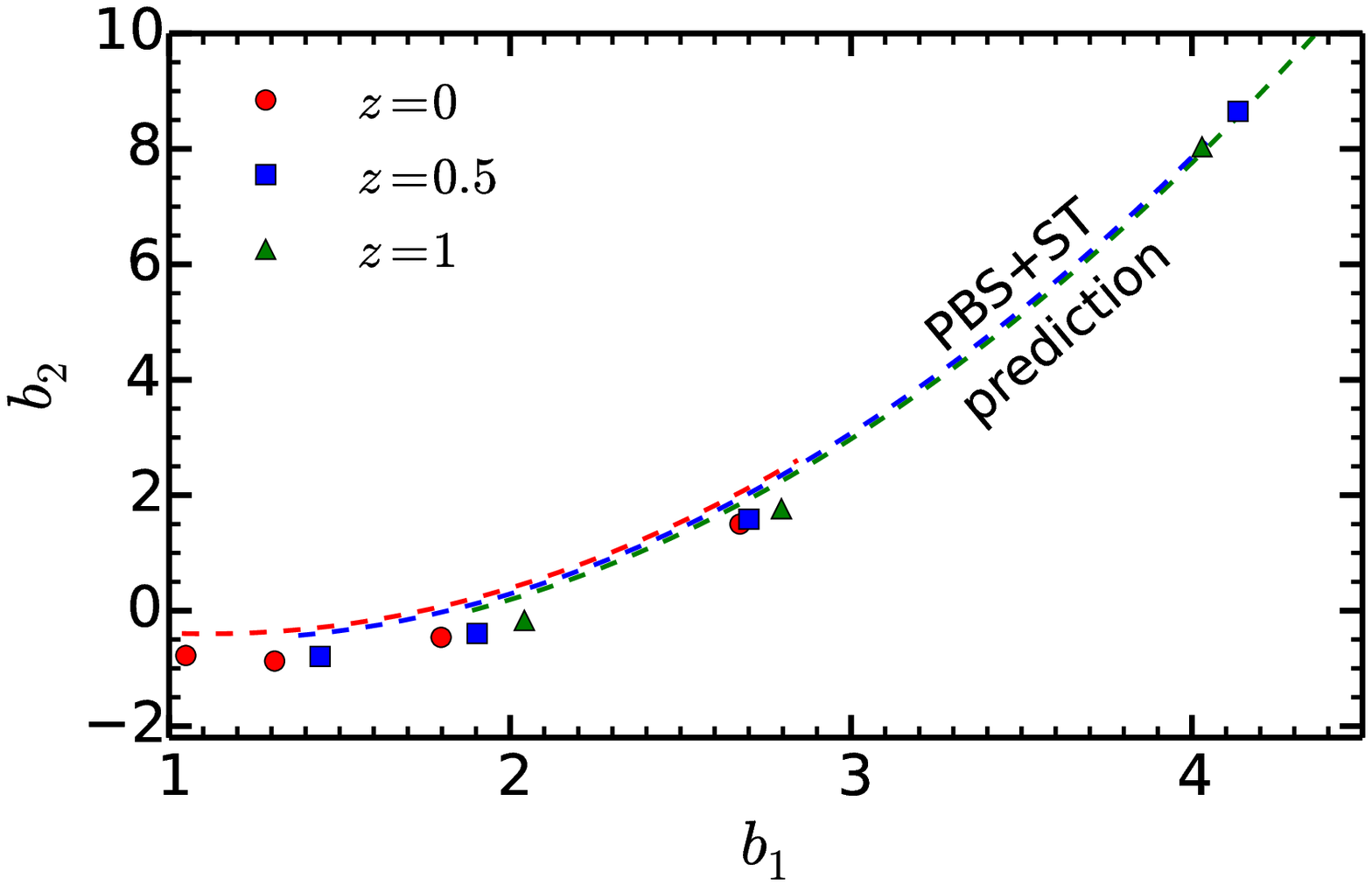}
\end{center}
\vspace*{-2em}
\caption{ 
({\it Left}) nonlocal bias values at second and third orders as a function of the linear bias parameter. 
Each point is taken from the result in the joint fit of the power spectrum and the bispectrum. 
Both of them are compared with the values expected from the local Lagrangian bias 
in the simple convolution picture ({\it dashed} lines). 
({\it Right}) second-order local bias bias $b_{2}$ against $b_{1}$. 
Again each point is obtained from our joint fit. For comparison, we draw theory lines 
which is based on the peak-background split with the universal mass function 
(see text in detail). Note that the range of these lines are limited since we here consider 
relevant halo mass range only ($0.6\ltsim M_{\rm halo}\,[10^{13}M_{\odot}/h] \ltsim 21$). 
}  
\label{fig: b1 vs bx}
\end{figure}

\section{Results}
\label{sec: results}

Now we show our measurements of the bias parameters from the simulated halo-matter power spectra 
combined together with the bispectrum.
In Fig.~\ref{fig: estimate b3nl}, we show the best-fitting values of $b_{3{\rm nl}}$ as a function of 
$k_{\rm max}$ for each mass bin at each redshift. 
First of all, the preferred values of $b_{3{\rm nl}}$ are nonzero generally for any halo mass bin at any redshift, 
at $k_{\rm max}\sim 0.1\,h/{\rm Mpc}$. Also, the best-fitting values of $b_{3{\rm nl}}$ from $P^{\rm hm}_{\,00}(k)$ are 
generally in a good agreement with those from $P^{\rm hm}_{\,01}(k)$, 
indicating that the third-order nonlocal bias term is important to explain both $P^{\rm hm}_{\,00}(k)$ and 
$P^{\rm hm}_{\,01}(k)$. If looking at smaller scales at $k\gtsim 0.1\,h/{\rm Mpc}$, we start to see 
a discrepancy between the two results, and the best-fitting values tend to vary as a function of $k_{\rm max}$. 
In addition, a goodness of fit, $\chi^{2}$/(d.o.f.) becomes worse at larger $k$.  
This clearly shows that our bias model fails to describe the halo-matter power spectra at such small scales, 
and higher-order contribution would start to kick in. 
Notice again that our values of the goodness of fit is somewhat small ($\mathcal{O}(0.1)$) simply because 
we adopt the nonlinear matter power spectra taken from the simulation itself, and we do not worry about 
unrealistic overfitting issues here. Since our $b_{3\rm nl}$ measurements look convergent up to 
a certain $k_{\rm max}$ but start to vary at larger $k_{\rm max}$, it is difficult to define 
the reliable range of the bias model which could depends on both redshift and halo mass. 
We here simply and conservatively quote the measured values of $b_{3{\rm nl}}$ at 
$k_{\rm max}= 0.08,\,0.1$, and 0.125 at $z=0,\,0.5$, and 1, respectively, which roughly 
correspond to valid range of the standard perturbation theory \cite{Carlson:2009kr,Nishimichi:2009uq}. \par 

We quantify contribution of the third-order nonlocal bias term to each power spectrum 
in Figs.~\ref{fig: estimate b3nl vs Pk z=1}, \ref{fig: estimate b3nl vs Pk z=0.5}, and 
\ref{fig: estimate b3nl vs Pk z=0} for $z=1$, 0.5 and 0, respectively. 
We plot $P^{\rm hm}_{X}(k)/(b_{1}P^{\rm mm}_{X}(k))-1$ with 
$X$ being `00' or `01', which manifests deviation from the linear bias term. 
The blue lines show the nonlinear contributions from local term only, i.e., the $b_{2}$ term, 
while the green lines show ones from second-order local plus nonlocal terms, i.e., the $b_{2}$ term plus 
the $b_{s^{2}}$ one. Our best-fitting results including the third-order nonlocal bias term is shown by
the red curves.  Clearly seen from the figures, the local bias model cannot explain the simulated halo-matter spectra, 
and even including second-order nonlocal bias terms does not drastically help in general.
Meanwhile, adding the third-order nonlocal bias term can apparently explain the power 
spectra very well. Within the valid range, the fractional differences between the simulated 
and fitted spectra are typically at a few percent level. 
This result is already expected from the behavior of the PT terms in Fig.~\ref{fig: PT comparison}. 
The reason why we obtain negative values of $b_{\rm 3nl}$ at mass bin IV at $z=0.5$ 
is obvious from the figures. At these bins, the bispectrum  
prefers large second-order bias parameters, especially $b_{2}$, whose contribution exceed 
the measured halo-matter power spectra. Therefore the negative $b_{\rm 3nl}$ is necessary 
to compensate with the second-order terms.
\par  

Given the fact that the contribution of the second-order terms are generally lower than that of 
the third-order nonlocal term, it is interesting to see to what extent we can simultaneously 
constrain four bias parameters only from the power spectra, i.e., without help of 
information on the second-order bias parameters from the bispectrum. 
We often encounter a similar situation in analyzing the actual galaxy survey if we only have 
the power spectrum measurement available, although we focus on the unobservable halo-matter 
power spectra throughout this work. 
Also, it is interesting to separate the information of the power spectrum out of 
that of the bispectrum and to understand the parameter degeneracy in the PT model. 
Figs.~\ref{fig: Pk only z0mbin1}-\ref{fig: Pk only z05mbin4} show 1D and 2D marginalized 
posterior distribution for constraints on the bias parameters. 
Generally speaking, the third-order nonlocal bias $b_{3\rm nl}$ is well constrained 
even only from the power spectra, while the second-order bias parameters cannot be 
tightly constrained only by the power spectra (green). The second-order nonlocal bias, 
$b_{s^{2}}$ cannot be constrained at all by the power spectra, since the amplitude of 
the $b_{s^{2}}$ term in the power spectra is fairly small compared to other terms 
as seen in Fig.~\ref{fig: PT comparison}. 
The second-order local bias $b_{2}$ can be constrained by the power spectra, but 
we confirm that the bispectrum is more sensitive to $b_{2}$. 
At low and intermediate mass bins (see Figs.~\ref{fig: Pk only z0mbin1} and 
\ref{fig: Pk only z05mbin2}), the preferred values of $b_{2}$ both from the power 
spectra and the bispectrum are consistent with each other, and hence the resultant 
values of $b_{\rm 3nl}$ in both cases of $P(k)$ and of $P(k)+B(k)$ become consistent 
as well. 
At massive bin (see Fig.~\ref{fig: Pk only z05mbin4}), this story seems a bit different. 
Since the preferred values of $b_{1}$ and $b_{2}$ from the bispectrum at at mass bin 
IV of $z=0.5$ are larger than those from the power spectra, the well-fitting $b_{\rm 3nl}$ 
from the combined case becomes lower than the one only from the power spectrum. 
Equivalently, the $b_{\rm 3nl}$ term become less important at higher mass bins, and 
the $b_{2}$ terms become dominant over the $b_{\rm 3nl}$ term. 
Furthermore, the linear bias value can be constrained solely by the bispectrum and 
its agreement with the power spectrum-only result becomes worse for more massive halos. 
The constraining power of the power spectrum on $b_{1}$ is weaker than  
what can be found in the literature. This is a consequence of an anti-correlation 
between $b_{1}$ and $b_{\rm 3nl}$. 
This fact implies that the $b_{\rm 3nl}$ term becomes important at fairly large scales, 
$k\ltsim 0.1\,h$/Mpc and has a non-negligible impact on 
determination of the linear bias value. We here do not investigate how these correlations 
affect estimation of cosmological parameters of interest and will be addressed in future work. 

Finally, we make a comparison between our $b_{3{\rm nl}}$ measurements with a theoretical 
prediction in order to make sure if our results are physically expected. For this purpose, we 
compare our results with the prediction, Eq.~(\ref{eq: LLB b3nl prediction}), in the simple 
coevolution picture (or the local Lagrangian bias) 
as discussed in \S.~\ref{subsec: co-evolution}. The cyan horizontal (vertical) line in each panel 
of Fig.~\ref{fig: estimate b3nl} (Figs.~\ref{fig: Pk only z0mbin1}-\ref{fig: Pk only z05mbin4}) 
is already drawn, and the left panel of Fig.~\ref{fig: b1 vs bx} summarizes 
such a comparison which includes both second- and third-order nonlocal bias parameters 
as a function of the linear bias $b_{1}$. Notice that it is not clear if our measured $b_{1}$ truly 
corresponds to $b_{1}^{\rm E}$ (see Eq.~(\ref{eq: LLB b1 prediction})) but we here simply assume 
$b_{1}\simeq b_{1}^{\rm E}$ for simplicity. As clearly seen in Fig.~\ref{fig: b1 vs bx}, overall agreement 
in third-order nonlocal bias is as good as that in second-order, although the agreement is 
apparently not perfect. 
Also, the $b_{\rm 3nl}$ value at mass bin IV of $z=0.5$ exceptionally 
deviates from the coevolution prediction. As we discussed above, however, the value preferred 
from the power spectrum only is more consistent with the coevolution prediction 
(see green dotted line in Fig.\ref{fig: Pk only z05mbin4}). Again, this difference comes from 
the fact that the mass bin IV of $z=0.5$ prefers larger $b_{2}$ which more affects the power spectrum 
and the bispectrum than the nonlocal bias terms. 
There are several sources which could make the prediction different from 
the local Lagrangian bias as we will discuss in the following section. 
However, it is worth mentioning that our $b_{\rm 3nl}$ measurement is not far from 
the coevolution prediction which is one of the simplest physical models one thinks of. 
This fact also implies an evidence of the third-order nonlocal bias term.
In the right panel of Fig.~\ref{fig: b1 vs bx}, we also compare our measurements of 
the second-order local bias $b_{2}$ from the joint fit with theoretical prediction 
that is based on the peak-background split (PBS) with the universal mass function 
(see Appendix.~\ref{sec: PBS-ST} in detail). 
Clearly seen from the figure, the measured $b_{2}$ values are systematically lower 
than the theoretical predictions at fixed $b_{1}$, while the characteristic dependence 
on $b_{1}$ is qualitatively similar. Note that it is a coincidence that two points around 
$b_{1}\sim 4$ look in a perfect agreement with the prediction, since they deviate 
from predictions in $(b_{1},M_{\rm halo})$ or $(b_{2},M_{\rm halo})$ plane. 


\section{Summary and discussion}
\label{sec: summary}
The nonlocality of halo bias is naturally induced by nonlinear gravitational evolution as suggested by recent studies. 
In this paper we study how well the PT model including nonlocal bias effects perform against the halo statistics 
simulated in $N$-body simulations in a $\Lambda$CDM universe. For this purpose we first revisit 
the bias renormalization scheme proposed by \cite{McDonald:2009lr} and show that, while the leading-order 
bispectrum requires only one second-order nonlocal bias term, $b_{s^{2}}$ (see Eq.~(\ref{eq: B^hmm_000})), 
the power spectrum at next-to-leading order demands an additional nonlocal bias term, $b_{\rm 3nl}$, 
associated with the third-order perturbation (see Eq.~(\ref{eq: dd_fin})). 
We extend this model to the power spectrum between halo density and matter momentum, and show that  
there is an exactly same correction of the $b_{\rm 3nl}$ term in this case as well (see Eq.~(\ref{eq: P^hm_01 fin})). 
The fact that we only need one additional nonlocal bias even at third order may sound surprising. However, 
we argue that this is actually expected since the symmetry in gravity basically restricts the allowed functional 
form of nonlocal terms. In order to confirm this, we show that the PT kernel in the $b_{3\rm nl}$ term 
exactly matches to the solution in a simple coevolution picture between dark matter and halo fluids 
(see discussion in \S.~\ref{subsec: co-evolution}). Also, this circumstance evidence becomes even much 
clearer when the solution in coevolution picture is found out to be consistent with that derived by the Galileon 
invariants (see Appendix~\ref{sec: Chuen-Chan} and similar discussions can be found 
in \cite{Chuen-Chan:2012fk}). Also we note that Ref.~\cite{Assassi:2014lr} readdress the bias renormalization 
in terms of the Effective Field Theory (EFT) language and drew the same conclusion. \par 

Then an inevitable question is whether the model can really well describe the halo statistics in $N$-body simulations. 
In particular, can the model simultaneously explain the halo power spectrum and the bispectrum which is never achieved 
in a simple local bias model \cite{Pollack:2012kh,Pollack:2013fk}? To answer this question, we fit the model including 
nonlocal bias terms to the power spectrum, combined with the bispectrum. 
We here focus on the cross spectra between halos and dark matter which are free from issues such as halo exclusion 
\cite{Baldauf:2013lr} or stochasticity \cite{Desjacques:2010qy,Hamaus:2012mz,Sato:2013fk}. 
A novel thing in this work is to compare the model for the 
cross spectrum between halo density and matter momentum. The momentum power spectrum is the essential 
ingredient in predicting RSD in the so-called Distribution Function approach as initiated by \cite{Seljak:2011dw}. 
We show that the fitting values of $b_{3\rm nl}$ up to a certain $k_{\rm max}$ (typically, $k_{\rm max}\ltsim 0.1h/$Mpc) 
are in a good agreement for two power spectra, saying that the model seems to be able to explain the power spectra 
and the bispectrum at the same time. We also explore if the derived values of $b_{3\rm nl}$ are consistent with predictions 
from the simple coevolution picture (or the local Lagrangian bias) and find as a good agreement as second order tidal 
bias, $b_{s^{2}}$, although the agreement is not perfect.\par 

Our study does indicate that there is no reason to ignore the nonlocal bias terms in predicting the halo statistics 
at a high accuracy. In fact there have been some evidences which suggests the third-order nonlocal bias term 
should be included in the literature. For instance, Refs.~\cite{Blazek:2013ys,Vlah:2013qy} find that they need to 
introduce two different second-order bias parameters for the halo density-density, $b_{2}^{00}$ 
and for the halo density-momentum, $b_{2}^{01}$ to explain the simulated halo power spectrum. 
As is already discussed in \cite{Vlah:2013qy}, the difference can be, at least qualitatively,  
explained by the $b_{3\rm nl}$ term. 
However, we need to be more careful to analyze the halo-halo statistics by properly taking stochasticity 
noise and velocity bias into account. Even though many improvements still need to be considered, 
Ref.~\cite{Beutler:2013kx} apply the model based on our study with nonlocal bias values fixed to be
the coevolution predisctions to the actual galaxy survey data. One of the reasons why it seems to work 
is that the authors primarily focus on the anisotropic clustering signal 
to extract RSD which has larger statistical errors (typically $\sim 10\%$) than the isotropic part (i.e., monopole, 
typically a few \%). Also, additional bias parameters such as the second-order local bias, $b_{2}$, and shot-noise 
like bias, $N$, are conservatively treated as free. In order to extract the shape information from the monopole, 
however, more refined analysis will be required. 
We leave it as our future work and hope to report it elsewhere in the near future.  
Also, there are extensions of the model considered here, which could make the fit and the comparisons better 
and extend to higher wavenumbers. Let us summarize the key assumptions of our simple coevolution picture again: 
local Lagrangian initial conditions, a continuity equation for the halo fluid, and no velocity bias. 
The local Lagrangian initial conditions will be likely to be modified by the presence of initial $b_{s^2}$ and $b_{\rm 3nl}$ 
due to e.g. ellipsoidal collapse \cite{Sheth:2013lr}. Since we are fitting for the amplitude of these terms, our inferred 
values are a combination of the initial and dynamical contributions and the agreement with 
the $b_{s^2}$, $b_{\rm 3nl}\propto (b_1-1)$ scaling tells us that the initial contributions are expected be fairly small. 
Furthermore, the peak model \cite{Bardeen:1986yq} and studies of proto-haloes in $N$-body simulations 
\cite{Elia:2011ds} suggest that there is an initial scale dependent linear bias $b_1(k)$, which arises from the 
dependence of the peak clustering on second derivatives of the field (see \cite{Desjacques:2008ry,Schmidt:2012fk} 
for a rigorous derivation, and also see \cite{Desjacques:2010fk,Matsubara:2011qy} for subsequent gravitational 
evolution taken into account). The same calculation also reveals that proto-halo velocities are likely statistically 
biased on small scales with respect to the underlying matter.
Simple considerations for the motions of peaks suggest that these effects are damped by gravitational evolution 
at linear level. In absence of a well tested description of these effects at the non-linear level, we refrain 
from taking these effects into account.\par 

Let us make a comment on a related work in Ref.~\cite{Biagetti:2014xx}. 
The authors in Ref.~\cite{Biagetti:2014xx} predict the halo-matter power spectrum by fixing 
bias parameters: the local bias parameters, $b_{1}$ and $b_{2}$, are calculated by 
the peak-background split combined with the non-universal mass function 
in the excursion set peak formalism \cite{Musso:2012uq,Paranjape:2013dd,Paranjape:2013qy}, 
and the nonlocal bias parameters are fixed with the results of the local Lagrangian bias 
(i.e., the same as our \S.~\ref{subsec: co-evolution}). In addition, a crucial difference is 
that they include $k^{2}$-type bias term based on the peak formalism. 
They claim that their predictions are in a good agreement with simulations including cosmology 
with massive neutrinos \cite{Villaescusa-Navarro:2014qy} at a few percent level, 
and the $k^2$-type term, which we ignored, 
is important. This sounds contradictory to our results, but we argue it is not actually the case:
in Fig.~\ref{fig: b1 vs bx}, we observe that our preferred $b_{\rm 3nl}$ values are sometimes 
larger than the coevolution prediction. This means that it is necessary to introduce another 
component (like $k^{2}$ term) to well fit to the simulated data, if the $b_{\rm 3nl}$ is fixed to 
the coevolution prediction. In addition, as is already pointed out in \cite{Baldauf:2012lr} and is shown 
in Fig.~\ref{fig: b1 vs bx}, the preferred values of the second-order bias, $b_{2}$ and $b_{s^{2}}$, 
are not in a perfect agreement with the simple theoretical predictions. 
It is interesting to clarify whether the source of this discrepancy comes truly from the $k^{2}$ bias 
or something different, which would require more careful investigation. 

As a final remark, we make a comment on future directions of our study. 
As shown in Fig.~\ref{fig: b1 vs bx}, our measurements suggests a characteristic dependence of 
the higher-order local and nonlocal biases on the linear bias $b_{1}$. 
This fact implies that there would be a possibility that we could model higher-order bias terms 
simply in terms of $b_{1}$ (or the halo mass $M_{\rm halo}$), which is an ultimate goal of modeling 
the halo bias. We believe that our results provide a hint toward a more refined modeling of the nonlinear 
halo bias without any free parameters. 
Another legitimate extension of our study is to investigate if the $b_{3\rm nl}$ term can explain the trispectrum 
simultaneously. However, Ref.~\cite{Assassi:2014lr} shows that there exists an additional nonlocal 
term even in the tree-level trispectrum. In addition, the trispectrum analysis requires a gigantic simulation 
volume to gain ample signal-to-noise ratio. Thus such an analysis would take a rigorous amount of work, 
even though it is straightforward to do.\par 

\begin{acknowledgments}
We would like to thank Roman Scoccimarro, Ravi Sheth , Kwan Chuen Chan, 
and Takahiko Matsubara for useful comments and discussions. 
We acknowledges Vincent Desjacques for correspondence regarding their work \cite{Biagetti:2014xx}.  
SS is supported by a Grant-in-Aid for Young Scientists (Start-up) from the Japan Society 
for the Promotion of Science (JSPS) (No. 25887012). 
TB gratefully acknowledges support from the Institute for Advanced Study 
through the W. M. Keck Foundation Fund.
\end{acknowledgments}

\appendix 

\section{Perturbation Theory basics}
\label{sec: PT basics}

In this appendix we summarize basic equations in perturbation theory. 

\subsection{Matter density}
\label{subsec: PT matter}
A matter density in Fourier space is perturbatively expanded into 
\ba
 \delta_{\rm m}(\bfk) & = & \delta_{0}(\bfk)\nonumber\\
 & & + \int\frac{d^{3}q}{(2\pi)^{3}}\,F^{(2)}_{\rm S}(\bfq,\bfk-\bfq)\delta_{0}(\bfq)\delta_{0}(\bfk-\bfq)\nonumber\\
 & & + \int\frac{d^{3}q_{1}}{(2\pi)^{3}}\frac{d^{3}q_{2}}{(2\pi)^{3}}\,F^{(3)}_{\rm S}(\bfq_{1},\bfq_{2},\bfk-\bfq_{1}-\bfq_{2})
            \delta_{0}(\bfq_{1}) \delta_{0}(\bfq_{2})\delta_{0}(\bfk-\bfq_{1}-\bfq_{2})\nonumber\\
 & &  + \calO({\delta_{0}}^{4}), 
\ea
where $\delta_{0}$ is the linear density perturbation and 
the symmetrized PT kernels are given by
\ba
 F^{(2)}_{\rm S}(\bfq_{1},\bfq_{2}) & = & \frac{1}{2}\left\{F^{(2)}(\bfq_{1},\bfq_{2})+F^{(2)}(\bfq_{2},\bfq_{1})\right\}\nonumber\\
 & = & \frac{5}{7}+\frac{1}{2}\frac{\bfq_{1}\cdot\bfq_{2}}{q_{1}q_{2}}\left(\frac{q_1}{q_2}+\frac{q_2}{q_1}\right)
           +\frac{2}{7}\left(\frac{\bfq_{1}\cdot\bfq_{2}}{q_{1}q_{2}}\right)^{2}, \\
 G^{(2)}_{\rm S}(\bfq_{1},\bfq_{2}) & = &  \frac{3}{7}+\frac{1}{2}\frac{\bfq_{1}\cdot\bfq_{2}}{q_{1}q_{2}}\left(\frac{q_1}{q_2}+\frac{q_2}{q_1}\right)
           +\frac{4}{7}\left(\frac{\bfq_{1}\cdot\bfq_{2}}{q_{1}q_{2}}\right)^{2}, \\
 F^{(3)}_{\rm S}(\bfq_{1},\bfq_{2},\bfq_{3}) & = & \frac{1}{3!}\left\{F^{(3)}(\bfq_{1},\bfq_{2},\bfq_{3})+\,{\rm cyclic}\right\}\nonumber\\
 & = & \frac{1}{6}\left[\frac{7}{9}\frac{\bfq_{123}\cdot\bfq_{3}}{q_{3}^{2}}F^{(2)}_{\rm S}(\bfq_{1},\bfq_{2})+\
 \left\{\frac{7}{9}\frac{\bfq_{123}\cdot(\bfq_{1}+\bfq_{2})}{|\bfq_{1}+\bfq_{2}|^{2}}+
 \frac{2}{9}\frac{q_{123}^{2}\bfq_{3}\cdot(\bfq_{1}+\bfq_{2})}{|\bfq_{1}+\bfq_{2}|^{2}\cdot \bfq_{3}^{2}}\right\}
 G^{(2)}_{\rm S}(\bfq_{1},\bfq_{2})\right]\nonumber\\
 & & + \,{\rm cyclic},\\
 G^{(3)}_{\rm S}(\bfq_{1},\bfq_{2},\bfq_{3})
 & = & \frac{1}{6}\left[\frac{1}{3}\frac{\bfq_{123}\cdot\bfq_{3}}{q_{3}^{2}}F^{(2)}_{\rm S}(\bfq_{1},\bfq_{2})+\
 \left\{\frac{1}{3}\frac{\bfq_{123}\cdot(\bfq_{1}+\bfq_{2})}{|\bfq_{1}+\bfq_{2}|^{2}}+
 \frac{2}{3}\frac{q_{123}^{2}\bfq_{3}\cdot(\bfq_{1}+\bfq_{2})}{|\bfq_{1}+\bfq_{2}|^{2}\cdot \bfq_{3}^{2}}\right\}
 G^{(2)}_{\rm S}(\bfq_{1},\bfq_{2})\right]\nonumber\\
 & & + \,{\rm cyclic},
\ea
where $\bfq_{123}=\bfq_{1}+\bfq_{2}+\bfq_{3}$. The unsymmetrized kernels are given by
\ba
 F^{(2)}(\bfq_{1},\bfq_{2}) & = & \frac{5}{7}\alpha(\bfq_{1},\bfq_{2}) + \frac{2}{7}\beta(\bfq_{1},\bfq_{2}),\\
 G^{(2)}(\bfq_{1},\bfq_{2}) & = & \frac{3}{7}\alpha(\bfq_{1},\bfq_{2}) + \frac{4}{7}\beta(\bfq_{1},\bfq_{2}),\\
 \alpha(\bfq_{1},\bfq_{2}) & = & \frac{(\bfq_{1}+\bfq_{2})\cdot \bfq_{1}}{q_{1}^{2}},\\
 \beta(\bfq_{1},\bfq_{2}) & = & \frac{1}{2}(\bfq_{1}+\bfq_{2})^{2}\frac{\bfq_{1}\cdot \bfq_{2}}{q_{1}^{2}q_{2}^{2}}. 
\ea

\subsection{Biased tracer's density}
\label{subsec: galaxy}
Following an ansatz in McDonald \& Roy (2010) \cite{McDonald:2009lr}, a halo density field 
(or generally biased tracer) is written as 
\ba
 \delta_{\rm h}(\bfx) & = & c_{\delta}\delta_{\rm m}(\bfx)\nonumber\\
 & & + \frac{1}{2}c_{\delta^{2}}\delta_{\rm m}(\bfx)^{2} + \frac{1}{2}c_{s^{2}}s(\bfx)^{2}\nonumber\\
 & & + \frac{1}{3!}c_{\delta^{3}}\delta_{\rm m}(\bfx)^{3} + \frac{1}{2}c_{\delta s^{2}}\delta_{\rm m}(\bfx)s(\bfx)^{2}
 + c_{\psi}\psi(\bfx) + c_{st}s(\bfx)t(\bfx)+\frac{1}{3!}c_{s^{3}}s(\bfx)^{3}\nonumber\\
 & & + c_{\epsilon}\epsilon + \dots, 
 \label{eq: MR ansatz}
\ea
where each independent variable is defined as 
\ba
 s_{ij}(\bfx) &\equiv& \partial_{i} \partial_{j}\phi(\bfx) -\frac{1}{3}\delta^{\rm K}_{ij}\delta_{\rm m}(\bfx) =  
 \left[\partial_{i} \partial_{j}\partial^{-2}-\frac{1}{3}\delta^{\rm K}_{ij}\right]\delta_{\rm m}(\bfx), \\
 t_{ij}(\bfx)  &\equiv& \partial_{i}v_{j} -\frac{1}{3}\delta^{\rm K}_{ij}\theta_{\rm m}(\bfx) - s_{ij}(\bfx) = 
 \left[\partial_{i} \partial_{j}\partial^{-2}-\frac{1}{3}\delta^{\rm K}_{ij}\right][\theta(\bfx)-\delta_{\rm m}(\bfx)],\\
 \psi(\bfx) &\equiv& [\theta(\bfx)-\delta_{\rm m}(\bfx)] -\frac{2}{7}s(\bfx)^{2}+\frac{4}{21}\delta_{\rm m}(\bfx)^{2}. 
\ea
Note that $t_{ij}$ is zero at first order, and $\psi$ is zero up to second order. 
In Fourier space, the halo density contrast is given by 
\ba
 \delta_{\rm h}(\bfk) & = & c_{\delta}\delta_{0}(\bfk)\nonumber\\
 && + c_{\delta}\int\frac{d^{3}q}{(2\pi)^{3}}F^{(2)}_{\rm S}(\bfq,\bfk-\bfq)\delta_{0}(\bfq)\delta_{0}(\bfk-\bfq)\nonumber\\
 && + \frac{1}{2}c_{\delta^{2}}\int\frac{d^{3}q}{(2\pi)^{3}}\delta_{0}(\bfq)\delta_{0}(\bfk-\bfq)\nonumber\\
 && + \frac{1}{2}c_{s^{2}}\int\frac{d^{3}q}{(2\pi)^{3}}\,S^{(2)}(\bfq,\bfk-\bfq)\delta_{0}(\bfq)\delta_{0}(\bfk-\bfq)\nonumber\\
 && + c_{\delta}\int\frac{d^{3}q_{1}}{(2\pi)^{3}}\frac{d^{3}q_{2}}{(2\pi)^{3}}\,F^{(3)}_{\rm S}(\bfq_{1},\bfq_{2},\bfk-\bfq_{1}-\bfq_{2})
            \delta_{0}(\bfq_{1}) \delta_{0}(\bfq_{2})\delta_{0}(\bfk-\bfq_{1}-\bfq_{2})\nonumber\\
 && + c_{\delta^{2}}\int\frac{d^{3}q_{1}}{(2\pi)^{3}}\frac{d^{3}q_{2}}{(2\pi)^{3}}\,F^{(2)}_{\rm S}(\bfq_{1},\bfk-\bfq_{1}-\bfq_{2})
            \delta_{0}(\bfq_{1}) \delta_{0}(\bfq_{2})\delta_{0}(\bfk-\bfq_{1}-\bfq_{2})\nonumber\\
 && + \frac{1}{3!}c_{\delta^{3}}\int\frac{d^{3}q_{1}}{(2\pi)^{3}}\frac{d^{3}q_{2}}{(2\pi)^{3}}\, 
            \delta_{0}(\bfq_{1}) \delta_{0}(\bfq_{2})\delta_{0}(\bfk-\bfq_{1}-\bfq_{2})\nonumber\\
 && + c_{s^{2}}\int\frac{d^{3}q_{1}}{(2\pi)^{3}}\frac{d^{3}q_{2}}{(2\pi)^{3}}\,S^{(2)}(\bfq_{1},\bfk-\bfq_{1})F^{(2)}_{\rm S}(\bfq_{2},\bfk-\bfq_{1}-\bfq_{2})
            \delta_{0}(\bfq_{1}) \delta_{0}(\bfq_{2})\delta_{0}(\bfk-\bfq_{1}-\bfq_{2})\nonumber\\
 &&  + \frac{1}{3!}c_{s^{3}}\int\frac{d^{3}q_{1}}{(2\pi)^{3}}\frac{d^{3}q_{2}}{(2\pi)^{3}} \,S^{(3)}(\bfq_{1},\bfq_{2},\bfk-\bfq_{1}-\bfq_{2})
            \delta_{0}(\bfq_{1}) \delta_{0}(\bfq_{2})\delta_{0}(\bfk-\bfq_{1}-\bfq_{2})\nonumber\\
 &&  + \frac{1}{2}c_{\delta s^{2}}\int\frac{d^{3}q_{1}}{(2\pi)^{3}}\frac{d^{3}q_{2}}{(2\pi)^{3}} \,S^{(2)}(\bfq_{2},\bfk-\bfq_{1}-\bfq_{2})
            \delta_{0}(\bfq_{1}) \delta_{0}(\bfq_{2})\delta_{0}(\bfk-\bfq_{1}-\bfq_{2})\nonumber\\
 &&  + c_{\psi}\int\frac{d^{3}q_{1}}{(2\pi)^{3}}\frac{d^{3}q_{2}}{(2\pi)^{3}}\, \left\{D^{(3)}_{\rm S}(\bfq_{1},\bfq_{2},\bfk-\bfq_{1}-\bfq_{2})
        -2F^{(2)}_{\rm S}(\bfq_{1},\bfk-\bfq_{1}-\bfq_{2})D^{(2)}_{\rm S}(\bfq_{2},\bfk-\bfq_{2})\right\}\nonumber\\
 &&   \;\;\;\;\;\;\;\; \times \delta_{0}(\bfq_{1}) \delta_{0}(\bfq_{2})\delta_{0}(\bfk-\bfq_{1}-\bfq_{2})\nonumber\\
 &&  + 2c_{st}\int\frac{d^{3}q_{1}}{(2\pi)^{3}}\frac{d^{3}q_{2}}{(2\pi)^{3}}\, S^{(2)}(\bfq_{1},\bfk-\bfq_{1})D^{(2)}_{\rm S}(\bfq_{2},\bfq_{1}-\bfq_{2})
        \delta_{0}(\bfq_{1}) \delta_{0}(\bfq_{2})\delta_{0}(\bfk-\bfq_{1}-\bfq_{2}), 
\label{eq: full expression for third order}
\ea
where 
\ba
 S^{(2)}(\bfq_{1},\bfq_{2}) & = & \left(\frac{\bfq_{1}\cdot\bfq_{2}}{q_{1}q_{2}}\right)^{2}-\frac{1}{3},\\
 S^{(3)}(\bfq_{1},\bfq_{2},\bfq_{3}) & = & \frac{(\bfq_{1}\cdot\bfq_{2})(\bfq_{2}\cdot\bfq_{3})(\bfq_{3}\cdot\bfq_{1})}{q_{1}^{2}q_{2}^{2}q_{3}^{2}}
 -\frac{1}{3}\frac{(\bfq_{1}\cdot\bfq_{2})^{2}}{q_{1}^{2}q_{2}^{2}}
 -\frac{1}{3}\frac{(\bfq_{2}\cdot\bfq_{3})^{2}}{q_{2}^{2}q_{3}^{2}}
 -\frac{1}{3}\frac{(\bfq_{3}\cdot\bfq_{1})^{2}}{q_{3}^{2}q_{1}^{2}} + \frac{2}{9},\\
 D^{(N)} & \equiv & G^{(N)}-F^{(N)}. 
\ea

\subsection{Distribution function approach}
\label{subsec: DF approach}
In the Distribution Function approach to model the redshift-space distortion proposed 
in Ref.~\cite{Seljak:2011dw}, the redshift-space 
power spectrum, $P^{\rm S}({\bf k})$, is expanded into infinite sum of momentum power spectrum, 
\ba
 P^{\rm S}(\bfk) = \sum_{LL'}\frac{(-1)^{L'}}{L!L'!}(ik_{\parallel})^{L+L'}P_{LL'}({\bfk}), 
\ea
where the momentum and its power spectrum are defined by
\ba
 && T_{\parallel}^{L}(\bfx) \equiv \left\{1+\delta(\bfx)\right\}v_{\parallel}(\bfx)^{L},\\
 && P_{LL'}({\bfk})(2\pi)^{3}\delta_{D}(\bfk+\bfk') \equiv \langle T_{\parallel}^{L}(\bfk)T_{\parallel}^{L'}(\bfk') \rangle.
\ea
Note that the velocity is defined in units of the Hubble velocity, and we define the velocity dispersion $\theta$ so that 
$\delta=\theta$ in linear regime. The velocity divergence $\theta$ is written in Fourier space as 
\be
 v_{\parallel}(\bfk) = -if\frac{k_{\parallel}}{k^{2}}\theta(\bfk).
\ee

\subsection{Halo-halo power spectrum}
\label{subsec: halo-halo}
The auto power spectrum of halo is similarly given by 
\ba
 P^{\rm hh}_{\,00}(k) & = & 
  b_{1}^{2}P^{\rm NL}_{\delta\delta}(k) + 2b_{1}b_{2}P_{b2,\delta}(k) 
 + 2b_{1}b_{s^{2}}P_{bs2,\delta}(k) + 2b_{1}b_{3{\rm nl}}\,\sigma^{2}_{3}(k)P(k)\nonumber\\
 && \ \ \ \
 +\, b_{2}^{2}P_{b22}(k) + 2b_{2}b_{s^{2}}P_{b2s2}(k) 
 + b_{s^2}^{2}P_{s22}(k)
 +N, 
 \label{eq: hh_00_fin}
\ea
where
\ba
 P_{b22}(k) & \equiv & \frac{1}{2}\int\frac{d^{3}q}{(2\pi)^{3}}\,P(q)
 \left\{P(|\bfk-\bfq|)-P(q)\right\},\\
 P_{b2s2}(k) & \equiv & \frac{1}{2}\int\frac{d^{3}q}{(2\pi)^{3}}\,P(q)
 \left\{P(|\bfk-\bfq|)S^{(2)}(\bfq,\bfk-\bfq)-\frac{2}{3}P(q)\right\},\\
 P_{bs22}(k) & \equiv & \frac{1}{2}\int\frac{d^{3}q}{(2\pi)^{3}}\,P(q)
 \left\{P(|\bfk-\bfq|)S^{(2)}(\bfq,\bfk-\bfq)^{2}-\frac{4}{9}P(q)\right\}. 
\ea
Here we subtract the constant terms like $\int d^{3}q\,P(q)^{2}$ to keep nonlinear corrections vanishing
in the limit of $k\to 0$. 
Also, cross spectrum between halo density and halo momentum is given by 
\ba
 P^{\rm hh}_{\,01}(k) & = & b_{1}f
 \left\{P^{\rm NL}_{\delta\theta}(k) + B_{b1}(k)\right\} +b_{1}(b_{1}-1)B_{b1}(k)
 + b_{2}\left\{P_{b2,\theta}(k) + b_{1}B_{b2}(k)\right\}\nonumber\\
 && \ \ \ \ \ \ 
 + b_{s^{2}}\left\{P_{bs2,\theta}(k) + b_{1}B_{bs2}(k)\right\} 
 + b_{3{\rm nl}}\,\sigma^{2}_{3}(k)P(k). 
 \label{eq: P^hh_01 fin}
\ea

\section{Fitting bias parameters only against the bispectrum}
\label{sec: bispectrum only}

\begin{figure}[t]
\begin{center}
\includegraphics[width=0.48\textwidth]{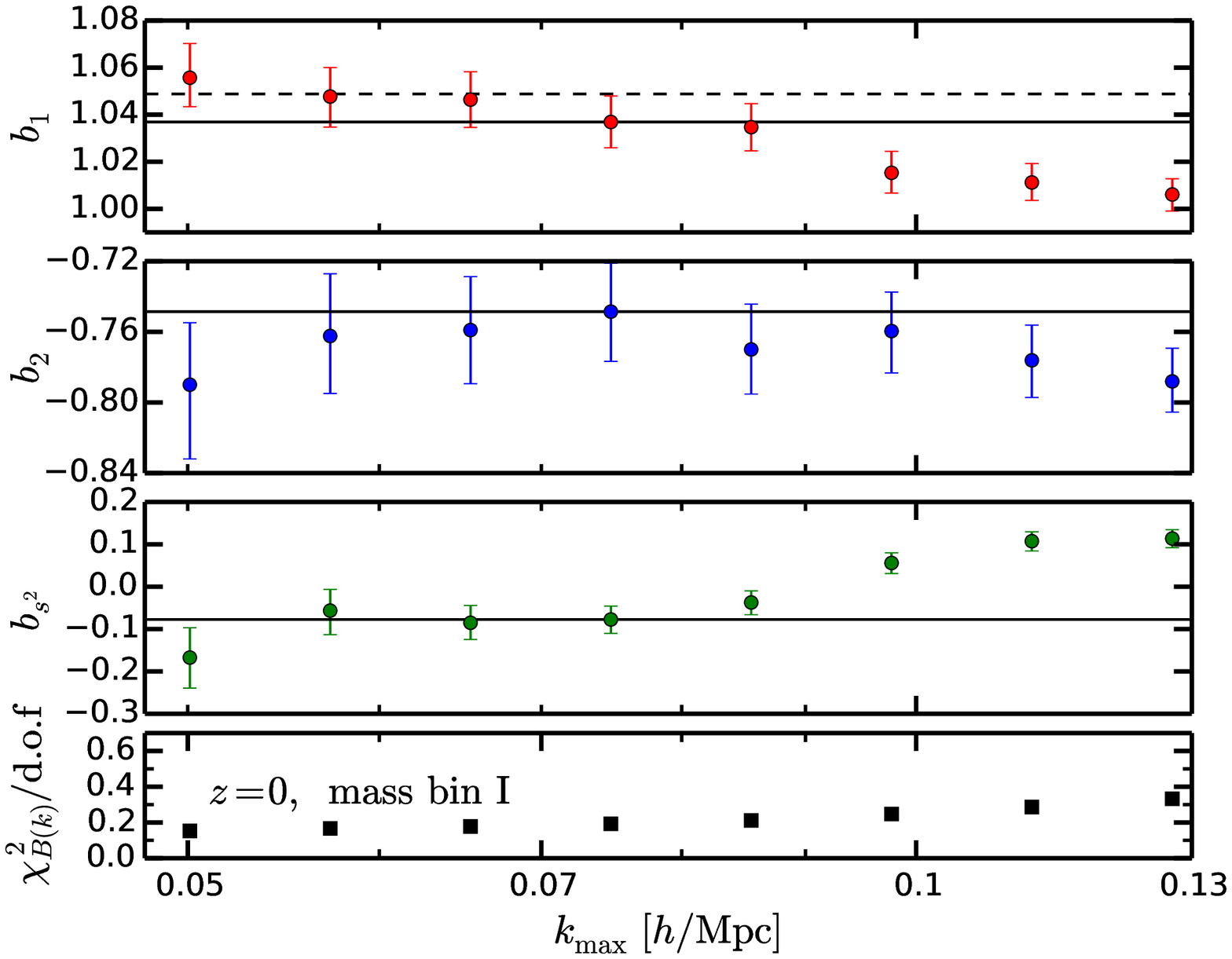}
\includegraphics[width=0.48\textwidth]{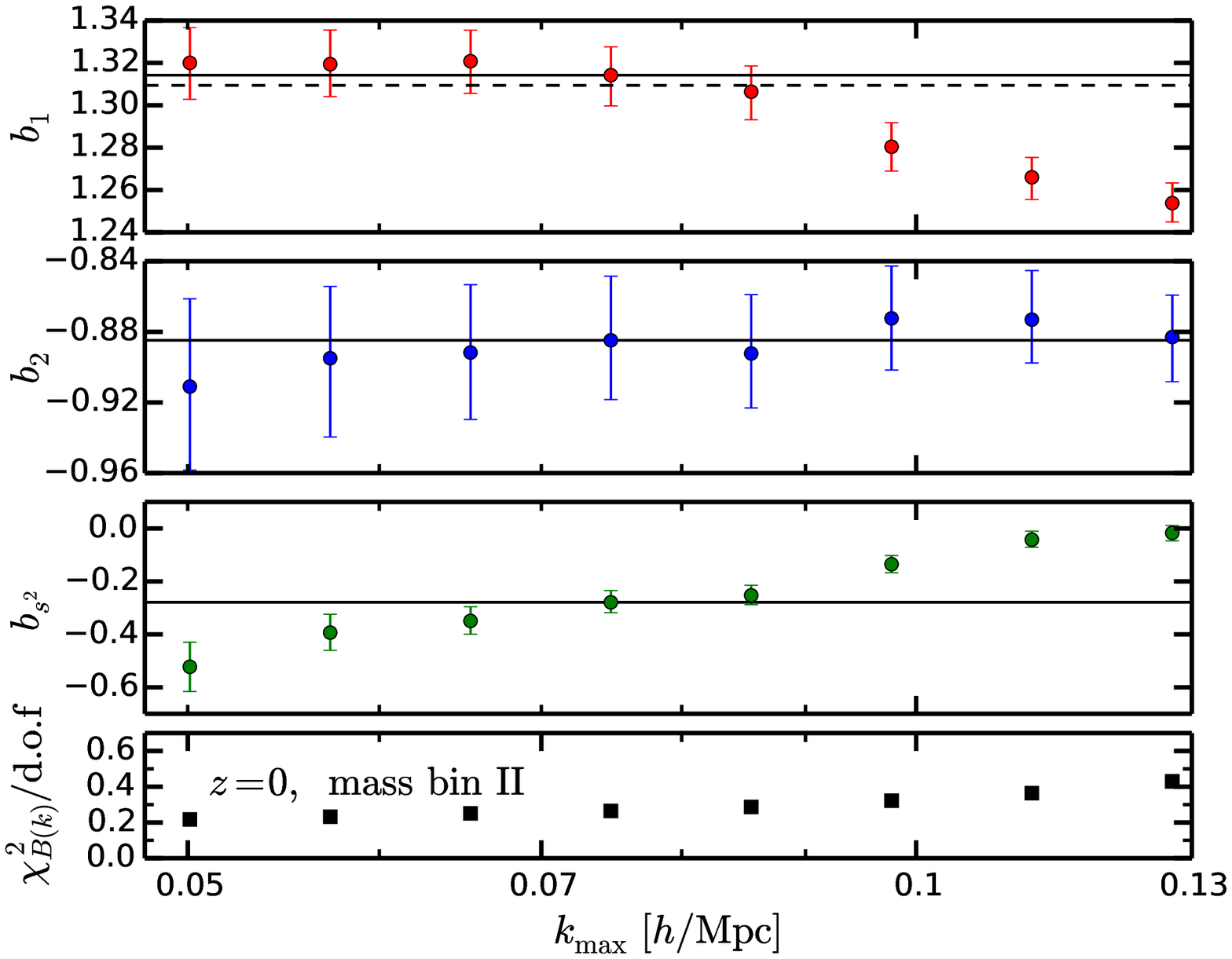}\\
\includegraphics[width=0.48\textwidth]{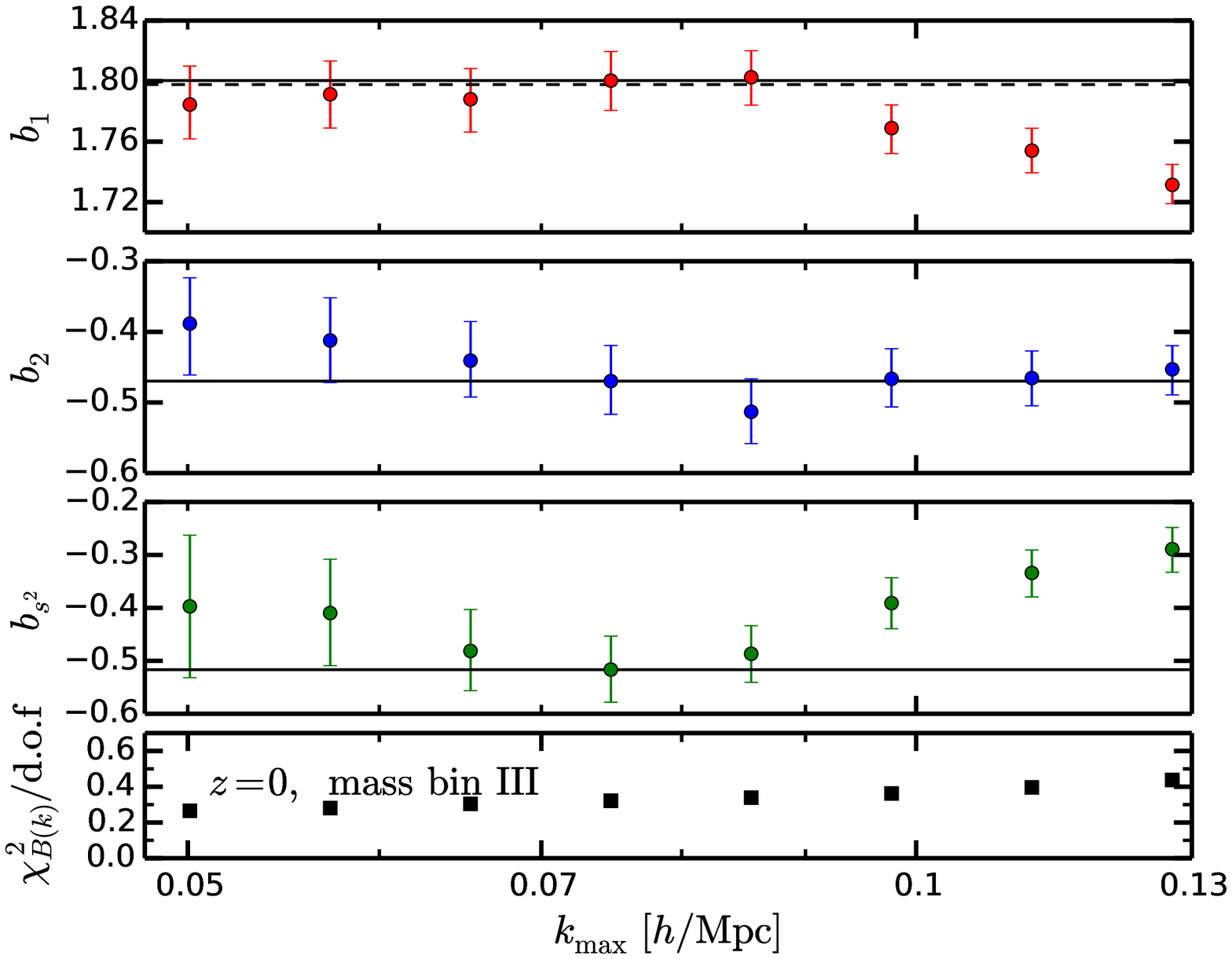}
\includegraphics[width=0.48\textwidth]{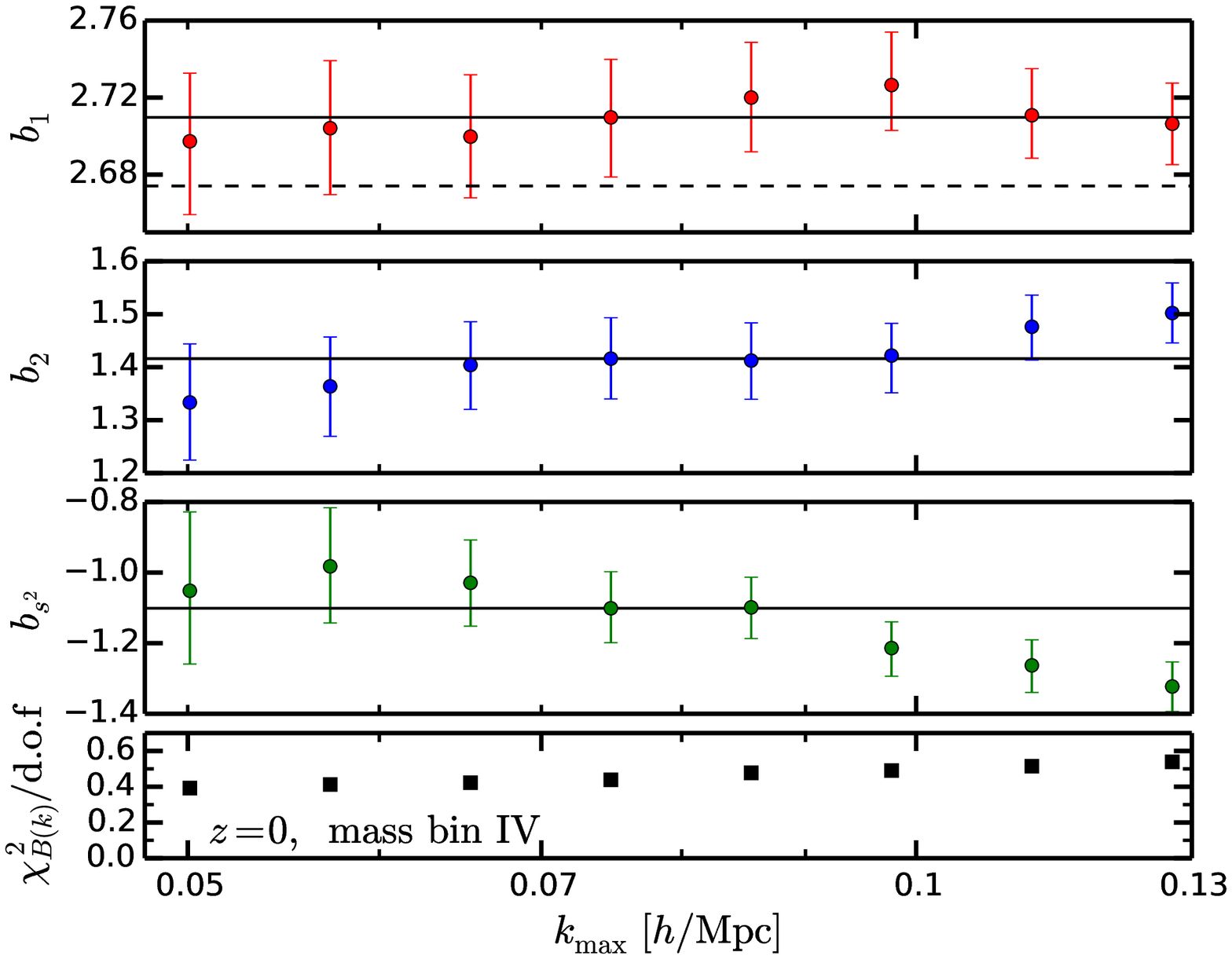}
\end{center}
\vspace*{-2em}
\caption{ 
The best-fitting values of the bias parameters only from the bispectrum at $z=0$ 
as a function $k_{\rm max}$. Our fiducial choice at $k_{{\rm max},B(k)}=0.065h/$Mpc 
is highlighted with a {\it black solid} line. As a reference, the value of $b_{1}$ preferred 
by joint fitting with the power spectrum is indicated with a {\it black dashed} line. 
}  
\label{fig: bkonly bias-kmax z=0}
\end{figure}

\begin{figure}[t]
\begin{center}
\includegraphics[width=0.48\textwidth]{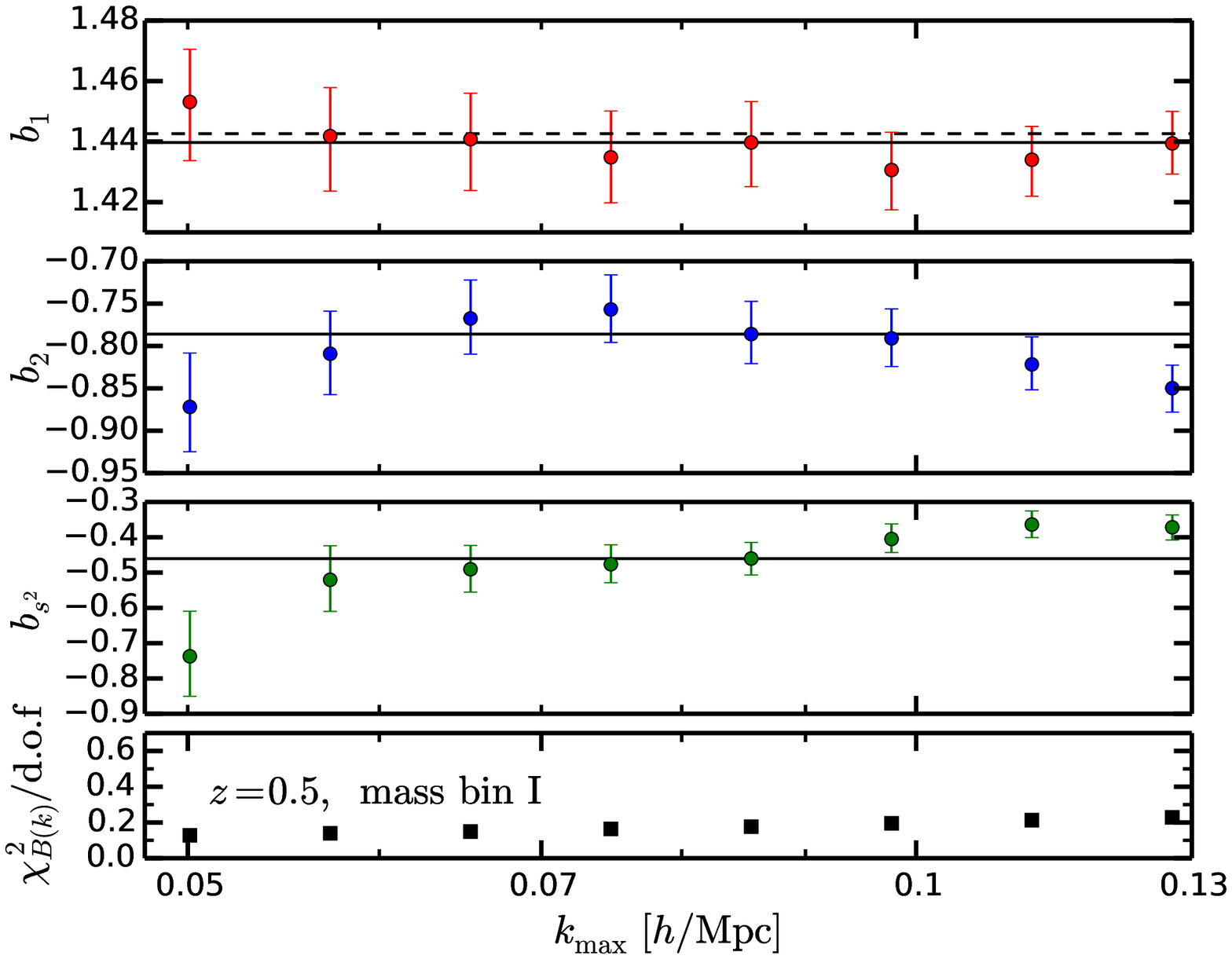}
\includegraphics[width=0.48\textwidth]{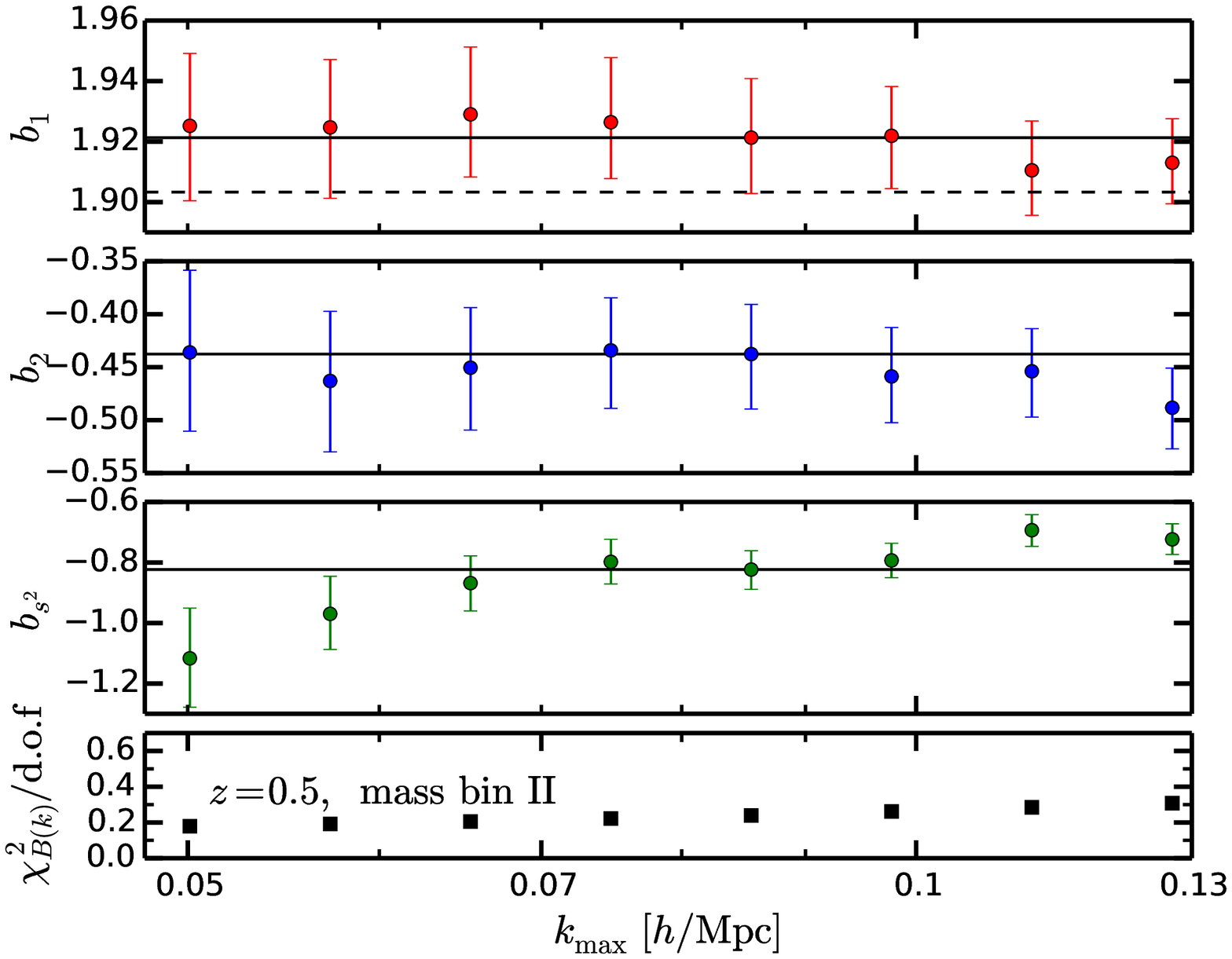}\\
\includegraphics[width=0.48\textwidth]{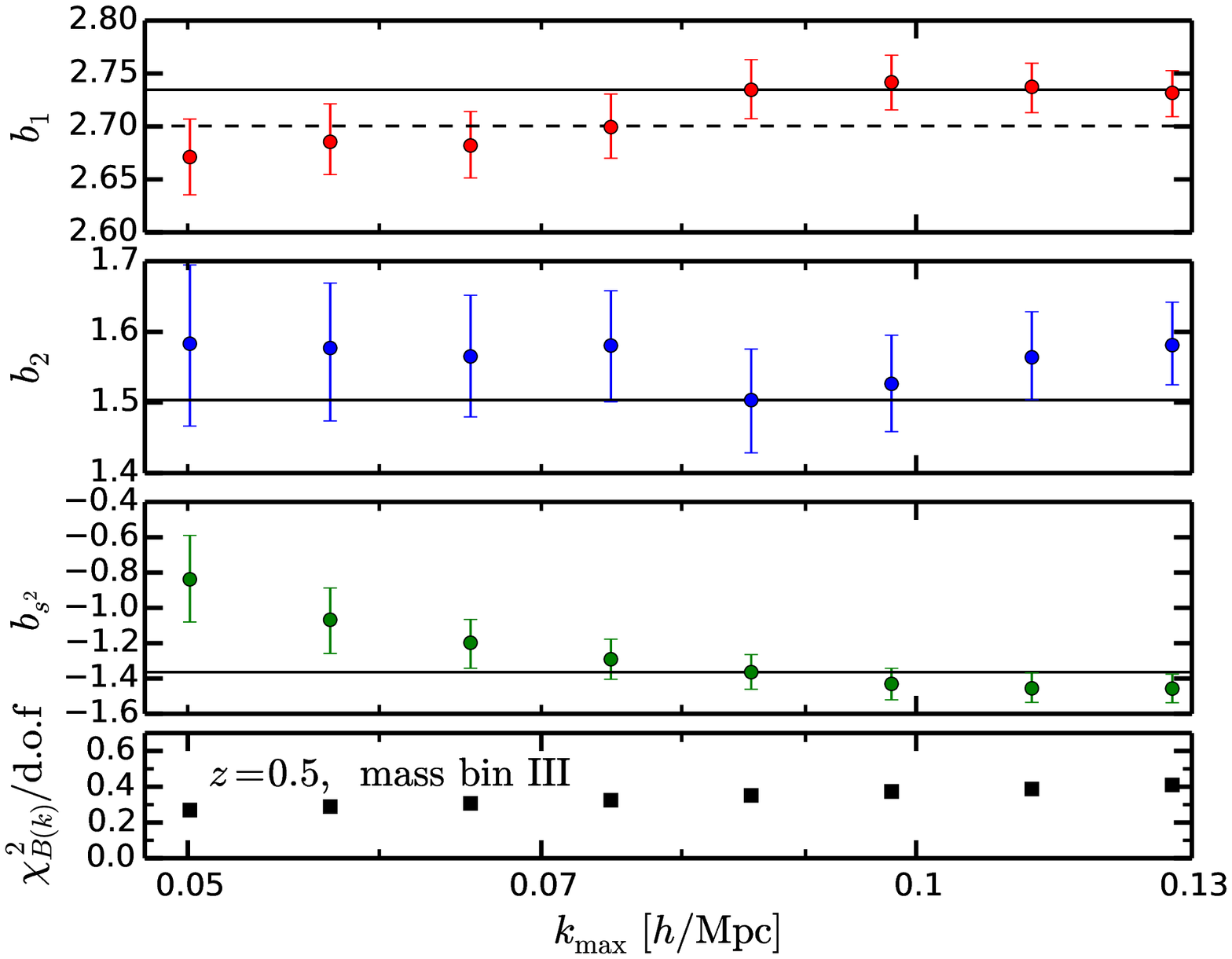}
\includegraphics[width=0.48\textwidth]{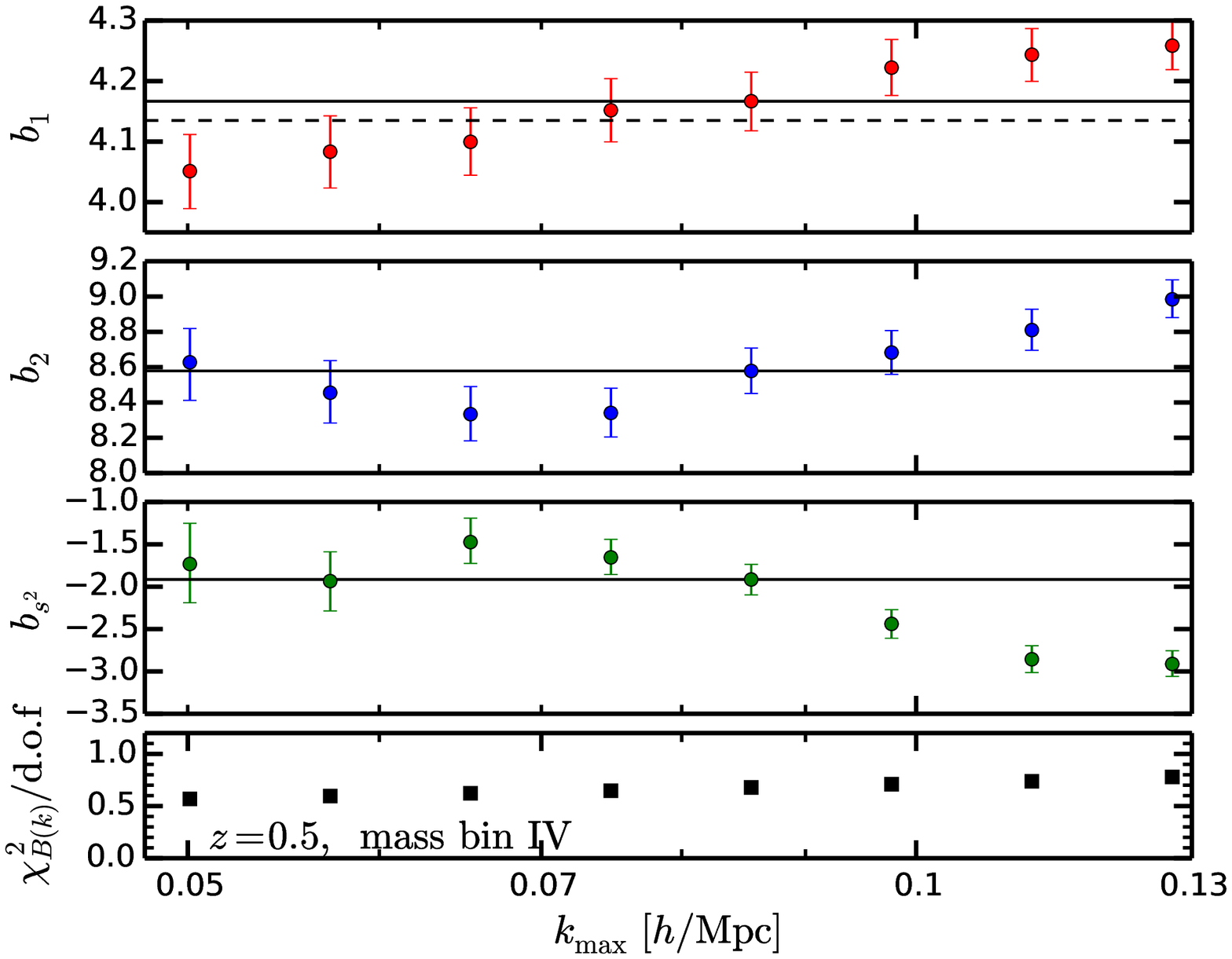}
\end{center}
\vspace*{-2em}
\caption{ 
Same as Fig.~\ref{fig: bkonly bias-kmax z=0}, but at $z=0.5$. 
}  
\label{fig: bkonly bias-kmax z=0.5}
\end{figure}

\begin{figure}[t]
\begin{center}
\includegraphics[width=0.48\textwidth]{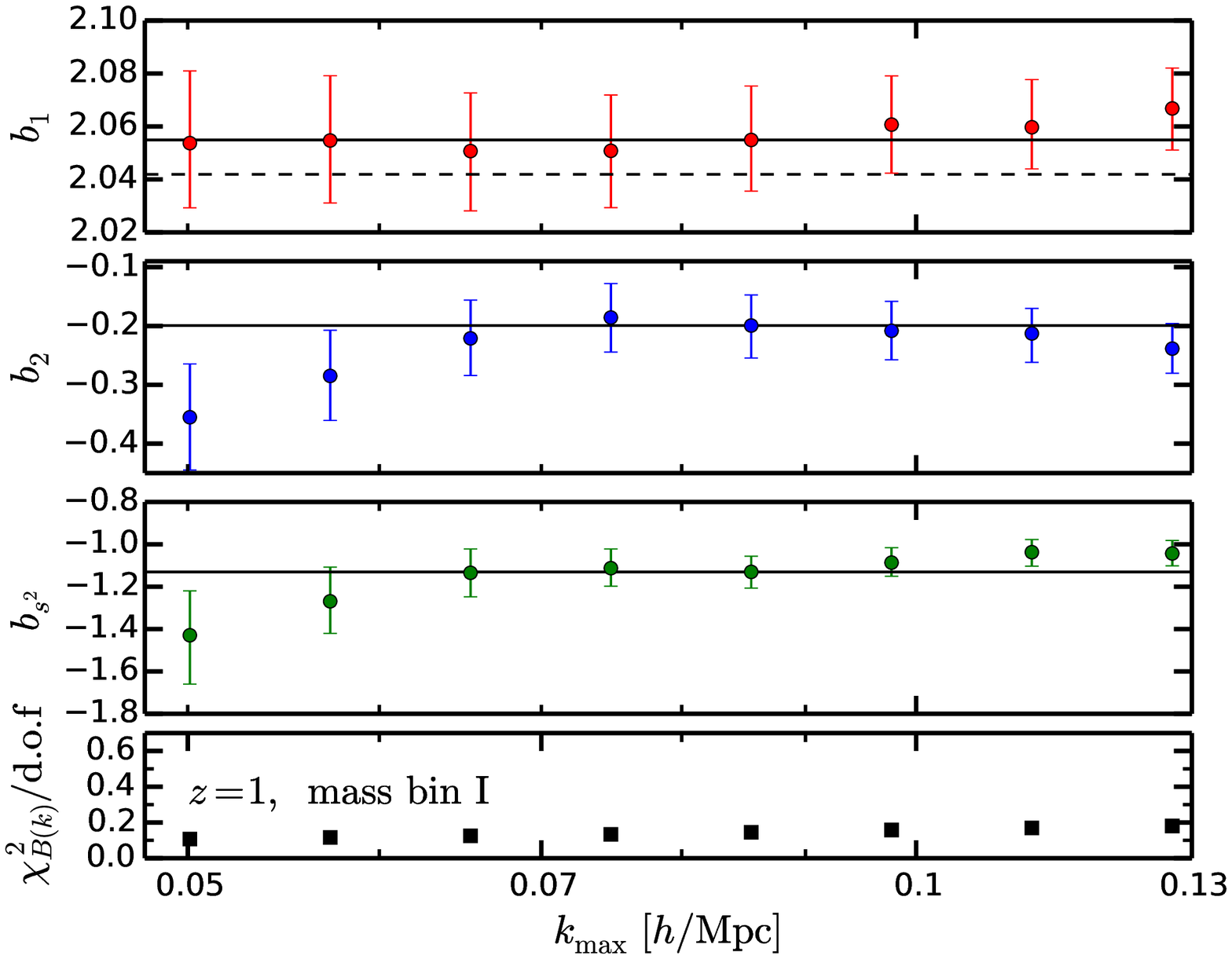}
\includegraphics[width=0.48\textwidth]{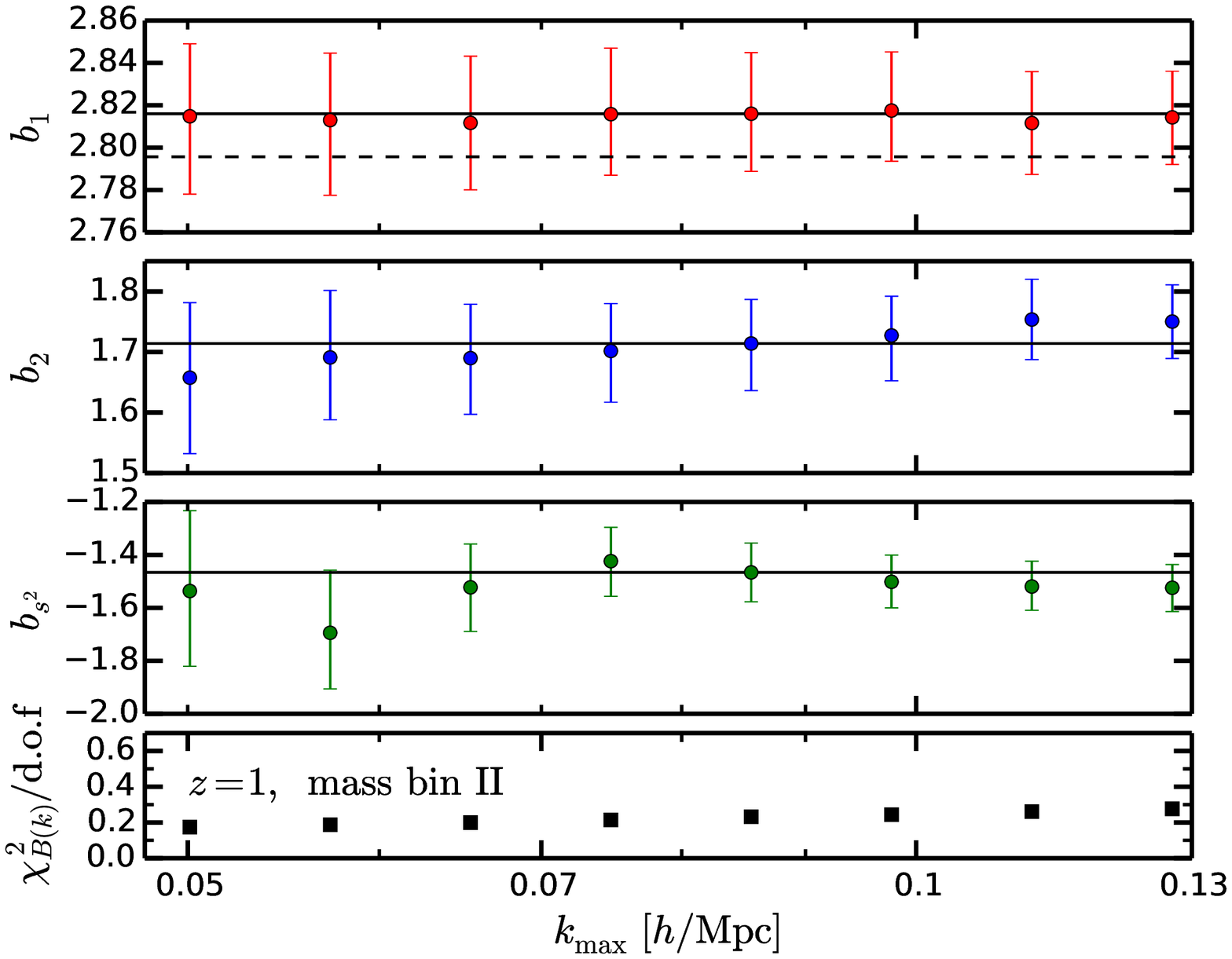}\\
\includegraphics[width=0.48\textwidth]{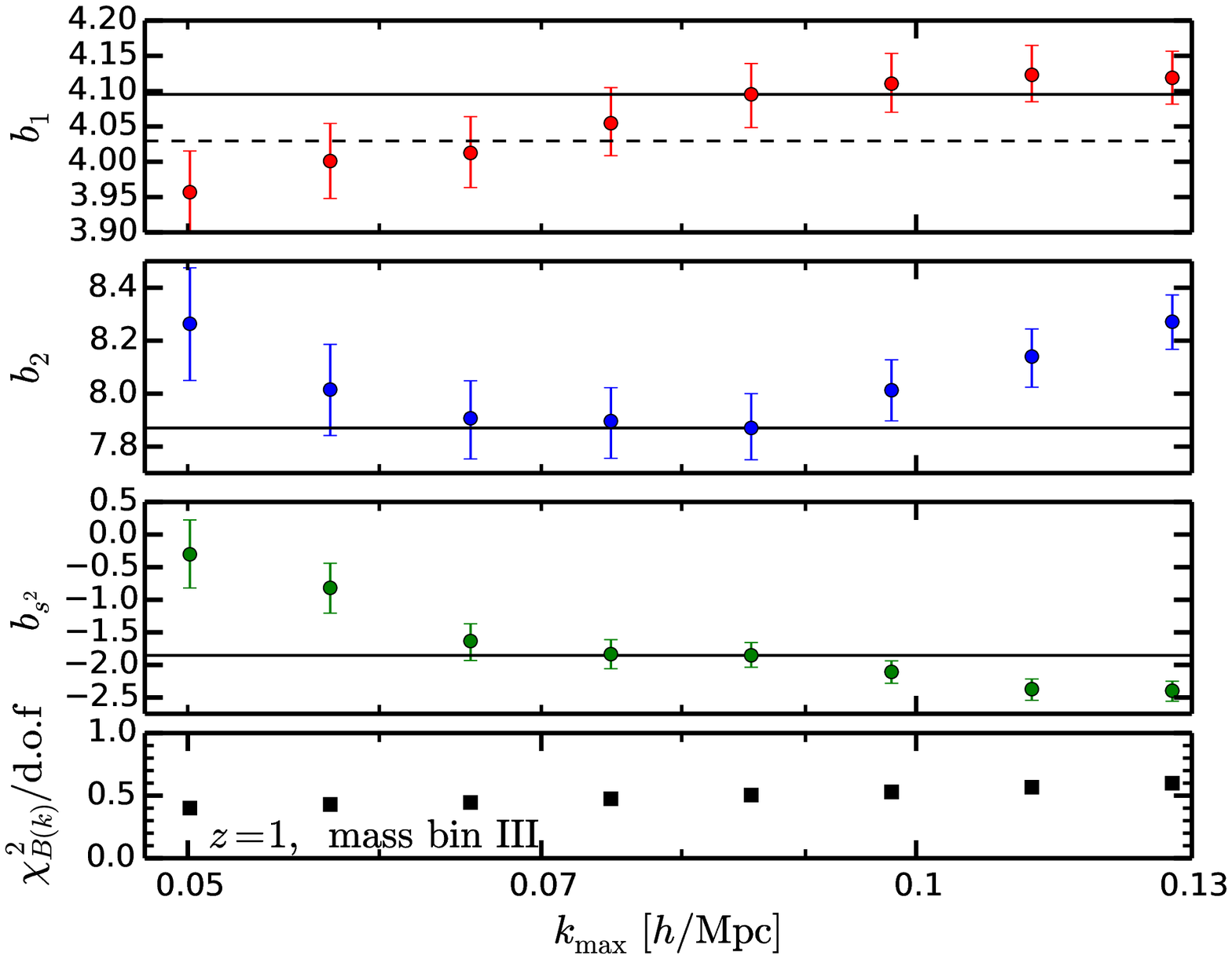}
\end{center}
\vspace*{-2em}
\caption{ 
Same as Fig.~\ref{fig: bkonly bias-kmax z=0}, but at $z=1$. 
}  
\label{fig: bkonly bias-kmax z=1}
\end{figure}

As we discussed in \S.~\ref{subsec: B^hmm_000}, the bispectrum is useful to access the second-order 
bias parameters, since the tree-level bispectrum depends only on the bias parameters up to second order. 
In other words, however, it is necessary to carefully investigate the valid range of the tree-level bispectrum. 
Here we show the fitting results using the bispectrum alone in our simulation. A set of free parameters is 
$(b_{1},b_{2},b_{s^{2}})$ in this case. Note that this analysis is slightly different from that in previous work 
\cite{Baldauf:2012lr}: we vary $b_{1}$ as a free parameter, while the authors in \cite{Baldauf:2012lr} fixed 
the value of $b_{1}$ taken from the halo-matter power spectrum. Since we intend to combine the power 
spectrum with the bispectrum and we have already seen that there exists an anti-correlation between 
$b_{1}$ and $b_{\rm 3nl}$ in the joint fit, it is helpful to isolate the information only from the bispectrum.\par 

Figs.~\ref{fig: bkonly bias-kmax z=0}-\ref{fig: bkonly bias-kmax z=1} show bias parameters derived 
at $z=0$, 0.5 and 1, respectively, from our MCMC fitting as a function of $k_{\rm max}$. 
As found in \cite{Baldauf:2012lr}, we see non-zero 
second-order tidal bias $b_{s^{2}}$ for a variety of halo mass bins and redshifts. 
In addition the figures show that larger $k_{\rm max}$ results in general deviate more from low $k_{\rm max}$ 
ones with higher $\chi^{2}$ values, implying the PT model certainly breaks down at such small scales. 
Based upon these considerations we choose the valid range of the maximum wavenumber in the bispectrum in 
a redshift-dependent way: $k_{{\rm max},B(k)}=0.065 (0.075) h$/Mpc at $z=0$ ($z=0.5$ or $1$). 
Note this choice is fully consistent with the $z=0$ result in Ref.~\cite{Baldauf:2012lr}. 
Interestingly, this is achieved without adding information on the linear bias $b_{1}$ from the power spectrum. 
In fact the preferred values of $b_{1}$ only from the bispectrum tend to more deviate from ones in the 
joint-fit results at higher mass bins at higher redshift.  This issue is also addressed in 
Figs.~\ref{fig: Pk only z0mbin1}-\ref{fig: Pk only z05mbin4}. When the preferred $b_{1}$ value from 
the bispectrum differs from that from the power spectrum, the bias values presented here could be different 
from those exhibited in the main text.

\section{Consistency check with the Galileon invariant approach}
\label{sec: Chuen-Chan}
In \S.~\ref{subsec: co-evolution} we derived solutions up to third order for the simple co-evolution equations 
of dark matter and halo fluids starting from initial condition with purely local bias (i.e., local Lagrangian bias). 
As a matter of fact such solutions have been already derived in Ref.~\cite{Chuen-Chan:2012fk} 
\footnote{Ref. \cite{Taruya:2000uq} 
has already derived such solutions in an exactly same way with ours, but investigated nonlocal terms in terms of 
stochastic bias in the halo-halo power spectrum}, 
but the authors took a different route which is based on Galileon symmetry in gravity. 
In this appendix we review the Galileon invariant approach 
and check that this approach is perfectly consistent with ours as expected.\par 

In the Lagrangian picture, gravitational evolution of displacement field is solely governed by the velocity potential, 
$\Phi_{\rm v}$, defined by $\theta=\nabla^{2} \Phi_{\rm v}$. Since the halo distribution is a scalar under 
translations and rotation in three dimensional space, it should be written down in terms of scalar invariants of 
$\nabla_{i}\nabla_{j}\Phi_{\rm v}$. It is known that there are only three such invariants in three dimensional space, 
so called `Galileons' \cite{Nicolis:2009fk}:
\ba
 \mathcal{G}_{1} & = & \nabla^{2}\Phi_{\rm v},\\
 \mathcal{G}_{2} & = & (\nabla_{i}\nabla_{j}\Phi_{\rm v})^{2}-(\nabla^{2}\Phi_{\rm v})^{2},\\
 \mathcal{G}_{3} & = & (\nabla^{2}\Phi_{\rm v})^{3}
 +2(\nabla_{i}\nabla_{j}\Phi_{\rm v})(\nabla_{j}\nabla_{k}\Phi_{\rm v})(\nabla_{k}\nabla_{i}\Phi_{\rm v})
 -3(\nabla_{i}\nabla_{j}\Phi_{\rm v})^{2}\nabla^{2}\Phi_{\rm v}. 
\ea
Ref.~\cite{Chuen-Chan:2012fk} rewrote the co-evolution equations and derived the solutions 
in terms of Galileons. Since their approach solves the exactly same gravity system, it is quite 
natural to achieve the consistent solution with what we derived in \S.~\ref{subsec: co-evolution}. 
Note that this approach does not hold if there exists a velocity bias, since relative motion 
between dark matter and halo fluids obviously breaks down the Galileon symmetry. 
Let us first begin with the second-order solution which is obtained as 
($\epsilon=1$ and $y\to \infty$ in Eq.~(95) in Ref.~\cite{Chuen-Chan:2012fk})
\be
 \delta^{(2)}_{\rm h} = b_{1}^{\rm E}\delta^{(2)} + \frac{b^{\rm L}_{2}}{2}{\delta^{(1)}}^{2}
 -\frac{2}{7}b^{\rm L}_{1}\mathcal{G}_{2}^{(2)}(\Phi_{\rm v}), 
 \label{eq: Galileon second solution}
\ee
where $\mathcal{G}_{2}^{(2)}(\Phi_{\rm v})$ in Fourier space is 
\be
 \mathcal{G}_{2}^{(2)}(\Phi_{\rm v})(\bfq_{1},\bfq_{2}) = \frac{\left(\bfq_{1}\cdot\bfq_{2}\right)^{2}}{q_{1}^{2}q_{2}^{2}}-1. 
\ee
Thus the Fourier-transformed version of Eq.~(\ref{eq: Galileon second solution}) matches
Eq.~(\ref{eq: co-evolution second solution}). Note that the simple relation between Eulerian and Lagrangian bias, 
$b^{\rm E}_{2}=b^{\rm L}_{2}+(8/21)b^{\rm L}_{1}$ is used. 
Likewise the third-order solution is given by 
($\epsilon=1$ and $y\to \infty$ in Eq.~(99) in Ref.~\cite{Chuen-Chan:2012fk})
\be
 \delta^{(3)}_{\rm h} = b_{1}^{\rm E}\delta^{(3)} + b^{\rm L}_{2}\delta^{(1)}\delta^{\rm L(2)}
 +\left(\frac{b^{\rm L}_{3}}{6}-\frac{b_{2}^{\rm L}}{2}\right){\delta^{(1)}}^{3}
 -\frac{2}{7}b^{\rm L}_{2}\delta^{(1)}\mathcal{G}_{2}^{(2)}(\Phi_{\rm v})
 -\frac{22}{126}b^{\rm L}_{1}\nabla[\mathcal{G}_{2}^{(2)}(\Phi_{\rm v}){\bf v}^{(1)}]
 -\frac{1}{9}b_{1}^{\rm L}\mathcal{G}_{2}^{(3)}(\Phi_{\rm v}),
\ee
where $\delta^{\rm L(2)}={\delta^{(1)}}^{2}+{\bf v}^{(1)}\cdot\nabla\delta^{(1)}+2\mathcal{G}^{(2)}_{2}/7$, 
and the (unsymmetrized) third-order part of the second-order Galileon is written in Fourier space as 
\ba
 \mathcal{G}_{2}^{(3)}(\Phi_{\rm v})(\bfq_{1},\bfq_{2},\bfq_{3})
 &=&2\left[
 \frac{\{\bfq_{1}\cdot(\bfq_{2}+\bfq_{3})\}^{2}}{q_{1}^{2}}\left\{
 \frac{1}{2}\frac{\bfq_{2}\cdot\bfq_{3}}{q_{2}^{2}q_{3}^{2}}-\frac{3}{7}\frac{1}{(\bfq_{2}+\bfq_{3})^{2}}
 \left[\frac{(\bfq_{2}\cdot\bfq_{3})^{2}}{q_{2}^{2}q_{3}^{2}}-1
 \right]
 \right\}\right.\nonumber\\
 &&\,\,\, 
 - (\bfq_{2}+\bfq_{3})^{2} \left\{
 \frac{1}{2}\frac{\bfq_{2}\cdot\bfq_{3}}{q_{2}^{2}q_{3}^{2}}-\frac{3}{7}\frac{1}{(\bfq_{2}+\bfq_{3})^{2}}
 \left[\frac{(\bfq_{2}\cdot\bfq_{3})^{2}}{q_{2}^{2}q_{3}^{2}}-1\right]
 \right\}
 \left.\right]\delta_{0}(\bfq_{1})\delta_{0}(\bfq_{2})\delta_{0}(\bfq_{3}).
\ea
A tedious and long calculation shows that this solution exactly matches Eq.~(\ref{eq: co-evolution third solution}). 
Here also $b^{\rm E}_{3}=-(708/567)b^{\rm L}_{1}-(13/7)b^{\rm L}_{2}+b^{\rm L}_{3}$ is helpful to find the match. 
As discussed in Ref.~\cite{Chuen-Chan:2012fk} (see also \cite{Assassi:2014lr}), it is not necessary to start with 
Eq.~(\ref{eq: full expression for third order}) and there are duplicated terms in third-order terms in 
Eq.~(\ref{eq: full expression for third order}). However, this fact does not alter our discussion since 
all the nonlocal third-order terms can be summarized into the $b_{\rm 3nl}$ term anyway as shown in 
\S.~\ref{subsec: co-evolution} or in Ref.~\cite{McDonald:2009lr}. 

\section{Predicting local bias parameters from the peak-background split with the universal mass function}
\label{sec: PBS-ST}
In this appendix, we summarize how to predict the local bias parameters, $b_{1}$ and $b_{2}$, 
on the basis of a simple peak-background split \cite{Bardeen:1986yq} combined with the universal 
halo mass function. For this purpose we here adopt the Sheth-Tormen (ST) fitting formula 
for the universal mass function \cite{Sheth:1999dq}. The similar contents can be found in the literature 
(see e.g., \cite{Saito:2009fk}) and, this appendix follows the notation in 
Refs.\cite{Baldauf:2011qp,Baldauf:2012lr,Slosar:2008aa}.\par 

The universal halo mass function basically assume that it depends only the peak hight $\nu$ defined as
\be
 \nu(R,z) = \frac{\delta_{\rm c}^{2}}{\sigma(R,z)^{2}}, 
\ee
where we set the density threshold $\delta_{\rm c}$ to be 1.686 based on the spherical collapse, 
and the variance of the matter fluctuation field smoothed over the scale $R$ is given by
\be
 \sigma^{2}(R,z) = \int \frac{k^{2}dk}{2\pi^{2}}\,P(k,z)|W(kR)|^{2}, 
\ee
with $W(kR)$ being the top-hat window function, i.e., $W(x)=3(\sin x - x\cos x)/x^{3}$. 
Here the Lagrangian radius $R$ is simply connected to 
the halo mass as $R=\{3M/(4\pi\overline{\rho}_{\rm m0})\}^{1/3}$. Note that $R$ does not depend on redshift. 
In the peak-background split, the local Lagrangian bias parameters are written down as 
\ba
 b_{1}^{\rm L}(M,z) & = & -\frac{1}{\overline{n}}\frac{2\nu}{\delta_{\rm c}}\frac{\partial n}{\partial \nu},\\
 b_{2}^{\rm L}(M,z) & = & \frac{4}{\overline{n}}\frac{\nu^{2}}{\delta^{2}_{\rm c}}\frac{\partial^{2} n}{\partial \nu^{2}}
 +\frac{2}{\overline{n}}\frac{\nu}{\delta^{2}_{\rm c}}\frac{\partial n}{\partial \nu}. 
\ea
In the case of the ST mass function, the derivatives are analytically expressed by
\ba
\frac{1}{\overline{n}}\frac{\partial n}{\partial \nu} & = & -\frac{q\nu -1}{2\nu}-\frac{p}{\nu\{1+(q\nu)^{p}\}},\\
\frac{1}{\overline{n}}\frac{\partial^{2} n}{\partial \nu^{2}} & = & \frac{p^{2}+\nu p q}{\nu^{2}\{1+(q\nu)^{p}\}}
+\frac{(q\nu)^{2}-2q\nu-1}{4\nu^{2}}, 
\ea
where we adopt $(p,q)=(0.15,0.75)$. Finally we obtain the Eulerian local bias parameters using 
Eqs.~(\ref{eq: b1L-b1E}) and (\ref{eq: b2L-b2E}). 

\bibliography{nlb}

\begin{thebibliography}{97}%
\makeatletter
\providecommand \@ifxundefined [1]{%
 \@ifx{#1\undefined}
}%
\providecommand \@ifnum [1]{%
 \ifnum #1\expandafter \@firstoftwo
 \else \expandafter \@secondoftwo
 \fi
}%
\providecommand \@ifx [1]{%
 \ifx #1\expandafter \@firstoftwo
 \else \expandafter \@secondoftwo
 \fi
}%
\providecommand \natexlab [1]{#1}%
\providecommand \enquote  [1]{``#1''}%
\providecommand \bibnamefont  [1]{#1}%
\providecommand \bibfnamefont [1]{#1}%
\providecommand \citenamefont [1]{#1}%
\providecommand \href@noop [0]{\@secondoftwo}%
\providecommand \href [0]{\begingroup \@sanitize@url \@href}%
\providecommand \@href[1]{\@@startlink{#1}\@@href}%
\providecommand \@@href[1]{\endgroup#1\@@endlink}%
\providecommand \@sanitize@url [0]{\catcode `\\12\catcode `\$12\catcode
  `\&12\catcode `\#12\catcode `\^12\catcode `\_12\catcode `\%12\relax}%
\providecommand \@@startlink[1]{}%
\providecommand \@@endlink[0]{}%
\providecommand \url  [0]{\begingroup\@sanitize@url \@url }%
\providecommand \@url [1]{\endgroup\@href {#1}{\urlprefix }}%
\providecommand \urlprefix  [0]{URL }%
\providecommand \Eprint [0]{\href }%
\providecommand \doibase [0]{http://dx.doi.org/}%
\providecommand \selectlanguage [0]{\@gobble}%
\providecommand \bibinfo  [0]{\@secondoftwo}%
\providecommand \bibfield  [0]{\@secondoftwo}%
\providecommand \translation [1]{[#1]}%
\providecommand \BibitemOpen [0]{}%
\providecommand \bibitemStop [0]{}%
\providecommand \bibitemNoStop [0]{.\EOS\space}%
\providecommand \EOS [0]{\spacefactor3000\relax}%
\providecommand \BibitemShut  [1]{\csname bibitem#1\endcsname}%
\let\auto@bib@innerbib\@empty
\bibitem [{\citenamefont {{Komatsu}}\ \emph {et~al.}(2009)\citenamefont
  {{Komatsu}}, \citenamefont {{Dunkley}}, \citenamefont {{Nolta}},
  \citenamefont {{Bennett}}, \citenamefont {{Gold}}, \citenamefont {{Hinshaw}},
  \citenamefont {{Jarosik}}, \citenamefont {{Larson}}, \citenamefont {{Limon}},
  \citenamefont {{Page}}, \citenamefont {{Spergel}}, \citenamefont {{Halpern}},
  \citenamefont {{Hill}}, \citenamefont {{Kogut}}, \citenamefont {{Meyer}},
  \citenamefont {{Tucker}}, \citenamefont {{Weiland}}, \citenamefont
  {{Wollack}},\ and\ \citenamefont {{Wright}}}]{Komatsu:2009qy}%
  \BibitemOpen
  \bibfield  {author} {\bibinfo {author} {\bibfnamefont {E.}~\bibnamefont
  {{Komatsu}}}, \bibinfo {author} {\bibfnamefont {J.}~\bibnamefont
  {{Dunkley}}}, \bibinfo {author} {\bibfnamefont {M.~R.}\ \bibnamefont
  {{Nolta}}}, \bibinfo {author} {\bibfnamefont {C.~L.}\ \bibnamefont
  {{Bennett}}}, \bibinfo {author} {\bibfnamefont {B.}~\bibnamefont {{Gold}}},
  \bibinfo {author} {\bibfnamefont {G.}~\bibnamefont {{Hinshaw}}}, \bibinfo
  {author} {\bibfnamefont {N.}~\bibnamefont {{Jarosik}}}, \bibinfo {author}
  {\bibfnamefont {D.}~\bibnamefont {{Larson}}}, \bibinfo {author}
  {\bibfnamefont {M.}~\bibnamefont {{Limon}}}, \bibinfo {author} {\bibfnamefont
  {L.}~\bibnamefont {{Page}}}, \bibinfo {author} {\bibfnamefont {D.~N.}\
  \bibnamefont {{Spergel}}}, \bibinfo {author} {\bibfnamefont {M.}~\bibnamefont
  {{Halpern}}}, \bibinfo {author} {\bibfnamefont {R.~S.}\ \bibnamefont
  {{Hill}}}, \bibinfo {author} {\bibfnamefont {A.}~\bibnamefont {{Kogut}}},
  \bibinfo {author} {\bibfnamefont {S.~S.}\ \bibnamefont {{Meyer}}}, \bibinfo
  {author} {\bibfnamefont {G.~S.}\ \bibnamefont {{Tucker}}}, \bibinfo {author}
  {\bibfnamefont {J.~L.}\ \bibnamefont {{Weiland}}}, \bibinfo {author}
  {\bibfnamefont {E.}~\bibnamefont {{Wollack}}}, \ and\ \bibinfo {author}
  {\bibfnamefont {E.~L.}\ \bibnamefont {{Wright}}},\ }\href {\doibase
  10.1088/0067-0049/180/2/330} {\bibfield  {journal} {\bibinfo  {journal}
  {\apjs}\ }\textbf {\bibinfo {volume} {180}},\ \bibinfo {pages} {330}
  (\bibinfo {year} {2009})},\ \Eprint {http://arxiv.org/abs/0803.0547}
  {arXiv:0803.0547} \BibitemShut {NoStop}%
\bibitem [{\citenamefont {{Komatsu}}\ \emph {et~al.}(2011)\citenamefont
  {{Komatsu}}, \citenamefont {{Smith}}, \citenamefont {{Dunkley}},
  \citenamefont {{Bennett}}, \citenamefont {{Gold}}, \citenamefont {{Hinshaw}},
  \citenamefont {{Jarosik}}, \citenamefont {{Larson}}, \citenamefont {{Nolta}},
  \citenamefont {{Page}}, \citenamefont {{Spergel}}, \citenamefont {{Halpern}},
  \citenamefont {{Hill}}, \citenamefont {{Kogut}}, \citenamefont {{Limon}},
  \citenamefont {{Meyer}}, \citenamefont {{Odegard}}, \citenamefont {{Tucker}},
  \citenamefont {{Weiland}}, \citenamefont {{Wollack}},\ and\ \citenamefont
  {{Wright}}}]{Komatsu:2011rs}%
  \BibitemOpen
  \bibfield  {author} {\bibinfo {author} {\bibfnamefont {E.}~\bibnamefont
  {{Komatsu}}}, \bibinfo {author} {\bibfnamefont {K.~M.}\ \bibnamefont
  {{Smith}}}, \bibinfo {author} {\bibfnamefont {J.}~\bibnamefont {{Dunkley}}},
  \bibinfo {author} {\bibfnamefont {C.~L.}\ \bibnamefont {{Bennett}}}, \bibinfo
  {author} {\bibfnamefont {B.}~\bibnamefont {{Gold}}}, \bibinfo {author}
  {\bibfnamefont {G.}~\bibnamefont {{Hinshaw}}}, \bibinfo {author}
  {\bibfnamefont {N.}~\bibnamefont {{Jarosik}}}, \bibinfo {author}
  {\bibfnamefont {D.}~\bibnamefont {{Larson}}}, \bibinfo {author}
  {\bibfnamefont {M.~R.}\ \bibnamefont {{Nolta}}}, \bibinfo {author}
  {\bibfnamefont {L.}~\bibnamefont {{Page}}}, \bibinfo {author} {\bibfnamefont
  {D.~N.}\ \bibnamefont {{Spergel}}}, \bibinfo {author} {\bibfnamefont
  {M.}~\bibnamefont {{Halpern}}}, \bibinfo {author} {\bibfnamefont {R.~S.}\
  \bibnamefont {{Hill}}}, \bibinfo {author} {\bibfnamefont {A.}~\bibnamefont
  {{Kogut}}}, \bibinfo {author} {\bibfnamefont {M.}~\bibnamefont {{Limon}}},
  \bibinfo {author} {\bibfnamefont {S.~S.}\ \bibnamefont {{Meyer}}}, \bibinfo
  {author} {\bibfnamefont {N.}~\bibnamefont {{Odegard}}}, \bibinfo {author}
  {\bibfnamefont {G.~S.}\ \bibnamefont {{Tucker}}}, \bibinfo {author}
  {\bibfnamefont {J.~L.}\ \bibnamefont {{Weiland}}}, \bibinfo {author}
  {\bibfnamefont {E.}~\bibnamefont {{Wollack}}}, \ and\ \bibinfo {author}
  {\bibfnamefont {E.~L.}\ \bibnamefont {{Wright}}},\ }\href {\doibase
  10.1088/0067-0049/192/2/18} {\bibfield  {journal} {\bibinfo  {journal}
  {\apjs}\ }\textbf {\bibinfo {volume} {192}},\ \bibinfo {eid} {18} (\bibinfo
  {year} {2011})},\ \Eprint {http://arxiv.org/abs/1001.4538} {arXiv:1001.4538
  [astro-ph.CO]} \BibitemShut {NoStop}%
\bibitem [{\citenamefont {{Hinshaw}}\ \emph {et~al.}(2013)\citenamefont
  {{Hinshaw}}, \citenamefont {{Larson}}, \citenamefont {{Komatsu}},
  \citenamefont {{Spergel}}, \citenamefont {{Bennett}}, \citenamefont
  {{Dunkley}}, \citenamefont {{Nolta}}, \citenamefont {{Halpern}},
  \citenamefont {{Hill}}, \citenamefont {{Odegard}}, \citenamefont {{Page}},
  \citenamefont {{Smith}}, \citenamefont {{Weiland}}, \citenamefont {{Gold}},
  \citenamefont {{Jarosik}}, \citenamefont {{Kogut}}, \citenamefont {{Limon}},
  \citenamefont {{Meyer}}, \citenamefont {{Tucker}}, \citenamefont
  {{Wollack}},\ and\ \citenamefont {{Wright}}}]{Hinshaw:2013rt}%
  \BibitemOpen
  \bibfield  {author} {\bibinfo {author} {\bibfnamefont {G.}~\bibnamefont
  {{Hinshaw}}}, \bibinfo {author} {\bibfnamefont {D.}~\bibnamefont {{Larson}}},
  \bibinfo {author} {\bibfnamefont {E.}~\bibnamefont {{Komatsu}}}, \bibinfo
  {author} {\bibfnamefont {D.~N.}\ \bibnamefont {{Spergel}}}, \bibinfo {author}
  {\bibfnamefont {C.~L.}\ \bibnamefont {{Bennett}}}, \bibinfo {author}
  {\bibfnamefont {J.}~\bibnamefont {{Dunkley}}}, \bibinfo {author}
  {\bibfnamefont {M.~R.}\ \bibnamefont {{Nolta}}}, \bibinfo {author}
  {\bibfnamefont {M.}~\bibnamefont {{Halpern}}}, \bibinfo {author}
  {\bibfnamefont {R.~S.}\ \bibnamefont {{Hill}}}, \bibinfo {author}
  {\bibfnamefont {N.}~\bibnamefont {{Odegard}}}, \bibinfo {author}
  {\bibfnamefont {L.}~\bibnamefont {{Page}}}, \bibinfo {author} {\bibfnamefont
  {K.~M.}\ \bibnamefont {{Smith}}}, \bibinfo {author} {\bibfnamefont {J.~L.}\
  \bibnamefont {{Weiland}}}, \bibinfo {author} {\bibfnamefont {B.}~\bibnamefont
  {{Gold}}}, \bibinfo {author} {\bibfnamefont {N.}~\bibnamefont {{Jarosik}}},
  \bibinfo {author} {\bibfnamefont {A.}~\bibnamefont {{Kogut}}}, \bibinfo
  {author} {\bibfnamefont {M.}~\bibnamefont {{Limon}}}, \bibinfo {author}
  {\bibfnamefont {S.~S.}\ \bibnamefont {{Meyer}}}, \bibinfo {author}
  {\bibfnamefont {G.~S.}\ \bibnamefont {{Tucker}}}, \bibinfo {author}
  {\bibfnamefont {E.}~\bibnamefont {{Wollack}}}, \ and\ \bibinfo {author}
  {\bibfnamefont {E.~L.}\ \bibnamefont {{Wright}}},\ }\href {\doibase
  10.1088/0067-0049/208/2/19} {\bibfield  {journal} {\bibinfo  {journal}
  {\apjs}\ }\textbf {\bibinfo {volume} {208}},\ \bibinfo {eid} {19} (\bibinfo
  {year} {2013})},\ \Eprint {http://arxiv.org/abs/1212.5226} {arXiv:1212.5226
  [astro-ph.CO]} \BibitemShut {NoStop}%
\bibitem [{\citenamefont {{Planck Collaboration}}\ \emph
  {et~al.}(2013)\citenamefont {{Planck Collaboration}}, \citenamefont {{Ade}},
  \citenamefont {{Aghanim}}, \citenamefont {{Armitage-Caplan}}, \citenamefont
  {{Arnaud}}, \citenamefont {{Ashdown}}, \citenamefont {{Atrio-Barandela}},
  \citenamefont {{Aumont}}, \citenamefont {{Baccigalupi}}, \citenamefont
  {{Banday}},\ and\ \citenamefont {et~al.}}]{Planck-Collaboration:2013fr}%
  \BibitemOpen
  \bibfield  {author} {\bibinfo {author} {\bibnamefont {{Planck
  Collaboration}}}, \bibinfo {author} {\bibfnamefont {P.~A.~R.}\ \bibnamefont
  {{Ade}}}, \bibinfo {author} {\bibfnamefont {N.}~\bibnamefont {{Aghanim}}},
  \bibinfo {author} {\bibfnamefont {C.}~\bibnamefont {{Armitage-Caplan}}},
  \bibinfo {author} {\bibfnamefont {M.}~\bibnamefont {{Arnaud}}}, \bibinfo
  {author} {\bibfnamefont {M.}~\bibnamefont {{Ashdown}}}, \bibinfo {author}
  {\bibfnamefont {F.}~\bibnamefont {{Atrio-Barandela}}}, \bibinfo {author}
  {\bibfnamefont {J.}~\bibnamefont {{Aumont}}}, \bibinfo {author}
  {\bibfnamefont {C.}~\bibnamefont {{Baccigalupi}}}, \bibinfo {author}
  {\bibfnamefont {A.~J.}\ \bibnamefont {{Banday}}}, \ and\ \bibinfo {author}
  {\bibnamefont {et~al.}},\ }\href@noop {} {\bibfield  {journal} {\bibinfo
  {journal} {ArXiv e-prints}\ } (\bibinfo {year} {2013})},\ \Eprint
  {http://arxiv.org/abs/1303.5076} {arXiv:1303.5076 [astro-ph.CO]} \BibitemShut
  {NoStop}%
\bibitem [{\citenamefont {{Weinberg}}\ \emph {et~al.}(2013)\citenamefont
  {{Weinberg}}, \citenamefont {{Mortonson}}, \citenamefont {{Eisenstein}},
  \citenamefont {{Hirata}}, \citenamefont {{Riess}},\ and\ \citenamefont
  {{Rozo}}}]{Weinberg:2012uq}%
  \BibitemOpen
  \bibfield  {author} {\bibinfo {author} {\bibfnamefont {D.~H.}\ \bibnamefont
  {{Weinberg}}}, \bibinfo {author} {\bibfnamefont {M.~J.}\ \bibnamefont
  {{Mortonson}}}, \bibinfo {author} {\bibfnamefont {D.~J.}\ \bibnamefont
  {{Eisenstein}}}, \bibinfo {author} {\bibfnamefont {C.}~\bibnamefont
  {{Hirata}}}, \bibinfo {author} {\bibfnamefont {A.~G.}\ \bibnamefont
  {{Riess}}}, \ and\ \bibinfo {author} {\bibfnamefont {E.}~\bibnamefont
  {{Rozo}}},\ }\href@noop {} {\bibfield  {journal} {\bibinfo  {journal}
  {Physics Reports}\ }\textbf {\bibinfo {volume} {530}},\ \bibinfo {pages} {87}
  (\bibinfo {year} {2013})},\ \Eprint {http://arxiv.org/abs/1201.2434}
  {arXiv:1201.2434 [astro-ph.CO]} \BibitemShut {NoStop}%
\bibitem [{\citenamefont {{Bernardeau}}\ \emph {et~al.}(2002)\citenamefont
  {{Bernardeau}}, \citenamefont {{Colombi}}, \citenamefont {{Gazta{\~n}aga}},\
  and\ \citenamefont {{Scoccimarro}}}]{Bernardeau:2002lr}%
  \BibitemOpen
  \bibfield  {author} {\bibinfo {author} {\bibfnamefont {F.}~\bibnamefont
  {{Bernardeau}}}, \bibinfo {author} {\bibfnamefont {S.}~\bibnamefont
  {{Colombi}}}, \bibinfo {author} {\bibfnamefont {E.}~\bibnamefont
  {{Gazta{\~n}aga}}}, \ and\ \bibinfo {author} {\bibfnamefont {R.}~\bibnamefont
  {{Scoccimarro}}},\ }\href {\doibase 10.1016/S0370-1573(02)00135-7} {\bibfield
   {journal} {\bibinfo  {journal} {\physrep}\ }\textbf {\bibinfo {volume}
  {367}},\ \bibinfo {pages} {1} (\bibinfo {year} {2002})},\ \Eprint
  {http://arxiv.org/abs/arXiv:astro-ph/0112551} {arXiv:astro-ph/0112551}
  \BibitemShut {NoStop}%
\bibitem [{\citenamefont {{Schlegel}}\ \emph {et~al.}(2009)\citenamefont
  {{Schlegel}}, \citenamefont {{White}},\ and\ \citenamefont
  {{Eisenstein}}}]{Schlegel:2009uq}%
  \BibitemOpen
  \bibfield  {author} {\bibinfo {author} {\bibfnamefont {D.}~\bibnamefont
  {{Schlegel}}}, \bibinfo {author} {\bibfnamefont {M.}~\bibnamefont {{White}}},
  \ and\ \bibinfo {author} {\bibfnamefont {D.}~\bibnamefont {{Eisenstein}}},\
  }in\ \href@noop {} {\emph {\bibinfo {booktitle} {astro2010: The Astronomy and
  Astrophysics Decadal Survey}}},\ \bibinfo {series} {ArXiv Astrophysics
  e-prints}, Vol.\ \bibinfo {volume} {2010}\ (\bibinfo {year} {2009})\ p.\
  \bibinfo {pages} {314},\ \Eprint {http://arxiv.org/abs/0902.4680}
  {arXiv:0902.4680 [astro-ph.CO]} \BibitemShut {NoStop}%
\bibitem [{\citenamefont {{Eisenstein}}\ \emph {et~al.}(2011)\citenamefont
  {{Eisenstein}}, \citenamefont {{Weinberg}}, \citenamefont {{Agol}},
  \citenamefont {{Aihara}}, \citenamefont {{Allende Prieto}}, \citenamefont
  {{Anderson}}, \citenamefont {{Arns}}, \citenamefont {{Aubourg}},
  \citenamefont {{Bailey}}, \citenamefont {{Balbinot}},\ and\ \citenamefont
  {et~al.}}]{Eisenstein:2011ve}%
  \BibitemOpen
  \bibfield  {author} {\bibinfo {author} {\bibfnamefont {D.~J.}\ \bibnamefont
  {{Eisenstein}}}, \bibinfo {author} {\bibfnamefont {D.~H.}\ \bibnamefont
  {{Weinberg}}}, \bibinfo {author} {\bibfnamefont {E.}~\bibnamefont {{Agol}}},
  \bibinfo {author} {\bibfnamefont {H.}~\bibnamefont {{Aihara}}}, \bibinfo
  {author} {\bibfnamefont {C.}~\bibnamefont {{Allende Prieto}}}, \bibinfo
  {author} {\bibfnamefont {S.~F.}\ \bibnamefont {{Anderson}}}, \bibinfo
  {author} {\bibfnamefont {J.~A.}\ \bibnamefont {{Arns}}}, \bibinfo {author}
  {\bibfnamefont {{\'E}.}~\bibnamefont {{Aubourg}}}, \bibinfo {author}
  {\bibfnamefont {S.}~\bibnamefont {{Bailey}}}, \bibinfo {author}
  {\bibfnamefont {E.}~\bibnamefont {{Balbinot}}}, \ and\ \bibinfo {author}
  {\bibnamefont {et~al.}},\ }\href {\doibase 10.1088/0004-6256/142/3/72}
  {\bibfield  {journal} {\bibinfo  {journal} {\aj}\ }\textbf {\bibinfo {volume}
  {142}},\ \bibinfo {eid} {72} (\bibinfo {year} {2011})},\ \Eprint
  {http://arxiv.org/abs/1101.1529} {arXiv:1101.1529 [astro-ph.IM]} \BibitemShut
  {NoStop}%
\bibitem [{\citenamefont {{Drinkwater}}\ \emph {et~al.}(2010)\citenamefont
  {{Drinkwater}}, \citenamefont {{Jurek}}, \citenamefont {{Blake}},
  \citenamefont {{Woods}}, \citenamefont {{Pimbblet}}, \citenamefont
  {{Glazebrook}}, \citenamefont {{Sharp}}, \citenamefont {{Pracy}},
  \citenamefont {{Brough}}, \citenamefont {{Colless}}, \citenamefont {{Couch}},
  \citenamefont {{Croom}}, \citenamefont {{Davis}}, \citenamefont {{Forbes}},
  \citenamefont {{Forster}}, \citenamefont {{Gilbank}}, \citenamefont
  {{Gladders}}, \citenamefont {{Jelliffe}}, \citenamefont {{Jones}},
  \citenamefont {{Li}}, \citenamefont {{Madore}}, \citenamefont {{Martin}},
  \citenamefont {{Poole}}, \citenamefont {{Small}}, \citenamefont
  {{Wisnioski}}, \citenamefont {{Wyder}},\ and\ \citenamefont
  {{Yee}}}]{Drinkwater:2010rz}%
  \BibitemOpen
  \bibfield  {author} {\bibinfo {author} {\bibfnamefont {M.~J.}\ \bibnamefont
  {{Drinkwater}}}, \bibinfo {author} {\bibfnamefont {R.~J.}\ \bibnamefont
  {{Jurek}}}, \bibinfo {author} {\bibfnamefont {C.}~\bibnamefont {{Blake}}},
  \bibinfo {author} {\bibfnamefont {D.}~\bibnamefont {{Woods}}}, \bibinfo
  {author} {\bibfnamefont {K.~A.}\ \bibnamefont {{Pimbblet}}}, \bibinfo
  {author} {\bibfnamefont {K.}~\bibnamefont {{Glazebrook}}}, \bibinfo {author}
  {\bibfnamefont {R.}~\bibnamefont {{Sharp}}}, \bibinfo {author} {\bibfnamefont
  {M.~B.}\ \bibnamefont {{Pracy}}}, \bibinfo {author} {\bibfnamefont
  {S.}~\bibnamefont {{Brough}}}, \bibinfo {author} {\bibfnamefont
  {M.}~\bibnamefont {{Colless}}}, \bibinfo {author} {\bibfnamefont {W.~J.}\
  \bibnamefont {{Couch}}}, \bibinfo {author} {\bibfnamefont {S.~M.}\
  \bibnamefont {{Croom}}}, \bibinfo {author} {\bibfnamefont {T.~M.}\
  \bibnamefont {{Davis}}}, \bibinfo {author} {\bibfnamefont {D.}~\bibnamefont
  {{Forbes}}}, \bibinfo {author} {\bibfnamefont {K.}~\bibnamefont {{Forster}}},
  \bibinfo {author} {\bibfnamefont {D.~G.}\ \bibnamefont {{Gilbank}}}, \bibinfo
  {author} {\bibfnamefont {M.}~\bibnamefont {{Gladders}}}, \bibinfo {author}
  {\bibfnamefont {B.}~\bibnamefont {{Jelliffe}}}, \bibinfo {author}
  {\bibfnamefont {N.}~\bibnamefont {{Jones}}}, \bibinfo {author} {\bibfnamefont
  {I.-H.}\ \bibnamefont {{Li}}}, \bibinfo {author} {\bibfnamefont
  {B.}~\bibnamefont {{Madore}}}, \bibinfo {author} {\bibfnamefont {D.~C.}\
  \bibnamefont {{Martin}}}, \bibinfo {author} {\bibfnamefont {G.~B.}\
  \bibnamefont {{Poole}}}, \bibinfo {author} {\bibfnamefont {T.}~\bibnamefont
  {{Small}}}, \bibinfo {author} {\bibfnamefont {E.}~\bibnamefont
  {{Wisnioski}}}, \bibinfo {author} {\bibfnamefont {T.}~\bibnamefont
  {{Wyder}}}, \ and\ \bibinfo {author} {\bibfnamefont {H.~K.~C.}\ \bibnamefont
  {{Yee}}},\ }\href {\doibase 10.1111/j.1365-2966.2009.15754.x} {\bibfield
  {journal} {\bibinfo  {journal} {\mnras}\ }\textbf {\bibinfo {volume} {401}},\
  \bibinfo {pages} {1429} (\bibinfo {year} {2010})},\ \Eprint
  {http://arxiv.org/abs/0911.4246} {arXiv:0911.4246 [astro-ph.CO]} \BibitemShut
  {NoStop}%
\bibitem [{\citenamefont {{Anderson}}\ \emph
  {et~al.}(2013{\natexlab{a}})\citenamefont {{Anderson}}, \citenamefont
  {{Aubourg}}, \citenamefont {{Bailey}}, \citenamefont {{Beutler}},
  \citenamefont {{Bolton}}, \citenamefont {{Brinkmann}}, \citenamefont
  {{Brownstein}}, \citenamefont {{Chuang}}, \citenamefont {{Cuesta}},
  \citenamefont {{Dawson}}, \citenamefont {{Eisenstein}}, \citenamefont
  {{Honscheid}}, \citenamefont {{Kazin}}, \citenamefont {{Kirkby}},
  \citenamefont {{Manera}}, \citenamefont {{McBride}}, \citenamefont {{Mena}},
  \citenamefont {{Nichol}}, \citenamefont {{Olmstead}}, \citenamefont
  {{Padmanabhan}}, \citenamefont {{Palanque-Delabrouille}}, \citenamefont
  {{Percival}}, \citenamefont {{Prada}}, \citenamefont {{Ross}}, \citenamefont
  {{Ross}}, \citenamefont {{Sanchez}}, \citenamefont {{Samushia}},
  \citenamefont {{Schlegel}}, \citenamefont {{Schneider}}, \citenamefont
  {{Seo}}, \citenamefont {{Strauss}}, \citenamefont {{Thomas}}, \citenamefont
  {{Tinker}}, \citenamefont {{Tojeiro}}, \citenamefont {{Verde}}, \citenamefont
  {{Weinberg}}, \citenamefont {{Xu}},\ and\ \citenamefont
  {{Yeche}}}]{Anderson:2013qy}%
  \BibitemOpen
  \bibfield  {author} {\bibinfo {author} {\bibfnamefont {L.}~\bibnamefont
  {{Anderson}}}, \bibinfo {author} {\bibfnamefont {E.}~\bibnamefont
  {{Aubourg}}}, \bibinfo {author} {\bibfnamefont {S.}~\bibnamefont {{Bailey}}},
  \bibinfo {author} {\bibfnamefont {F.}~\bibnamefont {{Beutler}}}, \bibinfo
  {author} {\bibfnamefont {A.~S.}\ \bibnamefont {{Bolton}}}, \bibinfo {author}
  {\bibfnamefont {J.}~\bibnamefont {{Brinkmann}}}, \bibinfo {author}
  {\bibfnamefont {J.~R.}\ \bibnamefont {{Brownstein}}}, \bibinfo {author}
  {\bibfnamefont {C.-H.}\ \bibnamefont {{Chuang}}}, \bibinfo {author}
  {\bibfnamefont {A.~J.}\ \bibnamefont {{Cuesta}}}, \bibinfo {author}
  {\bibfnamefont {K.~S.}\ \bibnamefont {{Dawson}}}, \bibinfo {author}
  {\bibfnamefont {D.~J.}\ \bibnamefont {{Eisenstein}}}, \bibinfo {author}
  {\bibfnamefont {K.}~\bibnamefont {{Honscheid}}}, \bibinfo {author}
  {\bibfnamefont {E.~A.}\ \bibnamefont {{Kazin}}}, \bibinfo {author}
  {\bibfnamefont {D.}~\bibnamefont {{Kirkby}}}, \bibinfo {author}
  {\bibfnamefont {M.}~\bibnamefont {{Manera}}}, \bibinfo {author}
  {\bibfnamefont {C.~K.}\ \bibnamefont {{McBride}}}, \bibinfo {author}
  {\bibfnamefont {O.}~\bibnamefont {{Mena}}}, \bibinfo {author} {\bibfnamefont
  {R.~C.}\ \bibnamefont {{Nichol}}}, \bibinfo {author} {\bibfnamefont {M.~D.}\
  \bibnamefont {{Olmstead}}}, \bibinfo {author} {\bibfnamefont
  {N.}~\bibnamefont {{Padmanabhan}}}, \bibinfo {author} {\bibfnamefont
  {N.}~\bibnamefont {{Palanque-Delabrouille}}}, \bibinfo {author}
  {\bibfnamefont {W.~J.}\ \bibnamefont {{Percival}}}, \bibinfo {author}
  {\bibfnamefont {F.}~\bibnamefont {{Prada}}}, \bibinfo {author} {\bibfnamefont
  {A.~J.}\ \bibnamefont {{Ross}}}, \bibinfo {author} {\bibfnamefont {N.~P.}\
  \bibnamefont {{Ross}}}, \bibinfo {author} {\bibfnamefont {A.~G.}\
  \bibnamefont {{Sanchez}}}, \bibinfo {author} {\bibfnamefont {L.}~\bibnamefont
  {{Samushia}}}, \bibinfo {author} {\bibfnamefont {D.~J.}\ \bibnamefont
  {{Schlegel}}}, \bibinfo {author} {\bibfnamefont {D.~P.}\ \bibnamefont
  {{Schneider}}}, \bibinfo {author} {\bibfnamefont {H.-J.}\ \bibnamefont
  {{Seo}}}, \bibinfo {author} {\bibfnamefont {M.~A.}\ \bibnamefont
  {{Strauss}}}, \bibinfo {author} {\bibfnamefont {D.}~\bibnamefont {{Thomas}}},
  \bibinfo {author} {\bibfnamefont {J.~L.}\ \bibnamefont {{Tinker}}}, \bibinfo
  {author} {\bibfnamefont {R.}~\bibnamefont {{Tojeiro}}}, \bibinfo {author}
  {\bibfnamefont {L.}~\bibnamefont {{Verde}}}, \bibinfo {author} {\bibfnamefont
  {D.~H.}\ \bibnamefont {{Weinberg}}}, \bibinfo {author} {\bibfnamefont
  {X.}~\bibnamefont {{Xu}}}, \ and\ \bibinfo {author} {\bibfnamefont
  {C.}~\bibnamefont {{Yeche}}},\ }\href@noop {} {\bibfield  {journal} {\bibinfo
   {journal} {ArXiv e-prints}\ } (\bibinfo {year} {2013}{\natexlab{a}})},\
  \Eprint {http://arxiv.org/abs/1303.4666} {arXiv:1303.4666 [astro-ph.CO]}
  \BibitemShut {NoStop}%
\bibitem [{\citenamefont {{Reid}}\ \emph {et~al.}(2012)\citenamefont {{Reid}},
  \citenamefont {{Samushia}}, \citenamefont {{White}}, \citenamefont
  {{Percival}}, \citenamefont {{Manera}}, \citenamefont {{Padmanabhan}},
  \citenamefont {{Ross}}, \citenamefont {{S{\'a}nchez}}, \citenamefont
  {{Bailey}}, \citenamefont {{Bizyaev}}, \citenamefont {{Bolton}},
  \citenamefont {{Brewington}}, \citenamefont {{Brinkmann}}, \citenamefont
  {{Brownstein}}, \citenamefont {{Cuesta}}, \citenamefont {{Eisenstein}},
  \citenamefont {{Gunn}}, \citenamefont {{Honscheid}}, \citenamefont
  {{Malanushenko}}, \citenamefont {{Malanushenko}}, \citenamefont {{Maraston}},
  \citenamefont {{McBride}}, \citenamefont {{Muna}}, \citenamefont {{Nichol}},
  \citenamefont {{Oravetz}}, \citenamefont {{Pan}}, \citenamefont {{de
  Putter}}, \citenamefont {{Roe}}, \citenamefont {{Ross}}, \citenamefont
  {{Schlegel}}, \citenamefont {{Schneider}}, \citenamefont {{Seo}},
  \citenamefont {{Shelden}}, \citenamefont {{Sheldon}}, \citenamefont
  {{Simmons}}, \citenamefont {{Skibba}}, \citenamefont {{Snedden}},
  \citenamefont {{Swanson}}, \citenamefont {{Thomas}}, \citenamefont
  {{Tinker}}, \citenamefont {{Tojeiro}}, \citenamefont {{Verde}}, \citenamefont
  {{Wake}}, \citenamefont {{Weaver}}, \citenamefont {{Weinberg}}, \citenamefont
  {{Zehavi}},\ and\ \citenamefont {{Zhao}}}]{Reid:2012ly}%
  \BibitemOpen
  \bibfield  {author} {\bibinfo {author} {\bibfnamefont {B.~A.}\ \bibnamefont
  {{Reid}}}, \bibinfo {author} {\bibfnamefont {L.}~\bibnamefont {{Samushia}}},
  \bibinfo {author} {\bibfnamefont {M.}~\bibnamefont {{White}}}, \bibinfo
  {author} {\bibfnamefont {W.~J.}\ \bibnamefont {{Percival}}}, \bibinfo
  {author} {\bibfnamefont {M.}~\bibnamefont {{Manera}}}, \bibinfo {author}
  {\bibfnamefont {N.}~\bibnamefont {{Padmanabhan}}}, \bibinfo {author}
  {\bibfnamefont {A.~J.}\ \bibnamefont {{Ross}}}, \bibinfo {author}
  {\bibfnamefont {A.~G.}\ \bibnamefont {{S{\'a}nchez}}}, \bibinfo {author}
  {\bibfnamefont {S.}~\bibnamefont {{Bailey}}}, \bibinfo {author}
  {\bibfnamefont {D.}~\bibnamefont {{Bizyaev}}}, \bibinfo {author}
  {\bibfnamefont {A.~S.}\ \bibnamefont {{Bolton}}}, \bibinfo {author}
  {\bibfnamefont {H.}~\bibnamefont {{Brewington}}}, \bibinfo {author}
  {\bibfnamefont {J.}~\bibnamefont {{Brinkmann}}}, \bibinfo {author}
  {\bibfnamefont {J.~R.}\ \bibnamefont {{Brownstein}}}, \bibinfo {author}
  {\bibfnamefont {A.~J.}\ \bibnamefont {{Cuesta}}}, \bibinfo {author}
  {\bibfnamefont {D.~J.}\ \bibnamefont {{Eisenstein}}}, \bibinfo {author}
  {\bibfnamefont {J.~E.}\ \bibnamefont {{Gunn}}}, \bibinfo {author}
  {\bibfnamefont {K.}~\bibnamefont {{Honscheid}}}, \bibinfo {author}
  {\bibfnamefont {E.}~\bibnamefont {{Malanushenko}}}, \bibinfo {author}
  {\bibfnamefont {V.}~\bibnamefont {{Malanushenko}}}, \bibinfo {author}
  {\bibfnamefont {C.}~\bibnamefont {{Maraston}}}, \bibinfo {author}
  {\bibfnamefont {C.~K.}\ \bibnamefont {{McBride}}}, \bibinfo {author}
  {\bibfnamefont {D.}~\bibnamefont {{Muna}}}, \bibinfo {author} {\bibfnamefont
  {R.~C.}\ \bibnamefont {{Nichol}}}, \bibinfo {author} {\bibfnamefont
  {D.}~\bibnamefont {{Oravetz}}}, \bibinfo {author} {\bibfnamefont
  {K.}~\bibnamefont {{Pan}}}, \bibinfo {author} {\bibfnamefont
  {R.}~\bibnamefont {{de Putter}}}, \bibinfo {author} {\bibfnamefont {N.~A.}\
  \bibnamefont {{Roe}}}, \bibinfo {author} {\bibfnamefont {N.~P.}\ \bibnamefont
  {{Ross}}}, \bibinfo {author} {\bibfnamefont {D.~J.}\ \bibnamefont
  {{Schlegel}}}, \bibinfo {author} {\bibfnamefont {D.~P.}\ \bibnamefont
  {{Schneider}}}, \bibinfo {author} {\bibfnamefont {H.-J.}\ \bibnamefont
  {{Seo}}}, \bibinfo {author} {\bibfnamefont {A.}~\bibnamefont {{Shelden}}},
  \bibinfo {author} {\bibfnamefont {E.~S.}\ \bibnamefont {{Sheldon}}}, \bibinfo
  {author} {\bibfnamefont {A.}~\bibnamefont {{Simmons}}}, \bibinfo {author}
  {\bibfnamefont {R.~A.}\ \bibnamefont {{Skibba}}}, \bibinfo {author}
  {\bibfnamefont {S.}~\bibnamefont {{Snedden}}}, \bibinfo {author}
  {\bibfnamefont {M.~E.~C.}\ \bibnamefont {{Swanson}}}, \bibinfo {author}
  {\bibfnamefont {D.}~\bibnamefont {{Thomas}}}, \bibinfo {author}
  {\bibfnamefont {J.}~\bibnamefont {{Tinker}}}, \bibinfo {author}
  {\bibfnamefont {R.}~\bibnamefont {{Tojeiro}}}, \bibinfo {author}
  {\bibfnamefont {L.}~\bibnamefont {{Verde}}}, \bibinfo {author} {\bibfnamefont
  {D.~A.}\ \bibnamefont {{Wake}}}, \bibinfo {author} {\bibfnamefont {B.~A.}\
  \bibnamefont {{Weaver}}}, \bibinfo {author} {\bibfnamefont {D.~H.}\
  \bibnamefont {{Weinberg}}}, \bibinfo {author} {\bibfnamefont
  {I.}~\bibnamefont {{Zehavi}}}, \ and\ \bibinfo {author} {\bibfnamefont
  {G.-B.}\ \bibnamefont {{Zhao}}},\ }\href {\doibase
  10.1111/j.1365-2966.2012.21779.x} {\bibfield  {journal} {\bibinfo  {journal}
  {\mnras}\ }\textbf {\bibinfo {volume} {426}},\ \bibinfo {pages} {2719}
  (\bibinfo {year} {2012})},\ \Eprint {http://arxiv.org/abs/1203.6641}
  {arXiv:1203.6641 [astro-ph.CO]} \BibitemShut {NoStop}%
\bibitem [{\citenamefont {{Anderson}}\ \emph
  {et~al.}(2013{\natexlab{b}})\citenamefont {{Anderson}}, \citenamefont
  {{Aubourg}}, \citenamefont {{Bailey}}, \citenamefont {{Beutler}},
  \citenamefont {{Bhardwaj}}, \citenamefont {{Blanton}}, \citenamefont
  {{Bolton}}, \citenamefont {{Brinkmann}}, \citenamefont {{Brownstein}},
  \citenamefont {{Burden}}, \citenamefont {{Chuang}}, \citenamefont {{Cuesta}},
  \citenamefont {{Dawson}}, \citenamefont {{Eisenstein}}, \citenamefont
  {{Escoffier}}, \citenamefont {{Gunn}}, \citenamefont {{Guo}}, \citenamefont
  {{Ho}}, \citenamefont {{Honscheid}}, \citenamefont {{Howlett}}, \citenamefont
  {{Kirkby}}, \citenamefont {{Lupton}}, \citenamefont {{Manera}}, \citenamefont
  {{Maraston}}, \citenamefont {{McBride}}, \citenamefont {{Mena}},
  \citenamefont {{Montesano}}, \citenamefont {{Nichol}}, \citenamefont
  {{Nuza}}, \citenamefont {{Olmstead}}, \citenamefont {{Padmanabhan}},
  \citenamefont {{Palanque-Delabrouille}}, \citenamefont {{Parejko}},
  \citenamefont {{Percival}}, \citenamefont {{Petitjean}}, \citenamefont
  {{Prada}}, \citenamefont {{Price-Whelan}}, \citenamefont {{Reid}},
  \citenamefont {{Roe}}, \citenamefont {{Ross}}, \citenamefont {{Ross}},
  \citenamefont {{Sabiu}}, \citenamefont {{Saito}}, \citenamefont {{Samushia}},
  \citenamefont {{Sanchez}}, \citenamefont {{Schlegel}}, \citenamefont
  {{Schneider}}, \citenamefont {{Scoccola}}, \citenamefont {{Seo}},
  \citenamefont {{Skibba}}, \citenamefont {{Strauss}}, \citenamefont
  {{Swanson}}, \citenamefont {{Thomas}}, \citenamefont {{Tinker}},
  \citenamefont {{Tojeiro}}, \citenamefont {{Magana}}, \citenamefont {{Verde}},
  \citenamefont {{Wake}}, \citenamefont {{Weaver}}, \citenamefont {{Weinberg}},
  \citenamefont {{White}}, \citenamefont {{Xu}}, \citenamefont {{Yeche}},
  \citenamefont {{Zehavi}},\ and\ \citenamefont {{Zhao}}}]{Anderson:2013qq}%
  \BibitemOpen
  \bibfield  {author} {\bibinfo {author} {\bibfnamefont {L.}~\bibnamefont
  {{Anderson}}}, \bibinfo {author} {\bibfnamefont {E.}~\bibnamefont
  {{Aubourg}}}, \bibinfo {author} {\bibfnamefont {S.}~\bibnamefont {{Bailey}}},
  \bibinfo {author} {\bibfnamefont {F.}~\bibnamefont {{Beutler}}}, \bibinfo
  {author} {\bibfnamefont {V.}~\bibnamefont {{Bhardwaj}}}, \bibinfo {author}
  {\bibfnamefont {M.}~\bibnamefont {{Blanton}}}, \bibinfo {author}
  {\bibfnamefont {A.~S.}\ \bibnamefont {{Bolton}}}, \bibinfo {author}
  {\bibfnamefont {J.}~\bibnamefont {{Brinkmann}}}, \bibinfo {author}
  {\bibfnamefont {J.~R.}\ \bibnamefont {{Brownstein}}}, \bibinfo {author}
  {\bibfnamefont {A.}~\bibnamefont {{Burden}}}, \bibinfo {author}
  {\bibfnamefont {C.-H.}\ \bibnamefont {{Chuang}}}, \bibinfo {author}
  {\bibfnamefont {A.~J.}\ \bibnamefont {{Cuesta}}}, \bibinfo {author}
  {\bibfnamefont {K.~S.}\ \bibnamefont {{Dawson}}}, \bibinfo {author}
  {\bibfnamefont {D.~J.}\ \bibnamefont {{Eisenstein}}}, \bibinfo {author}
  {\bibfnamefont {S.}~\bibnamefont {{Escoffier}}}, \bibinfo {author}
  {\bibfnamefont {J.~E.}\ \bibnamefont {{Gunn}}}, \bibinfo {author}
  {\bibfnamefont {H.}~\bibnamefont {{Guo}}}, \bibinfo {author} {\bibfnamefont
  {S.}~\bibnamefont {{Ho}}}, \bibinfo {author} {\bibfnamefont {K.}~\bibnamefont
  {{Honscheid}}}, \bibinfo {author} {\bibfnamefont {C.}~\bibnamefont
  {{Howlett}}}, \bibinfo {author} {\bibfnamefont {D.}~\bibnamefont {{Kirkby}}},
  \bibinfo {author} {\bibfnamefont {R.~H.}\ \bibnamefont {{Lupton}}}, \bibinfo
  {author} {\bibfnamefont {M.}~\bibnamefont {{Manera}}}, \bibinfo {author}
  {\bibfnamefont {C.}~\bibnamefont {{Maraston}}}, \bibinfo {author}
  {\bibfnamefont {C.~K.}\ \bibnamefont {{McBride}}}, \bibinfo {author}
  {\bibfnamefont {O.}~\bibnamefont {{Mena}}}, \bibinfo {author} {\bibfnamefont
  {F.}~\bibnamefont {{Montesano}}}, \bibinfo {author} {\bibfnamefont {R.~C.}\
  \bibnamefont {{Nichol}}}, \bibinfo {author} {\bibfnamefont {S.~E.}\
  \bibnamefont {{Nuza}}}, \bibinfo {author} {\bibfnamefont {M.~D.}\
  \bibnamefont {{Olmstead}}}, \bibinfo {author} {\bibfnamefont
  {N.}~\bibnamefont {{Padmanabhan}}}, \bibinfo {author} {\bibfnamefont
  {N.}~\bibnamefont {{Palanque-Delabrouille}}}, \bibinfo {author}
  {\bibfnamefont {J.}~\bibnamefont {{Parejko}}}, \bibinfo {author}
  {\bibfnamefont {W.~J.}\ \bibnamefont {{Percival}}}, \bibinfo {author}
  {\bibfnamefont {P.}~\bibnamefont {{Petitjean}}}, \bibinfo {author}
  {\bibfnamefont {F.}~\bibnamefont {{Prada}}}, \bibinfo {author} {\bibfnamefont
  {A.~M.}\ \bibnamefont {{Price-Whelan}}}, \bibinfo {author} {\bibfnamefont
  {B.}~\bibnamefont {{Reid}}}, \bibinfo {author} {\bibfnamefont {N.~A.}\
  \bibnamefont {{Roe}}}, \bibinfo {author} {\bibfnamefont {A.~J.}\ \bibnamefont
  {{Ross}}}, \bibinfo {author} {\bibfnamefont {N.~P.}\ \bibnamefont {{Ross}}},
  \bibinfo {author} {\bibfnamefont {C.~G.}\ \bibnamefont {{Sabiu}}}, \bibinfo
  {author} {\bibfnamefont {S.}~\bibnamefont {{Saito}}}, \bibinfo {author}
  {\bibfnamefont {L.}~\bibnamefont {{Samushia}}}, \bibinfo {author}
  {\bibfnamefont {A.~G.}\ \bibnamefont {{Sanchez}}}, \bibinfo {author}
  {\bibfnamefont {D.~J.}\ \bibnamefont {{Schlegel}}}, \bibinfo {author}
  {\bibfnamefont {D.~P.}\ \bibnamefont {{Schneider}}}, \bibinfo {author}
  {\bibfnamefont {C.~G.}\ \bibnamefont {{Scoccola}}}, \bibinfo {author}
  {\bibfnamefont {H.-J.}\ \bibnamefont {{Seo}}}, \bibinfo {author}
  {\bibfnamefont {R.~A.}\ \bibnamefont {{Skibba}}}, \bibinfo {author}
  {\bibfnamefont {M.~A.}\ \bibnamefont {{Strauss}}}, \bibinfo {author}
  {\bibfnamefont {M.~E.~C.}\ \bibnamefont {{Swanson}}}, \bibinfo {author}
  {\bibfnamefont {D.}~\bibnamefont {{Thomas}}}, \bibinfo {author}
  {\bibfnamefont {J.~L.}\ \bibnamefont {{Tinker}}}, \bibinfo {author}
  {\bibfnamefont {R.}~\bibnamefont {{Tojeiro}}}, \bibinfo {author}
  {\bibfnamefont {M.~V.}\ \bibnamefont {{Magana}}}, \bibinfo {author}
  {\bibfnamefont {L.}~\bibnamefont {{Verde}}}, \bibinfo {author} {\bibfnamefont
  {D.~A.}\ \bibnamefont {{Wake}}}, \bibinfo {author} {\bibfnamefont {B.~A.}\
  \bibnamefont {{Weaver}}}, \bibinfo {author} {\bibfnamefont {D.~H.}\
  \bibnamefont {{Weinberg}}}, \bibinfo {author} {\bibfnamefont
  {M.}~\bibnamefont {{White}}}, \bibinfo {author} {\bibfnamefont
  {X.}~\bibnamefont {{Xu}}}, \bibinfo {author} {\bibfnamefont {C.}~\bibnamefont
  {{Yeche}}}, \bibinfo {author} {\bibfnamefont {I.}~\bibnamefont {{Zehavi}}}, \
  and\ \bibinfo {author} {\bibfnamefont {G.-B.}\ \bibnamefont {{Zhao}}},\
  }\href@noop {} {\  (\bibinfo {year} {2013}{\natexlab{b}})},\ \Eprint
  {http://arxiv.org/abs/1312.4877v1} {arXiv:1312.4877v1 [astro-ph.CO]}
  \BibitemShut {NoStop}%
\bibitem [{\citenamefont {{Samushia}}\ \emph {et~al.}(2013)\citenamefont
  {{Samushia}}, \citenamefont {{Reid}}, \citenamefont {{White}}, \citenamefont
  {{Percival}}, \citenamefont {{Cuesta}}, \citenamefont {{Zhao}}, \citenamefont
  {{Ross}}, \citenamefont {{Manera}}, \citenamefont {{Aubourg}}, \citenamefont
  {{Beutler}}, \citenamefont {{Brinkmann}}, \citenamefont {{Brownstein}},
  \citenamefont {{Dawson}}, \citenamefont {{Eisenstein}}, \citenamefont {{Ho}},
  \citenamefont {{Honscheid}}, \citenamefont {{Maraston}}, \citenamefont
  {{Montesano}}, \citenamefont {{Nichol}}, \citenamefont {{Roe}}, \citenamefont
  {{Ross}}, \citenamefont {{S{\'a}nchez}}, \citenamefont {{Schlegel}},
  \citenamefont {{Schneider}}, \citenamefont {{Streblyanska}}, \citenamefont
  {{Thomas}}, \citenamefont {{Tinker}}, \citenamefont {{Wake}}, \citenamefont
  {{Weaver}},\ and\ \citenamefont {{Zehavi}}}]{Samushia:2013lr}%
  \BibitemOpen
  \bibfield  {author} {\bibinfo {author} {\bibfnamefont {L.}~\bibnamefont
  {{Samushia}}}, \bibinfo {author} {\bibfnamefont {B.~A.}\ \bibnamefont
  {{Reid}}}, \bibinfo {author} {\bibfnamefont {M.}~\bibnamefont {{White}}},
  \bibinfo {author} {\bibfnamefont {W.~J.}\ \bibnamefont {{Percival}}},
  \bibinfo {author} {\bibfnamefont {A.~J.}\ \bibnamefont {{Cuesta}}}, \bibinfo
  {author} {\bibfnamefont {G.-B.}\ \bibnamefont {{Zhao}}}, \bibinfo {author}
  {\bibfnamefont {A.~J.}\ \bibnamefont {{Ross}}}, \bibinfo {author}
  {\bibfnamefont {M.}~\bibnamefont {{Manera}}}, \bibinfo {author}
  {\bibfnamefont {{\'E}.}~\bibnamefont {{Aubourg}}}, \bibinfo {author}
  {\bibfnamefont {F.}~\bibnamefont {{Beutler}}}, \bibinfo {author}
  {\bibfnamefont {J.}~\bibnamefont {{Brinkmann}}}, \bibinfo {author}
  {\bibfnamefont {J.~R.}\ \bibnamefont {{Brownstein}}}, \bibinfo {author}
  {\bibfnamefont {K.~S.}\ \bibnamefont {{Dawson}}}, \bibinfo {author}
  {\bibfnamefont {D.~J.}\ \bibnamefont {{Eisenstein}}}, \bibinfo {author}
  {\bibfnamefont {S.}~\bibnamefont {{Ho}}}, \bibinfo {author} {\bibfnamefont
  {K.}~\bibnamefont {{Honscheid}}}, \bibinfo {author} {\bibfnamefont
  {C.}~\bibnamefont {{Maraston}}}, \bibinfo {author} {\bibfnamefont
  {F.}~\bibnamefont {{Montesano}}}, \bibinfo {author} {\bibfnamefont {R.~C.}\
  \bibnamefont {{Nichol}}}, \bibinfo {author} {\bibfnamefont {N.~A.}\
  \bibnamefont {{Roe}}}, \bibinfo {author} {\bibfnamefont {N.~P.}\ \bibnamefont
  {{Ross}}}, \bibinfo {author} {\bibfnamefont {A.~G.}\ \bibnamefont
  {{S{\'a}nchez}}}, \bibinfo {author} {\bibfnamefont {D.~J.}\ \bibnamefont
  {{Schlegel}}}, \bibinfo {author} {\bibfnamefont {D.~P.}\ \bibnamefont
  {{Schneider}}}, \bibinfo {author} {\bibfnamefont {A.}~\bibnamefont
  {{Streblyanska}}}, \bibinfo {author} {\bibfnamefont {D.}~\bibnamefont
  {{Thomas}}}, \bibinfo {author} {\bibfnamefont {J.~L.}\ \bibnamefont
  {{Tinker}}}, \bibinfo {author} {\bibfnamefont {D.~A.}\ \bibnamefont
  {{Wake}}}, \bibinfo {author} {\bibfnamefont {B.~A.}\ \bibnamefont
  {{Weaver}}}, \ and\ \bibinfo {author} {\bibfnamefont {I.}~\bibnamefont
  {{Zehavi}}},\ }\href@noop {} {\  (\bibinfo {year} {2013})},\ \Eprint
  {http://arxiv.org/abs/1312.4899v1} {arXiv:1312.4899v1 [astro-ph.CO]}
  \BibitemShut {NoStop}%
\bibitem [{\citenamefont {{Beutler}}\ \emph {et~al.}(2013)\citenamefont
  {{Beutler}}, \citenamefont {{Saito}}, \citenamefont {{Seo}}, \citenamefont
  {{Brinkmann}}, \citenamefont {{Dawson}}, \citenamefont {{Eisenstein}},
  \citenamefont {{Font-Ribera}}, \citenamefont {{Ho}}, \citenamefont
  {{McBride}}, \citenamefont {{Montesano}}, \citenamefont {{Percival}},
  \citenamefont {{Ross}}, \citenamefont {{Ross}}, \citenamefont {{Samushia}},
  \citenamefont {{Schlegel}}, \citenamefont {{S{\'a}nchez}}, \citenamefont
  {{Tinker}},\ and\ \citenamefont {{Weaver}}}]{Beutler:2013kx}%
  \BibitemOpen
  \bibfield  {author} {\bibinfo {author} {\bibfnamefont {F.}~\bibnamefont
  {{Beutler}}}, \bibinfo {author} {\bibfnamefont {S.}~\bibnamefont {{Saito}}},
  \bibinfo {author} {\bibfnamefont {H.-J.}\ \bibnamefont {{Seo}}}, \bibinfo
  {author} {\bibfnamefont {J.}~\bibnamefont {{Brinkmann}}}, \bibinfo {author}
  {\bibfnamefont {K.~S.}\ \bibnamefont {{Dawson}}}, \bibinfo {author}
  {\bibfnamefont {D.~J.}\ \bibnamefont {{Eisenstein}}}, \bibinfo {author}
  {\bibfnamefont {A.}~\bibnamefont {{Font-Ribera}}}, \bibinfo {author}
  {\bibfnamefont {S.}~\bibnamefont {{Ho}}}, \bibinfo {author} {\bibfnamefont
  {C.~K.}\ \bibnamefont {{McBride}}}, \bibinfo {author} {\bibfnamefont
  {F.}~\bibnamefont {{Montesano}}}, \bibinfo {author} {\bibfnamefont {W.~J.}\
  \bibnamefont {{Percival}}}, \bibinfo {author} {\bibfnamefont {A.~J.}\
  \bibnamefont {{Ross}}}, \bibinfo {author} {\bibfnamefont {N.~P.}\
  \bibnamefont {{Ross}}}, \bibinfo {author} {\bibfnamefont {L.}~\bibnamefont
  {{Samushia}}}, \bibinfo {author} {\bibfnamefont {D.~J.}\ \bibnamefont
  {{Schlegel}}}, \bibinfo {author} {\bibfnamefont {A.~G.}\ \bibnamefont
  {{S{\'a}nchez}}}, \bibinfo {author} {\bibfnamefont {J.~L.}\ \bibnamefont
  {{Tinker}}}, \ and\ \bibinfo {author} {\bibfnamefont {B.~A.}\ \bibnamefont
  {{Weaver}}},\ }\href@noop {} {\  (\bibinfo {year} {2013})},\ \Eprint
  {http://arxiv.org/abs/1312.4611v1} {arXiv:1312.4611v1 [astro-ph.CO]}
  \BibitemShut {NoStop}%
\bibitem [{\citenamefont {{Beutler}}\ \emph {et~al.}(2014)\citenamefont
  {{Beutler}}, \citenamefont {{Saito}}, \citenamefont {{Brownstein}},
  \citenamefont {{Chuang}}, \citenamefont {{Cuesta}}, \citenamefont
  {{Percival}}, \citenamefont {{Ross}}, \citenamefont {{Ross}}, \citenamefont
  {{Schneider}}, \citenamefont {{Samushia}}, \citenamefont {{S{\'a}nchez}},
  \citenamefont {{Seo}}, \citenamefont {{Tinker}}, \citenamefont {{Wagner}},\
  and\ \citenamefont {{Weaver}}}]{Beutler:2014ty}%
  \BibitemOpen
  \bibfield  {author} {\bibinfo {author} {\bibfnamefont {F.}~\bibnamefont
  {{Beutler}}}, \bibinfo {author} {\bibfnamefont {S.}~\bibnamefont {{Saito}}},
  \bibinfo {author} {\bibfnamefont {J.~R.}\ \bibnamefont {{Brownstein}}},
  \bibinfo {author} {\bibfnamefont {C.-H.}\ \bibnamefont {{Chuang}}}, \bibinfo
  {author} {\bibfnamefont {A.~J.}\ \bibnamefont {{Cuesta}}}, \bibinfo {author}
  {\bibfnamefont {W.~J.}\ \bibnamefont {{Percival}}}, \bibinfo {author}
  {\bibfnamefont {A.~J.}\ \bibnamefont {{Ross}}}, \bibinfo {author}
  {\bibfnamefont {N.~P.}\ \bibnamefont {{Ross}}}, \bibinfo {author}
  {\bibfnamefont {D.~P.}\ \bibnamefont {{Schneider}}}, \bibinfo {author}
  {\bibfnamefont {L.}~\bibnamefont {{Samushia}}}, \bibinfo {author}
  {\bibfnamefont {A.~G.}\ \bibnamefont {{S{\'a}nchez}}}, \bibinfo {author}
  {\bibfnamefont {H.-J.}\ \bibnamefont {{Seo}}}, \bibinfo {author}
  {\bibfnamefont {J.~L.}\ \bibnamefont {{Tinker}}}, \bibinfo {author}
  {\bibfnamefont {C.}~\bibnamefont {{Wagner}}}, \ and\ \bibinfo {author}
  {\bibfnamefont {B.~A.}\ \bibnamefont {{Weaver}}},\ }\href@noop {} {\
  (\bibinfo {year} {2014})},\ \Eprint {http://arxiv.org/abs/1403.4599v1}
  {arXiv:1403.4599v1 [astro-ph.CO]} \BibitemShut {NoStop}%
\bibitem [{\citenamefont {{Zhao}}\ \emph {et~al.}(2013)\citenamefont {{Zhao}},
  \citenamefont {{Saito}}, \citenamefont {{Percival}}, \citenamefont {{Ross}},
  \citenamefont {{Montesano}}, \citenamefont {{Viel}}, \citenamefont
  {{Schneider}}, \citenamefont {{Ernst}}, \citenamefont {{Manera}},
  \citenamefont {{Miralda-Escude}}, \citenamefont {{Ross}}, \citenamefont
  {{Samushia}}, \citenamefont {{Sanchez}}, \citenamefont {{Swanson}},
  \citenamefont {{Thomas}}, \citenamefont {{Tojeiro}}, \citenamefont
  {{Yeche}},\ and\ \citenamefont {{York}}}]{Zhao:2013fk}%
  \BibitemOpen
  \bibfield  {author} {\bibinfo {author} {\bibfnamefont {G.-B.}\ \bibnamefont
  {{Zhao}}}, \bibinfo {author} {\bibfnamefont {S.}~\bibnamefont {{Saito}}},
  \bibinfo {author} {\bibfnamefont {W.~J.}\ \bibnamefont {{Percival}}},
  \bibinfo {author} {\bibfnamefont {A.~J.}\ \bibnamefont {{Ross}}}, \bibinfo
  {author} {\bibfnamefont {F.}~\bibnamefont {{Montesano}}}, \bibinfo {author}
  {\bibfnamefont {M.}~\bibnamefont {{Viel}}}, \bibinfo {author} {\bibfnamefont
  {D.~P.}\ \bibnamefont {{Schneider}}}, \bibinfo {author} {\bibfnamefont
  {D.~J.}\ \bibnamefont {{Ernst}}}, \bibinfo {author} {\bibfnamefont
  {M.}~\bibnamefont {{Manera}}}, \bibinfo {author} {\bibfnamefont
  {J.}~\bibnamefont {{Miralda-Escude}}}, \bibinfo {author} {\bibfnamefont
  {N.~P.}\ \bibnamefont {{Ross}}}, \bibinfo {author} {\bibfnamefont
  {L.}~\bibnamefont {{Samushia}}}, \bibinfo {author} {\bibfnamefont {A.~G.}\
  \bibnamefont {{Sanchez}}}, \bibinfo {author} {\bibfnamefont {M.~E.~C.}\
  \bibnamefont {{Swanson}}}, \bibinfo {author} {\bibfnamefont {D.}~\bibnamefont
  {{Thomas}}}, \bibinfo {author} {\bibfnamefont {R.}~\bibnamefont {{Tojeiro}}},
  \bibinfo {author} {\bibfnamefont {C.}~\bibnamefont {{Yeche}}}, \ and\
  \bibinfo {author} {\bibfnamefont {D.~G.}\ \bibnamefont {{York}}},\
  }\href@noop {} {\bibfield  {journal} {\bibinfo  {journal} {\mnras}\ }\textbf
  {\bibinfo {volume} {436}},\ \bibinfo {pages} {2038} (\bibinfo {year}
  {2013})},\ \Eprint {http://arxiv.org/abs/1211.3741} {arXiv:1211.3741
  [astro-ph.CO]} \BibitemShut {NoStop}%
\bibitem [{\citenamefont {{Reid}}\ \emph {et~al.}(2014)\citenamefont {{Reid}},
  \citenamefont {{Seo}}, \citenamefont {{Leauthaud}}, \citenamefont
  {{Tinker}},\ and\ \citenamefont {{White}}}]{Reid:2014qy}%
  \BibitemOpen
  \bibfield  {author} {\bibinfo {author} {\bibfnamefont {B.~A.}\ \bibnamefont
  {{Reid}}}, \bibinfo {author} {\bibfnamefont {H.-J.}\ \bibnamefont {{Seo}}},
  \bibinfo {author} {\bibfnamefont {A.}~\bibnamefont {{Leauthaud}}}, \bibinfo
  {author} {\bibfnamefont {J.~L.}\ \bibnamefont {{Tinker}}}, \ and\ \bibinfo
  {author} {\bibfnamefont {M.}~\bibnamefont {{White}}},\ }\href@noop {}
  {\bibfield  {journal} {\bibinfo  {journal} {ArXiv e-prints}\ } (\bibinfo
  {year} {2014})},\ \Eprint {http://arxiv.org/abs/1404.3742} {arXiv:1404.3742}
  \BibitemShut {NoStop}%
\bibitem [{\citenamefont {{Blake}}\ \emph
  {et~al.}(2011{\natexlab{a}})\citenamefont {{Blake}}, \citenamefont
  {{Brough}}, \citenamefont {{Colless}}, \citenamefont {{Contreras}},
  \citenamefont {{Couch}}, \citenamefont {{Croom}}, \citenamefont {{Davis}},
  \citenamefont {{Drinkwater}}, \citenamefont {{Forster}}, \citenamefont
  {{Gilbank}}, \citenamefont {{Gladders}}, \citenamefont {{Glazebrook}},
  \citenamefont {{Jelliffe}}, \citenamefont {{Jurek}}, \citenamefont {{Li}},
  \citenamefont {{Madore}}, \citenamefont {{Martin}}, \citenamefont
  {{Pimbblet}}, \citenamefont {{Poole}}, \citenamefont {{Pracy}}, \citenamefont
  {{Sharp}}, \citenamefont {{Wisnioski}}, \citenamefont {{Woods}},
  \citenamefont {{Wyder}},\ and\ \citenamefont {{Yee}}}]{Blake:2011fj}%
  \BibitemOpen
  \bibfield  {author} {\bibinfo {author} {\bibfnamefont {C.}~\bibnamefont
  {{Blake}}}, \bibinfo {author} {\bibfnamefont {S.}~\bibnamefont {{Brough}}},
  \bibinfo {author} {\bibfnamefont {M.}~\bibnamefont {{Colless}}}, \bibinfo
  {author} {\bibfnamefont {C.}~\bibnamefont {{Contreras}}}, \bibinfo {author}
  {\bibfnamefont {W.}~\bibnamefont {{Couch}}}, \bibinfo {author} {\bibfnamefont
  {S.}~\bibnamefont {{Croom}}}, \bibinfo {author} {\bibfnamefont
  {T.}~\bibnamefont {{Davis}}}, \bibinfo {author} {\bibfnamefont {M.~J.}\
  \bibnamefont {{Drinkwater}}}, \bibinfo {author} {\bibfnamefont
  {K.}~\bibnamefont {{Forster}}}, \bibinfo {author} {\bibfnamefont
  {D.}~\bibnamefont {{Gilbank}}}, \bibinfo {author} {\bibfnamefont
  {M.}~\bibnamefont {{Gladders}}}, \bibinfo {author} {\bibfnamefont
  {K.}~\bibnamefont {{Glazebrook}}}, \bibinfo {author} {\bibfnamefont
  {B.}~\bibnamefont {{Jelliffe}}}, \bibinfo {author} {\bibfnamefont {R.~J.}\
  \bibnamefont {{Jurek}}}, \bibinfo {author} {\bibfnamefont {I.-H.}\
  \bibnamefont {{Li}}}, \bibinfo {author} {\bibfnamefont {B.}~\bibnamefont
  {{Madore}}}, \bibinfo {author} {\bibfnamefont {D.~C.}\ \bibnamefont
  {{Martin}}}, \bibinfo {author} {\bibfnamefont {K.}~\bibnamefont
  {{Pimbblet}}}, \bibinfo {author} {\bibfnamefont {G.~B.}\ \bibnamefont
  {{Poole}}}, \bibinfo {author} {\bibfnamefont {M.}~\bibnamefont {{Pracy}}},
  \bibinfo {author} {\bibfnamefont {R.}~\bibnamefont {{Sharp}}}, \bibinfo
  {author} {\bibfnamefont {E.}~\bibnamefont {{Wisnioski}}}, \bibinfo {author}
  {\bibfnamefont {D.}~\bibnamefont {{Woods}}}, \bibinfo {author} {\bibfnamefont
  {T.~K.}\ \bibnamefont {{Wyder}}}, \ and\ \bibinfo {author} {\bibfnamefont
  {H.~K.~C.}\ \bibnamefont {{Yee}}},\ }\href {\doibase
  10.1111/j.1365-2966.2011.18903.x} {\bibfield  {journal} {\bibinfo  {journal}
  {\mnras}\ }\textbf {\bibinfo {volume} {415}},\ \bibinfo {pages} {2876}
  (\bibinfo {year} {2011}{\natexlab{a}})},\ \Eprint
  {http://arxiv.org/abs/1104.2948} {arXiv:1104.2948 [astro-ph.CO]} \BibitemShut
  {NoStop}%
\bibitem [{\citenamefont {{Blake}}\ \emph
  {et~al.}(2011{\natexlab{b}})\citenamefont {{Blake}}, \citenamefont {{Davis}},
  \citenamefont {{Poole}}, \citenamefont {{Parkinson}}, \citenamefont
  {{Brough}}, \citenamefont {{Colless}}, \citenamefont {{Contreras}},
  \citenamefont {{Couch}}, \citenamefont {{Croom}}, \citenamefont
  {{Drinkwater}}, \citenamefont {{Forster}}, \citenamefont {{Gilbank}},
  \citenamefont {{Gladders}}, \citenamefont {{Glazebrook}}, \citenamefont
  {{Jelliffe}}, \citenamefont {{Jurek}}, \citenamefont {{Li}}, \citenamefont
  {{Madore}}, \citenamefont {{Martin}}, \citenamefont {{Pimbblet}},
  \citenamefont {{Pracy}}, \citenamefont {{Sharp}}, \citenamefont
  {{Wisnioski}}, \citenamefont {{Woods}}, \citenamefont {{Wyder}},\ and\
  \citenamefont {{Yee}}}]{Blake:2011jf}%
  \BibitemOpen
  \bibfield  {author} {\bibinfo {author} {\bibfnamefont {C.}~\bibnamefont
  {{Blake}}}, \bibinfo {author} {\bibfnamefont {T.}~\bibnamefont {{Davis}}},
  \bibinfo {author} {\bibfnamefont {G.~B.}\ \bibnamefont {{Poole}}}, \bibinfo
  {author} {\bibfnamefont {D.}~\bibnamefont {{Parkinson}}}, \bibinfo {author}
  {\bibfnamefont {S.}~\bibnamefont {{Brough}}}, \bibinfo {author}
  {\bibfnamefont {M.}~\bibnamefont {{Colless}}}, \bibinfo {author}
  {\bibfnamefont {C.}~\bibnamefont {{Contreras}}}, \bibinfo {author}
  {\bibfnamefont {W.}~\bibnamefont {{Couch}}}, \bibinfo {author} {\bibfnamefont
  {S.}~\bibnamefont {{Croom}}}, \bibinfo {author} {\bibfnamefont {M.~J.}\
  \bibnamefont {{Drinkwater}}}, \bibinfo {author} {\bibfnamefont
  {K.}~\bibnamefont {{Forster}}}, \bibinfo {author} {\bibfnamefont
  {D.}~\bibnamefont {{Gilbank}}}, \bibinfo {author} {\bibfnamefont
  {M.}~\bibnamefont {{Gladders}}}, \bibinfo {author} {\bibfnamefont
  {K.}~\bibnamefont {{Glazebrook}}}, \bibinfo {author} {\bibfnamefont
  {B.}~\bibnamefont {{Jelliffe}}}, \bibinfo {author} {\bibfnamefont {R.~J.}\
  \bibnamefont {{Jurek}}}, \bibinfo {author} {\bibfnamefont {I.-H.}\
  \bibnamefont {{Li}}}, \bibinfo {author} {\bibfnamefont {B.}~\bibnamefont
  {{Madore}}}, \bibinfo {author} {\bibfnamefont {D.~C.}\ \bibnamefont
  {{Martin}}}, \bibinfo {author} {\bibfnamefont {K.}~\bibnamefont
  {{Pimbblet}}}, \bibinfo {author} {\bibfnamefont {M.}~\bibnamefont {{Pracy}}},
  \bibinfo {author} {\bibfnamefont {R.}~\bibnamefont {{Sharp}}}, \bibinfo
  {author} {\bibfnamefont {E.}~\bibnamefont {{Wisnioski}}}, \bibinfo {author}
  {\bibfnamefont {D.}~\bibnamefont {{Woods}}}, \bibinfo {author} {\bibfnamefont
  {T.~K.}\ \bibnamefont {{Wyder}}}, \ and\ \bibinfo {author} {\bibfnamefont
  {H.~K.~C.}\ \bibnamefont {{Yee}}},\ }\href {\doibase
  10.1111/j.1365-2966.2011.19077.x} {\bibfield  {journal} {\bibinfo  {journal}
  {\mnras}\ }\textbf {\bibinfo {volume} {415}},\ \bibinfo {pages} {2892}
  (\bibinfo {year} {2011}{\natexlab{b}})},\ \Eprint
  {http://arxiv.org/abs/1105.2862} {arXiv:1105.2862 [astro-ph.CO]} \BibitemShut
  {NoStop}%
\bibitem [{\citenamefont {{Blake}}\ \emph
  {et~al.}(2011{\natexlab{c}})\citenamefont {{Blake}}, \citenamefont {{Kazin}},
  \citenamefont {{Beutler}}, \citenamefont {{Davis}}, \citenamefont
  {{Parkinson}}, \citenamefont {{Brough}}, \citenamefont {{Colless}},
  \citenamefont {{Contreras}}, \citenamefont {{Couch}}, \citenamefont
  {{Croom}}, \citenamefont {{Croton}}, \citenamefont {{Drinkwater}},
  \citenamefont {{Forster}}, \citenamefont {{Gilbank}}, \citenamefont
  {{Gladders}}, \citenamefont {{Glazebrook}}, \citenamefont {{Jelliffe}},
  \citenamefont {{Jurek}}, \citenamefont {{Li}}, \citenamefont {{Madore}},
  \citenamefont {{Martin}}, \citenamefont {{Pimbblet}}, \citenamefont
  {{Poole}}, \citenamefont {{Pracy}}, \citenamefont {{Sharp}}, \citenamefont
  {{Wisnioski}}, \citenamefont {{Woods}}, \citenamefont {{Wyder}},\ and\
  \citenamefont {{Yee}}}]{Blake:2011kc}%
  \BibitemOpen
  \bibfield  {author} {\bibinfo {author} {\bibfnamefont {C.}~\bibnamefont
  {{Blake}}}, \bibinfo {author} {\bibfnamefont {E.~A.}\ \bibnamefont
  {{Kazin}}}, \bibinfo {author} {\bibfnamefont {F.}~\bibnamefont {{Beutler}}},
  \bibinfo {author} {\bibfnamefont {T.~M.}\ \bibnamefont {{Davis}}}, \bibinfo
  {author} {\bibfnamefont {D.}~\bibnamefont {{Parkinson}}}, \bibinfo {author}
  {\bibfnamefont {S.}~\bibnamefont {{Brough}}}, \bibinfo {author}
  {\bibfnamefont {M.}~\bibnamefont {{Colless}}}, \bibinfo {author}
  {\bibfnamefont {C.}~\bibnamefont {{Contreras}}}, \bibinfo {author}
  {\bibfnamefont {W.}~\bibnamefont {{Couch}}}, \bibinfo {author} {\bibfnamefont
  {S.}~\bibnamefont {{Croom}}}, \bibinfo {author} {\bibfnamefont
  {D.}~\bibnamefont {{Croton}}}, \bibinfo {author} {\bibfnamefont {M.~J.}\
  \bibnamefont {{Drinkwater}}}, \bibinfo {author} {\bibfnamefont
  {K.}~\bibnamefont {{Forster}}}, \bibinfo {author} {\bibfnamefont
  {D.}~\bibnamefont {{Gilbank}}}, \bibinfo {author} {\bibfnamefont
  {M.}~\bibnamefont {{Gladders}}}, \bibinfo {author} {\bibfnamefont
  {K.}~\bibnamefont {{Glazebrook}}}, \bibinfo {author} {\bibfnamefont
  {B.}~\bibnamefont {{Jelliffe}}}, \bibinfo {author} {\bibfnamefont {R.~J.}\
  \bibnamefont {{Jurek}}}, \bibinfo {author} {\bibfnamefont {I.-H.}\
  \bibnamefont {{Li}}}, \bibinfo {author} {\bibfnamefont {B.}~\bibnamefont
  {{Madore}}}, \bibinfo {author} {\bibfnamefont {D.~C.}\ \bibnamefont
  {{Martin}}}, \bibinfo {author} {\bibfnamefont {K.}~\bibnamefont
  {{Pimbblet}}}, \bibinfo {author} {\bibfnamefont {G.~B.}\ \bibnamefont
  {{Poole}}}, \bibinfo {author} {\bibfnamefont {M.}~\bibnamefont {{Pracy}}},
  \bibinfo {author} {\bibfnamefont {R.}~\bibnamefont {{Sharp}}}, \bibinfo
  {author} {\bibfnamefont {E.}~\bibnamefont {{Wisnioski}}}, \bibinfo {author}
  {\bibfnamefont {D.}~\bibnamefont {{Woods}}}, \bibinfo {author} {\bibfnamefont
  {T.~K.}\ \bibnamefont {{Wyder}}}, \ and\ \bibinfo {author} {\bibfnamefont
  {H.~K.~C.}\ \bibnamefont {{Yee}}},\ }\href {\doibase
  10.1111/j.1365-2966.2011.19592.x} {\bibfield  {journal} {\bibinfo  {journal}
  {\mnras}\ }\textbf {\bibinfo {volume} {418}},\ \bibinfo {pages} {1707}
  (\bibinfo {year} {2011}{\natexlab{c}})},\ \Eprint
  {http://arxiv.org/abs/1108.2635} {arXiv:1108.2635 [astro-ph.CO]} \BibitemShut
  {NoStop}%
\bibitem [{\citenamefont {{Blake}}\ \emph
  {et~al.}(2011{\natexlab{d}})\citenamefont {{Blake}}, \citenamefont
  {{Glazebrook}}, \citenamefont {{Davis}}, \citenamefont {{Brough}},
  \citenamefont {{Colless}}, \citenamefont {{Contreras}}, \citenamefont
  {{Couch}}, \citenamefont {{Croom}}, \citenamefont {{Drinkwater}},
  \citenamefont {{Forster}}, \citenamefont {{Gilbank}}, \citenamefont
  {{Gladders}}, \citenamefont {{Jelliffe}}, \citenamefont {{Jurek}},
  \citenamefont {{Li}}, \citenamefont {{Madore}}, \citenamefont {{Martin}},
  \citenamefont {{Pimbblet}}, \citenamefont {{Poole}}, \citenamefont {{Pracy}},
  \citenamefont {{Sharp}}, \citenamefont {{Wisnioski}}, \citenamefont
  {{Woods}}, \citenamefont {{Wyder}},\ and\ \citenamefont
  {{Yee}}}]{Blake:2011zv}%
  \BibitemOpen
  \bibfield  {author} {\bibinfo {author} {\bibfnamefont {C.}~\bibnamefont
  {{Blake}}}, \bibinfo {author} {\bibfnamefont {K.}~\bibnamefont
  {{Glazebrook}}}, \bibinfo {author} {\bibfnamefont {T.~M.}\ \bibnamefont
  {{Davis}}}, \bibinfo {author} {\bibfnamefont {S.}~\bibnamefont {{Brough}}},
  \bibinfo {author} {\bibfnamefont {M.}~\bibnamefont {{Colless}}}, \bibinfo
  {author} {\bibfnamefont {C.}~\bibnamefont {{Contreras}}}, \bibinfo {author}
  {\bibfnamefont {W.}~\bibnamefont {{Couch}}}, \bibinfo {author} {\bibfnamefont
  {S.}~\bibnamefont {{Croom}}}, \bibinfo {author} {\bibfnamefont {M.~J.}\
  \bibnamefont {{Drinkwater}}}, \bibinfo {author} {\bibfnamefont
  {K.}~\bibnamefont {{Forster}}}, \bibinfo {author} {\bibfnamefont
  {D.}~\bibnamefont {{Gilbank}}}, \bibinfo {author} {\bibfnamefont
  {M.}~\bibnamefont {{Gladders}}}, \bibinfo {author} {\bibfnamefont
  {B.}~\bibnamefont {{Jelliffe}}}, \bibinfo {author} {\bibfnamefont {R.~J.}\
  \bibnamefont {{Jurek}}}, \bibinfo {author} {\bibfnamefont {I.-H.}\
  \bibnamefont {{Li}}}, \bibinfo {author} {\bibfnamefont {B.}~\bibnamefont
  {{Madore}}}, \bibinfo {author} {\bibfnamefont {D.~C.}\ \bibnamefont
  {{Martin}}}, \bibinfo {author} {\bibfnamefont {K.}~\bibnamefont
  {{Pimbblet}}}, \bibinfo {author} {\bibfnamefont {G.~B.}\ \bibnamefont
  {{Poole}}}, \bibinfo {author} {\bibfnamefont {M.}~\bibnamefont {{Pracy}}},
  \bibinfo {author} {\bibfnamefont {R.}~\bibnamefont {{Sharp}}}, \bibinfo
  {author} {\bibfnamefont {E.}~\bibnamefont {{Wisnioski}}}, \bibinfo {author}
  {\bibfnamefont {D.}~\bibnamefont {{Woods}}}, \bibinfo {author} {\bibfnamefont
  {T.~K.}\ \bibnamefont {{Wyder}}}, \ and\ \bibinfo {author} {\bibfnamefont
  {H.~K.~C.}\ \bibnamefont {{Yee}}},\ }\href {\doibase
  10.1111/j.1365-2966.2011.19606.x} {\bibfield  {journal} {\bibinfo  {journal}
  {\mnras}\ }\textbf {\bibinfo {volume} {418}},\ \bibinfo {pages} {1725}
  (\bibinfo {year} {2011}{\natexlab{d}})},\ \Eprint
  {http://arxiv.org/abs/1108.2637} {arXiv:1108.2637 [astro-ph.CO]} \BibitemShut
  {NoStop}%
\bibitem [{\citenamefont {{Blake}}\ \emph {et~al.}(2012)\citenamefont
  {{Blake}}, \citenamefont {{Brough}}, \citenamefont {{Colless}}, \citenamefont
  {{Contreras}}, \citenamefont {{Couch}}, \citenamefont {{Croom}},
  \citenamefont {{Croton}}, \citenamefont {{Davis}}, \citenamefont
  {{Drinkwater}}, \citenamefont {{Forster}}, \citenamefont {{Gilbank}},
  \citenamefont {{Gladders}}, \citenamefont {{Glazebrook}}, \citenamefont
  {{Jelliffe}}, \citenamefont {{Jurek}}, \citenamefont {{Li}}, \citenamefont
  {{Madore}}, \citenamefont {{Martin}}, \citenamefont {{Pimbblet}},
  \citenamefont {{Poole}}, \citenamefont {{Pracy}}, \citenamefont {{Sharp}},
  \citenamefont {{Wisnioski}}, \citenamefont {{Woods}}, \citenamefont
  {{Wyder}},\ and\ \citenamefont {{Yee}}}]{Blake:2012nx}%
  \BibitemOpen
  \bibfield  {author} {\bibinfo {author} {\bibfnamefont {C.}~\bibnamefont
  {{Blake}}}, \bibinfo {author} {\bibfnamefont {S.}~\bibnamefont {{Brough}}},
  \bibinfo {author} {\bibfnamefont {M.}~\bibnamefont {{Colless}}}, \bibinfo
  {author} {\bibfnamefont {C.}~\bibnamefont {{Contreras}}}, \bibinfo {author}
  {\bibfnamefont {W.}~\bibnamefont {{Couch}}}, \bibinfo {author} {\bibfnamefont
  {S.}~\bibnamefont {{Croom}}}, \bibinfo {author} {\bibfnamefont
  {D.}~\bibnamefont {{Croton}}}, \bibinfo {author} {\bibfnamefont {T.~M.}\
  \bibnamefont {{Davis}}}, \bibinfo {author} {\bibfnamefont {M.~J.}\
  \bibnamefont {{Drinkwater}}}, \bibinfo {author} {\bibfnamefont
  {K.}~\bibnamefont {{Forster}}}, \bibinfo {author} {\bibfnamefont
  {D.}~\bibnamefont {{Gilbank}}}, \bibinfo {author} {\bibfnamefont
  {M.}~\bibnamefont {{Gladders}}}, \bibinfo {author} {\bibfnamefont
  {K.}~\bibnamefont {{Glazebrook}}}, \bibinfo {author} {\bibfnamefont
  {B.}~\bibnamefont {{Jelliffe}}}, \bibinfo {author} {\bibfnamefont {R.~J.}\
  \bibnamefont {{Jurek}}}, \bibinfo {author} {\bibfnamefont {I.-h.}\
  \bibnamefont {{Li}}}, \bibinfo {author} {\bibfnamefont {B.}~\bibnamefont
  {{Madore}}}, \bibinfo {author} {\bibfnamefont {D.~C.}\ \bibnamefont
  {{Martin}}}, \bibinfo {author} {\bibfnamefont {K.}~\bibnamefont
  {{Pimbblet}}}, \bibinfo {author} {\bibfnamefont {G.~B.}\ \bibnamefont
  {{Poole}}}, \bibinfo {author} {\bibfnamefont {M.}~\bibnamefont {{Pracy}}},
  \bibinfo {author} {\bibfnamefont {R.}~\bibnamefont {{Sharp}}}, \bibinfo
  {author} {\bibfnamefont {E.}~\bibnamefont {{Wisnioski}}}, \bibinfo {author}
  {\bibfnamefont {D.}~\bibnamefont {{Woods}}}, \bibinfo {author} {\bibfnamefont
  {T.~K.}\ \bibnamefont {{Wyder}}}, \ and\ \bibinfo {author} {\bibfnamefont
  {H.~K.~C.}\ \bibnamefont {{Yee}}},\ }\href {\doibase
  10.1111/j.1365-2966.2012.21473.x} {\bibfield  {journal} {\bibinfo  {journal}
  {\mnras}\ }\textbf {\bibinfo {volume} {425}},\ \bibinfo {pages} {405}
  (\bibinfo {year} {2012})},\ \Eprint {http://arxiv.org/abs/1204.3674}
  {arXiv:1204.3674 [astro-ph.CO]} \BibitemShut {NoStop}%
\bibitem [{\citenamefont {{Contreras}}\ \emph {et~al.}(2013)\citenamefont
  {{Contreras}}, \citenamefont {{Blake}}, \citenamefont {{Poole}},
  \citenamefont {{Marin}}, \citenamefont {{Brough}}, \citenamefont {{Colless}},
  \citenamefont {{Couch}}, \citenamefont {{Croom}}, \citenamefont {{Croton}},
  \citenamefont {{Davis}}, \citenamefont {{Drinkwater}}, \citenamefont
  {{Forster}}, \citenamefont {{Gilbank}}, \citenamefont {{Gladders}},
  \citenamefont {{Glazebrook}}, \citenamefont {{Jelliffe}}, \citenamefont
  {{Jurek}}, \citenamefont {{Li}}, \citenamefont {{Madore}}, \citenamefont
  {{Martin}}, \citenamefont {{Pimbblet}}, \citenamefont {{Pracy}},
  \citenamefont {{Sharp}}, \citenamefont {{Wisnioski}}, \citenamefont
  {{Woods}}, \citenamefont {{Wyder}},\ and\ \citenamefont
  {{Yee}}}]{Contreras:2013ys}%
  \BibitemOpen
  \bibfield  {author} {\bibinfo {author} {\bibfnamefont {C.}~\bibnamefont
  {{Contreras}}}, \bibinfo {author} {\bibfnamefont {C.}~\bibnamefont
  {{Blake}}}, \bibinfo {author} {\bibfnamefont {G.~B.}\ \bibnamefont
  {{Poole}}}, \bibinfo {author} {\bibfnamefont {F.}~\bibnamefont {{Marin}}},
  \bibinfo {author} {\bibfnamefont {S.}~\bibnamefont {{Brough}}}, \bibinfo
  {author} {\bibfnamefont {M.}~\bibnamefont {{Colless}}}, \bibinfo {author}
  {\bibfnamefont {W.}~\bibnamefont {{Couch}}}, \bibinfo {author} {\bibfnamefont
  {S.}~\bibnamefont {{Croom}}}, \bibinfo {author} {\bibfnamefont
  {D.}~\bibnamefont {{Croton}}}, \bibinfo {author} {\bibfnamefont {T.~M.}\
  \bibnamefont {{Davis}}}, \bibinfo {author} {\bibfnamefont {M.~J.}\
  \bibnamefont {{Drinkwater}}}, \bibinfo {author} {\bibfnamefont
  {K.}~\bibnamefont {{Forster}}}, \bibinfo {author} {\bibfnamefont
  {D.}~\bibnamefont {{Gilbank}}}, \bibinfo {author} {\bibfnamefont
  {M.}~\bibnamefont {{Gladders}}}, \bibinfo {author} {\bibfnamefont
  {K.}~\bibnamefont {{Glazebrook}}}, \bibinfo {author} {\bibfnamefont
  {B.}~\bibnamefont {{Jelliffe}}}, \bibinfo {author} {\bibfnamefont {R.~J.}\
  \bibnamefont {{Jurek}}}, \bibinfo {author} {\bibfnamefont {I.-h.}\
  \bibnamefont {{Li}}}, \bibinfo {author} {\bibfnamefont {B.}~\bibnamefont
  {{Madore}}}, \bibinfo {author} {\bibfnamefont {D.~C.}\ \bibnamefont
  {{Martin}}}, \bibinfo {author} {\bibfnamefont {K.}~\bibnamefont
  {{Pimbblet}}}, \bibinfo {author} {\bibfnamefont {M.}~\bibnamefont {{Pracy}}},
  \bibinfo {author} {\bibfnamefont {R.}~\bibnamefont {{Sharp}}}, \bibinfo
  {author} {\bibfnamefont {E.}~\bibnamefont {{Wisnioski}}}, \bibinfo {author}
  {\bibfnamefont {D.}~\bibnamefont {{Woods}}}, \bibinfo {author} {\bibfnamefont
  {T.~K.}\ \bibnamefont {{Wyder}}}, \ and\ \bibinfo {author} {\bibfnamefont
  {H.~K.~C.}\ \bibnamefont {{Yee}}},\ }\href {\doibase 10.1093/mnras/sts608}
  {\bibfield  {journal} {\bibinfo  {journal} {\mnras}\ }\textbf {\bibinfo
  {volume} {430}},\ \bibinfo {pages} {924} (\bibinfo {year} {2013})},\ \Eprint
  {http://arxiv.org/abs/1302.5178} {arXiv:1302.5178 [astro-ph.CO]} \BibitemShut
  {NoStop}%
\bibitem [{\citenamefont {{Scrimgeour}}\ \emph {et~al.}(2012)\citenamefont
  {{Scrimgeour}}, \citenamefont {{Davis}}, \citenamefont {{Blake}},
  \citenamefont {{James}}, \citenamefont {{Poole}}, \citenamefont
  {{Staveley-Smith}}, \citenamefont {{Brough}}, \citenamefont {{Colless}},
  \citenamefont {{Contreras}}, \citenamefont {{Couch}}, \citenamefont
  {{Croom}}, \citenamefont {{Croton}}, \citenamefont {{Drinkwater}},
  \citenamefont {{Forster}}, \citenamefont {{Gilbank}}, \citenamefont
  {{Gladders}}, \citenamefont {{Glazebrook}}, \citenamefont {{Jelliffe}},
  \citenamefont {{Jurek}}, \citenamefont {{Li}}, \citenamefont {{Madore}},
  \citenamefont {{Martin}}, \citenamefont {{Pimbblet}}, \citenamefont
  {{Pracy}}, \citenamefont {{Sharp}}, \citenamefont {{Wisnioski}},
  \citenamefont {{Woods}}, \citenamefont {{Wyder}},\ and\ \citenamefont
  {{Yee}}}]{Scrimgeour:2012pd}%
  \BibitemOpen
  \bibfield  {author} {\bibinfo {author} {\bibfnamefont {M.~I.}\ \bibnamefont
  {{Scrimgeour}}}, \bibinfo {author} {\bibfnamefont {T.}~\bibnamefont
  {{Davis}}}, \bibinfo {author} {\bibfnamefont {C.}~\bibnamefont {{Blake}}},
  \bibinfo {author} {\bibfnamefont {J.~B.}\ \bibnamefont {{James}}}, \bibinfo
  {author} {\bibfnamefont {G.~B.}\ \bibnamefont {{Poole}}}, \bibinfo {author}
  {\bibfnamefont {L.}~\bibnamefont {{Staveley-Smith}}}, \bibinfo {author}
  {\bibfnamefont {S.}~\bibnamefont {{Brough}}}, \bibinfo {author}
  {\bibfnamefont {M.}~\bibnamefont {{Colless}}}, \bibinfo {author}
  {\bibfnamefont {C.}~\bibnamefont {{Contreras}}}, \bibinfo {author}
  {\bibfnamefont {W.}~\bibnamefont {{Couch}}}, \bibinfo {author} {\bibfnamefont
  {S.}~\bibnamefont {{Croom}}}, \bibinfo {author} {\bibfnamefont
  {D.}~\bibnamefont {{Croton}}}, \bibinfo {author} {\bibfnamefont {M.~J.}\
  \bibnamefont {{Drinkwater}}}, \bibinfo {author} {\bibfnamefont
  {K.}~\bibnamefont {{Forster}}}, \bibinfo {author} {\bibfnamefont
  {D.}~\bibnamefont {{Gilbank}}}, \bibinfo {author} {\bibfnamefont
  {M.}~\bibnamefont {{Gladders}}}, \bibinfo {author} {\bibfnamefont
  {K.}~\bibnamefont {{Glazebrook}}}, \bibinfo {author} {\bibfnamefont
  {B.}~\bibnamefont {{Jelliffe}}}, \bibinfo {author} {\bibfnamefont {R.~J.}\
  \bibnamefont {{Jurek}}}, \bibinfo {author} {\bibfnamefont {I.-h.}\
  \bibnamefont {{Li}}}, \bibinfo {author} {\bibfnamefont {B.}~\bibnamefont
  {{Madore}}}, \bibinfo {author} {\bibfnamefont {D.~C.}\ \bibnamefont
  {{Martin}}}, \bibinfo {author} {\bibfnamefont {K.}~\bibnamefont
  {{Pimbblet}}}, \bibinfo {author} {\bibfnamefont {M.}~\bibnamefont {{Pracy}}},
  \bibinfo {author} {\bibfnamefont {R.}~\bibnamefont {{Sharp}}}, \bibinfo
  {author} {\bibfnamefont {E.}~\bibnamefont {{Wisnioski}}}, \bibinfo {author}
  {\bibfnamefont {D.}~\bibnamefont {{Woods}}}, \bibinfo {author} {\bibfnamefont
  {T.~K.}\ \bibnamefont {{Wyder}}}, \ and\ \bibinfo {author} {\bibfnamefont
  {H.~K.~C.}\ \bibnamefont {{Yee}}},\ }\href {\doibase
  10.1111/j.1365-2966.2012.21402.x} {\bibfield  {journal} {\bibinfo  {journal}
  {\mnras}\ }\textbf {\bibinfo {volume} {425}},\ \bibinfo {pages} {116}
  (\bibinfo {year} {2012})},\ \Eprint {http://arxiv.org/abs/1205.6812}
  {arXiv:1205.6812 [astro-ph.CO]} \BibitemShut {NoStop}%
\bibitem [{\citenamefont {{Takada}}\ \emph {et~al.}(2012)\citenamefont
  {{Takada}}, \citenamefont {{Ellis}}, \citenamefont {{Chiba}}, \citenamefont
  {{Greene}}, \citenamefont {{Aihara}}, \citenamefont {{Arimoto}},
  \citenamefont {{Bundy}}, \citenamefont {{Cohen}}, \citenamefont {{Dor{\'e}}},
  \citenamefont {{Graves}}, \citenamefont {{Gunn}}, \citenamefont {{Heckman}},
  \citenamefont {{Hirata}}, \citenamefont {{Ho}}, \citenamefont {{Kneib}},
  \citenamefont {{Le F{\`e}vre}}, \citenamefont {{Lin}}, \citenamefont
  {{More}}, \citenamefont {{Murayama}}, \citenamefont {{Nagao}}, \citenamefont
  {{Ouchi}}, \citenamefont {{Seiffert}}, \citenamefont {{Silverman}},
  \citenamefont {{Sodr{\'e}}}, \citenamefont {{Spergel}}, \citenamefont
  {{Strauss}}, \citenamefont {{Sugai}}, \citenamefont {{Suto}}, \citenamefont
  {{Takami}},\ and\ \citenamefont {{Wyse}}}]{Takada:2012ww}%
  \BibitemOpen
  \bibfield  {author} {\bibinfo {author} {\bibfnamefont {M.}~\bibnamefont
  {{Takada}}}, \bibinfo {author} {\bibfnamefont {R.}~\bibnamefont {{Ellis}}},
  \bibinfo {author} {\bibfnamefont {M.}~\bibnamefont {{Chiba}}}, \bibinfo
  {author} {\bibfnamefont {J.~E.}\ \bibnamefont {{Greene}}}, \bibinfo {author}
  {\bibfnamefont {H.}~\bibnamefont {{Aihara}}}, \bibinfo {author}
  {\bibfnamefont {N.}~\bibnamefont {{Arimoto}}}, \bibinfo {author}
  {\bibfnamefont {K.}~\bibnamefont {{Bundy}}}, \bibinfo {author} {\bibfnamefont
  {J.}~\bibnamefont {{Cohen}}}, \bibinfo {author} {\bibfnamefont
  {O.}~\bibnamefont {{Dor{\'e}}}}, \bibinfo {author} {\bibfnamefont
  {G.}~\bibnamefont {{Graves}}}, \bibinfo {author} {\bibfnamefont {J.~E.}\
  \bibnamefont {{Gunn}}}, \bibinfo {author} {\bibfnamefont {T.}~\bibnamefont
  {{Heckman}}}, \bibinfo {author} {\bibfnamefont {C.}~\bibnamefont {{Hirata}}},
  \bibinfo {author} {\bibfnamefont {P.}~\bibnamefont {{Ho}}}, \bibinfo {author}
  {\bibfnamefont {J.-P.}\ \bibnamefont {{Kneib}}}, \bibinfo {author}
  {\bibfnamefont {O.}~\bibnamefont {{Le F{\`e}vre}}}, \bibinfo {author}
  {\bibfnamefont {L.}~\bibnamefont {{Lin}}}, \bibinfo {author} {\bibfnamefont
  {S.}~\bibnamefont {{More}}}, \bibinfo {author} {\bibfnamefont
  {H.}~\bibnamefont {{Murayama}}}, \bibinfo {author} {\bibfnamefont
  {T.}~\bibnamefont {{Nagao}}}, \bibinfo {author} {\bibfnamefont
  {M.}~\bibnamefont {{Ouchi}}}, \bibinfo {author} {\bibfnamefont
  {M.}~\bibnamefont {{Seiffert}}}, \bibinfo {author} {\bibfnamefont
  {J.}~\bibnamefont {{Silverman}}}, \bibinfo {author} {\bibfnamefont
  {L.}~\bibnamefont {{Sodr{\'e}}}, \bibfnamefont {Jr}}, \bibinfo {author}
  {\bibfnamefont {D.~N.}\ \bibnamefont {{Spergel}}}, \bibinfo {author}
  {\bibfnamefont {M.~A.}\ \bibnamefont {{Strauss}}}, \bibinfo {author}
  {\bibfnamefont {H.}~\bibnamefont {{Sugai}}}, \bibinfo {author} {\bibfnamefont
  {Y.}~\bibnamefont {{Suto}}}, \bibinfo {author} {\bibfnamefont
  {H.}~\bibnamefont {{Takami}}}, \ and\ \bibinfo {author} {\bibfnamefont
  {R.}~\bibnamefont {{Wyse}}},\ }\href@noop {} {\bibfield  {journal} {\bibinfo
  {journal} {ArXiv e-prints}\ } (\bibinfo {year} {2012})},\ \Eprint
  {http://arxiv.org/abs/1206.0737} {arXiv:1206.0737 [astro-ph.CO]} \BibitemShut
  {NoStop}%
\bibitem [{\citenamefont {{Adams}}\ \emph {et~al.}(2011)\citenamefont
  {{Adams}}, \citenamefont {{Blanc}}, \citenamefont {{Hill}}, \citenamefont
  {{Gebhardt}}, \citenamefont {{Drory}}, \citenamefont {{Hao}}, \citenamefont
  {{Bender}}, \citenamefont {{Byun}}, \citenamefont {{Ciardullo}},
  \citenamefont {{Cornell}}, \citenamefont {{Finkelstein}}, \citenamefont
  {{Fry}}, \citenamefont {{Gawiser}}, \citenamefont {{Gronwall}}, \citenamefont
  {{Hopp}}, \citenamefont {{Jeong}}, \citenamefont {{Kelz}}, \citenamefont
  {{Kelzenberg}}, \citenamefont {{Komatsu}}, \citenamefont {{MacQueen}},
  \citenamefont {{Murphy}}, \citenamefont {{Odoms}}, \citenamefont {{Roth}},
  \citenamefont {{Schneider}}, \citenamefont {{Tufts}},\ and\ \citenamefont
  {{Wilkinson}}}]{Adams:2011ak}%
  \BibitemOpen
  \bibfield  {author} {\bibinfo {author} {\bibfnamefont {J.~J.}\ \bibnamefont
  {{Adams}}}, \bibinfo {author} {\bibfnamefont {G.~A.}\ \bibnamefont
  {{Blanc}}}, \bibinfo {author} {\bibfnamefont {G.~J.}\ \bibnamefont {{Hill}}},
  \bibinfo {author} {\bibfnamefont {K.}~\bibnamefont {{Gebhardt}}}, \bibinfo
  {author} {\bibfnamefont {N.}~\bibnamefont {{Drory}}}, \bibinfo {author}
  {\bibfnamefont {L.}~\bibnamefont {{Hao}}}, \bibinfo {author} {\bibfnamefont
  {R.}~\bibnamefont {{Bender}}}, \bibinfo {author} {\bibfnamefont
  {J.}~\bibnamefont {{Byun}}}, \bibinfo {author} {\bibfnamefont
  {R.}~\bibnamefont {{Ciardullo}}}, \bibinfo {author} {\bibfnamefont {M.~E.}\
  \bibnamefont {{Cornell}}}, \bibinfo {author} {\bibfnamefont {S.~L.}\
  \bibnamefont {{Finkelstein}}}, \bibinfo {author} {\bibfnamefont
  {A.}~\bibnamefont {{Fry}}}, \bibinfo {author} {\bibfnamefont
  {E.}~\bibnamefont {{Gawiser}}}, \bibinfo {author} {\bibfnamefont
  {C.}~\bibnamefont {{Gronwall}}}, \bibinfo {author} {\bibfnamefont
  {U.}~\bibnamefont {{Hopp}}}, \bibinfo {author} {\bibfnamefont
  {D.}~\bibnamefont {{Jeong}}}, \bibinfo {author} {\bibfnamefont
  {A.}~\bibnamefont {{Kelz}}}, \bibinfo {author} {\bibfnamefont
  {R.}~\bibnamefont {{Kelzenberg}}}, \bibinfo {author} {\bibfnamefont
  {E.}~\bibnamefont {{Komatsu}}}, \bibinfo {author} {\bibfnamefont {P.~J.}\
  \bibnamefont {{MacQueen}}}, \bibinfo {author} {\bibfnamefont
  {J.}~\bibnamefont {{Murphy}}}, \bibinfo {author} {\bibfnamefont {P.~S.}\
  \bibnamefont {{Odoms}}}, \bibinfo {author} {\bibfnamefont {M.}~\bibnamefont
  {{Roth}}}, \bibinfo {author} {\bibfnamefont {D.~P.}\ \bibnamefont
  {{Schneider}}}, \bibinfo {author} {\bibfnamefont {J.~R.}\ \bibnamefont
  {{Tufts}}}, \ and\ \bibinfo {author} {\bibfnamefont {C.~P.}\ \bibnamefont
  {{Wilkinson}}},\ }\href {\doibase 10.1088/0067-0049/192/1/5} {\bibfield
  {journal} {\bibinfo  {journal} {\apjs}\ }\textbf {\bibinfo {volume} {192}},\
  \bibinfo {eid} {5} (\bibinfo {year} {2011})},\ \Eprint
  {http://arxiv.org/abs/1011.0426} {arXiv:1011.0426 [astro-ph.CO]} \BibitemShut
  {NoStop}%
\bibitem [{\citenamefont {{Levi}}\ \emph {et~al.}(2013)\citenamefont {{Levi}},
  \citenamefont {{Bebek}}, \citenamefont {{Beers}}, \citenamefont {{Blum}},
  \citenamefont {{Cahn}}, \citenamefont {{Eisenstein}}, \citenamefont
  {{Flaugher}}, \citenamefont {{Honscheid}}, \citenamefont {{Kron}},
  \citenamefont {{Lahav}}, \citenamefont {{McDonald}}, \citenamefont {{Roe}},
  \citenamefont {{Schlegel}},\ and\ \citenamefont {{representing the DESI
  collaboration}}}]{Levi:2013ly}%
  \BibitemOpen
  \bibfield  {author} {\bibinfo {author} {\bibfnamefont {M.}~\bibnamefont
  {{Levi}}}, \bibinfo {author} {\bibfnamefont {C.}~\bibnamefont {{Bebek}}},
  \bibinfo {author} {\bibfnamefont {T.}~\bibnamefont {{Beers}}}, \bibinfo
  {author} {\bibfnamefont {R.}~\bibnamefont {{Blum}}}, \bibinfo {author}
  {\bibfnamefont {R.}~\bibnamefont {{Cahn}}}, \bibinfo {author} {\bibfnamefont
  {D.}~\bibnamefont {{Eisenstein}}}, \bibinfo {author} {\bibfnamefont
  {B.}~\bibnamefont {{Flaugher}}}, \bibinfo {author} {\bibfnamefont
  {K.}~\bibnamefont {{Honscheid}}}, \bibinfo {author} {\bibfnamefont
  {R.}~\bibnamefont {{Kron}}}, \bibinfo {author} {\bibfnamefont
  {O.}~\bibnamefont {{Lahav}}}, \bibinfo {author} {\bibfnamefont
  {P.}~\bibnamefont {{McDonald}}}, \bibinfo {author} {\bibfnamefont
  {N.}~\bibnamefont {{Roe}}}, \bibinfo {author} {\bibfnamefont
  {D.}~\bibnamefont {{Schlegel}}}, \ and\ \bibinfo {author} {\bibnamefont
  {{representing the DESI collaboration}}},\ }\href@noop {} {\bibfield
  {journal} {\bibinfo  {journal} {ArXiv e-prints}\ } (\bibinfo {year}
  {2013})},\ \Eprint {http://arxiv.org/abs/1308.0847} {arXiv:1308.0847
  [astro-ph.CO]} \BibitemShut {NoStop}%
\bibitem [{\citenamefont {{Laureijs}}\ \emph {et~al.}(2011)\citenamefont
  {{Laureijs}}, \citenamefont {{Amiaux}}, \citenamefont {{Arduini}},
  \citenamefont {{Augu{\`e}res}}, \citenamefont {{Brinchmann}}, \citenamefont
  {{Cole}}, \citenamefont {{Cropper}}, \citenamefont {{Dabin}}, \citenamefont
  {{Duvet}}, \citenamefont {{Ealet}},\ and\ \citenamefont
  {et~al.}}]{Laureijs:2011jw}%
  \BibitemOpen
  \bibfield  {author} {\bibinfo {author} {\bibfnamefont {R.}~\bibnamefont
  {{Laureijs}}}, \bibinfo {author} {\bibfnamefont {J.}~\bibnamefont
  {{Amiaux}}}, \bibinfo {author} {\bibfnamefont {S.}~\bibnamefont {{Arduini}}},
  \bibinfo {author} {\bibfnamefont {J.~.}\ \bibnamefont {{Augu{\`e}res}}},
  \bibinfo {author} {\bibfnamefont {J.}~\bibnamefont {{Brinchmann}}}, \bibinfo
  {author} {\bibfnamefont {R.}~\bibnamefont {{Cole}}}, \bibinfo {author}
  {\bibfnamefont {M.}~\bibnamefont {{Cropper}}}, \bibinfo {author}
  {\bibfnamefont {C.}~\bibnamefont {{Dabin}}}, \bibinfo {author} {\bibfnamefont
  {L.}~\bibnamefont {{Duvet}}}, \bibinfo {author} {\bibfnamefont
  {A.}~\bibnamefont {{Ealet}}}, \ and\ \bibinfo {author} {\bibnamefont
  {et~al.}},\ }\href@noop {} {\bibfield  {journal} {\bibinfo  {journal} {ArXiv
  e-prints}\ } (\bibinfo {year} {2011})},\ \Eprint
  {http://arxiv.org/abs/1110.3193} {arXiv:1110.3193 [astro-ph.CO]} \BibitemShut
  {NoStop}%
\bibitem [{\citenamefont {{Cooray}}\ and\ \citenamefont
  {{Sheth}}(2002)}]{Cooray:2002lr}%
  \BibitemOpen
  \bibfield  {author} {\bibinfo {author} {\bibfnamefont {A.}~\bibnamefont
  {{Cooray}}}\ and\ \bibinfo {author} {\bibfnamefont {R.}~\bibnamefont
  {{Sheth}}},\ }\href {\doibase 10.1016/S0370-1573(02)00276-4} {\bibfield
  {journal} {\bibinfo  {journal} {\physrep}\ }\textbf {\bibinfo {volume}
  {372}},\ \bibinfo {pages} {1} (\bibinfo {year} {2002})},\ \Eprint
  {http://arxiv.org/abs/arXiv:astro-ph/0206508} {arXiv:astro-ph/0206508}
  \BibitemShut {NoStop}%
\bibitem [{\citenamefont {{Seljak}}(2000)}]{Seljak:2000mz}%
  \BibitemOpen
  \bibfield  {author} {\bibinfo {author} {\bibfnamefont {U.}~\bibnamefont
  {{Seljak}}},\ }\href {\doibase 10.1046/j.1365-8711.2000.03715.x} {\bibfield
  {journal} {\bibinfo  {journal} {\mnras}\ }\textbf {\bibinfo {volume} {318}},\
  \bibinfo {pages} {203} (\bibinfo {year} {2000})},\ \Eprint
  {http://arxiv.org/abs/astro-ph/0001493} {astro-ph/0001493} \BibitemShut
  {NoStop}%
\bibitem [{\citenamefont {{Behroozi}}\ \emph {et~al.}(2010)\citenamefont
  {{Behroozi}}, \citenamefont {{Conroy}},\ and\ \citenamefont
  {{Wechsler}}}]{Behroozi:2010lr}%
  \BibitemOpen
  \bibfield  {author} {\bibinfo {author} {\bibfnamefont {P.~S.}\ \bibnamefont
  {{Behroozi}}}, \bibinfo {author} {\bibfnamefont {C.}~\bibnamefont
  {{Conroy}}}, \ and\ \bibinfo {author} {\bibfnamefont {R.~H.}\ \bibnamefont
  {{Wechsler}}},\ }\href {\doibase 10.1088/0004-637X/717/1/379} {\bibfield
  {journal} {\bibinfo  {journal} {\apj}\ }\textbf {\bibinfo {volume} {717}},\
  \bibinfo {pages} {379} (\bibinfo {year} {2010})},\ \Eprint
  {http://arxiv.org/abs/1001.0015} {arXiv:1001.0015 [astro-ph.CO]} \BibitemShut
  {NoStop}%
\bibitem [{\citenamefont {{Reddick}}\ \emph {et~al.}(2012)\citenamefont
  {{Reddick}}, \citenamefont {{Wechsler}}, \citenamefont {{Tinker}},\ and\
  \citenamefont {{Behroozi}}}]{Reddick:2012qy}%
  \BibitemOpen
  \bibfield  {author} {\bibinfo {author} {\bibfnamefont {R.~M.}\ \bibnamefont
  {{Reddick}}}, \bibinfo {author} {\bibfnamefont {R.~H.}\ \bibnamefont
  {{Wechsler}}}, \bibinfo {author} {\bibfnamefont {J.~L.}\ \bibnamefont
  {{Tinker}}}, \ and\ \bibinfo {author} {\bibfnamefont {P.~S.}\ \bibnamefont
  {{Behroozi}}},\ }\href@noop {} {\bibfield  {journal} {\bibinfo  {journal}
  {ArXiv e-prints}\ } (\bibinfo {year} {2012})},\ \Eprint
  {http://arxiv.org/abs/1207.2160} {arXiv:1207.2160 [astro-ph.CO]} \BibitemShut
  {NoStop}%
\bibitem [{\citenamefont {{Hearin}}\ and\ \citenamefont
  {{Watson}}(2013)}]{Hearin:2013vn}%
  \BibitemOpen
  \bibfield  {author} {\bibinfo {author} {\bibfnamefont {A.~P.}\ \bibnamefont
  {{Hearin}}}\ and\ \bibinfo {author} {\bibfnamefont {D.~F.}\ \bibnamefont
  {{Watson}}},\ }\href {\doibase 10.1093/mnras/stt1374} {\bibfield  {journal}
  {\bibinfo  {journal} {\mnras}\ }\textbf {\bibinfo {volume} {435}},\ \bibinfo
  {pages} {1313} (\bibinfo {year} {2013})},\ \Eprint
  {http://arxiv.org/abs/1304.5557} {arXiv:1304.5557 [astro-ph.CO]} \BibitemShut
  {NoStop}%
\bibitem [{\citenamefont {{Zentner}}\ \emph {et~al.}(2013)\citenamefont
  {{Zentner}}, \citenamefont {{Hearin}},\ and\ \citenamefont
  {{Bosch}}}]{Zentner:2013lq}%
  \BibitemOpen
  \bibfield  {author} {\bibinfo {author} {\bibfnamefont {A.~R.}\ \bibnamefont
  {{Zentner}}}, \bibinfo {author} {\bibfnamefont {A.~P.}\ \bibnamefont
  {{Hearin}}}, \ and\ \bibinfo {author} {\bibfnamefont {F.~C.~v.~d.}\
  \bibnamefont {{Bosch}}},\ }\href@noop {} {\  (\bibinfo {year} {2013})},\
  \Eprint {http://arxiv.org/abs/1311.1818v1} {arXiv:1311.1818v1 [astro-ph.CO]}
  \BibitemShut {NoStop}%
\bibitem [{\citenamefont {{Leauthaud}}\ \emph {et~al.}(2011)\citenamefont
  {{Leauthaud}}, \citenamefont {{Tinker}}, \citenamefont {{Behroozi}},
  \citenamefont {{Busha}},\ and\ \citenamefont
  {{Wechsler}}}]{Leauthaud:2011fk}%
  \BibitemOpen
  \bibfield  {author} {\bibinfo {author} {\bibfnamefont {A.}~\bibnamefont
  {{Leauthaud}}}, \bibinfo {author} {\bibfnamefont {J.}~\bibnamefont
  {{Tinker}}}, \bibinfo {author} {\bibfnamefont {P.~S.}\ \bibnamefont
  {{Behroozi}}}, \bibinfo {author} {\bibfnamefont {M.~T.}\ \bibnamefont
  {{Busha}}}, \ and\ \bibinfo {author} {\bibfnamefont {R.~H.}\ \bibnamefont
  {{Wechsler}}},\ }\href {\doibase 10.1088/0004-637X/738/1/45} {\bibfield
  {journal} {\bibinfo  {journal} {\apj}\ }\textbf {\bibinfo {volume} {738}},\
  \bibinfo {eid} {45} (\bibinfo {year} {2011})},\ \Eprint
  {http://arxiv.org/abs/1103.2077} {arXiv:1103.2077 [astro-ph.CO]} \BibitemShut
  {NoStop}%
\bibitem [{\citenamefont {{Tinker}}\ \emph {et~al.}(2013)\citenamefont
  {{Tinker}}, \citenamefont {{Leauthaud}}, \citenamefont {{Bundy}},
  \citenamefont {{George}}, \citenamefont {{Behroozi}}, \citenamefont
  {{Massey}}, \citenamefont {{Rhodes}},\ and\ \citenamefont
  {{Wechsler}}}]{Tinker:2013fk}%
  \BibitemOpen
  \bibfield  {author} {\bibinfo {author} {\bibfnamefont {J.~L.}\ \bibnamefont
  {{Tinker}}}, \bibinfo {author} {\bibfnamefont {A.}~\bibnamefont
  {{Leauthaud}}}, \bibinfo {author} {\bibfnamefont {K.}~\bibnamefont
  {{Bundy}}}, \bibinfo {author} {\bibfnamefont {M.~R.}\ \bibnamefont
  {{George}}}, \bibinfo {author} {\bibfnamefont {P.}~\bibnamefont
  {{Behroozi}}}, \bibinfo {author} {\bibfnamefont {R.}~\bibnamefont
  {{Massey}}}, \bibinfo {author} {\bibfnamefont {J.}~\bibnamefont {{Rhodes}}},
  \ and\ \bibinfo {author} {\bibfnamefont {R.}~\bibnamefont {{Wechsler}}},\
  }\href@noop {} {\  (\bibinfo {year} {2013})},\ \Eprint
  {http://arxiv.org/abs/1308.2974v1} {arXiv:1308.2974v1 [astro-ph.CO]}
  \BibitemShut {NoStop}%
\bibitem [{\citenamefont {{Guo}}\ \emph {et~al.}(2014)\citenamefont {{Guo}},
  \citenamefont {{Zheng}}, \citenamefont {{Zehavi}}, \citenamefont {{Xu}},
  \citenamefont {{Eisenstein}}, \citenamefont {{Weinberg}}, \citenamefont
  {{Bahcall}}, \citenamefont {{Berlind}}, \citenamefont {{Comparat}},
  \citenamefont {{McBride}}, \citenamefont {{Ross}}, \citenamefont
  {{Schneider}}, \citenamefont {{Skibba}}, \citenamefont {{Swanson}},
  \citenamefont {{Tinker}}, \citenamefont {{Tojeiro}},\ and\ \citenamefont
  {{Wake}}}]{Guo:2014fj}%
  \BibitemOpen
  \bibfield  {author} {\bibinfo {author} {\bibfnamefont {H.}~\bibnamefont
  {{Guo}}}, \bibinfo {author} {\bibfnamefont {Z.}~\bibnamefont {{Zheng}}},
  \bibinfo {author} {\bibfnamefont {I.}~\bibnamefont {{Zehavi}}}, \bibinfo
  {author} {\bibfnamefont {H.}~\bibnamefont {{Xu}}}, \bibinfo {author}
  {\bibfnamefont {D.~J.}\ \bibnamefont {{Eisenstein}}}, \bibinfo {author}
  {\bibfnamefont {D.~H.}\ \bibnamefont {{Weinberg}}}, \bibinfo {author}
  {\bibfnamefont {N.~A.}\ \bibnamefont {{Bahcall}}}, \bibinfo {author}
  {\bibfnamefont {A.~A.}\ \bibnamefont {{Berlind}}}, \bibinfo {author}
  {\bibfnamefont {J.}~\bibnamefont {{Comparat}}}, \bibinfo {author}
  {\bibfnamefont {C.~K.}\ \bibnamefont {{McBride}}}, \bibinfo {author}
  {\bibfnamefont {A.~J.}\ \bibnamefont {{Ross}}}, \bibinfo {author}
  {\bibfnamefont {D.~P.}\ \bibnamefont {{Schneider}}}, \bibinfo {author}
  {\bibfnamefont {R.~A.}\ \bibnamefont {{Skibba}}}, \bibinfo {author}
  {\bibfnamefont {M.~E.~C.}\ \bibnamefont {{Swanson}}}, \bibinfo {author}
  {\bibfnamefont {J.~L.}\ \bibnamefont {{Tinker}}}, \bibinfo {author}
  {\bibfnamefont {R.}~\bibnamefont {{Tojeiro}}}, \ and\ \bibinfo {author}
  {\bibfnamefont {D.~A.}\ \bibnamefont {{Wake}}},\ }\href@noop {} {\bibfield
  {journal} {\bibinfo  {journal} {ArXiv e-prints}\ } (\bibinfo {year}
  {2014})},\ \Eprint {http://arxiv.org/abs/1401.3009} {arXiv:1401.3009
  [astro-ph.CO]} \BibitemShut {NoStop}%
\bibitem [{\citenamefont {{Cole}}\ \emph {et~al.}(2005)\citenamefont {{Cole}},
  \citenamefont {{Percival}}, \citenamefont {{Peacock}}, \citenamefont
  {{Norberg}}, \citenamefont {{Baugh}}, \citenamefont {{Frenk}}, \citenamefont
  {{Baldry}}, \citenamefont {{Bland-Hawthorn}}, \citenamefont {{Bridges}},
  \citenamefont {{Cannon}}, \citenamefont {{Colless}}, \citenamefont
  {{Collins}}, \citenamefont {{Couch}}, \citenamefont {{Cross}}, \citenamefont
  {{Dalton}}, \citenamefont {{Eke}}, \citenamefont {{De Propris}},
  \citenamefont {{Driver}}, \citenamefont {{Efstathiou}}, \citenamefont
  {{Ellis}}, \citenamefont {{Glazebrook}}, \citenamefont {{Jackson}},
  \citenamefont {{Jenkins}}, \citenamefont {{Lahav}}, \citenamefont {{Lewis}},
  \citenamefont {{Lumsden}}, \citenamefont {{Maddox}}, \citenamefont
  {{Madgwick}}, \citenamefont {{Peterson}}, \citenamefont {{Sutherland}},\ and\
  \citenamefont {{Taylor}}}]{Cole:2005fp}%
  \BibitemOpen
  \bibfield  {author} {\bibinfo {author} {\bibfnamefont {S.}~\bibnamefont
  {{Cole}}}, \bibinfo {author} {\bibfnamefont {W.~J.}\ \bibnamefont
  {{Percival}}}, \bibinfo {author} {\bibfnamefont {J.~A.}\ \bibnamefont
  {{Peacock}}}, \bibinfo {author} {\bibfnamefont {P.}~\bibnamefont
  {{Norberg}}}, \bibinfo {author} {\bibfnamefont {C.~M.}\ \bibnamefont
  {{Baugh}}}, \bibinfo {author} {\bibfnamefont {C.~S.}\ \bibnamefont
  {{Frenk}}}, \bibinfo {author} {\bibfnamefont {I.}~\bibnamefont {{Baldry}}},
  \bibinfo {author} {\bibfnamefont {J.}~\bibnamefont {{Bland-Hawthorn}}},
  \bibinfo {author} {\bibfnamefont {T.}~\bibnamefont {{Bridges}}}, \bibinfo
  {author} {\bibfnamefont {R.}~\bibnamefont {{Cannon}}}, \bibinfo {author}
  {\bibfnamefont {M.}~\bibnamefont {{Colless}}}, \bibinfo {author}
  {\bibfnamefont {C.}~\bibnamefont {{Collins}}}, \bibinfo {author}
  {\bibfnamefont {W.}~\bibnamefont {{Couch}}}, \bibinfo {author} {\bibfnamefont
  {N.~J.~G.}\ \bibnamefont {{Cross}}}, \bibinfo {author} {\bibfnamefont
  {G.}~\bibnamefont {{Dalton}}}, \bibinfo {author} {\bibfnamefont {V.~R.}\
  \bibnamefont {{Eke}}}, \bibinfo {author} {\bibfnamefont {R.}~\bibnamefont
  {{De Propris}}}, \bibinfo {author} {\bibfnamefont {S.~P.}\ \bibnamefont
  {{Driver}}}, \bibinfo {author} {\bibfnamefont {G.}~\bibnamefont
  {{Efstathiou}}}, \bibinfo {author} {\bibfnamefont {R.~S.}\ \bibnamefont
  {{Ellis}}}, \bibinfo {author} {\bibfnamefont {K.}~\bibnamefont
  {{Glazebrook}}}, \bibinfo {author} {\bibfnamefont {C.}~\bibnamefont
  {{Jackson}}}, \bibinfo {author} {\bibfnamefont {A.}~\bibnamefont
  {{Jenkins}}}, \bibinfo {author} {\bibfnamefont {O.}~\bibnamefont {{Lahav}}},
  \bibinfo {author} {\bibfnamefont {I.}~\bibnamefont {{Lewis}}}, \bibinfo
  {author} {\bibfnamefont {S.}~\bibnamefont {{Lumsden}}}, \bibinfo {author}
  {\bibfnamefont {S.}~\bibnamefont {{Maddox}}}, \bibinfo {author}
  {\bibfnamefont {D.}~\bibnamefont {{Madgwick}}}, \bibinfo {author}
  {\bibfnamefont {B.~A.}\ \bibnamefont {{Peterson}}}, \bibinfo {author}
  {\bibfnamefont {W.}~\bibnamefont {{Sutherland}}}, \ and\ \bibinfo {author}
  {\bibfnamefont {K.}~\bibnamefont {{Taylor}}},\ }\href {\doibase
  10.1111/j.1365-2966.2005.09318.x} {\bibfield  {journal} {\bibinfo  {journal}
  {\mnras}\ }\textbf {\bibinfo {volume} {362}},\ \bibinfo {pages} {505}
  (\bibinfo {year} {2005})},\ \Eprint {http://arxiv.org/abs/astro-ph/0501174}
  {astro-ph/0501174} \BibitemShut {NoStop}%
\bibitem [{\citenamefont {{Tegmark}}\ \emph {et~al.}(2006)\citenamefont
  {{Tegmark}}, \citenamefont {{Eisenstein}}, \citenamefont {{Strauss}},
  \citenamefont {{Weinberg}}, \citenamefont {{Blanton}}, \citenamefont
  {{Frieman}}, \citenamefont {{Fukugita}}, \citenamefont {{Gunn}},
  \citenamefont {{Hamilton}}, \citenamefont {{Knapp}}, \citenamefont
  {{Nichol}}, \citenamefont {{Ostriker}}, \citenamefont {{Padmanabhan}},
  \citenamefont {{Percival}}, \citenamefont {{Schlegel}}, \citenamefont
  {{Schneider}}, \citenamefont {{Scoccimarro}}, \citenamefont {{Seljak}},
  \citenamefont {{Seo}}, \citenamefont {{Swanson}}, \citenamefont {{Szalay}},
  \citenamefont {{Vogeley}}, \citenamefont {{Yoo}}, \citenamefont {{Zehavi}},
  \citenamefont {{Abazajian}}, \citenamefont {{Anderson}}, \citenamefont
  {{Annis}}, \citenamefont {{Bahcall}}, \citenamefont {{Bassett}},
  \citenamefont {{Berlind}}, \citenamefont {{Brinkmann}}, \citenamefont
  {{Budavari}}, \citenamefont {{Castander}}, \citenamefont {{Connolly}},
  \citenamefont {{Csabai}}, \citenamefont {{Doi}}, \citenamefont
  {{Finkbeiner}}, \citenamefont {{Gillespie}}, \citenamefont {{Glazebrook}},
  \citenamefont {{Hennessy}}, \citenamefont {{Hogg}}, \citenamefont
  {{Ivezi{\'c}}}, \citenamefont {{Jain}}, \citenamefont {{Johnston}},
  \citenamefont {{Kent}}, \citenamefont {{Lamb}}, \citenamefont {{Lee}},
  \citenamefont {{Lin}}, \citenamefont {{Loveday}}, \citenamefont {{Lupton}},
  \citenamefont {{Munn}}, \citenamefont {{Pan}}, \citenamefont {{Park}},
  \citenamefont {{Peoples}}, \citenamefont {{Pier}}, \citenamefont {{Pope}},
  \citenamefont {{Richmond}}, \citenamefont {{Rockosi}}, \citenamefont
  {{Scranton}}, \citenamefont {{Sheth}}, \citenamefont {{Stebbins}},
  \citenamefont {{Stoughton}}, \citenamefont {{Szapudi}}, \citenamefont
  {{Tucker}}, \citenamefont {{vanden Berk}}, \citenamefont {{Yanny}},\ and\
  \citenamefont {{York}}}]{Tegmark:2006pl}%
  \BibitemOpen
  \bibfield  {author} {\bibinfo {author} {\bibfnamefont {M.}~\bibnamefont
  {{Tegmark}}}, \bibinfo {author} {\bibfnamefont {D.~J.}\ \bibnamefont
  {{Eisenstein}}}, \bibinfo {author} {\bibfnamefont {M.~A.}\ \bibnamefont
  {{Strauss}}}, \bibinfo {author} {\bibfnamefont {D.~H.}\ \bibnamefont
  {{Weinberg}}}, \bibinfo {author} {\bibfnamefont {M.~R.}\ \bibnamefont
  {{Blanton}}}, \bibinfo {author} {\bibfnamefont {J.~A.}\ \bibnamefont
  {{Frieman}}}, \bibinfo {author} {\bibfnamefont {M.}~\bibnamefont
  {{Fukugita}}}, \bibinfo {author} {\bibfnamefont {J.~E.}\ \bibnamefont
  {{Gunn}}}, \bibinfo {author} {\bibfnamefont {A.~J.~S.}\ \bibnamefont
  {{Hamilton}}}, \bibinfo {author} {\bibfnamefont {G.~R.}\ \bibnamefont
  {{Knapp}}}, \bibinfo {author} {\bibfnamefont {R.~C.}\ \bibnamefont
  {{Nichol}}}, \bibinfo {author} {\bibfnamefont {J.~P.}\ \bibnamefont
  {{Ostriker}}}, \bibinfo {author} {\bibfnamefont {N.}~\bibnamefont
  {{Padmanabhan}}}, \bibinfo {author} {\bibfnamefont {W.~J.}\ \bibnamefont
  {{Percival}}}, \bibinfo {author} {\bibfnamefont {D.~J.}\ \bibnamefont
  {{Schlegel}}}, \bibinfo {author} {\bibfnamefont {D.~P.}\ \bibnamefont
  {{Schneider}}}, \bibinfo {author} {\bibfnamefont {R.}~\bibnamefont
  {{Scoccimarro}}}, \bibinfo {author} {\bibfnamefont {U.}~\bibnamefont
  {{Seljak}}}, \bibinfo {author} {\bibfnamefont {H.}~\bibnamefont {{Seo}}},
  \bibinfo {author} {\bibfnamefont {M.}~\bibnamefont {{Swanson}}}, \bibinfo
  {author} {\bibfnamefont {A.~S.}\ \bibnamefont {{Szalay}}}, \bibinfo {author}
  {\bibfnamefont {M.~S.}\ \bibnamefont {{Vogeley}}}, \bibinfo {author}
  {\bibfnamefont {J.}~\bibnamefont {{Yoo}}}, \bibinfo {author} {\bibfnamefont
  {I.}~\bibnamefont {{Zehavi}}}, \bibinfo {author} {\bibfnamefont
  {K.}~\bibnamefont {{Abazajian}}}, \bibinfo {author} {\bibfnamefont {S.~F.}\
  \bibnamefont {{Anderson}}}, \bibinfo {author} {\bibfnamefont
  {J.}~\bibnamefont {{Annis}}}, \bibinfo {author} {\bibfnamefont {N.~A.}\
  \bibnamefont {{Bahcall}}}, \bibinfo {author} {\bibfnamefont {B.}~\bibnamefont
  {{Bassett}}}, \bibinfo {author} {\bibfnamefont {A.}~\bibnamefont
  {{Berlind}}}, \bibinfo {author} {\bibfnamefont {J.}~\bibnamefont
  {{Brinkmann}}}, \bibinfo {author} {\bibfnamefont {T.}~\bibnamefont
  {{Budavari}}}, \bibinfo {author} {\bibfnamefont {F.}~\bibnamefont
  {{Castander}}}, \bibinfo {author} {\bibfnamefont {A.}~\bibnamefont
  {{Connolly}}}, \bibinfo {author} {\bibfnamefont {I.}~\bibnamefont
  {{Csabai}}}, \bibinfo {author} {\bibfnamefont {M.}~\bibnamefont {{Doi}}},
  \bibinfo {author} {\bibfnamefont {D.~P.}\ \bibnamefont {{Finkbeiner}}},
  \bibinfo {author} {\bibfnamefont {B.}~\bibnamefont {{Gillespie}}}, \bibinfo
  {author} {\bibfnamefont {K.}~\bibnamefont {{Glazebrook}}}, \bibinfo {author}
  {\bibfnamefont {G.~S.}\ \bibnamefont {{Hennessy}}}, \bibinfo {author}
  {\bibfnamefont {D.~W.}\ \bibnamefont {{Hogg}}}, \bibinfo {author}
  {\bibfnamefont {{\v Z}.}~\bibnamefont {{Ivezi{\'c}}}}, \bibinfo {author}
  {\bibfnamefont {B.}~\bibnamefont {{Jain}}}, \bibinfo {author} {\bibfnamefont
  {D.}~\bibnamefont {{Johnston}}}, \bibinfo {author} {\bibfnamefont
  {S.}~\bibnamefont {{Kent}}}, \bibinfo {author} {\bibfnamefont {D.~Q.}\
  \bibnamefont {{Lamb}}}, \bibinfo {author} {\bibfnamefont {B.~C.}\
  \bibnamefont {{Lee}}}, \bibinfo {author} {\bibfnamefont {H.}~\bibnamefont
  {{Lin}}}, \bibinfo {author} {\bibfnamefont {J.}~\bibnamefont {{Loveday}}},
  \bibinfo {author} {\bibfnamefont {R.~H.}\ \bibnamefont {{Lupton}}}, \bibinfo
  {author} {\bibfnamefont {J.~A.}\ \bibnamefont {{Munn}}}, \bibinfo {author}
  {\bibfnamefont {K.}~\bibnamefont {{Pan}}}, \bibinfo {author} {\bibfnamefont
  {C.}~\bibnamefont {{Park}}}, \bibinfo {author} {\bibfnamefont
  {J.}~\bibnamefont {{Peoples}}}, \bibinfo {author} {\bibfnamefont {J.~R.}\
  \bibnamefont {{Pier}}}, \bibinfo {author} {\bibfnamefont {A.}~\bibnamefont
  {{Pope}}}, \bibinfo {author} {\bibfnamefont {M.}~\bibnamefont {{Richmond}}},
  \bibinfo {author} {\bibfnamefont {C.}~\bibnamefont {{Rockosi}}}, \bibinfo
  {author} {\bibfnamefont {R.}~\bibnamefont {{Scranton}}}, \bibinfo {author}
  {\bibfnamefont {R.~K.}\ \bibnamefont {{Sheth}}}, \bibinfo {author}
  {\bibfnamefont {A.}~\bibnamefont {{Stebbins}}}, \bibinfo {author}
  {\bibfnamefont {C.}~\bibnamefont {{Stoughton}}}, \bibinfo {author}
  {\bibfnamefont {I.}~\bibnamefont {{Szapudi}}}, \bibinfo {author}
  {\bibfnamefont {D.~L.}\ \bibnamefont {{Tucker}}}, \bibinfo {author}
  {\bibfnamefont {D.~E.}\ \bibnamefont {{vanden Berk}}}, \bibinfo {author}
  {\bibfnamefont {B.}~\bibnamefont {{Yanny}}}, \ and\ \bibinfo {author}
  {\bibfnamefont {D.~G.}\ \bibnamefont {{York}}},\ }\href {\doibase
  10.1103/PhysRevD.74.123507} {\bibfield  {journal} {\bibinfo  {journal}
  {\prd}\ }\textbf {\bibinfo {volume} {74}},\ \bibinfo {pages} {123507}
  (\bibinfo {year} {2006})},\ \Eprint
  {http://arxiv.org/abs/arXiv:astro-ph/0608632} {arXiv:astro-ph/0608632}
  \BibitemShut {NoStop}%
\bibitem [{\citenamefont {{Yamamoto}}\ \emph {et~al.}(2010)\citenamefont
  {{Yamamoto}}, \citenamefont {{Nakamura}}, \citenamefont {{H{\"u}tsi}},
  \citenamefont {{Narikawa}},\ and\ \citenamefont {{Sato}}}]{Yamamoto:2010kj}%
  \BibitemOpen
  \bibfield  {author} {\bibinfo {author} {\bibfnamefont {K.}~\bibnamefont
  {{Yamamoto}}}, \bibinfo {author} {\bibfnamefont {G.}~\bibnamefont
  {{Nakamura}}}, \bibinfo {author} {\bibfnamefont {G.}~\bibnamefont
  {{H{\"u}tsi}}}, \bibinfo {author} {\bibfnamefont {T.}~\bibnamefont
  {{Narikawa}}}, \ and\ \bibinfo {author} {\bibfnamefont {T.}~\bibnamefont
  {{Sato}}},\ }\href {\doibase 10.1103/PhysRevD.81.103517} {\bibfield
  {journal} {\bibinfo  {journal} {\prd}\ }\textbf {\bibinfo {volume} {81}},\
  \bibinfo {eid} {103517} (\bibinfo {year} {2010})},\ \Eprint
  {http://arxiv.org/abs/1004.3231} {arXiv:1004.3231 [astro-ph.CO]} \BibitemShut
  {NoStop}%
\bibitem [{\citenamefont {{Oka}}\ \emph {et~al.}(2014)\citenamefont {{Oka}},
  \citenamefont {{Saito}}, \citenamefont {{Nishimichi}}, \citenamefont
  {{Taruya}},\ and\ \citenamefont {{Yamamoto}}}]{Oka:2014aa}%
  \BibitemOpen
  \bibfield  {author} {\bibinfo {author} {\bibfnamefont {A.}~\bibnamefont
  {{Oka}}}, \bibinfo {author} {\bibfnamefont {S.}~\bibnamefont {{Saito}}},
  \bibinfo {author} {\bibfnamefont {T.}~\bibnamefont {{Nishimichi}}}, \bibinfo
  {author} {\bibfnamefont {A.}~\bibnamefont {{Taruya}}}, \ and\ \bibinfo
  {author} {\bibfnamefont {K.}~\bibnamefont {{Yamamoto}}},\ }\href {\doibase
  10.1093/mnras/stu111} {\bibfield  {journal} {\bibinfo  {journal} {\mnras}\ }
  (\bibinfo {year} {2014}),\ 10.1093/mnras/stu111},\ \Eprint
  {http://arxiv.org/abs/1310.2820} {arXiv:1310.2820 [astro-ph.CO]} \BibitemShut
  {NoStop}%
\bibitem [{\citenamefont {{Kaiser}}(1984)}]{Kaiser:1984bh}%
  \BibitemOpen
  \bibfield  {author} {\bibinfo {author} {\bibfnamefont {N.}~\bibnamefont
  {{Kaiser}}},\ }\href {\doibase 10.1086/184341} {\bibfield  {journal}
  {\bibinfo  {journal} {\apjl}\ }\textbf {\bibinfo {volume} {284}},\ \bibinfo
  {pages} {L9} (\bibinfo {year} {1984})}\BibitemShut {NoStop}%
\bibitem [{\citenamefont {{Fry}}\ and\ \citenamefont
  {{Gaztanaga}}(1993)}]{Fry:1993lq}%
  \BibitemOpen
  \bibfield  {author} {\bibinfo {author} {\bibfnamefont {J.~N.}\ \bibnamefont
  {{Fry}}}\ and\ \bibinfo {author} {\bibfnamefont {E.}~\bibnamefont
  {{Gaztanaga}}},\ }\href {\doibase 10.1086/173015} {\bibfield  {journal}
  {\bibinfo  {journal} {\apj}\ }\textbf {\bibinfo {volume} {413}},\ \bibinfo
  {pages} {447} (\bibinfo {year} {1993})},\ \Eprint
  {http://arxiv.org/abs/astro-ph/9302009} {astro-ph/9302009} \BibitemShut
  {NoStop}%
\bibitem [{\citenamefont {{Sheth}}\ and\ \citenamefont
  {{Tormen}}(1999)}]{Sheth:1999dq}%
  \BibitemOpen
  \bibfield  {author} {\bibinfo {author} {\bibfnamefont {R.~K.}\ \bibnamefont
  {{Sheth}}}\ and\ \bibinfo {author} {\bibfnamefont {G.}~\bibnamefont
  {{Tormen}}},\ }\href {\doibase 10.1046/j.1365-8711.1999.02692.x} {\bibfield
  {journal} {\bibinfo  {journal} {\mnras}\ }\textbf {\bibinfo {volume} {308}},\
  \bibinfo {pages} {119} (\bibinfo {year} {1999})},\ \Eprint
  {http://arxiv.org/abs/astro-ph/9901122} {astro-ph/9901122} \BibitemShut
  {NoStop}%
\bibitem [{\citenamefont {{Tinker}}\ \emph {et~al.}(2008)\citenamefont
  {{Tinker}}, \citenamefont {{Kravtsov}}, \citenamefont {{Klypin}},
  \citenamefont {{Abazajian}}, \citenamefont {{Warren}}, \citenamefont
  {{Yepes}}, \citenamefont {{Gottl{\"o}ber}},\ and\ \citenamefont
  {{Holz}}}]{Tinker:2008fk}%
  \BibitemOpen
  \bibfield  {author} {\bibinfo {author} {\bibfnamefont {J.}~\bibnamefont
  {{Tinker}}}, \bibinfo {author} {\bibfnamefont {A.~V.}\ \bibnamefont
  {{Kravtsov}}}, \bibinfo {author} {\bibfnamefont {A.}~\bibnamefont
  {{Klypin}}}, \bibinfo {author} {\bibfnamefont {K.}~\bibnamefont
  {{Abazajian}}}, \bibinfo {author} {\bibfnamefont {M.}~\bibnamefont
  {{Warren}}}, \bibinfo {author} {\bibfnamefont {G.}~\bibnamefont {{Yepes}}},
  \bibinfo {author} {\bibfnamefont {S.}~\bibnamefont {{Gottl{\"o}ber}}}, \ and\
  \bibinfo {author} {\bibfnamefont {D.~E.}\ \bibnamefont {{Holz}}},\ }\href
  {\doibase 10.1086/591439} {\bibfield  {journal} {\bibinfo  {journal} {\apj}\
  }\textbf {\bibinfo {volume} {688}},\ \bibinfo {pages} {709} (\bibinfo {year}
  {2008})},\ \Eprint {http://arxiv.org/abs/0803.2706} {arXiv:0803.2706}
  \BibitemShut {NoStop}%
\bibitem [{\citenamefont {{Pollack}}\ \emph {et~al.}(2012)\citenamefont
  {{Pollack}}, \citenamefont {{Smith}},\ and\ \citenamefont
  {{Porciani}}}]{Pollack:2012kh}%
  \BibitemOpen
  \bibfield  {author} {\bibinfo {author} {\bibfnamefont {J.~E.}\ \bibnamefont
  {{Pollack}}}, \bibinfo {author} {\bibfnamefont {R.~E.}\ \bibnamefont
  {{Smith}}}, \ and\ \bibinfo {author} {\bibfnamefont {C.}~\bibnamefont
  {{Porciani}}},\ }\href {\doibase 10.1111/j.1365-2966.2011.20279.x} {\bibfield
   {journal} {\bibinfo  {journal} {\mnras}\ }\textbf {\bibinfo {volume}
  {420}},\ \bibinfo {pages} {3469} (\bibinfo {year} {2012})},\ \Eprint
  {http://arxiv.org/abs/1109.3458} {arXiv:1109.3458 [astro-ph.CO]} \BibitemShut
  {NoStop}%
\bibitem [{\citenamefont {{Pollack}}\ \emph {et~al.}(2013)\citenamefont
  {{Pollack}}, \citenamefont {{Smith}},\ and\ \citenamefont
  {{Porciani}}}]{Pollack:2013fk}%
  \BibitemOpen
  \bibfield  {author} {\bibinfo {author} {\bibfnamefont {J.~E.}\ \bibnamefont
  {{Pollack}}}, \bibinfo {author} {\bibfnamefont {R.~E.}\ \bibnamefont
  {{Smith}}}, \ and\ \bibinfo {author} {\bibfnamefont {C.}~\bibnamefont
  {{Porciani}}},\ }\href@noop {} {\  (\bibinfo {year} {2013})},\ \Eprint
  {http://arxiv.org/abs/1309.0504v1} {arXiv:1309.0504v1 [astro-ph.CO]}
  \BibitemShut {NoStop}%
\bibitem [{\citenamefont {{Jeong}}\ and\ \citenamefont
  {{Komatsu}}(2009)}]{Jeong:2009xe}%
  \BibitemOpen
  \bibfield  {author} {\bibinfo {author} {\bibfnamefont {D.}~\bibnamefont
  {{Jeong}}}\ and\ \bibinfo {author} {\bibfnamefont {E.}~\bibnamefont
  {{Komatsu}}},\ }\href {\doibase 10.1088/0004-637X/691/1/569} {\bibfield
  {journal} {\bibinfo  {journal} {\apj}\ }\textbf {\bibinfo {volume} {691}},\
  \bibinfo {pages} {569} (\bibinfo {year} {2009})},\ \Eprint
  {http://arxiv.org/abs/0805.2632} {arXiv:0805.2632} \BibitemShut {NoStop}%
\bibitem [{\citenamefont {{Saito}}\ \emph {et~al.}(2011)\citenamefont
  {{Saito}}, \citenamefont {{Takada}},\ and\ \citenamefont
  {{Taruya}}}]{Saito:2011yq}%
  \BibitemOpen
  \bibfield  {author} {\bibinfo {author} {\bibfnamefont {S.}~\bibnamefont
  {{Saito}}}, \bibinfo {author} {\bibfnamefont {M.}~\bibnamefont {{Takada}}}, \
  and\ \bibinfo {author} {\bibfnamefont {A.}~\bibnamefont {{Taruya}}},\ }\href
  {\doibase 10.1103/PhysRevD.83.043529} {\bibfield  {journal} {\bibinfo
  {journal} {\prd}\ }\textbf {\bibinfo {volume} {83}},\ \bibinfo {eid} {043529}
  (\bibinfo {year} {2011})},\ \Eprint {http://arxiv.org/abs/1006.4845}
  {arXiv:1006.4845 [astro-ph.CO]} \BibitemShut {NoStop}%
\bibitem [{\citenamefont {{Nishizawa}}\ \emph {et~al.}(2012)\citenamefont
  {{Nishizawa}}, \citenamefont {{Takada}},\ and\ \citenamefont
  {{Nishimichi}}}]{Nishizawa:2012lr}%
  \BibitemOpen
  \bibfield  {author} {\bibinfo {author} {\bibfnamefont {A.~J.}\ \bibnamefont
  {{Nishizawa}}}, \bibinfo {author} {\bibfnamefont {M.}~\bibnamefont
  {{Takada}}}, \ and\ \bibinfo {author} {\bibfnamefont {T.}~\bibnamefont
  {{Nishimichi}}},\ }\href@noop {} {\bibfield  {journal} {\bibinfo  {journal}
  {ArXiv e-prints}\ } (\bibinfo {year} {2012})},\ \Eprint
  {http://arxiv.org/abs/1212.4025} {arXiv:1212.4025 [astro-ph.CO]} \BibitemShut
  {NoStop}%
\bibitem [{\citenamefont {{McDonald}}\ and\ \citenamefont
  {{Roy}}(2009)}]{McDonald:2009lr}%
  \BibitemOpen
  \bibfield  {author} {\bibinfo {author} {\bibfnamefont {P.}~\bibnamefont
  {{McDonald}}}\ and\ \bibinfo {author} {\bibfnamefont {A.}~\bibnamefont
  {{Roy}}},\ }\href {\doibase 10.1088/1475-7516/2009/08/020} {\bibfield
  {journal} {\bibinfo  {journal} {\jcap}\ }\textbf {\bibinfo {volume} {8}},\
  \bibinfo {pages} {20} (\bibinfo {year} {2009})},\ \Eprint
  {http://arxiv.org/abs/0902.0991} {arXiv:0902.0991 [astro-ph.CO]} \BibitemShut
  {NoStop}%
\bibitem [{\citenamefont {{Matsubara}}(2011)}]{Matsubara:2011qy}%
  \BibitemOpen
  \bibfield  {author} {\bibinfo {author} {\bibfnamefont {T.}~\bibnamefont
  {{Matsubara}}},\ }\href {\doibase 10.1103/PhysRevD.83.083518} {\bibfield
  {journal} {\bibinfo  {journal} {\prd}\ }\textbf {\bibinfo {volume} {83}},\
  \bibinfo {eid} {083518} (\bibinfo {year} {2011})},\ \Eprint
  {http://arxiv.org/abs/1102.4619} {arXiv:1102.4619 [astro-ph.CO]} \BibitemShut
  {NoStop}%
\bibitem [{\citenamefont {{Baldauf}}\ \emph {et~al.}(2012)\citenamefont
  {{Baldauf}}, \citenamefont {{Seljak}}, \citenamefont {{Desjacques}},\ and\
  \citenamefont {{McDonald}}}]{Baldauf:2012lr}%
  \BibitemOpen
  \bibfield  {author} {\bibinfo {author} {\bibfnamefont {T.}~\bibnamefont
  {{Baldauf}}}, \bibinfo {author} {\bibfnamefont {U.}~\bibnamefont {{Seljak}}},
  \bibinfo {author} {\bibfnamefont {V.}~\bibnamefont {{Desjacques}}}, \ and\
  \bibinfo {author} {\bibfnamefont {P.}~\bibnamefont {{McDonald}}},\
  }\href@noop {} {\bibfield  {journal} {\bibinfo  {journal} {ArXiv e-prints}\ }
  (\bibinfo {year} {2012})},\ \Eprint {http://arxiv.org/abs/1201.4827}
  {arXiv:1201.4827 [astro-ph.CO]} \BibitemShut {NoStop}%
\bibitem [{\citenamefont {{Chuen Chan}}\ \emph {et~al.}(2012)\citenamefont
  {{Chuen Chan}}, \citenamefont {{Scoccimarro}},\ and\ \citenamefont
  {{Sheth}}}]{Chuen-Chan:2012fk}%
  \BibitemOpen
  \bibfield  {author} {\bibinfo {author} {\bibfnamefont {K.}~\bibnamefont
  {{Chuen Chan}}}, \bibinfo {author} {\bibfnamefont {R.}~\bibnamefont
  {{Scoccimarro}}}, \ and\ \bibinfo {author} {\bibfnamefont {R.~K.}\
  \bibnamefont {{Sheth}}},\ }\href@noop {} {\bibfield  {journal} {\bibinfo
  {journal} {ArXiv e-prints}\ } (\bibinfo {year} {2012})},\ \Eprint
  {http://arxiv.org/abs/1201.3614} {arXiv:1201.3614 [astro-ph.CO]} \BibitemShut
  {NoStop}%
\bibitem [{\citenamefont {{McDonald}}(2006)}]{McDonald:2006lr}%
  \BibitemOpen
  \bibfield  {author} {\bibinfo {author} {\bibfnamefont {P.}~\bibnamefont
  {{McDonald}}},\ }\href {\doibase 10.1103/PhysRevD.74.103512} {\bibfield
  {journal} {\bibinfo  {journal} {\prd}\ }\textbf {\bibinfo {volume} {74}},\
  \bibinfo {pages} {103512} (\bibinfo {year} {2006})},\ \Eprint
  {http://arxiv.org/abs/arXiv:astro-ph/0609413} {arXiv:astro-ph/0609413}
  \BibitemShut {NoStop}%
\bibitem [{\citenamefont {{Assassi}}\ \emph {et~al.}(2014)\citenamefont
  {{Assassi}}, \citenamefont {{Baumann}}, \citenamefont {{Green}},\ and\
  \citenamefont {{Zaldarriaga}}}]{Assassi:2014lr}%
  \BibitemOpen
  \bibfield  {author} {\bibinfo {author} {\bibfnamefont {V.}~\bibnamefont
  {{Assassi}}}, \bibinfo {author} {\bibfnamefont {D.}~\bibnamefont
  {{Baumann}}}, \bibinfo {author} {\bibfnamefont {D.}~\bibnamefont {{Green}}},
  \ and\ \bibinfo {author} {\bibfnamefont {M.}~\bibnamefont {{Zaldarriaga}}},\
  }\href@noop {} {\bibfield  {journal} {\bibinfo  {journal} {ArXiv e-prints}\ }
  (\bibinfo {year} {2014})},\ \Eprint {http://arxiv.org/abs/1402.5916}
  {arXiv:1402.5916 [astro-ph.CO]} \BibitemShut {NoStop}%
\bibitem [{\citenamefont {{Kehagias}}\ \emph {et~al.}(2013)\citenamefont
  {{Kehagias}}, \citenamefont {{Nore{\~n}a}}, \citenamefont {{Perrier}},\ and\
  \citenamefont {{Riotto}}}]{Kehagias:2013gf}%
  \BibitemOpen
  \bibfield  {author} {\bibinfo {author} {\bibfnamefont {A.}~\bibnamefont
  {{Kehagias}}}, \bibinfo {author} {\bibfnamefont {J.}~\bibnamefont
  {{Nore{\~n}a}}}, \bibinfo {author} {\bibfnamefont {H.}~\bibnamefont
  {{Perrier}}}, \ and\ \bibinfo {author} {\bibfnamefont {A.}~\bibnamefont
  {{Riotto}}},\ }\href@noop {} {\  (\bibinfo {year} {2013})},\ \Eprint
  {http://arxiv.org/abs/1311.0786v1} {arXiv:1311.0786v1 [astro-ph.CO]}
  \BibitemShut {NoStop}%
\bibitem [{\citenamefont {{Seljak}}\ \emph {et~al.}(2009)\citenamefont
  {{Seljak}}, \citenamefont {{Hamaus}},\ and\ \citenamefont
  {{Desjacques}}}]{Seljak:2009nl}%
  \BibitemOpen
  \bibfield  {author} {\bibinfo {author} {\bibfnamefont {U.}~\bibnamefont
  {{Seljak}}}, \bibinfo {author} {\bibfnamefont {N.}~\bibnamefont {{Hamaus}}},
  \ and\ \bibinfo {author} {\bibfnamefont {V.}~\bibnamefont {{Desjacques}}},\
  }\href {\doibase 10.1103/PhysRevLett.103.091303} {\bibfield  {journal}
  {\bibinfo  {journal} {Physical Review Letters}\ }\textbf {\bibinfo {volume}
  {103}},\ \bibinfo {eid} {091303} (\bibinfo {year} {2009})},\ \Eprint
  {http://arxiv.org/abs/0904.2963} {arXiv:0904.2963 [astro-ph.CO]} \BibitemShut
  {NoStop}%
\bibitem [{\citenamefont {{Percival}}\ and\ \citenamefont
  {{Sch{\"a}fer}}(2008)}]{Percival:2008fk}%
  \BibitemOpen
  \bibfield  {author} {\bibinfo {author} {\bibfnamefont {W.~J.}\ \bibnamefont
  {{Percival}}}\ and\ \bibinfo {author} {\bibfnamefont {B.~M.}\ \bibnamefont
  {{Sch{\"a}fer}}},\ }\href {\doibase 10.1111/j.1745-3933.2008.00437.x}
  {\bibfield  {journal} {\bibinfo  {journal} {\mnras}\ }\textbf {\bibinfo
  {volume} {385}},\ \bibinfo {pages} {L78} (\bibinfo {year} {2008})},\ \Eprint
  {http://arxiv.org/abs/0712.2729} {arXiv:0712.2729} \BibitemShut {NoStop}%
\bibitem [{\citenamefont {{Desjacques}}\ and\ \citenamefont
  {{Sheth}}(2010)}]{Desjacques:2010qy}%
  \BibitemOpen
  \bibfield  {author} {\bibinfo {author} {\bibfnamefont {V.}~\bibnamefont
  {{Desjacques}}}\ and\ \bibinfo {author} {\bibfnamefont {R.~K.}\ \bibnamefont
  {{Sheth}}},\ }\href {\doibase 10.1103/PhysRevD.81.023526} {\bibfield
  {journal} {\bibinfo  {journal} {\prd}\ }\textbf {\bibinfo {volume} {81}},\
  \bibinfo {eid} {023526} (\bibinfo {year} {2010})},\ \Eprint
  {http://arxiv.org/abs/0909.4544} {arXiv:0909.4544 [astro-ph.CO]} \BibitemShut
  {NoStop}%
\bibitem [{\citenamefont {{Desjacques}}\ \emph {et~al.}(2010)\citenamefont
  {{Desjacques}}, \citenamefont {{Crocce}}, \citenamefont {{Scoccimarro}},\
  and\ \citenamefont {{Sheth}}}]{Desjacques:2010fk}%
  \BibitemOpen
  \bibfield  {author} {\bibinfo {author} {\bibfnamefont {V.}~\bibnamefont
  {{Desjacques}}}, \bibinfo {author} {\bibfnamefont {M.}~\bibnamefont
  {{Crocce}}}, \bibinfo {author} {\bibfnamefont {R.}~\bibnamefont
  {{Scoccimarro}}}, \ and\ \bibinfo {author} {\bibfnamefont {R.~K.}\
  \bibnamefont {{Sheth}}},\ }\href {\doibase 10.1103/PhysRevD.82.103529}
  {\bibfield  {journal} {\bibinfo  {journal} {\prd}\ }\textbf {\bibinfo
  {volume} {82}},\ \bibinfo {eid} {103529} (\bibinfo {year} {2010})},\ \Eprint
  {http://arxiv.org/abs/1009.3449} {arXiv:1009.3449 [astro-ph.CO]} \BibitemShut
  {NoStop}%
\bibitem [{\citenamefont {{Seljak}}\ and\ \citenamefont
  {{McDonald}}(2011)}]{Seljak:2011dw}%
  \BibitemOpen
  \bibfield  {author} {\bibinfo {author} {\bibfnamefont {U.}~\bibnamefont
  {{Seljak}}}\ and\ \bibinfo {author} {\bibfnamefont {P.}~\bibnamefont
  {{McDonald}}},\ }\href {\doibase 10.1088/1475-7516/2011/11/039} {\bibfield
  {journal} {\bibinfo  {journal} {\jcap}\ }\textbf {\bibinfo {volume} {11}},\
  \bibinfo {eid} {039} (\bibinfo {year} {2011})},\ \Eprint
  {http://arxiv.org/abs/1109.1888} {arXiv:1109.1888 [astro-ph.CO]} \BibitemShut
  {NoStop}%
\bibitem [{\citenamefont {{Okumura}}\ \emph
  {et~al.}(2012{\natexlab{a}})\citenamefont {{Okumura}}, \citenamefont
  {{Seljak}},\ and\ \citenamefont {{Desjacques}}}]{Okumura:2012gf}%
  \BibitemOpen
  \bibfield  {author} {\bibinfo {author} {\bibfnamefont {T.}~\bibnamefont
  {{Okumura}}}, \bibinfo {author} {\bibfnamefont {U.}~\bibnamefont {{Seljak}}},
  \ and\ \bibinfo {author} {\bibfnamefont {V.}~\bibnamefont {{Desjacques}}},\
  }\href {\doibase 10.1088/1475-7516/2012/11/014} {\bibfield  {journal}
  {\bibinfo  {journal} {\jcap}\ }\textbf {\bibinfo {volume} {11}},\ \bibinfo
  {eid} {014} (\bibinfo {year} {2012}{\natexlab{a}})},\ \Eprint
  {http://arxiv.org/abs/1206.4070} {arXiv:1206.4070 [astro-ph.CO]} \BibitemShut
  {NoStop}%
\bibitem [{\citenamefont {{Okumura}}\ \emph
  {et~al.}(2012{\natexlab{b}})\citenamefont {{Okumura}}, \citenamefont
  {{Seljak}}, \citenamefont {{McDonald}},\ and\ \citenamefont
  {{Desjacques}}}]{Okumura:2012rq}%
  \BibitemOpen
  \bibfield  {author} {\bibinfo {author} {\bibfnamefont {T.}~\bibnamefont
  {{Okumura}}}, \bibinfo {author} {\bibfnamefont {U.}~\bibnamefont {{Seljak}}},
  \bibinfo {author} {\bibfnamefont {P.}~\bibnamefont {{McDonald}}}, \ and\
  \bibinfo {author} {\bibfnamefont {V.}~\bibnamefont {{Desjacques}}},\ }\href
  {\doibase 10.1088/1475-7516/2012/02/010} {\bibfield  {journal} {\bibinfo
  {journal} {\jcap}\ }\textbf {\bibinfo {volume} {2}},\ \bibinfo {eid} {010}
  (\bibinfo {year} {2012}{\natexlab{b}})},\ \Eprint
  {http://arxiv.org/abs/1109.1609} {arXiv:1109.1609 [astro-ph.CO]} \BibitemShut
  {NoStop}%
\bibitem [{\citenamefont {{Vlah}}\ \emph {et~al.}(2012)\citenamefont {{Vlah}},
  \citenamefont {{Seljak}}, \citenamefont {{McDonald}}, \citenamefont
  {{Okumura}},\ and\ \citenamefont {{Baldauf}}}]{Vlah:2012fr}%
  \BibitemOpen
  \bibfield  {author} {\bibinfo {author} {\bibfnamefont {Z.}~\bibnamefont
  {{Vlah}}}, \bibinfo {author} {\bibfnamefont {U.}~\bibnamefont {{Seljak}}},
  \bibinfo {author} {\bibfnamefont {P.}~\bibnamefont {{McDonald}}}, \bibinfo
  {author} {\bibfnamefont {T.}~\bibnamefont {{Okumura}}}, \ and\ \bibinfo
  {author} {\bibfnamefont {T.}~\bibnamefont {{Baldauf}}},\ }\href {\doibase
  10.1088/1475-7516/2012/11/009} {\bibfield  {journal} {\bibinfo  {journal}
  {\jcap}\ }\textbf {\bibinfo {volume} {11}},\ \bibinfo {eid} {009} (\bibinfo
  {year} {2012})},\ \Eprint {http://arxiv.org/abs/1207.0839} {arXiv:1207.0839
  [astro-ph.CO]} \BibitemShut {NoStop}%
\bibitem [{\citenamefont {{Vlah}}\ \emph {et~al.}(2013)\citenamefont {{Vlah}},
  \citenamefont {{Seljak}}, \citenamefont {{Okumura}},\ and\ \citenamefont
  {{Desjacques}}}]{Vlah:2013qy}%
  \BibitemOpen
  \bibfield  {author} {\bibinfo {author} {\bibfnamefont {Z.}~\bibnamefont
  {{Vlah}}}, \bibinfo {author} {\bibfnamefont {U.}~\bibnamefont {{Seljak}}},
  \bibinfo {author} {\bibfnamefont {T.}~\bibnamefont {{Okumura}}}, \ and\
  \bibinfo {author} {\bibfnamefont {V.}~\bibnamefont {{Desjacques}}},\ }\href
  {\doibase 10.1088/1475-7516/2013/10/053} {\bibfield  {journal} {\bibinfo
  {journal} {\jcap}\ }\textbf {\bibinfo {volume} {10}},\ \bibinfo {eid} {053}
  (\bibinfo {year} {2013})},\ \Eprint {http://arxiv.org/abs/1308.6294}
  {arXiv:1308.6294 [astro-ph.CO]} \BibitemShut {NoStop}%
\bibitem [{\citenamefont {{Blazek}}\ \emph {et~al.}(2013)\citenamefont
  {{Blazek}}, \citenamefont {{Seljak}}, \citenamefont {{Vlah}},\ and\
  \citenamefont {{Okumura}}}]{Blazek:2013ys}%
  \BibitemOpen
  \bibfield  {author} {\bibinfo {author} {\bibfnamefont {J.}~\bibnamefont
  {{Blazek}}}, \bibinfo {author} {\bibfnamefont {U.}~\bibnamefont {{Seljak}}},
  \bibinfo {author} {\bibfnamefont {Z.}~\bibnamefont {{Vlah}}}, \ and\ \bibinfo
  {author} {\bibfnamefont {T.}~\bibnamefont {{Okumura}}},\ }\href@noop {}
  {\bibfield  {journal} {\bibinfo  {journal} {ArXiv e-prints}\ } (\bibinfo
  {year} {2013})},\ \Eprint {http://arxiv.org/abs/1311.5563} {arXiv:1311.5563
  [astro-ph.CO]} \BibitemShut {NoStop}%
\bibitem [{\citenamefont {{Matsubara}}(2008)}]{Matsubara:2008vp}%
  \BibitemOpen
  \bibfield  {author} {\bibinfo {author} {\bibfnamefont {T.}~\bibnamefont
  {{Matsubara}}},\ }\href {\doibase 10.1103/PhysRevD.77.063530} {\bibfield
  {journal} {\bibinfo  {journal} {\prd}\ }\textbf {\bibinfo {volume} {77}},\
  \bibinfo {pages} {063530} (\bibinfo {year} {2008})},\ \Eprint
  {http://arxiv.org/abs/0711.2521} {arXiv:0711.2521} \BibitemShut {NoStop}%
\bibitem [{\citenamefont {{Taruya}}\ \emph {et~al.}(2010)\citenamefont
  {{Taruya}}, \citenamefont {{Nishimichi}},\ and\ \citenamefont
  {{Saito}}}]{Taruya:2010lr}%
  \BibitemOpen
  \bibfield  {author} {\bibinfo {author} {\bibfnamefont {A.}~\bibnamefont
  {{Taruya}}}, \bibinfo {author} {\bibfnamefont {T.}~\bibnamefont
  {{Nishimichi}}}, \ and\ \bibinfo {author} {\bibfnamefont {S.}~\bibnamefont
  {{Saito}}},\ }\href {\doibase 10.1103/PhysRevD.82.063522} {\bibfield
  {journal} {\bibinfo  {journal} {\prd}\ }\textbf {\bibinfo {volume} {82}},\
  \bibinfo {pages} {063522} (\bibinfo {year} {2010})},\ \Eprint
  {http://arxiv.org/abs/1006.0699} {arXiv:1006.0699 [astro-ph.CO]} \BibitemShut
  {NoStop}%
\bibitem [{\citenamefont {{Fry}}(1996)}]{Fry:1996lr}%
  \BibitemOpen
  \bibfield  {author} {\bibinfo {author} {\bibfnamefont {J.~N.}\ \bibnamefont
  {{Fry}}},\ }\href {\doibase 10.1086/310006} {\bibfield  {journal} {\bibinfo
  {journal} {\apjl}\ }\textbf {\bibinfo {volume} {461}},\ \bibinfo {pages}
  {L65} (\bibinfo {year} {1996})}\BibitemShut {NoStop}%
\bibitem [{\citenamefont {{Taruya}}(2000)}]{Taruya:2000uq}%
  \BibitemOpen
  \bibfield  {author} {\bibinfo {author} {\bibfnamefont {A.}~\bibnamefont
  {{Taruya}}},\ }\href {\doibase 10.1086/309007} {\bibfield  {journal}
  {\bibinfo  {journal} {\apj}\ }\textbf {\bibinfo {volume} {537}},\ \bibinfo
  {pages} {37} (\bibinfo {year} {2000})},\ \Eprint
  {http://arxiv.org/abs/arXiv:astro-ph/9909124} {arXiv:astro-ph/9909124}
  \BibitemShut {NoStop}%
\bibitem [{\citenamefont {{Hui}}\ and\ \citenamefont
  {{Parfrey}}(2008)}]{Hui:2008qy}%
  \BibitemOpen
  \bibfield  {author} {\bibinfo {author} {\bibfnamefont {L.}~\bibnamefont
  {{Hui}}}\ and\ \bibinfo {author} {\bibfnamefont {K.~P.}\ \bibnamefont
  {{Parfrey}}},\ }\href {\doibase 10.1103/PhysRevD.77.043527} {\bibfield
  {journal} {\bibinfo  {journal} {\prd}\ }\textbf {\bibinfo {volume} {77}},\
  \bibinfo {eid} {043527} (\bibinfo {year} {2008})},\ \Eprint
  {http://arxiv.org/abs/0712.1162} {arXiv:0712.1162} \BibitemShut {NoStop}%
\bibitem [{\citenamefont {{Catelan}}\ \emph {et~al.}(1998)\citenamefont
  {{Catelan}}, \citenamefont {{Lucchin}}, \citenamefont {{Matarrese}},\ and\
  \citenamefont {{Porciani}}}]{Catelan:1998qy}%
  \BibitemOpen
  \bibfield  {author} {\bibinfo {author} {\bibfnamefont {P.}~\bibnamefont
  {{Catelan}}}, \bibinfo {author} {\bibfnamefont {F.}~\bibnamefont
  {{Lucchin}}}, \bibinfo {author} {\bibfnamefont {S.}~\bibnamefont
  {{Matarrese}}}, \ and\ \bibinfo {author} {\bibfnamefont {C.}~\bibnamefont
  {{Porciani}}},\ }\href {\doibase 10.1046/j.1365-8711.1998.01455.x} {\bibfield
   {journal} {\bibinfo  {journal} {\mnras}\ }\textbf {\bibinfo {volume}
  {297}},\ \bibinfo {pages} {692} (\bibinfo {year} {1998})},\ \Eprint
  {http://arxiv.org/abs/astro-ph/9708067} {astro-ph/9708067} \BibitemShut
  {NoStop}%
\bibitem [{\citenamefont {{Springel}}(2005)}]{Springel:2005uq}%
  \BibitemOpen
  \bibfield  {author} {\bibinfo {author} {\bibfnamefont {V.}~\bibnamefont
  {{Springel}}},\ }\href {\doibase 10.1111/j.1365-2966.2005.09655.x} {\bibfield
   {journal} {\bibinfo  {journal} {\mnras}\ }\textbf {\bibinfo {volume}
  {364}},\ \bibinfo {pages} {1105} (\bibinfo {year} {2005})},\ \Eprint
  {http://arxiv.org/abs/arXiv:astro-ph/0505010} {arXiv:astro-ph/0505010}
  \BibitemShut {NoStop}%
\bibitem [{\citenamefont {{Desjacques}}\ \emph {et~al.}(2009)\citenamefont
  {{Desjacques}}, \citenamefont {{Seljak}},\ and\ \citenamefont
  {{Iliev}}}]{Desjacques:2009qy}%
  \BibitemOpen
  \bibfield  {author} {\bibinfo {author} {\bibfnamefont {V.}~\bibnamefont
  {{Desjacques}}}, \bibinfo {author} {\bibfnamefont {U.}~\bibnamefont
  {{Seljak}}}, \ and\ \bibinfo {author} {\bibfnamefont {I.~T.}\ \bibnamefont
  {{Iliev}}},\ }\href {\doibase 10.1111/j.1365-2966.2009.14721.x} {\bibfield
  {journal} {\bibinfo  {journal} {\mnras}\ }\textbf {\bibinfo {volume} {396}},\
  \bibinfo {pages} {85} (\bibinfo {year} {2009})},\ \Eprint
  {http://arxiv.org/abs/0811.2748} {arXiv:0811.2748} \BibitemShut {NoStop}%
\bibitem [{\citenamefont {{Percival}}\ \emph {et~al.}(2013)\citenamefont
  {{Percival}}, \citenamefont {{Ross}}, \citenamefont {{Sanchez}},
  \citenamefont {{Samushia}}, \citenamefont {{Burden}}, \citenamefont
  {{Crittenden}}, \citenamefont {{Cuesta}}, \citenamefont {{Magana}},
  \citenamefont {{Manera}}, \citenamefont {{Beutler}}, \citenamefont
  {{Chuang}}, \citenamefont {{Eisenstein}}, \citenamefont {{Ho}}, \citenamefont
  {{McBride}}, \citenamefont {{Montesano}}, \citenamefont {{Padmanabhan}},
  \citenamefont {{Reid}}, \citenamefont {{Saito}}, \citenamefont {{Schneider}},
  \citenamefont {{Seo}}, \citenamefont {{Tojeiro}},\ and\ \citenamefont
  {{Weaver}}}]{Percival:2013fj}%
  \BibitemOpen
  \bibfield  {author} {\bibinfo {author} {\bibfnamefont {W.~J.}\ \bibnamefont
  {{Percival}}}, \bibinfo {author} {\bibfnamefont {A.~J.}\ \bibnamefont
  {{Ross}}}, \bibinfo {author} {\bibfnamefont {A.~G.}\ \bibnamefont
  {{Sanchez}}}, \bibinfo {author} {\bibfnamefont {L.}~\bibnamefont
  {{Samushia}}}, \bibinfo {author} {\bibfnamefont {A.}~\bibnamefont
  {{Burden}}}, \bibinfo {author} {\bibfnamefont {R.}~\bibnamefont
  {{Crittenden}}}, \bibinfo {author} {\bibfnamefont {A.~J.}\ \bibnamefont
  {{Cuesta}}}, \bibinfo {author} {\bibfnamefont {M.~V.}\ \bibnamefont
  {{Magana}}}, \bibinfo {author} {\bibfnamefont {M.}~\bibnamefont {{Manera}}},
  \bibinfo {author} {\bibfnamefont {F.}~\bibnamefont {{Beutler}}}, \bibinfo
  {author} {\bibfnamefont {C.-H.}\ \bibnamefont {{Chuang}}}, \bibinfo {author}
  {\bibfnamefont {D.~J.}\ \bibnamefont {{Eisenstein}}}, \bibinfo {author}
  {\bibfnamefont {S.}~\bibnamefont {{Ho}}}, \bibinfo {author} {\bibfnamefont
  {C.~K.}\ \bibnamefont {{McBride}}}, \bibinfo {author} {\bibfnamefont
  {F.}~\bibnamefont {{Montesano}}}, \bibinfo {author} {\bibfnamefont
  {N.}~\bibnamefont {{Padmanabhan}}}, \bibinfo {author} {\bibfnamefont
  {B.}~\bibnamefont {{Reid}}}, \bibinfo {author} {\bibfnamefont
  {S.}~\bibnamefont {{Saito}}}, \bibinfo {author} {\bibfnamefont {D.~P.}\
  \bibnamefont {{Schneider}}}, \bibinfo {author} {\bibfnamefont {H.-J.}\
  \bibnamefont {{Seo}}}, \bibinfo {author} {\bibfnamefont {R.}~\bibnamefont
  {{Tojeiro}}}, \ and\ \bibinfo {author} {\bibfnamefont {B.~A.}\ \bibnamefont
  {{Weaver}}},\ }\href@noop {} {\  (\bibinfo {year} {2013})},\ \Eprint
  {http://arxiv.org/abs/1312.4841v1} {arXiv:1312.4841v1 [astro-ph.CO]}
  \BibitemShut {NoStop}%
\bibitem [{\citenamefont {{Lewis}}\ and\ \citenamefont
  {{Bridle}}(2002)}]{Lewis:2002lr}%
  \BibitemOpen
  \bibfield  {author} {\bibinfo {author} {\bibfnamefont {A.}~\bibnamefont
  {{Lewis}}}\ and\ \bibinfo {author} {\bibfnamefont {S.}~\bibnamefont
  {{Bridle}}},\ }\href {\doibase 10.1103/PhysRevD.66.103511} {\bibfield
  {journal} {\bibinfo  {journal} {\prd}\ }\textbf {\bibinfo {volume} {66}},\
  \bibinfo {pages} {103511} (\bibinfo {year} {2002})},\ \Eprint
  {http://arxiv.org/abs/arXiv:astro-ph/0205436} {arXiv:astro-ph/0205436}
  \BibitemShut {NoStop}%
\bibitem [{\citenamefont {{Carlson}}\ \emph {et~al.}(2009)\citenamefont
  {{Carlson}}, \citenamefont {{White}},\ and\ \citenamefont
  {{Padmanabhan}}}]{Carlson:2009kr}%
  \BibitemOpen
  \bibfield  {author} {\bibinfo {author} {\bibfnamefont {J.}~\bibnamefont
  {{Carlson}}}, \bibinfo {author} {\bibfnamefont {M.}~\bibnamefont {{White}}},
  \ and\ \bibinfo {author} {\bibfnamefont {N.}~\bibnamefont {{Padmanabhan}}},\
  }\href {\doibase 10.1103/PhysRevD.80.043531} {\bibfield  {journal} {\bibinfo
  {journal} {\prd}\ }\textbf {\bibinfo {volume} {80}},\ \bibinfo {eid} {043531}
  (\bibinfo {year} {2009})},\ \Eprint {http://arxiv.org/abs/0905.0479}
  {arXiv:0905.0479 [astro-ph.CO]} \BibitemShut {NoStop}%
\bibitem [{\citenamefont {{Nishimichi}}\ \emph {et~al.}(2009)\citenamefont
  {{Nishimichi}}, \citenamefont {{Shirata}}, \citenamefont {{Taruya}},
  \citenamefont {{Yahata}}, \citenamefont {{Saito}}, \citenamefont {{Suto}},
  \citenamefont {{Takahashi}}, \citenamefont {{Yoshida}}, \citenamefont
  {{Matsubara}}, \citenamefont {{Sugiyama}}, \citenamefont {{Kayo}},
  \citenamefont {{Jing}},\ and\ \citenamefont
  {{Yoshikawa}}}]{Nishimichi:2009uq}%
  \BibitemOpen
  \bibfield  {author} {\bibinfo {author} {\bibfnamefont {T.}~\bibnamefont
  {{Nishimichi}}}, \bibinfo {author} {\bibfnamefont {A.}~\bibnamefont
  {{Shirata}}}, \bibinfo {author} {\bibfnamefont {A.}~\bibnamefont {{Taruya}}},
  \bibinfo {author} {\bibfnamefont {K.}~\bibnamefont {{Yahata}}}, \bibinfo
  {author} {\bibfnamefont {S.}~\bibnamefont {{Saito}}}, \bibinfo {author}
  {\bibfnamefont {Y.}~\bibnamefont {{Suto}}}, \bibinfo {author} {\bibfnamefont
  {R.}~\bibnamefont {{Takahashi}}}, \bibinfo {author} {\bibfnamefont
  {N.}~\bibnamefont {{Yoshida}}}, \bibinfo {author} {\bibfnamefont
  {T.}~\bibnamefont {{Matsubara}}}, \bibinfo {author} {\bibfnamefont
  {N.}~\bibnamefont {{Sugiyama}}}, \bibinfo {author} {\bibfnamefont
  {I.}~\bibnamefont {{Kayo}}}, \bibinfo {author} {\bibfnamefont
  {Y.}~\bibnamefont {{Jing}}}, \ and\ \bibinfo {author} {\bibfnamefont
  {K.}~\bibnamefont {{Yoshikawa}}},\ }\href@noop {} {\bibfield  {journal}
  {\bibinfo  {journal} {\pasj}\ }\textbf {\bibinfo {volume} {61}},\ \bibinfo
  {pages} {321} (\bibinfo {year} {2009})},\ \Eprint
  {http://arxiv.org/abs/0810.0813} {arXiv:0810.0813} \BibitemShut {NoStop}%
\bibitem [{\citenamefont {{Baldauf}}\ \emph {et~al.}(2013)\citenamefont
  {{Baldauf}}, \citenamefont {{Seljak}}, \citenamefont {{Smith}}, \citenamefont
  {{Hamaus}},\ and\ \citenamefont {{Desjacques}}}]{Baldauf:2013lr}%
  \BibitemOpen
  \bibfield  {author} {\bibinfo {author} {\bibfnamefont {T.}~\bibnamefont
  {{Baldauf}}}, \bibinfo {author} {\bibfnamefont {U.}~\bibnamefont {{Seljak}}},
  \bibinfo {author} {\bibfnamefont {R.~E.}\ \bibnamefont {{Smith}}}, \bibinfo
  {author} {\bibfnamefont {N.}~\bibnamefont {{Hamaus}}}, \ and\ \bibinfo
  {author} {\bibfnamefont {V.}~\bibnamefont {{Desjacques}}},\ }\href@noop {}
  {\bibfield  {journal} {\bibinfo  {journal} {ArXiv e-prints}\ } (\bibinfo
  {year} {2013})},\ \Eprint {http://arxiv.org/abs/1305.2917} {arXiv:1305.2917
  [astro-ph.CO]} \BibitemShut {NoStop}%
\bibitem [{\citenamefont {{Hamaus}}\ \emph {et~al.}(2013)\citenamefont
  {{Hamaus}}, \citenamefont {{Wandelt}}, \citenamefont {{Sutter}},
  \citenamefont {{Lavaux}},\ and\ \citenamefont {{Warren}}}]{Hamaus:2013lr}%
  \BibitemOpen
  \bibfield  {author} {\bibinfo {author} {\bibfnamefont {N.}~\bibnamefont
  {{Hamaus}}}, \bibinfo {author} {\bibfnamefont {B.~D.}\ \bibnamefont
  {{Wandelt}}}, \bibinfo {author} {\bibfnamefont {P.~M.}\ \bibnamefont
  {{Sutter}}}, \bibinfo {author} {\bibfnamefont {G.}~\bibnamefont {{Lavaux}}},
  \ and\ \bibinfo {author} {\bibfnamefont {M.~S.}\ \bibnamefont {{Warren}}},\
  }\href@noop {} {\bibfield  {journal} {\bibinfo  {journal} {ArXiv e-prints}\ }
  (\bibinfo {year} {2013})},\ \Eprint {http://arxiv.org/abs/1307.2571}
  {arXiv:1307.2571 [astro-ph.CO]} \BibitemShut {NoStop}%
\bibitem [{\citenamefont {{Sato}}\ and\ \citenamefont
  {{Matsubara}}(2013)}]{Sato:2013fk}%
  \BibitemOpen
  \bibfield  {author} {\bibinfo {author} {\bibfnamefont {M.}~\bibnamefont
  {{Sato}}}\ and\ \bibinfo {author} {\bibfnamefont {T.}~\bibnamefont
  {{Matsubara}}},\ }\href@noop {} {\bibfield  {journal} {\bibinfo  {journal}
  {ArXiv e-prints}\ } (\bibinfo {year} {2013})},\ \Eprint
  {http://arxiv.org/abs/1304.4228} {arXiv:1304.4228 [astro-ph.CO]} \BibitemShut
  {NoStop}%
\bibitem [{\citenamefont {{Sheth}}\ \emph {et~al.}(2013)\citenamefont
  {{Sheth}}, \citenamefont {{Chan}},\ and\ \citenamefont
  {{Scoccimarro}}}]{Sheth:2013lr}%
  \BibitemOpen
  \bibfield  {author} {\bibinfo {author} {\bibfnamefont {R.~K.}\ \bibnamefont
  {{Sheth}}}, \bibinfo {author} {\bibfnamefont {K.~C.}\ \bibnamefont {{Chan}}},
  \ and\ \bibinfo {author} {\bibfnamefont {R.}~\bibnamefont {{Scoccimarro}}},\
  }\href {\doibase 10.1103/PhysRevD.87.083002} {\bibfield  {journal} {\bibinfo
  {journal} {\prd}\ }\textbf {\bibinfo {volume} {87}},\ \bibinfo {eid} {083002}
  (\bibinfo {year} {2013})},\ \Eprint {http://arxiv.org/abs/1207.7117}
  {arXiv:1207.7117 [astro-ph.CO]} \BibitemShut {NoStop}%
\bibitem [{\citenamefont {{Bardeen}}\ \emph {et~al.}(1986)\citenamefont
  {{Bardeen}}, \citenamefont {{Bond}}, \citenamefont {{Kaiser}},\ and\
  \citenamefont {{Szalay}}}]{Bardeen:1986yq}%
  \BibitemOpen
  \bibfield  {author} {\bibinfo {author} {\bibfnamefont {J.~M.}\ \bibnamefont
  {{Bardeen}}}, \bibinfo {author} {\bibfnamefont {J.~R.}\ \bibnamefont
  {{Bond}}}, \bibinfo {author} {\bibfnamefont {N.}~\bibnamefont {{Kaiser}}}, \
  and\ \bibinfo {author} {\bibfnamefont {A.~S.}\ \bibnamefont {{Szalay}}},\
  }\href {\doibase 10.1086/164143} {\bibfield  {journal} {\bibinfo  {journal}
  {\apj}\ }\textbf {\bibinfo {volume} {304}},\ \bibinfo {pages} {15} (\bibinfo
  {year} {1986})}\BibitemShut {NoStop}%
\bibitem [{\citenamefont {{Elia}}\ \emph {et~al.}(2012)\citenamefont {{Elia}},
  \citenamefont {{Ludlow}},\ and\ \citenamefont {{Porciani}}}]{Elia:2011ds}%
  \BibitemOpen
  \bibfield  {author} {\bibinfo {author} {\bibfnamefont {A.}~\bibnamefont
  {{Elia}}}, \bibinfo {author} {\bibfnamefont {A.~D.}\ \bibnamefont
  {{Ludlow}}}, \ and\ \bibinfo {author} {\bibfnamefont {C.}~\bibnamefont
  {{Porciani}}},\ }\href {\doibase 10.1111/j.1365-2966.2012.20572.x} {\bibfield
   {journal} {\bibinfo  {journal} {\mnras}\ }\textbf {\bibinfo {volume}
  {421}},\ \bibinfo {pages} {3472} (\bibinfo {year} {2012})},\ \Eprint
  {http://arxiv.org/abs/1111.4211} {arXiv:1111.4211 [astro-ph.CO]} \BibitemShut
  {NoStop}%
\bibitem [{\citenamefont {{Desjacques}}(2008)}]{Desjacques:2008ry}%
  \BibitemOpen
  \bibfield  {author} {\bibinfo {author} {\bibfnamefont {V.}~\bibnamefont
  {{Desjacques}}},\ }\href {\doibase 10.1103/PhysRevD.78.103503} {\bibfield
  {journal} {\bibinfo  {journal} {\prd}\ }\textbf {\bibinfo {volume} {78}},\
  \bibinfo {eid} {103503} (\bibinfo {year} {2008})},\ \Eprint
  {http://arxiv.org/abs/0806.0007} {arXiv:0806.0007} \BibitemShut {NoStop}%
\bibitem [{\citenamefont {{Schmidt}}\ \emph {et~al.}(2012)\citenamefont
  {{Schmidt}}, \citenamefont {{Jeong}},\ and\ \citenamefont
  {{Desjacques}}}]{Schmidt:2012fk}%
  \BibitemOpen
  \bibfield  {author} {\bibinfo {author} {\bibfnamefont {F.}~\bibnamefont
  {{Schmidt}}}, \bibinfo {author} {\bibfnamefont {D.}~\bibnamefont {{Jeong}}},
  \ and\ \bibinfo {author} {\bibfnamefont {V.}~\bibnamefont {{Desjacques}}},\
  }\href@noop {} {\bibfield  {journal} {\bibinfo  {journal} {ArXiv e-prints}\ }
  (\bibinfo {year} {2012})},\ \Eprint {http://arxiv.org/abs/1212.0868}
  {arXiv:1212.0868 [astro-ph.CO]} \BibitemShut {NoStop}%
\bibitem [{\citenamefont {{Biagetti}}\ \emph {et~al.}(2014)\citenamefont
  {{Biagetti}}, \citenamefont {{Desjacques}}, \citenamefont {{Kehagias}},\ and\
  \citenamefont {{Riotto}}}]{Biagetti:2014xx}%
  \BibitemOpen
  \bibfield  {author} {\bibinfo {author} {\bibfnamefont {M.}~\bibnamefont
  {{Biagetti}}}, \bibinfo {author} {\bibfnamefont {V.}~\bibnamefont
  {{Desjacques}}}, \bibinfo {author} {\bibfnamefont {A.}~\bibnamefont
  {{Kehagias}}}, \ and\ \bibinfo {author} {\bibfnamefont {A.}~\bibnamefont
  {{Riotto}}},\ }\href@noop {} {\bibfield  {journal} {\bibinfo  {journal}
  {ArXiv e-prints}\ } (\bibinfo {year} {2014})},\ \Eprint
  {http://arxiv.org/abs/1405.1435} {1405.1435 [astro-ph.CO]} \BibitemShut
  {NoStop}%
\bibitem [{\citenamefont {{Musso}}\ \emph {et~al.}(2012)\citenamefont
  {{Musso}}, \citenamefont {{Paranjape}},\ and\ \citenamefont
  {{Sheth}}}]{Musso:2012uq}%
  \BibitemOpen
  \bibfield  {author} {\bibinfo {author} {\bibfnamefont {M.}~\bibnamefont
  {{Musso}}}, \bibinfo {author} {\bibfnamefont {A.}~\bibnamefont
  {{Paranjape}}}, \ and\ \bibinfo {author} {\bibfnamefont {R.~K.}\ \bibnamefont
  {{Sheth}}},\ }\href {\doibase 10.1111/j.1365-2966.2012.21903.x} {\bibfield
  {journal} {\bibinfo  {journal} {\mnras}\ }\textbf {\bibinfo {volume} {427}},\
  \bibinfo {pages} {3145} (\bibinfo {year} {2012})},\ \Eprint
  {http://arxiv.org/abs/1205.3401} {arXiv:1205.3401 [astro-ph.CO]} \BibitemShut
  {NoStop}%
\bibitem [{\citenamefont {{Paranjape}}\ \emph
  {et~al.}(2013{\natexlab{a}})\citenamefont {{Paranjape}}, \citenamefont
  {{Sheth}},\ and\ \citenamefont {{Desjacques}}}]{Paranjape:2013dd}%
  \BibitemOpen
  \bibfield  {author} {\bibinfo {author} {\bibfnamefont {A.}~\bibnamefont
  {{Paranjape}}}, \bibinfo {author} {\bibfnamefont {R.~K.}\ \bibnamefont
  {{Sheth}}}, \ and\ \bibinfo {author} {\bibfnamefont {V.}~\bibnamefont
  {{Desjacques}}},\ }\href {\doibase 10.1093/mnras/stt267} {\bibfield
  {journal} {\bibinfo  {journal} {\mnras}\ }\textbf {\bibinfo {volume} {431}},\
  \bibinfo {pages} {1503} (\bibinfo {year} {2013}{\natexlab{a}})},\ \Eprint
  {http://arxiv.org/abs/1210.1483} {arXiv:1210.1483 [astro-ph.CO]} \BibitemShut
  {NoStop}%
\bibitem [{\citenamefont {{Paranjape}}\ \emph
  {et~al.}(2013{\natexlab{b}})\citenamefont {{Paranjape}}, \citenamefont
  {{Sefusatti}}, \citenamefont {{Chan}}, \citenamefont {{Desjacques}},
  \citenamefont {{Monaco}},\ and\ \citenamefont {{Sheth}}}]{Paranjape:2013qy}%
  \BibitemOpen
  \bibfield  {author} {\bibinfo {author} {\bibfnamefont {A.}~\bibnamefont
  {{Paranjape}}}, \bibinfo {author} {\bibfnamefont {E.}~\bibnamefont
  {{Sefusatti}}}, \bibinfo {author} {\bibfnamefont {K.~C.}\ \bibnamefont
  {{Chan}}}, \bibinfo {author} {\bibfnamefont {V.}~\bibnamefont
  {{Desjacques}}}, \bibinfo {author} {\bibfnamefont {P.}~\bibnamefont
  {{Monaco}}}, \ and\ \bibinfo {author} {\bibfnamefont {R.~K.}\ \bibnamefont
  {{Sheth}}},\ }\href {\doibase 10.1093/mnras/stt1578} {\bibfield  {journal}
  {\bibinfo  {journal} {\mnras}\ }\textbf {\bibinfo {volume} {436}},\ \bibinfo
  {pages} {449} (\bibinfo {year} {2013}{\natexlab{b}})},\ \Eprint
  {http://arxiv.org/abs/1305.5830} {arXiv:1305.5830 [astro-ph.CO]} \BibitemShut
  {NoStop}%
\bibitem [{\citenamefont {{Villaescusa-Navarro}}\ \emph
  {et~al.}(2014)\citenamefont {{Villaescusa-Navarro}}, \citenamefont
  {{Marulli}}, \citenamefont {{Viel}}, \citenamefont {{Branchini}},
  \citenamefont {{Castorina}}, \citenamefont {{Sefusatti}},\ and\ \citenamefont
  {{Saito}}}]{Villaescusa-Navarro:2014qy}%
  \BibitemOpen
  \bibfield  {author} {\bibinfo {author} {\bibfnamefont {F.}~\bibnamefont
  {{Villaescusa-Navarro}}}, \bibinfo {author} {\bibfnamefont {F.}~\bibnamefont
  {{Marulli}}}, \bibinfo {author} {\bibfnamefont {M.}~\bibnamefont {{Viel}}},
  \bibinfo {author} {\bibfnamefont {E.}~\bibnamefont {{Branchini}}}, \bibinfo
  {author} {\bibfnamefont {E.}~\bibnamefont {{Castorina}}}, \bibinfo {author}
  {\bibfnamefont {E.}~\bibnamefont {{Sefusatti}}}, \ and\ \bibinfo {author}
  {\bibfnamefont {S.}~\bibnamefont {{Saito}}},\ }\href {\doibase
  10.1088/1475-7516/2014/03/011} {\bibfield  {journal} {\bibinfo  {journal}
  {\jcap}\ }\textbf {\bibinfo {volume} {3}},\ \bibinfo {eid} {011} (\bibinfo
  {year} {2014})},\ \Eprint {http://arxiv.org/abs/1311.0866} {arXiv:1311.0866
  [astro-ph.CO]} \BibitemShut {NoStop}%
\bibitem [{Note1()}]{Note1}%
  \BibitemOpen
  \bibinfo {note} {Ref. \cite {Taruya:2000uq} has already derived such
  solutions in an exactly same way with ours, but investigated nonlocal terms
  in terms of stochastic bias in the halo-halo power spectrum}\BibitemShut
  {NoStop}%
\bibitem [{\citenamefont {{Nicolis}}\ \emph {et~al.}(2009)\citenamefont
  {{Nicolis}}, \citenamefont {{Rattazzi}},\ and\ \citenamefont
  {{Trincherini}}}]{Nicolis:2009fk}%
  \BibitemOpen
  \bibfield  {author} {\bibinfo {author} {\bibfnamefont {A.}~\bibnamefont
  {{Nicolis}}}, \bibinfo {author} {\bibfnamefont {R.}~\bibnamefont
  {{Rattazzi}}}, \ and\ \bibinfo {author} {\bibfnamefont {E.}~\bibnamefont
  {{Trincherini}}},\ }\href {\doibase 10.1103/PhysRevD.79.064036} {\bibfield
  {journal} {\bibinfo  {journal} {\prd}\ }\textbf {\bibinfo {volume} {79}},\
  \bibinfo {eid} {064036} (\bibinfo {year} {2009})},\ \Eprint
  {http://arxiv.org/abs/0811.2197} {arXiv:0811.2197 [hep-th]} \BibitemShut
  {NoStop}%
\bibitem [{\citenamefont {{Saito}}\ \emph {et~al.}(2009)\citenamefont
  {{Saito}}, \citenamefont {{Takada}},\ and\ \citenamefont
  {{Taruya}}}]{Saito:2009fk}%
  \BibitemOpen
  \bibfield  {author} {\bibinfo {author} {\bibfnamefont {S.}~\bibnamefont
  {{Saito}}}, \bibinfo {author} {\bibfnamefont {M.}~\bibnamefont {{Takada}}}, \
  and\ \bibinfo {author} {\bibfnamefont {A.}~\bibnamefont {{Taruya}}},\ }\href
  {\doibase 10.1103/PhysRevD.80.083528} {\bibfield  {journal} {\bibinfo
  {journal} {\prd}\ }\textbf {\bibinfo {volume} {80}},\ \bibinfo {pages}
  {083528} (\bibinfo {year} {2009})},\ \Eprint {http://arxiv.org/abs/0907.2922}
  {arXiv:0907.2922} \BibitemShut {NoStop}%
\bibitem [{\citenamefont {{Baldauf}}\ \emph {et~al.}(2011)\citenamefont
  {{Baldauf}}, \citenamefont {{Seljak}},\ and\ \citenamefont
  {{Senatore}}}]{Baldauf:2011qp}%
  \BibitemOpen
  \bibfield  {author} {\bibinfo {author} {\bibfnamefont {T.}~\bibnamefont
  {{Baldauf}}}, \bibinfo {author} {\bibfnamefont {U.}~\bibnamefont {{Seljak}}},
  \ and\ \bibinfo {author} {\bibfnamefont {L.}~\bibnamefont {{Senatore}}},\
  }\href {\doibase 10.1088/1475-7516/2011/04/006} {\bibfield  {journal}
  {\bibinfo  {journal} {\jcap}\ }\textbf {\bibinfo {volume} {4}},\ \bibinfo
  {pages} {6} (\bibinfo {year} {2011})},\ \Eprint
  {http://arxiv.org/abs/1011.1513} {arXiv:1011.1513 [astro-ph.CO]} \BibitemShut
  {NoStop}%
\bibitem [{\citenamefont {{Slosar}}\ \emph {et~al.}(2008)\citenamefont
  {{Slosar}}, \citenamefont {{Hirata}}, \citenamefont {{Seljak}}, \citenamefont
  {{Ho}},\ and\ \citenamefont {{Padmanabhan}}}]{Slosar:2008aa}%
  \BibitemOpen
  \bibfield  {author} {\bibinfo {author} {\bibfnamefont {A.}~\bibnamefont
  {{Slosar}}}, \bibinfo {author} {\bibfnamefont {C.}~\bibnamefont {{Hirata}}},
  \bibinfo {author} {\bibfnamefont {U.}~\bibnamefont {{Seljak}}}, \bibinfo
  {author} {\bibfnamefont {S.}~\bibnamefont {{Ho}}}, \ and\ \bibinfo {author}
  {\bibfnamefont {N.}~\bibnamefont {{Padmanabhan}}},\ }\href {\doibase
  10.1088/1475-7516/2008/08/031} {\bibfield  {journal} {\bibinfo  {journal}
  {\jcap}\ }\textbf {\bibinfo {volume} {8}},\ \bibinfo {eid} {031} (\bibinfo
  {year} {2008})},\ \Eprint {http://arxiv.org/abs/0805.3580} {arXiv:0805.3580}
  \BibitemShut {NoStop}%
\end{thebibliography}%

\end{document}